%% file: Q_N_form.tex
\documentclass[12pt,twoside,a4paper]{article}

\usepackage{amsmath}
\usepackage{a4wide}
\usepackage[dvips]{graphicx}
\usepackage{dsfont}
\usepackage{amsthm}
\usepackage{amssymb}
\usepackage{float}

\input defs

\newcommand{\dof}{DoF}

\newcommand{\nhim}{{S_{\text{NHIM}}^{2d-3}(E)}}
\newcommand{\ts}{{S_{\text{ds}}^{2d-2}(E)}}

\newcommand{\tsf}{{B_{\text{ds,\,f}}^{2d-2}(E)}}
\newcommand{\tsb}{{B_{\text{ds,\,b}}^{2d-2}(E)}}

 \normalsize

\newcommand{\capsty}{\footnotesize}

\newcommand{\rem}[1]{}
\newcommand{\todo}[1]{}
\newcommand{\New}[1]{ {#1}} 

\newcommand{\Capts}[1]{}
\newcommand{\FIGo}[3]{\begin{figure}%
#3
\caption[]{\capsty #2}%
\label{#1}%
\end{figure}}

\numberwithin{equation}{section}

\title{Wigner's Dynamical Transition State Theory in Phase Space: Classical and Quantum}
\author{
        Holger Waalkens$^{1,2}$,
        Roman Schubert$^1$,
and Stephen Wiggins$^1$
   }

\begin{document}
\maketitle

\noindent
{\small $^1$ School of Mathematics, University of Bristol, University Walk, Bristol BS8 1TW, UK}\\
{\small $^2$ Department of Mathematics, University of Groningen, Nijenborgh 9, 9747 AG Groningen, The Netherlands}\\[2ex]
E-mail: h.waalkens@math.rug.nl, roman.schubert@bristol.ac.uk, s.wiggins@bristol.ac.uk

\abstract{We develop Wigner's approach to a dynamical transition
state theory in phase space  in both the classical and quantum
mechanical settings. The key to our development is the
construction of a normal form for describing the dynamics 
in the neighborhood of a specific type of saddle point that
governs the evolution from reactants to products in high
dimensional systems. In the classical case this is  the
standard Poincar\'e-Birkhoff normal form. In the quantum case we
develop a normal form based on the Weyl 
calculus and an explicit algorithm for computing this quantum
normal form. 
The classical normal form
allows us to discover and compute the  phase space structures that
govern classical reaction dynamics. From this knowledge we are able to
provide a direct construction of an energy dependent dividing
surface in phase space having the properties that trajectories do
not locally ``re-cross'' the surface and the directional flux
across the surface is minimal. Using this, we are able to give a
formula for the directional flux through the dividing surface 
that goes beyond the harmonic
approximation. We relate this construction to the flux-flux
autocorrelation function which is a standard ingredient in the
expression for the reaction rate in the chemistry community. We
also give a classical mechanical interpretation of the activated
complex as a normally hyperbolic invariant manifold (NHIM), and
further describe the structure of the NHIM.
The quantum normal form provides us with an efficient algorithm to 
compute quantum reaction rates and we relate this algorithm to the quantum version of the flux-flux autocorrelation function formalism. 
The significance of the classical phase space
structures for the quantum mechanics of reactions is elucidated by studying 
the phase space distribution of scattering states. 
The quantum normal form also provides  an efficient
way of computing  Gamov-Siegert resonances. We relate these
resonances to the lifetimes of the quantum activated complex. 
We consider several one, two, and three
degree-of-freedom systems and show explicitly how calculations of
the above quantities can be carried out. Our theoretical framework
is valid for Hamiltonian systems with an arbitrary number of
degrees of freedom and we demonstrate that in several situations
it gives rise to algorithms that are computationally more
efficient than existing methods.}

\tableofcontents

\input introduction

\input classical_NF

\input semiclassical_quantum_NF

\input class_phase_space

\input smatrix

\input resonances

\input examples

\input conclusions

\input appendix


\section*{Acknowledgments}

H.W. is grateful to Andrew Burbanks for many helpful discussions
on the numerical implementation of the normal form
computation in the programming language {\sf C}$++$. Furthermore,
H.W. and S. W. would like to thank Peter Collins for comparing the
results for the normal forms presented in this paper with the
results of his numerical computations. 
R.S., H.W. and S.W. acknowledge individual support by EPSRC.
S.W. also acknowledges support by   ONR (Grant No.~N00014-01-1-0769). 
\newpage

\appendix





\newcommand{\etalchar}[1]{$^{#1}$}
\def\cprime{$'$}
\providecommand{\bysame}{\leavevmode\hbox to3em{\hrulefill}\thinspace}
\providecommand{\MR}{\relax\ifhmode\unskip\space\fi MR }
\providecommand{\MRhref}[2]{%
  \href{http://www.ams.org/mathscinet-getitem?mr=#1}{#2}
}
\providecommand{\href}[2]{#2}

\end{document}

%% file: defs.tex
\newcommand{\ud}{\mathrm{d}}
\newcommand{\ui}{\mathrm{i}}
\newcommand{\ue}{\mathrm{e}}

\newcommand{\T}{\mathds{T}}
\newcommand{\R}{\mathds{R}}
\newcommand{\Z}{\mathds{Z}}
\newcommand{\N}{\mathds{N}}
\newcommand{\C}{\mathds{C}}

\newcommand{\M}{\mathcal{M}}

\newcommand{\cS}{{\mathcal S}}

\newcommand{\cH}{{\widehat{H}}}
\newcommand{\ccH}{{\mathcal H}}

\newcommand{\cW}{{\mathcal W}_{\mathrm{ qm}}}

\newcommand{\cWl}{{\mathcal W}_{\mathrm{ qm; \,loc}}}

\newcommand{\ccW}{{\mathcal W}_{\mathrm{ cl}}}

\newcommand{\flow}{\Phi}
\newcommand{\flowparam}{\epsilon}
\newcommand{\Hess}{\text{D}^2}

\newcommand{\Hcnf}{H_{\text{CNF}}}
\newcommand{\Hqnf}{K_{\text{QNF}}}
\newcommand{\Hloc}{H_{\text{loc}}}

\newcommand{\Nweyl}{N_{\text{Weyl}}}
\newcommand{\flux}{f}

\newcommand{\Ihat}{\hat{I}}
\newcommand{\Jhat}{\hat{J}}
\newcommand{\nscatt}{{n_{\text{sca}}}}

\newcommand{\mscatt}{{m_{\text{sca}}}}
\newcommand{\psiI}{\psi_{I}}

\providecommand{\abs}[1]{\lvert#1\rvert}
\providecommand{\biggabs}[1]{\bigg\lvert#1\bigg\rvert}

\newcommand{\pa}{\partial}
\newcommand{\la}{\langle}
\newcommand{\ra}{\rangle}

\newcommand{\adM}{\operatorname{Mad}}
\newcommand{\ad}{\operatorname{ad}}
\newcommand{\Op}{\operatorname{Op}}
\newcommand{\Tr}{\operatorname{Tr}}

\newcommand{\opD}{{\cal D}}

\renewcommand{\Im}{\operatorname{Im}}
\renewcommand{\Re}{\operatorname{Re}}

\newtheorem{thm}{Theorem}
\newtheorem{lem}{Lemma}
\newtheorem{prop}{Proposition}

\newtheorem{Def}{Definition}

%% file: introduction.tex
\section{Introduction}

The subject of this paper is {\em transition state
theory} -- classical and quantum. Transition state theory (TST)
(sometimes also referred to as  ``activated complex theory'' or
the ``theory of absolute reaction rates'') is widely regarded as
the most important theoretical and computational approach to
analyzing chemical reactions, both from a qualitative and
quantitative point of view. The central ideas of TST are so
fundamental that in recent years TST has been recognized as a very
natural and fruitful approach in areas  far beyond its origin of
conception in chemistry. For example, it has been used in atomic
physics \cite{JaffeFarellyUzer2000}, studies of the rearrangements
of clusters \cite{KomaBerry1999,KomatsuzakiBerry2002}, solid state
and semi-conductor physics
\cite{JacucciTollerDeLorenziFlynn84,Eckhardt1995}, diffusion
dynamics in materials \cite{vmg}, cosmology
\cite{DeOlivieiraAlmeidaSoaresTonini2002}, and celestial mechanics
\cite{JaffeRossLoMarsdenFarrellyUzer2002,WaalkensBurbanksWiggins05b}.

The literature on TST is vast, which befits the importance,
utility,  breadth, scope, and success  of the theory. Searching
ISI Web of Knowledge on the phrase ``transition state theory''
yields more than 17,600 hits. Searching Google with the same
phrase gives more than 41,000,000 hits. There have been numerous
reviews of TST, and the relatively recent review of
\cite{TruhlarGarrettKlippenstein96} is an excellent source for
earlier reviews, historical accounts, books, pedagogical articles,
and handbook chapters dealing with TST. Moreover,
\cite{TruhlarGarrettKlippenstein96} is notable from the point of
view of that in little more that 10 years it has attracted more
than 458 citations (and it also contains 844 references)!

Certainly the existence of this vast literature begs the question
``why does there need to be yet another paper on the theoretical
foundations of TST, what new could it possibly add?'' The one word
answer to this questions is, ``dynamics''.  Advances in
experimental techniques over the past twenty years, such as, e.g.,
femtosecond laser spectroscopy, transition state spectroscopy, and
single molecule techniques (\cite{Neumark,PolanyiZewail95,Zewail})
now provide us with "real time" dynamical information on the
progress of a chemical reaction from "reactants" to "products". At
the same time, these new experimental techniques, as well as
advances in computational capabilities,  have resulted in a
growing   realization among chemists of the ubiquity of
non-ergodic behaviour  in complex molecular systems, see, e.g.
\cite{SkodjeYang04,bhc1, bhc2, Carpenter}. All of these results
point to a need to develop a framework for studying and
understanding dynamics in high dimensional dynamical systems and
recently developed tools in computational and applied dynamical
systems theory are giving new insights and results in the study of
the dynamics of molecular systems with three or more degrees of
freedom. In particular, we will show how these recent advances in
analytical and computational techniques can enable us to realize
Wigner's {\em dynamical} picture of transition state theory in
{\em phase space} for  systems with three or more
degrees-of-freedom. However,  to set this in context we first need
to describe a bit of the historical background and setting of TST.

Transition state theory was created in the 1930's, with most of
the credit being given to Eyring, Polanyi, and Wigner, who are
referred to as the ``founding trinity of TST'' in Miller's
important review on chemical reaction rates (\cite{Miller1}).
Nevertheless, important contributions were also made by Evans,
Farkas, Szilard,  Horiuti, Pelzer, and Marcelin, and these are
described in the discussions of the historical development of the
subject given in \cite{LaidlerKing, PollakTalkner}.

The approach to TST taken by Eyring \cite{Eyring35} emphasized
thermodynamics (see the perspective article of \cite{Petersson}).
The approach of Wigner \cite{Wigner38} on the other hand is 
based on classical mechanics
(see the perspective article of \cite{Garrett}). It is the
dynamical approach of Wigner  that is the focus of this paper.
Despite the fact that  the original framework of TST is
classical mechanics, it is natural to consider quantum mechanical
versions of this approach to reaction dynamics. We will first
describe the classical mechanical setting, and then consider the
quantum mechanical version, and we will emphasize how much of the
structure and philosophy of the classical approach influences the
quantum approach.

\subsection{Transition State Theory: Classical Dynamics}

To begin with, we first  examine the assumptions of classical TST,
as set out by Wigner. Wigner begins by stating that he considers
chemical reactions in a setting where  the equilibrium
Maxwell-Boltzmann velocity and energy distributions are maintained
(see \cite{Mahan} for a detailed discussion of this point) and for
which the potential energy surface is known (\cite{Garrett}). He
then gives the following assumptions from which he derives TST:

\begin{enumerate}

\item the motion of the nuclei occurs on the Born-Oppenheimer
potential energy surface (``electronic adiabaticity'' of the
reaction)

\item classical mechanics adequately describes the  motion of the
nuclei

\item there exists a hypersurface in phase space dividing the
energy surface into a region of reactants and a region of products
having the property that all trajectories that pass from reactants
to products must cross this dividing surface precisely once.

\end{enumerate}

\noindent It is important to note that Wigner clearly developed
his ideas in {\em phase space}, the arena for dynamics. It is
important to keep this in mind since a great deal of later
developments occur in {\em configuration space}, in which certain
dynamical properties are obscured.

From the modelling point of view, the first two assumptions are of
a very different nature than the third. The first two are central
to developing the model, or {\em dynamical system} (i.e.
determining the potential energy surface and Hamiltonian
function). As a result, once a dynamical system describing the
reaction has been developed the third ``assumption'' cannot really
have the status of an assumption. Rather, such a hypersurface
satisfying these properties must be shown to exist for the
dynamical system. Of course, in practice this is exactly how the
theory is utilized. One starts with a dynamical system describing
the reaction, and then one attempts to construct a ``dividing
surface'' having the required characteristics. It is precisely
this third ``assumption'' that is at the heart of this paper and
from which, as we shall see, many dynamical consequences flow.

We will be concerned with dynamics on a fixed energy surface.  In
this paper ``energy'' means the total energy of the system, e.g
the sum of the kinetic and potential energies.  More
mathematically, the energy surface is the level set of the
Hamiltonian function.\footnote{Hamiltonian functions can be more
general than the sum of the kinetic and potential energy terms.
They could contain magnetic terms or Coriolis terms, for example.
Nevertheless, we will still refer to the level set of the
Hamiltonian function as the ``energy surface''.} This is important
to keep in mind because in not an insignificant portion of the
relevant literature the meaning of the phrase ``energy surface''
is actually the ``potential energy surface'', and a great deal of
effort is expended in attempting to infer dynamical phenomena from
the ``topography'' of the potential energy surface. Certainly for
one degree-of-freedom (\dof) Hamiltonian systems (i.e. one
configuration space coordinate and one associated momentum) one can
understand all possible dynamics from the shape of the potential
energy surface. This is definitely not true for more than one \dof
(or else dynamical phenomena such as ``chaos'' would have been
discovered many years earlier). However, two \dof Hamiltonian
systems, where the Hamiltonian is the sum of the kinetic and
potential energies, does allow for certain constructions based solely
on the potential energy surface that imply important dynamical
phenomena. We will survey these later in this introduction. We
emphasize that similar constructions using the potential energy
surface for systems having more than 2 \dof simply do not work in
the same way as they do for 2 \dof.

Now to realize assumption 3, on a fixed energy surface, we need to
choose a dividing surface that will ``separate'' the energy
surface into two parts (``two parts'' is a bit too simplistic, but
we will come back to that later) -- one part corresponding the
reactants and the other to products. The dividing surface would
have the additional (dynamical) property that trajectories
evolving from reactants to products cross it only once. Again,
these reactant and product regions are typically defined via the
potential energy surface. They are often interpreted as
``potential wells'' (i.e., local minima of the potential energy
function) that are ``separated'' by a ``saddle point'' and a   surface
(in configuration space) passing through a neighborhood of the
saddle point serves as the  dividing surface (\cite{Pechukas81}).
We will show that for systems with three or more \dof such a
configuration space approach, in several different ways,  does not
allow one to realize Wigner's original construction of TST. In
fact, this is a central message of this paper. It can be
misleading, and even wrong, to attempt to infer dynamical
phenomena from the topography of the potential energy surface.

In the series of papers
\cite{WWJU01,UJPYW01,WaalkensBurbanksWiggins04,WaalkensWiggins04,WaalkensBurbanksWigginsb04,WaalkensBurbanksWiggins05,WaalkensBurbanksWiggins05c}
the fundamental framework for phase space TST is developed. The
starting point is  classical mechanics and a  Hamiltonian function
describing the system (the same as \cite{Wigner38}). The
Hamiltonian can be expressed in any convenient set of coordinates,
have any number, $d$, degrees of freedom (\dof), and does
\emph{not} have to be of the form ``kinetic plus potential
energy'', e.g., it can include rotational or magnetic terms.

With the Hamiltonian function in hand, the next step is to locate
particular saddle-like equilibrium points of the associated
Hamilton's equations that are of a certain type. Namely, the
matrix associated with the linearization of Hamilton's equations
about the equilibrium has a pair of real eigenvalues of opposite
signs ($\pm \lambda$) and $2d-2$ purely imaginary eigenvalues
occurring in complex conjugate pairs ($\pm \ui \omega_k, \, k=2,
\ldots, d$). Such equilibria are called
saddle-centre-$\cdots$-centres, and structures associated with
these equilibria provide the fundamental mechanism for
``transformation'' in a large, and diverse, number of applications
(some listed at the beginning of this introduction), whose {\em
dynamical} consequences have remained a mystery.

Of course, locating saddles is in the spirit of classical
transition state theory, but there is an important difference
here. We are concerned with the {\em dynamical consequences} of
certain types of saddles of Hamilton's equations in phase space.
The usual approach is to consider saddles of the potential energy
surface (the setting of the ``landscape paradigm''
\cite{Wales04}). However, if the Hamiltonian has the form of the
sum of the kinetic energy and the potential energy, then there is
a correspondence between the rank one saddles of the potential
energy surface and the saddle-centre-$\cdots$-centre type
equilibria of Hamilton's equations. But here we emphasize the
phase space setting and the influence of this saddle in the
dynamical arena of phase space. We reiterate that a central point
of ours is that it is difficult, and often misleading, to try to
infer dynamics from properties of configuration space.

Next we seek to understand the phase space geometry  near this
saddle-centre-$\cdots$-centre (henceforth referred to as a
``saddle'') equilibrium point of Hamilton's equations. 
An understanding of the geometry will give
rise to a set of coordinates that will enable us to explicitly
compute the phase space structures that govern transport and to
quantify their influence on trajectories. This set of coordinates
is realized in an algorithmic manner through the use of the
Poincar\'e-Birkhoff normal form procedure.
These {\em normal form coordinates} are central to
our theory and the resulting  analytical and computational
techniques.  In particular, they enable us to show that ``near''
the saddle the energy surface has what we call the ``bottleneck
property'' which facilitates the construction of an energy
dependent dividing surface. This dividing surface has the
``no-recrossing'' property and the flux across the dividing
surface is ``minimal'' (in a sense that we will make precise).
Moreover, the coordinates also naturally give rise to a
``dynamical reaction path''.  We want
to describe these notions in a bit more detail and place them in
the context of the chemistry literature.

Further, we note that historically it has been well-recognized
that the computation of quantities associated with chemical
reactions is greatly facilitated by adopting a ``good'' set of
coordinates (\cite{jr,em,Miller76,Miller77}). 
\New{
In particular, if
the Hamiltonian is separable, i.e. there is a set of \emph{configuration space coordinates} in terms of which the equations of motion decouple,  then the choice of a dividing surface with the no recrossing
property is trivial (\cite{Garrett}). This situation is extremely special and therefore almost irrelevant for chemical reactions.  
However, the normal form method shows that such a decoupling can always be obtained in the neighbourhood of the dividing surface through the symplectic (``canonical'') transformation of the full phase space coordinates (i.e. a symplectic transformation mixing the configuration space coordinates and the conjugate momenta).
The normal form thus is a 
constructive way of obtaining ``good'' coordinates in phase space. 
}

\paragraph{The Bottleneck Property of the Energy Surface and the  Energy
Dependent Dividing Surface:}

The geometry or ``shape'' of a fixed energy surface has received
little attention, as opposed to
consideration of the geometry or ``shape'' of \emph{potential} energy
surfaces. This is unfortunate since an understanding of the
geometry of the energy surface is essential for constraining and
interpreting the possible global dynamics. Nevertheless, the lack
of attention to this issue is
understandable since such considerations give rise to extremely
difficult mathematical problems. As an example, the importance of
an understanding of the topology of the energy surface for an
understanding of the dynamics of the three body problem was
emphasized by Poincar\'e (\cite{poinc1, poinc2, poinc3}), and work
on this problem has involved some of the giants of mathematics of
the $20^{\scriptstyle th}$ century and has resulted in the
creation of many new areas of mathematical research. Very recent
results on the three body problem, as well as a discussion of the
history of the subject, can be found in \cite{mmw}, and a
discussion of the developments of an appropriate computational
framework for studying such questions for general Hamiltonian
systems can be found in \cite{kmm}. We would expect that similar
studies of the structure of the energy surfaces for standard
Hamiltonian's arising in studies of reaction dynamics will be
similarly fruitful and lead to new global dynamical insights.

However, there are ``local'' results that describe the geometry of
the energy surface that are very relevant to studies of reaction
dynamics and TST. In particular, for a range of energies above
that of the saddle,
the $(2d-1)$-dimensional energy surface has locally the structure
of the product of a $(2d-2)$-dimensional sphere with the real line,
 $S^{2d-2} \times \mathbb R$. 
We say that in this region of the
phase space the energy surface has the ``bottleneck property''
because it  is (locally) separated into two pieces: $S^{2d-2}
\times \mathbb R^+$  and $S^{2d-2} \times \mathbb R^-$, and
$S^{2d-2} \times \{0 \}$ is the dividing surface that separates
these two pieces of the energy surface, and we identify the two
pieces separated by this dividing surface as ``reactants'' and
``products''. It will turn out 
that
$\mathbb R$ corresponds to a natural (energy dependent) ``reaction
coordinate'' and $S^{2d-2}$ will correspond to (energy dependent)
unstable bath modes, or vibrations ``normal'' to the reaction
coordinate.

It should be clear that the geometry of the energy surface, as the
energy varies, is an important feature of reaction dynamics. In
particular, the geometry changes with energy and the
``bottleneck'', $S^{2d-2} \times \mathbb R$, may deform into a
more complicated shape as the energy is further increased above
that of the saddle. This can  lead to the ``breakdown'' of the
validity of transition state theory in the sense that we are not
able to construct a dividing surface separating reactants from
products that is not recrossed. We note that an ``energy limit'' for TST has been
discussed in \cite{Grimmel, Sverdlik}. Looking at it another way,
the energy surface deforms in such a way that the distinction
between reactants and products becomes unclear. This is one way in
which TST can ``break down''. We will mention one other way after
we have introduced the notion of a normally hyperbolic invariant
manifold.

\paragraph{The ``No-Recrossing'' Property and Minimal Flux:}

The dividing surface described above can be realized through the
normal form computations and transformations
(\cite{WWJU01,UJPYW01}).
The high dimensional spherical geometry, $S^{2d-2}$, is
significant in several ways.  A sphere is separated into two parts
along its equator, which in this high dimensional case is given by
$S^{2d-3}$, the $(2d-3)$-dimensional sphere.\footnote{Think of the
familiar, and easily visualizable, case of the $2$-dimensional sphere, $S^2$.
It is separated into two hemispheres by its equator, a sphere of
one less dimension, $S^1$.} The Hamiltonian vector  field is
transverse to each hemisphere, but in an opposite sense for each
hemisphere. This indicates the evolution from reactants to
products through one hemisphere, and the evolution from products
to reactants through the other hemisphere. Transversality of the
Hamiltonian vector field to a hemisphere is the mathematical
property one needs to show that there are ``no local recrossing
of trajectories'', as is shown in \cite{UJPYW01} and in this paper.
The Hamiltonian vector field is tangent to
the equator of the sphere.\footnote{If the Hamiltonian vector
field is transverse to one hemisphere, transverse to the other
hemisphere in the opposite directional sense, and it varies
smoothly in phase space, then we can view the equator as where the
Hamiltonian vector field ``changes direction''.} Mathematically,
this is the condition for the equator,
$S^{2d-3}$, to be an invariant manifold. More precisely, it is
saddle like in stability and an example of a {\em normally
hyperbolic invariant manifold}, or NHIM (\cite{Wiggins94, WWJU01,
UJPYW01}). The NHIM has the physical interpretation as the
``activated complex''
-- an unstable super molecule poised between reactants
and  products. 

Except for the equator, $S^{2d-3}$, (which is a
normally hyperbolic invariant manifold), the dividing surface thus is
locally a ``surface of no return'' in the sense that all
trajectories that start on the dividing
surface exit a neighborhood of the dividing surface \cite{UJPYW01}. 
Most importantly for reaction dynamics, the
energy surface has the ``bottleneck property''. That is, our
dividing surface locally divides the energy surface into two,
disjoint components, which correspond to reactants and products.
Therefore the {\em only} way a trajectory can pass from one of
these components of the energy surface to the other is to pass
through the dividing surface. The issue of ``recrossing'' is an important
part of the choice of the dividing surface. Truhlar \cite{Truhlar2}
distinguishes two type of recrossing: {\em local} and {\em global}
recrossing. Local recrossing cannot occur with our choice of
dividing surface. However, global recrossing is a very different
matter. If the energy surface is compact (i.e. closed and bounded,
for our purposes) then the Poincar\'e recurrence theorem
(\cite{Arnold78}) implies that global recrossing must occur for
almost all trajectories crossing the dividing surface. Moreover,
the existence of homoclinic orbits and heteroclinic cycles may
also be an intrinsic feature of the dynamics
(\cite{WaalkensBurbanksWiggins04,WaalkensBurbanksWigginsb04,WaalkensBurbanksWiggins05,WaalkensBurbanksWiggins05b,WaalkensBurbanksWiggins05c}).
Their existence also implies that global recrossing cannot be
avoided regardless of the choice of transition state; in other
words, {\em global} recrossing is a fundamental property of the
dynamics and its presence does not therefore indicate the
limitations of any particular method for constructing a dividing surface.

Wigner \cite{Wigner38} pointed out that  the effect of
trajectories recrossing the dividing surface would result in ``too
high values of the reaction rate''. This observation naturally
leads to the notion of {\em variational transition state theory},
where the idea is to vary the choice of choice of dividing surface
in such a way that the flux across the dividing surface attains a
minimum value (see \cite{Keck1967} and the review paper of
\cite{TruhlarGarrett84}).  The latter review paper contains 206
references and has more than 390 citations, which is indicative of the
fact that variational transition state theory is a huge subject in
its own right. Much of the work that falls under the heading of
``variational transition state theory'' involves dividing surfaces
in configuration space (see \cite{jj,bj} for a systematic
development of this approach).  These beautiful results obtained
by our predecessors can with modern day mathematical tools be
formulated differently, leading to more general results. The
beginnings of a general framework for such an approach was first
given by \cite{MacKay1991}, and this was used in
\cite{WaalkensWiggins04} to show that the dividing surfaces
computed by the normal form approach described in \cite{WWJU01,
UJPYW01} have ``minimal flux''.  It is worth re-emphasizing again,
that our dividing surface construction and our flux calculations
are carried out in phase space, not configuration space. The work
in \cite{WaalkensWiggins04} implies that one cannot find a surface
in configuration space for systems with more than 2 \dof that is
free of {\em local} recrossing, and therefore has minimal flux
({\em unless} the system is given in coordinates in which the
Hamiltonian is separable \New{or has some very special symmetries}).

It is worth pointing out here that flux across a dividing surface
is a ``local property'' with respect to the given surface, i.e. it
does not  require integration of trajectories for its computation.
If one makes a ``bad'' choice of dividing surface that is not free
of {\em local} recrossing then one must compute trajectories to
correct for the local recrossing effect (in the chemistry
literature these are referred to as ``dynamical corrections'' to
the rate, see \cite{mm,Pritchard} for specific examples of the
effect of recrossing and how it is treated). This is particularly
apparent when one carefully examines a standard ingredient in the
reaction rate in use in the chemistry community -- the flux-flux
autocorrelation function
for which we show that
the use of our dividing surface and phase space approach allows
the computation of this function without the long time integration
of trajectories.

In summary, our work on the geometry of  reaction dynamics allows
for a careful analysis and realization of Wigner's \cite{Wigner38}
dynamical version for transition state theory. The
dynamical foundations of Wigner's transition state theory did
receive a great deal of attention in the 1970's in a series of
seminal papers by Child, McLafferty, Pechukas and Pollak
\cite{PechukasMcLafferty73,PechukasPollak77,PechukasPollak78,PechukasPollak79,PollakPechukas79,PollakChildPechukas80,PollakChild80,ChildPollak80,Pechukas81}, and
there is a wealth of dynamical ideas in these works. However, it is
important to realize that these works focus  almost entirely on 2 \dof,
and most of the results have not been generalized to 3 or more
\dof. Nevertheless, for 2 \dof they show how to construct a dividing
surface without recrossing from the projection of a periodic
orbit, the Lyapunov orbit associated with a saddle equilibrium
point, to configuration space -- the so called \emph{periodic
orbit dividing surface} (PODS)
\cite{PechukasMcLafferty73,PechukasPollak78}. In addition to this
construction being limited to 2 \dof systems, the
Hamiltonian must  be of type `kinetic plus potential' -- Coriolis
terms due to a rotating coordinate system or a magnetic field are not
allowed.

The generalization to more than two degrees of freedom and to more
general Hamiltonians has posed a major problem for decades. The
reasons for the problems are twofold. On  the one hand, a construction
based on configuration space, as in the case of the PODS, simply
does not work for systems with more than two degrees of freedom,
as discussed in \cite{WaalkensWiggins04}. On the other hand, it
was not clear what replaces the periodic orbit in higher
dimension. For more than two degrees of freedom a periodic orbit
lacks sufficient dimensionality  to serve as a building block for
the construction of a dividing surface. In fact, a completely new
object, a so called {\em normally hyperbolic invariant manifold}
(NHIM)\cite{Wiggins94} takes the place of the periodic orbit in
two degrees of freedom. 
It is interesting to recall a remark of Pechukas from his
influential review paper \cite{Pechukas81}:

\smallskip
\begin{quotation} \em
\noindent It is easy to guess that generalized transition states
in problems with more degrees of freedom must be unstable
invariant classical manifolds of the appropriate dimension, but to
our knowledge no calculations have been done.
\end{quotation}
\smallskip

\noindent Our work gives a precise characterization of these
invariant manifolds in terms of the NHIM, as well as shows exactly what calculations
are required to realize them in specific systems.\footnote{It is
perhaps worth pointing out that when reading the chemistry
literature mathematicians might experience  some confusion
surrounding  the phrases ``transition state'' and ``dividing
surface''. In some parts of the literature they are used
synonymously. In other parts, they have a very different meaning,
as can be seen from the above quote of Pechukas.  A dividing
surface {\em cannot} be an invariant manifold, or else
trajectories could not cross the surface (trajectories on an
invariant manifold remain on that manifold for all time). The
confusion probably arose out of the PODS theory.
\New{ 
In that situation the dividing surface and the invariant manifold (the periodic orbit) project to the same line in configuration space. The projection of a reactive trajectory to configuration space intersects this line in configuration space.  In the three-dimensional energy surface however, the trajectory intersects the dividing surface and not the periodic orbit. The dividing surface is a 2-dimensional sphere, $S^2$, in this case (i.e. it is of one dimension less than the three-dimensional energy surface) and  
the periodic orbit is an invariant one-dimensional sphere, $S^1$, that forms the equator of the sphere. 
The same situation holds for more than two \dof. The
equator of our dividing surface is a normally hyperbolic
invariant manifold (but a periodic orbit does not have sufficient
dimensions to satisfy this requirement).}}
The NHIM is not only the building block
for the construction of a dividing surface in arbitrary dimension,
but it also forms the basis for  locating  the transition pathways
for reactions in terms of the stable and unstable manifolds of the
NHIM \cite{WaalkensBurbanksWigginsb04}.

Finally, we remarked earlier that one way in which TST can ``break
down'' is through deformation of the energy surface. Another way
in which it may break down is through {\em bifurcation of the
NHIM}. For 2 \dof systems the NHIM is a periodic orbit and
bifurcation theory for periodic orbits in Hamiltonian systems is
well-developed (\cite{Meyer70, MH92}). Bifurcation of the NHIM in
2 \dof systems can lead to stable motions that ``trap''
trajectories in the transition region. This has been observed in
\cite{ChildPollak80,mm}. At present there exists no general
bifurcation theory for NHIMs in systems with $d$ \dof, $d \ge 3$,
and this poses a limitation to the range of validity of our
approach. In this case the relevant NHIMs are $(2d-3)$-dimensional and
contain their own nontrivial dynamics. The development of
bifurcation theory for such objects promises to be a challenging
and interesting mathematical problem that should yield new
insights into reaction dynamics.

\paragraph{The Dynamical Reaction Path:}

Thus far we have described the geometry of the energy surface near
a saddle and the nature of the dividing surface that separates the
energy surface near the saddle into two regions corresponding to
reactants and products. Now we want to describe in more detail how
trajectories approach, and move away from, the dividing surface.
For this purpose the notion of the {\em reaction path} arises.

Traditionally, the {\em reaction path} of a polyatomic molecule is
the steepest descent path on the potential energy surface (if
mass-weighted Cartesian coordinates are used) connecting saddle
points and minima (\cite{Miller2}). Hence, it is a configuration
space notion derived from properties of the potential energy
surface that is used to describe a specific dynamical phenomenon.
Similarly to TST, the literature related to reaction paths is
vast. Searching ISI Web of Knowledge on the phrase ``reaction
path'' yields more than 6,600 hits. Searching Google with the same
phrase gives more than 7,900,000 hits.  It is often assumed that
 a reacting trajectory, when projected into configuration
space, will be ``close'' to this reaction path, and much work is
concerned with developing configuration space coordinates (and
their associated conjugate momenta) in which the dynamical
equations that describe evolution ``close'' to this reaction path
can be expressed (see, e.g., \cite{Marcus2,Marcus3,Marcus4,Miller2,Miller83,Natanson1,Natanson2,Natanson3,Natanson4,Natanson5,ggb2,ggb,gb}).
However, despite its fundamental importance in the historical
development of the subject of reaction dynamics, one might
question the relationship of this configuration based reaction
path to the actual path taken in the course of the dynamical evolution
from reactants to products. In fact, in recent years numerous
experiments have shown that the actual dynamics may exhibit
significant deviations from the ``classical reaction path''
\cite{pccachz,Sun,Ammal,lczbs,tllcszrhb,bowman,hk,pmo}.

In this paper
we show that the coordinates given by
normal form theory also give rise to an intrinsic {\em dynamical
reaction path}, which is a trajectory on the energy surface. Its
construction follows from the dynamical properties associated with
the NHIM (``activated complex''). The NHIM has stable and unstable manifolds which as we will explain in detail have the structure of spherical cylinders, $S^{2d-3}\times \R$,  and  form the phase space conduits for reaction in the sense that they enclose the reactive trajectories. Our dynamical reaction path forms  the centre line of these spherical cylinders.
 and gives rise to a phase
space description of an invariant ``modal partitioning'' along the
reaction path corresponding to energy in the reacting mode and
energies in the (nonlinear) vibrational modes normal to the
reaction path.

\subsection{Transition State Theory: Quantum Dynamics}

Historically a great deal of effort -- mostly in the chemistry
community -- has been devoted to developing a quantum mechanical
version of transition state theory (see the work by Miller and
coworkers \cite{Miller98}). Nonetheless, a quantum mechanical
formulation of transition state theory is still considered an open
problem (see the recent review by Pollak and Talkner
\cite{PollakTalkner05}). The nature of the difficulties are summed
up succinctly by Miller \cite{Miller1}:

\smallskip
\begin{quotation}
\noindent {\em
--- the conclusion of it all is that there is no uniquely
well-defined quantum version of TST in the sense that there is in
classical mechanics. This is because tunnelling along the reaction
coordinate necessarily requires one to solve the (quantum)
dynamics for some finite region about the TS dividing surface, and
if one does this fully quantum mechanically there is no `theory'
left, i.e., one has a full dimensional quantum treatment which is
ipso facto exact, a quantum simulation.}
\end{quotation}
\smallskip

Part of the problem leading to this statement originates from
(classical) transition state theory where the necessary  theoretical 
framework to realise transition state theory for multi-dimensional systems as described  in this paper has been developed only very recently. In particular, this realisation of classical TST requires one to work in \emph{phase space}  (as opposed to configuration space). This also has consequences for the development of a quantum version of transition state theory (which should reduce to classical TST in the classical limit and have all the computational benefits of a ``local'' theory like in the classical case).  
Again due to the lack of a theoretical framework, most approaches to developing a quantum version of transition state theory  involve attempts to achieve a separation of the Schr\"odinger equation that describes the chemical reaction. However, like in the classical case this separation does not exist.    
In contrast to this, we will develop a quantum version of TST 
which is  built in a systematic way on the classical theory presented in this paper. 



\noindent 
In the classical 
case the key idea to realise TST is to 
transform the Hamilton function describing the reaction to normal form.
In the quantum case we will establish a quantum version of the classical normal
form theory, and from this all of the quantum reaction dynamics
quantities will flow. In particular, the classical phase space
structures that we found will play a  central role in the
computation of quantum mechanical reaction quantities. Quantum
mechanical computations are notable for suffering from the ``curse
of dimensionality.'' We will see that the property of integrability 
which follows from the normal form in the classical case 
will have a quantum manifestation that
renders computations of ``local'' reaction quantities tractable
for  high dimensional systems. This leads to very efficient
algorithms for computing, e.g., quantum mechanical cumulative
reaction probabilities and resonances.

Classical normal form theory is a standard technique of dynamical
systems theory, and there are many textbooks and tutorial articles
that describe the subject. However, quantum normal form theory is
probably much less familiar in both the dynamical systems
community as well as the chemistry community. It is therefore 
useful to provide a discussion of the background, context, and
historical development of the subject.

Symplectic transformations like those involved in the classical normal form theory  also 
have  a long history in the
study of Partial Differential Equations.
In the theory of
\emph{microlocal analysis} they form one of the core
techniques  introduced in the late 60's and early
70's in the fundamental papers by Egorov, H{\"o}rmander
and Duistermaat, \cite{Ego69,Hoe71,DuiHoe72}.
These ideas lead naturally to the consideration of normal forms
for partial differential equations, and these were used to study
the solvability and the singularities of solutions.
The basic idea is the
following. One can associate with a linear partial differential
operator  a function on phase space  by substituting momenta
for the partial differentials. The resulting  function is called the
\emph{symbol} of the operator. One can now use
a symplectic transformation to find coordinates
in which the symbol has a particularly simple form.
The crucial point now is that the tools from microlocal analysis
allow one to quantise such a symplectic transformation. The result  is a unitary operator  which is called a 
\emph{Fourier integral operator} and yields the transformation of the original partial differential operator corresponding 
to the symplectic transformation of its symbol (plus small error terms). 
This is the content of Egorov's Theorem, \cite{Ego69}.  
If the transformed symbol assumes a simple form, then the 
the transformed operator assumes a simple form too and
its properties can be studied more easily. 
This construction
was the basis for many developments in the theory of linear
partial differential equations in the 70's, such as the study of
the solvability and the propagation of singularities (see, e.g.,
the compendium by H{\"o}rmander \cite{Hoe85III,Hoe85IV}).

In quantum mechanics
the relation between operators and symbols mentioned above is the relation  
between the Hamilton operator,
which defines a quantum mechanical system, and
the corresponding classical Hamilton function, which defines
the classical dynamical system corresponding to the
quantum system.  The operator thus is the quantisation of
the symbol, and microlocal analysis provides us with a
powerful set of tools to analyse quantum systems. 
These  ideas were applied,  e.g., in the seminal work
by Colin de Verdi{\`e}re on modes and quasimodes \cite{CdV77} where he
constructed classical and quantum normal forms around invariant
tori in phase space which still is  an active area of research
(see, e.g., the recent work by Cargo et al.
\cite{Cargoetal05}).

In transition state theory the classical Hamiltonian relevant for reaction type dynamics 
has an
equilibrium point, and as we have discussed in the first part of
the introduction one can use symplectic transformations to bring
the Hamilton function to a normal form in the neighbourhood of the
equilibrium point. The tools from microlocal analysis will
allow us to quantize this symplectic transformation and
bring the Hamilton operator into a normal form, too.
The problem of quantum normal forms near
equilibrium points of the symbol has been
studied quite extensively already. But most of this work concerns \emph{stable}
equilibrium points (see \cite{BelVit90,EspGraHer91,Sjo92,BamGraPau99}).
%
Here the aim is to construct a quantum normal form in order to
study energy spectra and eigenfunctions with very high precision.
In the physics literature
\cite{Robnik84b,Ali85,Eckhardt86,FriedEzra88,FriedEzra88b}
the same question was studied based on
the Lie approach to classical normal forms. In the
early works there has
been some confusion about the ordering problem in quantisation,
but these problems have been resolved  by Crehan \cite{Cre90}.

The case of an unstable equilibrium point (or more precisely
an equilibrium of saddle-center-$\cdots$-center type), which occurs in
transition state theory, has received much less attention
in the literature so far. In this case one expects the operator
to have continuous spectrum, and so instead of computing
eigenvalues one is looking for resonances.
Resonances are complex eigenvalues. Their imaginary parts
are related to the finite lifetime of quantum states in the neighbourhood
of the unstable equilibrium point. 
Since the problem is no longer selfadjoint, the
determination of resonances
is in general a much more difficult problem than that of eigenvalues
(see \cite{Zworski99} for a review).
The case of a Hamilton operator for which  the symbol has an unstable
equilibrium is one of the few cases where resonances can be computed
to high accuracy using a complex Bohr-Sommerfeld quantisation.
For 2 \dof systems, this was developed in \cite{GerSjo87,Sjo03} 
(for references in the chemistry literature, see, e.g.,
 \cite{SeidemanMiller91,Moiseyev98}  where
resonances are known as Gamov-Siegert eigenvalues).
The methods were then extended to systems with more \dof by
Sj{\"o}strand in \cite{Sjo87}, and building on this work more complete
results were obtained
by Kaidi and Kerdelhu{\'e} \cite{KaiKer00} who derived
quantisation conditions for the resonances which are valid to
all orders in the semiclassical parameter $\hbar$ and
are based on a quantum normal form. In \cite{IanSjo02} this was embedded into the study of
more general normal forms for Fourier integral operators.

The development and study of the quantum normal form near an equilibrium point of
saddle-center-$\cdots$-center type is one of the mains aims in the
quantum part of this paper. As  mentioned above,  the quantum
normal form has already been used to study resonances in the literature before.
We will see that the quantum normal form provides us with much more information 
which includes cumulative reaction probabilities and a detailed 
understanding of the dynamical mechanism of quantum reactions.
To this end we will relate the quantum states described by the quantum normal form to the  
phase space structures that control classical reaction dynamics.
In the classical case  the NHIM is the manifestation of the activated complex.   
Due to the
Heisenberg uncertainty relation,  there is no such invariant structure in quantum mechanics.
In fact the resonances will describe how the quantum activated complex decays. 

In order to use the quantum normal form  to study
concrete chemical reactions we have to be able to
compute it explicitly, i.e., we need an explicit algorithm
analogously to the classical normal form. 
The mathematical treatments in \cite{Sjo87,KaiKer00} do not give
us such an algorithm. Therefore we develop a quantized version
of the classical normal form algorithm which is similar to  quantum normal form
for stable equilibrium points in \cite{Cre90,EspGraHer91,BamGraPau99}.
We give a complete exposition of our algorithm to compute the quantum
normal form. 
At the level of symbols, the  classical and quantum normal form algorithms are almost identical.  The
essential differences are that the Poisson bracket is replaced by
the Moyal bracket, and rather than dealing with polynomial functions of the
phase space coordinates, we deal with polynomial functions of the phase space coordinates and $\hbar$.

\New{
The outline of this paper is as follows.
In Sec.~\ref{sec:CNF} we will start by reviewing classical normal form theory. 
We will show in detail how to construct symplectic transformations from the flows of Hamiltonian vector fields. The theory is presented in such a way that it allows for a direct comparison to the quantum normal form that we develop  in  
Sec.~\ref{sec:SNF}. This section includes a careful review of the necessary tools from the symbol calculus which are required to quantize symplectic transformations.
In Sec.~\ref{sec:classical}  we discuss the phase space structures which govern classical reaction dynamics and show how these phase space structures can be realised with the help of the classical normal form. This includes the construction of a dividing surface, the role of the NHIM and its stable and unstable manifolds, the foliation of the NHIM by invariant tori and its relation to the activated complex, the definition of dynamical reaction paths and a formula for the directional flux through the dividing surface. In this section we will also relate the theory presented to the flux-flux autocorrelation function formalism that can be found in the chemistry literature.
Sec.~\ref{sec:smatrix}  is the quantum mechanical analogue of Sec.~\ref{sec:classical}. 
We here use the  quantum normal form to study quantum  reaction dynamics. We show how to construct a local S-matrix from the quantum normal form and how this leads to an efficient algorithm to compute the cumulative reaction probability (the quantum analogue of the classical flux).
We study the distributions of the scattering states in phase space and relate them to the phase space structures governing classical reaction dynamics.  We here will also relate the quantum normal form computation of the cumulative reaction probability to the quantum version of the flux-flux autocorrelation function formalism. 
In Sec.~\ref{sec:resonances} we study quantum resonances that correspond to the (classical) activated complex. We will show how the resonances describe the quantum mechanical lifetimes of 
the  activated complex.  
We will study the phase space distributions of the corresponding resonance states and interpret these distributions in terms of the  phase space structures associated with the classical dynamics of reactions. 
%
In Sec. \ref{sec:examples} we illustrate the efficiency of the classical and quantum normal form algorithms for computing fluxes, cumulative reaction probabilities and resonances by applying the theory presented to several examples with  one, two and three degrees of freedom. 
} 

%% file: classical_NF.tex
\section{Classical Normal Form Theory}
\label{sec:CNF}

In this section we summarise the main elements of classical
Poincar\'e-Birkhoff normal form theory for Hamiltonian functions.
This is a well-known theory and has been the subject of many
review papers and books \cite{Deprit69,DragtFinn76,AKN88,MH92,Murdock03}. The
main reason for summarising the essential results here is so that
the reader can clearly see the classical and quantum normal form
theories ``side-by-side''. In this way the classical-quantum
correspondence is most apparent. This is explicitly illustrated by
developing the classical normal form theory in a way that is
rather different than that found in the literature.  This
difference allows us to explicitly show that the structure of the
classical and the quantum normal form theories is very similar. At
the same time, we emphasize that the classical normal form theory
is an essential tool for both discovering and computing the
necessary geometric structures in phase space with which we
construct our phase space transition state theory in Section
\ref{sec:classical}.

This section is organised as follows. In Sec.~\ref{sec:conj_symp}
we show how functions on phase space transform under symplectic
coordinate transformations, which are constructed as Hamiltonian
flows. In Sec.~\ref{sec:defcompcnf} we define what a  (classical)
normal form is and show how the formalism developed  in
Sec.~\ref{sec:conj_symp} can be used to transform a Hamiltonian
function into normal form to any desired order of its Taylor
expansion about an equilibrium point. The general scheme is
discussed in detail in Sec.~\ref{sec:examp_comp_cnf} for the case
of a saddle-centre-$\cdots$-centre equilibrium point.


\subsection{Transformation of Phase Space Functions through Symplectic
  Coordinate Transformations} 
\label{sec:conj_symp}

The essence of classical normal form theory is to find a new set
of coordinates, i.e., a change of variables, that transforms the
Hamiltonian to a ``simpler'' form (and we will explicitly define
what we mean by ``simpler'' shortly).  Since we are dealing with
Hamiltonian functions we want the coordinate transformation to
preserve the Hamiltonian structure, and this will be accomplished
if the transformation is {\em symplectic} (\cite{Arnold78,AM78}).
A standard approach to constructing  symplectic transformations is
through the use of Lie transforms (see, e.g., \cite{Murdock03}),
which we now review. Before proceeding we note that there are
issues related to differentiability of functions, existence and
uniqueness of solutions of ordinary differential equations, etc.
However, we will proceed  formally and assume that our functions
have as many derivatives as required and that solutions of
ordinary differential equations exist, and are sufficiently
differentiable, on domains of interest. Our purpose here is to
develop methods and an algorithm. Its applicability must be
verified for specific problems.

A function $W$ on phase space $\R^d\times\R^d$ defines a Hamiltonian vector field

\begin{equation}
X_W = \sum_{k=1}^d \bigg(\frac{\partial W}{\partial p_k}\frac{\partial }{\partial q_k}
- \frac{\partial W}{\partial q_k}\frac{\partial }{\partial p_k} \bigg)\,,
\end{equation}

\noindent
 and at a point $z=(q,p)=(q_1,\dots,q_d,p_1,\dots,p_d)$ in phase space
 this vector field takes the value

\begin{equation}
X_W(z) =\left(\frac{\partial W(z)}{\partial p_1}, \ldots ,\frac{\partial W(z)}{\partial p_d},
- \frac{\partial W(z)}{\partial q_1}, \ldots ,-\frac{\partial W(z)}{\partial q_d}  \right)\,.
\end{equation}

The  solutions of the ordinary differential equation (``Hamilton's
equations'')

\begin{equation}\label{eq:liouville}
\frac{\ud}{\ud \flowparam} z(\flowparam) = X_W (z(\flowparam))
\end{equation}

\noindent
 defines a Hamiltonian flow, $z\mapsto z(\epsilon):=\flow_W^\flowparam(z)$, which
 satisfies the properties

 \begin{itemize}

 \item $\flow_W^{\flowparam_1} \circ \flow_W^{\flowparam_2} = \flow_W^{\flowparam_1 + \flowparam_2}$,

 \item $\flow_W^\flowparam \circ  \flow_W^{-\flowparam} = {\rm id}$,

 \item $\flow_W^{0}= {\rm id}$,

 \end{itemize}

\noindent where id denotes the identity map, and
\begin{equation}
\frac{\ud}{\ud \flowparam} \flow_W^\flowparam (z)=
X_W(\flow^\flowparam(z))\,.
\end{equation}

Most importantly
for us, the Hamiltonian flow $\flow_W^\flowparam$
defines a symplectic, or `canonical', coordinate transformation of
the phase space onto itself \cite{Arnold78}.
This is significant because symplectic coordinate transformations
preserve the Hamiltonian structure. The Hamiltonian $W$ is referred to as the {\em
generating function} for the symplectic transformation
$\flow_W^\flowparam$.

We now consider the transformation of a (scalar valued) function
on phase space
under such a symplectic transformation. More precisely, for a
phase space function $A$ and a symplectic coordinate
transformation defined from the flow generated by Hamilton's
equations $z(\flowparam)=\flow_W^\flowparam (z)$, the
transformation of the function under this symplectic
transformation is given by

\begin{equation}
\label{eq:class_flow} A(\flowparam) = A \circ
\flow_W^{-\flowparam} \,,
\end{equation}

\noindent or, in coordinates,

\begin{equation}
A(\flowparam)\big(z(\flowparam)\big) = A(z)\,.
\end{equation}

For our purposes we want to develop $A(\flowparam)$ as a
(formal) power series in $\flowparam$. We begin by computing the
first derivative of $A(\flowparam)$ with respect to $\flowparam$ giving

\begin{equation} \label{eq:dAdepsilon_classical1}
\frac{\ud}{\ud \flowparam} A(\flowparam) =
- \langle \nabla  A,X_W \rangle \circ \flow_W^{-\flowparam}
= \{W,A\} \circ \flow_W^{-\flowparam}\,,
\end{equation}

\noindent
 where $\nabla A \equiv \left(\frac{\partial A}{\partial q_1}, \ldots ,\frac{\partial A}{\partial q_d},
 \frac{\partial A}{\partial p_1}, \ldots ,\frac{\partial A}{\partial p_d}  \right)$ is the gradient of $A$, $\langle
\cdot,\cdot \rangle$ is the standard scalar product in $\R^{2d}$,
and

\begin{equation}
\{W,A\} = \sum_{k=1}^d \bigg(\frac{\partial W}{\partial
q_k}\frac{\partial A }{\partial p_k} - \frac{\partial W}{\partial
p_k}\frac{\partial A}{\partial q_k} \bigg) = -\{ A, W \},
\label{pb_def}
\end{equation}

\noindent
 is the Poisson bracket of $W$ and $A$. Using the
fact that $W$ is invariant under the flow $\flow_W^\flowparam$ (
i.e., $W\left( \flow_W^{-\flowparam} \right) = W \left(
\flow_W^{0}\right)$, or, in other words, the Hamiltonian $W$ is
constant along trajectories of the vector field $X_W$ generated by $W$) we can rewrite
\eqref{eq:dAdepsilon_classical1} as

\begin{equation} \label{eq:dAdepsilon_classical}
\frac{\ud}{\ud \flowparam} A(\flowparam) = \{W,A(\flowparam)\}\,.
\end{equation}

The Poisson bracket gives us a convenient way of representing the
derivatives of a function along trajectories of Hamilton's
equations. We simplify the notation further by defining the
adjoint operator

\begin{equation}
\label{eq:ad_op_class}
 \ad_W: A \mapsto \ad_W A := \{ W,A \}
\end{equation}

\noindent
 associated with a generating function $W$. We can now
differentiate  \eqref{eq:dAdepsilon_classical} again
to obtain the second order derivative with respect to $\flowparam$,

\begin{equation}
\label{eq:2deriv_class}
\frac{\ud^2}{\ud \flowparam^2} A(\flowparam) =
 \frac{\ud}{\ud \flowparam} \left(\frac{\ud}{\ud \flowparam}  A(\flowparam)
 \right) =
\{W,\frac{\ud}{\ud \flowparam}A(\flowparam)\} =
 \{W,\{W,A(\flowparam)\}\} =: \big[\ad_W \big]^2 A(\flowparam) \,.
\end{equation}

\noindent Continuing this procedure for higher order derivatives
gives

\begin{equation}\label{eq:ho_deriv_class}
\begin{split}
\frac{\ud^n}{\ud \flowparam^n} A(\flowparam) =
\frac{\ud}{\ud \flowparam} \left(\frac{\ud^{n-1}}{\ud \flowparam^{n-1}} A(\flowparam)  \right)
&=  \{ W, \{ \cdots \{ W, \frac{\ud}{\ud \flowparam}A(\flowparam)\} \cdots \}\}  \\
&=  \{ W, \{ \cdots \{W, \{W,A(\flowparam)\} \} \cdots \}\}\\
& =: \big[\ad_W \big]^n A(\flowparam) \,.
\end{split}
\end{equation}

\noindent Using these results, we obtain the Taylor expansion of
$A(\flowparam)$ about $\epsilon =0$,

\begin{equation} \label{eq:Atransf_classical}
A(\flowparam) = \sum_{n=0}^{\infty} \frac{\flowparam^n}{n!}
\frac{\ud^n}{\ud \flowparam^n} A(\flowparam)\big|_{\flowparam = 0}
= \sum_{n=0}^{\infty} \frac{\flowparam^n}{n!} \big[ \ad_W \big]^n
A \,,
\end{equation}

\noindent where $A(0)=A$ and  $\big[ \ad_W \big]^n A$ are defined
as in Equations~\eqref{eq:dAdepsilon_classical}--\eqref{eq:ho_deriv_class} with $\big[
\ad_W \big]^0 A=A$.

\rem{
\begin{equation}
\big[ \ad_W \big]^0 =\text{id}\,, \qquad \big[ \ad_W \big]^n =
\ad_W  \circ \big[ \ad_W \big]^{n-1}\,,\quad n\ge 1\,.
\end{equation}
}

Equation~\eqref{eq:Atransf_classical} gives the Taylor expansion
with respect to the flow parameter or `time' $\epsilon$ for a
phase space function $A$ that is transformed by a symplectic
transformation defined by the Hamiltonian flow generated by
the function $W$. It will form the basis of the classical normal form
method where the idea is to ``simplify'' (or ``normalise'') a
function which, for us, will be a specific Hamiltonian through the
choice of an ``appropriately chosen''  sequence of symplectic
transformations that simplify the Hamiltonian ``order by order''
of its Taylor expansion with respect to the phase space
coordinates $z=(q,p)$. First, we need to make clear that  the
normal form procedure that we develop here is valid  {\em in a
neighborhood of an equilibrium point}. This means  that the normal
form is a local object whose dynamics have meaning  for the
original Hamiltonian only in a neighborhood of an equilibrium
point. In order to describe the terms in the Taylor expansion of a
given order in the phase space coordinates more precisely  we
introduce the vector spaces $\ccW^s$, $s\in \N_0$,  of polynomials 
which are homogeneous of order $s$. The space  $\ccW^s$ is  spanned (over
$\C$) by all monomials of the form

\begin{equation} \label{eq:classicalmonomials}
q^{\alpha}p^{\beta}:=\prod_{k=1}^d q_k^{\alpha_k} p_k^{\beta_k}
\,\, ,\quad \text{where} \quad \abs{\alpha}+\abs{\beta}:=
\sum_{k=1}^d \alpha_k + \beta_k =s\,\, .
\end{equation}

The following two lemmata are the key tools used in the
computation of the classical normal form.
\begin{lem} \label{lem:classical}
Let $W\in \ccW^{s'}$, $A\in \ccW^{s}$ with  $s,s'\ge 1$, then
\begin{equation}
\{W,A\}\in \ccW^{s+s'-2}\,,
\end{equation}
and for $n\ge 0$,
\begin{equation}
\big[\ad_W\big]^nA\in \ccW^{n(s'-2)+s}
\end{equation}
if $n(s'-2)+s\ge0$ and $\big[\ad_W\big]^nA=0$ otherwise.
\end{lem}

\begin{proof}
This Lemma can be proven by direct calculation.
\end{proof}

This lemma is key to the proof of

\begin{lem} \label{lem:classical2}

Let $W\in \ccW^{s'}$ with $s'\ge 3$ and
\begin{equation}
A= \sum_{s=0}^\infty A_s \label{ser_A}
\end{equation}
with $A_s\in \ccW^s$. Then
\begin{equation} \label{eq:trans_classical}
A' := A\circ \flow^{-1}_W
= \sum_{n=0}^{\infty} \frac{1}{n!} \big[ \ad_W \big]^n A
= \sum_{s=0}^\infty A'_s\,\,,
\end{equation}
where
\begin{equation}
A'_s =\sum_{n=0}^{[\frac{s}{s'-2}]} \frac{1}{n!} [\ad_{W}]^n
A_{s-n(s'-2)}\,, \label{order_s}
\end{equation}
where $[\frac{s}{s'-2}]$ denotes the integer part of
$\frac{s}{s'-2}$.
\end{lem}

\begin{proof}

Using \eqref{ser_A}, we write out the next to last term in
\eqref{eq:trans_classical} as a series of series as follows (where
we have also changed the summation index from $s$ to $j$ in order to
avoid possible confusion):

\begin{eqnarray}
\sum_{n=0}^{\infty} \frac{1}{n!} \big[ \ad_W \big]^n A & = &
\sum_{n=0}^{\infty} \frac{1}{n!} \big[ \ad_W \big]^n \sum_{j=0}^\infty A_j \nonumber \\
& = & \sum_{j=0}^\infty A_j + \sum_{j=0}^\infty
  \ad_W A_j + \sum_{j=0}^\infty \frac{1}{2} \big[ \ad_W \big]^2
 A_j\nonumber \\
 & + &  \sum_{j=0}^\infty
 \frac{1}{3!}\big[ \ad_W \big]^3 A_j + \ldots + \sum_{j=0}^\infty
 \frac{1}{n!}\big[ \ad_W \big]^n A_j + \ldots\,.
\end{eqnarray}

\noindent We now want to inspect each series in the series and
extract the order $s$ term from each one. Then summing these terms
will give the series \eqref{order_s}. Using Lemma
\ref{lem:classical}, we find

\begin{equation}
\big[ \ad_W \big]^n A_j  \in \ccW^{n(s'-2)+j}. \label{j1}
\end{equation}

\noindent Now we wish to choose $j$ such that

\begin{equation}
\big[ \ad_W \big]^n A_j \in \ccW^{s}. \label{j2}
\end{equation}

\noindent Comparing \eqref{j1} and \eqref{j2}, this is true for

\begin{equation}
j = s-n(s' -2).
\end{equation}

\noindent Hence it follows that

\begin{equation}  \label{order_s2}
A'_s =\sum_{n=0}^{[\frac{s}{s'-2}]} \frac{1}{n!} [\ad_{W}]^n
A_{s-n(s'-2)}\,.
\end{equation}

\end{proof}

\subsection{Definition and Computation of the Classical Normal Form}
\label{sec:defcompcnf}

We will now define when a Hamilton function is in classical normal
form. Here we use the adjective `classical' to distinguish the normal form in
the case of classical mechanics from the normal form that we will define for the
case of  quantum mechanics in Sec.~\ref{sec:SNF}.
As we will see, in general a Hamilton function is not in normal form. However,
as we will show in detail, the formalism reviewed in the previous section can
be used to construct an explicit algorithm which allows one to transform a
Hamilton function to normal form
to any desired order of its Taylor expansion.

The starting point is a Hamilton function with an equilibrium point at $z=z_0$, i.e., $\nabla H(z_0)=0$. Let
$H_2(z) := \frac12\la z-z_0,\Hess H(z_0)(z-z_0) \ra$ be the quadratic part of the Taylor expansion of $H$
about $z_0$.\footnote{Here $,\Hess H(z_0)$ denotes the Hessian of $H$ at $z_0$, i.e., the matrix of second derivatives $(\partial_{z_i}\partial_{z_j} H(z_0))_{ij}$} We then make the following
\begin{Def} \label{def:cnf}
We say that $H$ is in {\bf classical normal form} with respect to the equilibrium point $z_0$ if
\begin{equation}
\ad_{H_2} H  \equiv \{ H_2,H \} = 0\,.
\end{equation}
\end{Def}
It follows from  this definition that if $H$ is in normal form
then $H_2$ will be an integral of the motion generated by the
Hamilton function $H$ and moreover, as we will see below,
depending on the structure of $H_2$, further integrals of the
motion will exist. A consequence of the existence of integrals of
motion is the structuring, or \emph{foliation}, of the phase space
by lower dimensional surfaces or \emph{manifolds} that are
invariant under the dynamics. If we choose initial conditions for
Hamilton's equations then these initial conditions will determine
values of the integrals of motion. The full solution of Hamilton's
equation will then be contained in the manifold given by the
common level set of the integrals corresponding to the initial
values. This way the integrals of the motion confine the possible
dynamics. Moreover,  the existence of integrals of the motion
significantly simplifies the study of the dynamics.

In general a Hamilton function is not in normal form. However, we
will use the formalism and results developed in the previous
section to transform a Hamiltonian to normal form in a
neighbourhood of the equilibrium point to a certain order of its
Taylor expansion about the equilibrium point. As we will see, the
transformed Hamiltonian function truncated at this order will lead
to a very accurate description of the motion in the neighbourhood
of the equilibrium point. What we mean by ``accurate description'' is discussed in 
Sec~\ref{sec:accuracy_NF}).

We develop the following procedure.
We begin with our
``original Hamiltonian''
\begin{equation}
H= H^{(0)}  \label{orig_Ham}\,,
\end{equation}
and we construct a consecutive sequence of symplectic
transformations
\begin{equation} \label{eq:transformation-sequence}
H^{(0)}\to H^{(1)}\to H^{(2)}\to H^{(3)}\to\cdots\to H^{(N)}\,,
\end{equation}
where $N$ is a sufficiently large integer which will be the order
at which we will truncate the normal form series.

The first step in the sequence~\eqref{eq:transformation-sequence}
is obtained by shifting the critical point $z_0$ to the origin of
a new  coordinate system. We set
\begin{equation} \label{eq:shift_origin_classical}
z^{(1)} = z - z_0\,\,.
\end{equation}
The Hamiltonian function $H^{(1)}$ is the representation of
$H^{(0)}$ in terms of the new coordinates $z^{(1)}$, i.e.,
\begin{equation} \label{eq:H1_z1_classical}
H^{(1)}(z^{(1)}) = H^{(0)}(z^{(1)}+z_0) \,.
\end{equation}

Once the equilibrium point is shifted to the origin, our normal
form procedure will require us to work with the Taylor expansion
of the Hamiltonian $H^{(1)}$  about the origin in a ``term-by-term'' fashion. Let

\begin{equation}\label{eq:H-exp_classical}
H^{(1)}=E_0 + \sum_{s=2}^\infty H_s^{(1)}\,,
\end{equation}
where

\begin{equation}
H_s^{(1)}(q,p):=\sum_{\abs{\alpha}+\abs{\beta}=s}\frac{1}{\alpha!\beta!}
\pa_q^{\alpha}\pa_p^{\beta}H^{(1)}(0, 0)q^{\alpha}p^{\beta} 
\end{equation}
are the terms of order $s$. Here we employ the usual
multi-index notation; for $\alpha \equiv (\alpha_1, \ldots,
\alpha_d)\in N_0^d$  we have $\vert \alpha \vert \equiv \alpha_1 + \ldots +
\alpha_d$, $\alpha! \equiv \alpha_1! \alpha_2! \cdots \alpha_d!$,
$q^\alpha \equiv  q_1^{\alpha_1} q_2^{\alpha_2} \cdots
q_d^{\alpha_d}$ and $\partial_q^{\alpha} \equiv
\frac{\partial^{\alpha_1}}{\partial q_1^{\alpha_1}}
\cdots\frac{\partial^{\alpha_d}}{\partial q_d^{\alpha_d}}$ (for $\beta\in
N_0^d$ and $p\in\R^d$, the notation is analogous).
Since \eqref{eq:H-exp_classical}  is a Taylor
expansion of a Hamiltonian about an equilibrium point at the
origin it follows that $H^{(1)}_1=0$.
In particular, $H^{(1)}_0 \equiv E_0$ is the ``energy'' of the equilibrium point.


At the next step in the sequence~\eqref{eq:transformation-sequence} we choose
a linear symplectic transformation
such that $H^{(2)}_2$ assumes a
``simple form''.
In other words, we seek a transformation
that simplifies the quadratic part of the Hamiltonian or,
equivalently, the linear part of the Hamiltonian vector field. This is
accomplished by choosing an appropriate symplectic $2d\times 2d$ matrix $M$,
i.e., a matrix statisfying $M^T\,J\,M=J$,
where $J$ is the standard $2d\times 2d$ symplectic matrix
\begin{equation} \label{eq:def_stand_sympl_matr}
J= \left(
\begin{array}{rr}
0 & \text{id}\\
-\text{id} & 0
\end{array}
\right)
\end{equation}
whose blocks consist of $d\times d$ zero matrices and $d\times d$ identity
matrices.
We then set
\begin{equation}\label{eq:simplify_quadratic_classic}
z^{(2)} = M \, z^{(1)}\,,
\end{equation}
and the corresponding transformed Hamiltonian is given by
\begin{equation} \label{eq:H2_z2_classical}
H^{(2)} (z^{(2)}) = H^{(1)} (M^{-1}\,z^{(2)})\,.
\end{equation}

Which form of $H^{(2)}_2$ can be considered to be ``simple''
depends on the nature of the particular
equilibrium point (i.e., the eigenvalues and eigenvectors
associated with the matrix obtained by linearising Hamilton's
equations about the origin).
The main benefit of having $H^{(2)}_2$ in a ``simple'' form
is that this will simplify the explicit implementation of the
algorithm to normalise the higher order terms, $n\ge 3$, i.e., how
to choose the next steps in the
sequence~\eqref{eq:transformation-sequence}.
Therefore, ``simplify'' could mean that we would seek a
transformation that would diagonalise the linear part of
Hamilton's equations, or transform it to ``real Jordan canonical
form'' in the case of complex eigenvalues. Clearly, constructing
such a transformation is a problem in linear algebra for which
there is a large literature. However, the symplectic case tends to
bring with it new difficulties, both in the analytical and
computational areas (see, e.g., \cite{ChurchillKummer99}). In the
next section we will see how to simplify the linear part of
Hamilton's equations for our particular case of interest, i.e., a
saddle-centre-$\cdots$-centre equilibrium point satisfying a
certain ``nonresonance'' condition. However, it is important to
realise that the normal form {\em algorithm} does not depend on
the specific form taken by the linear part of Hamilton's
equations.

Up to this point we have located an equilibrium point of interest,
translated it to the origin, Taylor expanded the resulting transformed
Hamiltonian $H^{(1)}$ about the origin (for which $H_1^{(1)} =
0$), and constructed a linear symplectic transformation in such a way that the
quadratic part of the resulting transformed Hamiltonian,
$H_2^{(2)}$, is ``simple''.  Now we are ready to describe
how to normalise the terms of order three and higher, i.e., how to
define the next steps in the sequence~\eqref{eq:transformation-sequence}.
To accomplish these transformations we will use the formalism reviewed in
Sec.~\ref{sec:conj_symp} and successively transform the Hamiltonian by
the time one
maps of the flows generated by Hamiltonian vector fields. More
precisely, for $n\ge3$, $H^{(n)}$ is computed from $H^{(n-1)}$ according to
\begin{equation} \label{n_trans_Ham}
H^{(n)} = H^{(n-1)}\circ \flow^{-1}_{W_n} = \sum_{k=0}^{\infty}
\frac{1}{k!} \big[ \ad_{W_n} \big]^k H^{(n-1)}
\end{equation}
with a generating function $W_n\in \ccW^n$. The order $s$
term of the Taylor expansion of $H^{(n)}$ expressed as a series
involving terms in the Taylor expansion of $H^{(n-1)}$ and $W_n$
is obtained by substituting the Taylor expansion of $H^{(n-1)}$
into \eqref{n_trans_Ham} and  using Lemma \ref{lem:classical2}.
This gives
\begin{equation} \label{n_trans_Ham_s}
 H_s^{(n)} = \sum_{k=0}^{\left[ \frac{s}{n-2} \right]} \frac{1}{k!} \big[ \ad_{W_n}
 \big]^k H_{s-k(n-2)}^{(n-1)}\,,\qquad n\ge3\,\,.
\end{equation}
The corresponding transformation of phase space
coordinates is then given by
\begin{equation}
z^{(n)} = \flow^1_{W_n} (z^{(n-1)}) \,,\qquad n\ge3\,\,.
\end{equation}
%
%


We note that in fact also the affine linear symplectic coordinate transformations
\eqref{eq:shift_origin_classical} and \eqref{eq:simplify_quadratic_classic}
which formed the first two steps in the sequence
\eqref{eq:transformation-sequence}
can be formally expressed as time one maps of Hamiltonian flows with
generating functions $W_1\in \ccW^1$ and $W_2\in \ccW^2$, respectively.
A generating functions  $W_1$ whose time one map achieves the translation
\eqref{eq:shift_origin_classical}
is given by
\begin{equation}
W_1(z) = -\la z_0,J z \ra, \label{gen_trans}
\end{equation}
where $J$ is the standard $2d\times2d$ symplectic matrix defined in Equation~\eqref{eq:def_stand_sympl_matr}.
This gives
\begin{equation}
z^{(1)} = \flow^1_{W_1}(z) = z-z_0\,.
\end{equation}
In this case the upper limit of the sum in \eqref{n_trans_Ham_s}
is infinity. It is in general not straightforward to explicitly give an
expression for a generating function $W_2\in \ccW^2$ whose time
one map achieves the linear symplectic transformation
\eqref{eq:simplify_quadratic_classic} for a given symplectic
matrix $M$. But such a $W_2$ always exists\footnote{This follows from two facts. 
Firstly, the group of linear symplectic transformations is  connected, 
and therefore the image of the exponentiation of its Lie algebra is connected, too. 
Secondly this Lie algebra is isomorphic to
the vector space of quadratic polynomials endowed with
the Poisson bracket \cite{Fol89}. Therefore 
the set of all time one maps generated by quadratic elements of $\ccW^2$ 
is the whole symplectic group.}.
%
For $n=2$ in
Equation~\eqref{n_trans_Ham_s} the upper limit of the sum is again
infinity. In the next section we will provide a matrix $M$ which
achieves the simplification of the quadratic part of the
Hamiltonian function for the case of a saddle-centre-$\cdots$-centre
equilibrium point satisfying a nonresonance condition without
specifying the corresponding $W_2$. Note however that it is  $M$
and not necessarily $W_2$ which is required for our normalisation
procedure.

Let us now proceed with the nonlinear symplectic transformations
generated by polynomials $W_n\in\ccW^n$ with $n\ge3$ to achieve
the third and higher steps in the
sequence~\eqref{eq:transformation-sequence}. The first thing to
note is that these transformations will not alter the zeroth order
term, $E_0$, and we will also have $H_1^{(1)} = H_1^{(n)} = 0$,
$n \ge 3$. The zeroth order term is unaltered since the  upper limit in
the sum \eqref{n_trans_Ham_s} is zero for $s=0$. The first order term stays zero because
 for
$s\le1$ in combination with $n> 3$ and $s=0$ in combination with $n=3$,
the upper limit in the sum \eqref{n_trans_Ham_s} is again zero. For $n=3$ in combination with $s=1$, the upper limit is 1. However, the $k=1$ term, $\ad_{W_3}H^{(2)}_0$, in 
the sum \eqref{n_trans_Ham_s} is zero because $H^{(2)}_0$ is the constant $E_0$ and hence vanishes when $ad_{W_3}$ is applied to it.

Moreover, the quadratic part of the Hamiltonian $H_2^{(2)}$ will
not be modified by the transformations generated by $W_n, \, n \ge
3$. We will show this directly from our formalism.

 \begin{lem} \label{lem:H2unchanged}
 \label{lem:order2}
 $H_2^{(n)} = H_2^{(2)}, \quad n \ge 3.$
 \end{lem}

 \begin{proof}

 The idea is to use \eqref{n_trans_Ham} to transform from
 $H^{(n-1)}$ to $H^{(n)}$, and then to show that $H_2^{(n)}=
 H_2^{(n-1)}$ for $n \ge 3$.

 We separate out the constant and quadratic parts of $H^{(n-1)}$
 as

 \begin{equation}
 H^{(n-1)} = E_0 + H_2^{(n-1)} + \sum_{s=3}^{\infty} H_s^{(n-1)},
 \end{equation}

 \noindent
 and then we substitute this into \eqref{n_trans_Ham} to obtain

 \begin{eqnarray}
 H^{(n)} = \sum_{k=0}^{\infty} \frac{1}{k!} \big[ \ad_{W_n}
 \big]^k E_0 + \sum_{k=0}^{\infty} \frac{1}{k!} \big[ \ad_{W_n}
 \big]^k H_2^{(n-1)} + \sum_{k=0}^{\infty} \frac{1}{k!} \big[ \ad_{W_n}
 \big]^k \sum_{s=3}^{\infty} H_s^{(n-1)}. \nonumber \\
 \end{eqnarray}

 \noindent Note that the first series in this expression only admits the $k=0$ term, $E_0$. We consider the case $n \ge 3$.
 In this case,  the third series, using Lemma
 \ref{lem:classical}, only admits terms of order larger than or
 equal to three. Hence, all of the quadratic terms must be in the
 second series. Using Lemma \ref{lem:classical}, the $k^{\text{th}}$ term
 in that series is contained in $\ccW^{k(n-2) + 2}$.  Therefore
 the only quadratic term occurs for $k=0$, which is $H_2^{(n-1)}$.

 \end{proof}

Lemma~\ref{lem:H2unchanged} motivates the definition of the operator
\begin{equation}
\opD := \ad_{H_2^{(2)}} = \{ H_2^{(2)}, \cdot \}.
\end{equation}
In fact, $\opD$ will simply be a convenient shorthand notation for
the operator $\ad_{H_2}=\{ H_2 , \cdot \}$ in the definition of
the the normal form in Definition~\ref{def:cnf} in terms of the
coordinates $z^{(2)}$. The operator  $\opD$ plays a crucial role
in the computation of the normal form transformation.

The other important point to realise when transforming
$H^{(n-1)}$ to  $H^{(n)}$ with  $\flow^{-1}_{W_n}$, $W_n\in \ccW^n$, is that all terms of
order smaller than $n$ are unchanged (however, the terms of order larger
than $n$ are modified by the $n^{\rm th}$ order normalisation
transformation). This is essential for the
success of the iterative process and we provide a proof of this result now.

 \begin{lem}
 \label{lem:lower_order}
For $n\ge3$ and $ 0 \le s < n$, $H_s^{(n)} = H_s^{(n-1)}$.
 \end{lem}

 \begin{proof}

 First, it is important to consider the upper limit of the sum \eqref{n_trans_Ham_s}.
 For $0 \le s \le n-3$ it is zero, which indicates that for these
 values of $s$ only the $k=0$ term is nonzero. Hence, we have

 \begin{equation}
 H_s^{(n)} =  H_s^{(n-1)}, \quad 0 \le s \le n-3.
 \end{equation}

 \noindent
 Next we  separately consider the cases $s=n-2$ and $s=n-1$.
 Using \eqref{n_trans_Ham_s} we find  for $s=n-2$, 
   
 \begin{equation}
 H_{n-2}^{(n)} =  H_{n-2}^{(n-1)}+\ad_{W_n}H_{0}^{(n-1)}=H_{n-2}^{(n-1)}
 \end{equation}
since $H_{0}^{(n-1)}=E_0=const.$. For $s=n-1$, \eqref{n_trans_Ham_s} gives 

\begin{equation}
 H_{n-1}^{(n)} =  H_{n-1}^{(n-1)}+\ad_{W_n}H_{1}^{(n-1)}+\delta_{n,3}\frac{1}{2}\big[\ad_{W_n}\big]^2H_{0}^{(n-1)}=H_{n-1}^{(n-1)}
 \end{equation}
since $H_1^{(n-1)} =0$ and $H_{0}^{(n-1)}=E_0=const.$. The Kronecker symbol  in the last term of the second expression shows that this term only occurs for $n=3$. 
\end{proof}

Now if we consider the $n^{\text{th}}$ order term in $H^{(n)}$ this
will show us how to choose $W_n, \, n \ge 3$.

\begin{lem}[Homological Equation]
\label{lem:hom_eq}
For $s=n\ge 3$,
\begin{equation}\label{eq:homological_classical}
H_n^{(n)} = H_n^{(n-1)}-   \opD \, W_n\,\,.
\end{equation}
\end{lem}

 \begin{proof}

 This result is also obtained from \eqref{n_trans_Ham_s}, with a careful consideration of the upper limit of the sum. The case
 $n \ge 5$ is the most straightforward. In this case only $k=0$ and
 $k=1$ contribute in the sum, and using \eqref{pb_def}, we obtain immediately that

 \begin{equation} \label{eq:classicalhomologicalequation}
 H_n^{(n)} = H_n^{(n-1)} + \ad_{W_n}
  H_2^{(2)} = H_n^{(n-1)} - \ad_{H_2^{(2)}}
  W_n = H_n^{(n-1)}-   \opD \, W_n \,.
 \end{equation}

 \noindent
 The special cases $s=n=4$ and $s=n=3$ must be considered. These will
 give rise to some additional terms in \eqref{n_trans_Ham_s}.
 However, as for Lemma \ref{lem:lower_order}, these will be zero if we take  into account $H_1^{(n-1)} =0$ and
 $\big[ \ad_{W_n} \big]^k  E_0 =0$ for integers $k>0$, $n\ge 3$.

\end{proof}

Equation~\eqref{eq:classicalhomologicalequation} is known as the {\em
 homological equation}.
We want to solve the homological equation, i.e., find a function $W_n \in \ccW^n$,  in such a way that $H^{(n)}$ is in normal form
 up to order $n$. To this end note that it follows from Lemma
\ref{lem:classical} that $\opD$ defines a {\em linear} map of $\ccW^n $  into
$\ccW^n $, i.e., {\em for each n},
\begin{equation}
\opD : \ccW^n \rightarrow \ccW^n.
\label{lin_op}
\end{equation}
In order to have $H^{(n)}$ in normal form up to order $n$ we have to  require $\opD\, H_n^{(n)} =
0$. Looking at the homological equation~\eqref{eq:classicalhomologicalequation} this means we need to find a function
$W_n \in \ccW^n$ such that $H_n^{(n)} = H_n^{(n-1)}-   \opD \, W_n$ is in the
kernel of the restriction of $\opD$ to   $\ccW^n$, i.e.,
\begin{equation}
H_n^{(n)} = H_n^{(n-1)}-   \opD \, W_n \in \text{ Ker } \opD  \big|_{\ccW^n}\,.
\end{equation}

\begin{Def}\label{def:solvable}
We will call the homological equation \eqref{eq:homological_classical} \emph{\bf solvable} if for any $n\geq 3$ there 
exist for any $H_n\in \ccW^n$ an $W_n\in \ccW^n$ such that 
\begin{equation}
H_n-   \opD \, W_n \in \mathrm{Ker } \opD  \big|_{\ccW^n}\,\, .
\end{equation}
\end{Def}

Whether   the Homological equation is solvable and how such a  
$W_n$ can be found depends on the structure of $\opD$, i.e., on the
structure of the matrix associated with the linearisation of the vector field
about the equilibrium point. In the next subsection we will show that the homological equation is 
solvable in the case of a saddle-centre-$\cdots$-centre equilibrium point and explain how $W_n$
can be found.

We summarise the results of this section  in the following

\begin{thm} \label{theorem_classicalNF}

Assume that a Hamiltonian function $H$ has an equilibrium point at $z_0\in
\R^d\times\R^d$, and that the homological equation is solvable.  Then for every $N\in\N$
there is a symplectic transformation $\flow_N$ such that
\begin{equation}\label{eq:defPhiN}
H\circ\flow^{-1}_N = H_{\text{CNF}}^{(N)} + O_{N+1}\,,
\end{equation}
where $H_{\text{CNF}}^{(N)}$ is in normal form (with respect to $z=(0,0)$) and
$O_{N+1}$ is of order $N+1$, i.e., there exists an open neighbourhood $U$ of
$z=(0,0)$ and a constant $c>0$ such that
\begin{equation}
|O_{N+1}(\epsilon z)| < c \epsilon^{N+1}
\end{equation}
for $z\in U$ and $\epsilon<1$.
\end{thm}

\begin{proof}

\New{

Following the scheme described in this section we normalise
the Hamilton function $H$ order by order according to the
sequence~\eqref{eq:transformation-sequence}.
We start by  choosing a new coordinate system $z^{(1)}=z-z_0$ which has the
equilibrium point $z_0$ at the origin (see
\eqref{eq:shift_origin_classical}), and Taylor expand the Hamilton
function $H^{(1)}$, which we obtain from expressing $H$ in the new coordinates
$z^{(1)}$ (see Equation~\eqref{eq:H1_z1_classical}), about $z^{(1)}=0$ to 
order $N$. The remainder which we denote by $R_{N+1}^{(1)}$ is then  of order $N+1$. 

We then choose a symplectic $2d\times2d$ matrix $M$
to define a linear symplectic transformation to new coordinates $z^{(2)}=M\, z^{(1)}$ in
terms of which the
quadratic part of the transformed Hamilton function
$H^{(2)}$ (see Equation~\eqref{eq:H2_z2_classical}) assumes a simple form.
As mentioned above the choice of
$M$ depends on the nature of the equilibrium point and will simplify
the calculation of the steps for $n\ge3$ in the sequence~\eqref{eq:transformation-sequence}.
Apart from this however, the choice of the symplectic matrix $M$ is not
important. We thus get
\begin{equation}
H^{(2)} = E_0 + \sum_{s=2}^N H^{(2)}_s + R_{N+1}^{(2)}\,,
\end{equation}
where $H^{(2)}_s(z^{(2)}) = H^{(1)}_s(M^{-1}\, z^{(2)}) $, i.e. $H^{(2)}_s \in \ccW^s$ for $s=2,\ldots,N$, and the remainder term $R_{N+1}^{(2)}$ given by  $R_{N+1}^{(2)}(z^{(2)})=R_{N+1}^{(1)}(M^{-1}\, z^{(2)})$ is again of order $N+1$.

Having simplified the quadratic part, we proceed  inductively by subsequently
choosing generating functions $W_n\in\ccW^n$, which at each order $n$, $n=3,\ldots,N$, solve the homological
equation~\eqref{eq:homological_classical} and determing $H^{(n)}$ from $H^{(n-1)}$ as follows. 
For $n\ge 3$,
 $H^{(n-1)}$ is of the form
\begin{equation}
H^{(n-1)} = \sum_{s=0}^N H^{(n-1)}_s + R_{N+1}^{(n-1)}  \,,,
\end{equation}
where $H^{(n-1)}_s \in \ccW^s$ and  $R_{N+1}^{(n-1)} $ is of order $N+1$.
Using this decomposition of $H^{(n-1)}$ we can write for 
$H^{(n)} = H^{(n-1)} \circ \flow^{-1}_{W_n}$,
\begin{eqnarray} 
H^{(n)}&=& \sum_{s=0}^N H^{(n-1)}_s \circ \flow^{-1}_{W_n} + R_{N+1}^{(n-1)} \circ \flow^{-1}_{W_n}  \label{eq:thm1_trafo_1} \\
&=& \sum_{s=0}^N \sum_{k=0}^\infty \frac{1}{k!} \big[ \ad_{W_n} \big]^k     H^{(n-1)}_s    + R_{N+1}^{(n-1)} \circ \flow^{-1}_{W_n}  \label{eq:thm1_trafo_2} \\
&=&  \sum_{s=0}^N \sum_{k=0}^{\big[ \frac{N-s}{n-2} \big]} \frac{1}{k!}  \big[ \ad_{W_n} \big]^k     H^{(n-1)}_s    + R_{N+1}^{(n)}   \label{eq:thm1_trafo_3} \,,      
\end{eqnarray}
where
\begin{equation}
R_{N+1}^{(n)}   =  R_{N+1}^{(n-1)} \circ \flow^{-1}_{W_n} + 
\sum_{s=0}^N \sum_{k= \big[ \frac{N-s}{n-2} \big]+1 }^\infty \frac{1}{k!}  \big[ \ad_{W_n} \big]^k     H^{(n-1)}_s \label{eq:thm1_trafo_4} \,.
\end{equation}
We here used Eq.~\eqref{n_trans_Ham} to get \eqref{eq:thm1_trafo_2}. To obtain \eqref{eq:thm1_trafo_3} from  \eqref{eq:thm1_trafo_2}  we removed all those terms from the double sum in 
\eqref{eq:thm1_trafo_2}  contained in the $\ccW^s$ with $s\ge N+1$ and absorbed them in the new remainder term 
$R_{N+1}^{(n)} $ in \eqref{eq:thm1_trafo_4}. Since the symplectic transformations $\flow^{1}_{W_n}$ are near identity transformations for $n\ge 3$ the remainder term $R_{N+1}^{(n)}$   is again  of order $N+1$.

After the step $n=N$
the terms of order less than or equal to $N$ of the
Hamilton function $H^{(N)}$ are then in normal form
(with respect to $z=(0,0)$).
The symplectic transformation $\flow_N$ in Eq.~\eqref{eq:defPhiN} and the
corresponding new coordinates $z^{(N)}$
are then given by
\begin{equation} \label{eq:Nthordernormalformcoordinates}
z^{(N)} \equiv \flow_N(z) =\flow_{W_N}^{1} \circ \cdots \circ \flow_{W_3}^{1} (z^{(2)})\,,
\quad z^{(2)} = M z^{(1)}\,,\quad z^{(1)} = z-z_0\,.
\end{equation}

} 
\end{proof}

From the point of view of applications the definition of the normal
form in Definition~\ref{def:cnf} is not very practical since it
requires one to carry out the procedure described in the proof of Theorem~\ref{theorem_classicalNF} for $N\rightarrow \infty$.
In general,
it is well known that such normal form transformations do not
converge, except in special cases
\cite{Sieg71,Bruno71,Ruess67,Perez-Marco03}. For applications  it
is more practical to consider the {\em truncated normal form}.

\begin{Def}[$N^\text{th}$ Order Classical Normal Form]
\label{def:tcnf} Consider a Hamilton function $H$ with an equilibrium point
$z_0\in\R^d\times\R^d$ which, for $N\in \N$, we normalise as described in
Theorem~\ref{theorem_classicalNF}.
Then we refer to $H_{\text{CNF}}^{(N)}$ in Equation~\eqref{eq:defPhiN}
\rem{
\begin{equation}
H^{(N)} = E_0 + H_2^{(2)} + H_3^{(N)} + \ldots + H_{N-1}^{(N)} +
H_N^{(N)} + \ldots,
\end{equation}
where $H_s^{(N)}, \, 3 \le s \le N$ have all been
normalised, i.e., $\opD  H_s^{(N)} =\{
H_2^{(2)},H_s^{(N)} \} =0, \,3 \le s \le N$.
%
%
Then we
refer to the Hamiltonian function
\begin{equation}
\Hcnf^{(N)} = E_0 + H_2^{(2)} + H_3^{(N)} + \ldots + H_{N-1}^{(N)}
+ H_N^{(N)}
\end{equation}
} 
as the \emph{$N^\text{th}$ order  classical normal
form (CNF)} of $H$.

\rem{
The corresponding normal form coordinates,
denoted  $z^{(N)}$, are related to the original coordinates by

\begin{equation} \label{eq:Nthordernormalformcoordinates}
z^{(N)} =\flow_{W_N}^{1} \circ \cdots \circ \flow_{W_3}^{1} (z^{(2)})\,,
\quad z^{(2)} = M z^{(1)}\,,\quad z^{(1)} = z-z_0\,.
\end{equation}
}
\end{Def}

\New{
Note that in order to compute the $N^{\text{th}}$  order normal form 
it is sufficient to  carry out the Taylor expansion of the Hamiltonian up to order $N$. The remainder term can be neglected immediately since
the procedure described in the proof of Theorem~\ref{theorem_classicalNF} shows that no terms from the remainder term will enter the $N^{\text{th}}$ order normal form.  
} 

Of course, the normal form procedure presented in this section
raises questions like ``what is the error associated with
truncating the normal form at some finite order?''  After all, one
is interested in the dynamics associated with the full, original
Hamiltonian. Another obvious question is ``what is the optimum
order at which to truncate the normal form so that errors are
minimised?'' There is no general theory that can be used to answer
such questions. They must be addressed on a problem-by-problem
basis. Fortunately, truncating the normal form does give extremely
accurate results in a number of problems
\cite{WaalkensBurbanksWiggins04,WaalkensBurbanksWigginsb04,WaalkensBurbanksWiggins05b},
and we will consider this in more detail in Section
\ref{sec:accuracy_NF}.


\subsection{Nature and Computation of the Normal Form in a Neighborhood of an
  Equilibrium Point of Saddle-Centre-$\cdots$-Centre Stability Type}
\label{sec:examp_comp_cnf}

We now describe the computation of the normal form in the
classical situation of interest to us; in the neighborhood of an
equilibrium point of saddle-centre-$\cdots$-centre stability type.
This means that the matrix associated with the linearisation of
Hamilton's equations about the equilibrium point has two real
eigenvalues, $\pm \lambda$, and $d-1$ complex conjugate pairs of
pure imaginary eigenvalues, $\pm \ui\, \omega_k$, $k=2, \ldots,d$. Moreover, we will assume that the $\omega_k$, $k=2, \ldots,d$, are nonresonant in the sense that they are linearly
independent over the integers, i.e.,  $k_2\,\omega_2 +  \ldots +
k_d\, \omega_d \ne 0$ for all $(k_2, \ldots, k_d) \in \mathbb Z^{d-1} - \{0 \}$ (note that the more stringent diophantine
condition for nonresonance (\cite{AKN88}) is not required for our
work).

But first, we locate the equilibrium point of interest, denote it by 
$z_0= (q_0, p_0)$, and translate it to the origin using the
generating function given in \eqref{gen_trans}. The Taylor series
of the corresponding Hamiltonian then has the form

\begin{equation}
H^{(1)}(z^{(1)}) = E_0 + H_2^{(1)}(z^{(1)}) + \sum_{s=3}^{\infty}
H_s^{(1)}(z^{(1)}).
\end{equation}

We next construct a linear symplectic transformation $M:\R^{2d}\mapsto \R^{2d}$ such that for
$z^{(2)}=M\, z^{(1)} $, we have
\begin{equation} \label{eq:saddlequadratic-gen}
H^{(2)}_2 (z^{(2)}) = \lambda p_1^{(2)} q_1^{(2)} + \sum_{k=2}^d
\frac{\omega_k}{2} \big( (p^{(2)}_k)^2 + q^{(2)}_k)^2 \big)\,.
\end{equation}
We note that for some purposes it is convenient to consider also a slightly modified version
of the coordinates $z^{(2)}=(q^{(2)},p^{(2)})$ which for later reference we will we denote by
$(Q^{(2)},P^{(2)})$.
The coordinates $(q^{(2)},p^{(2)})$ and $(Q^{(2)},P^{(2)})$ agree in the
centre components, i.e., $Q^{(2)}_k = q^{(2)}_k$ and $P^{(2)}_k = p^{(2)}_k$
for $k=2,\ldots,d$, but are rotated versus each other by an angle of 45$^\circ$ in the saddle
plane, i.e.,
\begin{equation} \label{eq:def_Q_P_2}
Q^{(2)}_1 = \frac{1}{\sqrt{2}}(q^{(2)}_1-p^{(2)}_1)\,,
\quad P^{(2)}_1 = \frac{1}{\sqrt{2}}(q^{(2)}_1+p^{(2)}_1)\,.
\end{equation}
Note that the tranformation from $(q^{(2)},p^{(2)})$ to $(Q^{(2)},P^{(2)})$ is symplectic.
In terms of $(Q^{(2)},P^{(2)})$ the quadratic part of the Hamiltonian assumes
the form
\begin{equation} \label{eq:H22_QP_classical}
H^{(2)}_2 (Q^{(2)},P^{(2)}) = \frac{\lambda}{2} \big( (P^{(2)}_1)^2 - Q^{(2)}_1)^2 \big) + \sum_{k=2}^d
\frac{\omega_k}{2} \big( (P^{(2)}_k)^2 + Q^{(2)}_k)^2 \big)\,.
\end{equation}
The quadratic part then consists of the sum of one inverted harmononic
oscillator (or ``parabolic barrier'') and $d-1$ harmonic oscillators.

In order to construct the  $2d\times2d$ matrix $M$ above we  label the eigenvalues
of   $J\,\Hess H(z_0)$ \New{(which is the matrix corresponding to the  linearisation of Hamilton's vector field around the equilibrium point)} in such a way that
\begin{equation}
e_1 = -e_{1+d} = \lambda\,,\qquad e_k = -e_{k+d} = \ui \omega_k
\,,\quad k=2,\dots,d\,,
\end{equation}
and then use the corresponding eigenvectors
$v_1,\dots,v_{2d}$ to form the columns of the  matrix
$M$ according to
\begin{equation}
M = \left( c_1 v_1,     c_2 \Re v_2,      \dots, c_d \Re v_{d},
           c_1 v_{1+d}, c_2 \Im v_{2},  \dots, c_d \Im v_{d} \right)\,,
\end{equation}
where $c_1,\ldots,c_d$ are scalars defined as
\begin{equation}
c_1^{-2} := \la v_1, J v_{1+d} \ra\,,\quad c_k^{-2} := \la \Re
v_k, J \Im v_{k} \ra\,,\quad k=2,\dots,d\,.
\end{equation}
The constants $c_1,\ldots,c_d$ guarantee that the matrix
$M$ will be symplectic, i.e.,
$M$ will satisfy $M^T J M =J$.  Here we have assumed  that the eigenvectors $v_1$ and
$v_{1+d}$ have been chosen in such a way that  $\la v_1, J
v_{1+d}\ra$ is positive (if $\la v_1, J v_{1+d}\ra<0$ then we
multiply $v_{1+d}$ by -1). It is not difficult to see that
$c_k^{-2}$, $k=2,\dots,d$, are automatically positive if the
frequencies $\omega_k$ are positive \footnote{ In fact, if one of
the $d_k^{-2}$ is negative then this means that the corresponding
frequency is negative; this is a case which we have excluded,
although it can be dealt with in a way that is similar to the
procedure described in this paper. }. Using the fact that $\la
v_n,J v_k \ra=0$ for $n$ and $k$ from the distinct sets
$\{1,1+d\}$, $\{2,2+d\}$, \dots, $\{d,2d\}$, it is easily verified
that the matrix $M$ satisfies $M^T \,J \, M =J$.

\subsubsection{Solution of the homological equation}
\label{sec:solvhomequ}

Given a Hamiltonian function whose quadratic part is of the form
\eqref{eq:saddlequadratic-gen}, the solution of  the homological
equation derived in Lemma \ref{lem:hom_eq}  for any order $n\ge 3$
is extremely simple and transparent if we first perform the
following symplectic complex linear change of coordinates
$z^{(n)}=(q^{(n)},p^{(n)})\mapsto(x,\xi)$ which has the components
$x_1=q^{(n)}_1$, $\xi_1=p^{(n)}_1$ and

\begin{equation}\label{eq:comlex-rot_classical}
x_k   :=\frac{1}{\sqrt{2}}(q^{(n)}_k-\ui p^{(n)}_k) \,\, ,\quad
\xi_k :=\frac{1}{\sqrt{2}}(p^{(n)}_k- \ui q^{(n)}_k)  \,\, ,\qquad k=2,\ldots ,d\,\,.
\end{equation}
Here, and for the rest of this section, we omit the superscript $(n)$ for $x$ and $\xi$ for
the sake of a simpler and less cumbersome notation.

In terms of the phase space coordinates $(x,\xi)$, the linear
map $\opD$ takes the form

\begin{equation} \label{eq:def_D_coord}
\opD =  \lambda (\xi_1\pa_{\xi_1}-x_1\pa_{x_1})  +
\sum_{k=2}^{d}\ui \omega_k(\xi_k\pa_{\xi_k}-x_k\pa_{x_k})\,.
\end{equation}
The form of \eqref{eq:def_D_coord} is significant for
two reasons. One is that when the monomials of order $n$ defined in
\eqref{eq:classicalmonomials} are expressed in terms of the
coordinates $(x,\xi)$ they form a basis for $\ccW^n$. We have
\begin{equation}
\ccW^n = \text{span }  \bigg\{ x^{\alpha}\xi^{\beta}:=\prod_{k=1}^d x_k^{\alpha_k} \xi_k^{\beta_k}
\,:\,  \quad \abs{\alpha}+\abs{\beta}:=
\sum_{k=1}^d \alpha_k + \beta_k =n\bigg\}\,.
\end{equation}
 Secondly, in
this basis the linear map \eqref{eq:def_D_coord} is diagonal. In fact,
using  \eqref{eq:def_D_coord}, we see that
the image under $\opD $ of a monomial $x^\alpha \xi^\beta \in
\ccW^n$ is
\begin{equation} \label{eq:Dmonomial}
\opD \, \prod_{k=1}^d
x_k^{\alpha_k} \xi_k^{\beta_k} = \bigg(\lambda ( \beta_1 -
\alpha_1) + \sum_{k=2}^{d}\ui \omega_k(\beta_k-\alpha_k)\bigg)\,
\prod_{k=1}^d x_k^{\alpha_k} \xi_k^{\beta_k}\,.
\end{equation}
These monomials thus are eigenvectors of  \eqref{eq:def_D_coord}.

Since the map $\opD$ can be diagonalised it follows in a trivial way that
 $\ccW^n$ can be represented as the direct sum of the kernel of $\opD$ acting on $\ccW^n$,
 $\text{Ker } \opD  \big|_{\ccW^n}$, and the image of $\opD$ acting on
 $\ccW^n$, $\text{Im } \opD  \big|_{\ccW^n}$, i.e.,

\begin{equation}
\ccW^n =\text{Ker }  \opD \big|_{\ccW^n}
\oplus \text{Im }
\opD \big|_{\ccW^n}.
\label{ds_decomp}
\end{equation}
Now we can express $H_n^{(n-1)}$ as
\begin{equation}
H_n^{(n-1)} = H_{n; \text{Ker}}^{(n-1)} + H_{n; \text{Im}}^{(n-1)},
\label{order_decomp}
\end{equation}
where $H_{n; \text{Ker}}^{(n-1)}\in \text{Ker }  \opD
\big|_{\ccW^n}$ and  $H_{n; \text{Im}}^{(n-1)}\in \text{Im } \opD
\big|_{\ccW^n}$. We can then choose $W_n$ such that
\begin{equation}
 \opD W_n = H_{n; \text{Im}}^{(n-1)}\,\,,
\label{trans_sol}
 \end{equation}
and therefore by \eqref{eq:homological_classical}
\begin{equation}
 H_n^{(n)} =H_{n; \text{Ker}}^{(n-1)}.
\label{n_order_nform}
 \end{equation}
The choice of $W_n$ is not unique since one can always add terms from the
kernel of $\opD\big|_{\ccW^n}$. However, we will require $W_n \in \text{Im }
\opD  \big|_{\ccW^n}$, i.e., we will invert  $\opD$ on its
image $\text{Im } \opD  \big|_{\ccW^n}$, which renders the choice of $W_n$ unique.

Using our assumption  that the frequencies $\omega_2,\dots,\omega_d$ are nonresonant, i.e.,
linearly independent over $\Z$, we see from \eqref{eq:Dmonomial}
that a monomial $x^\alpha \xi^\beta$ is mapped to zero if and only
if $\alpha_k=\beta_k$ for all $k=1,\dots,d$.
In particular $ \text{Ker }
\opD  \big|_{\ccW^s} = \{0\} $ if $s$  is odd. This implies that coordinate transformations can
be constructed such that all odd order terms are eliminated. Moreover, for $s$ even, the terms that {\em cannot} be
eliminated are those which are sums of monomials for which $x_k$ and $\xi_k$
have equal integer exponents for all $k=1, \ldots, d$.

Concretely, we can compute  $W_{n}$ according to \eqref{trans_sol} as follows. We
assume that $H^{(n-1)}_{n;\text{ Im}}$ is the linear combination of
$L$ monomials of order $n$,

\begin{equation}
H^{(n-1)}_{n;\text{ Im}} = \sum_{l=1}^L h_l \prod_{k=1}^d
x_{k}^{\alpha_{k;l}} \xi_{k}^{\beta_{k;l}},
\label{sc_n_order_image}
\end{equation}

\noindent with $\sum_{k=1}^d \alpha_{k;l} + \beta_{k;l} = n$ for
all $l=1,\dots,L$, and for all $l=1,\dots,L$, there is at least
one $k=1,\dots,d$ for which  $\alpha_{k;l} \ne \beta_{k;l}$ (i.e.,
the vectors $(\alpha_{1;l},\ldots,\alpha_{d;l})$ and
$(\beta_{1;l},\ldots,\beta_{d;l})$ are different for all
$l=1,\dots,L$).  Upon inspecting  \eqref{eq:Dmonomial},  and using
\eqref{trans_sol}, we see that a generating function $W_n$ that
solves the homological equation is given by

\begin{equation}
W_{n} = \sum_{l=1}^L \frac{h_l}{\lambda(\beta_{1;l}-\alpha_{1;l}) +
\sum_{k=2}^{d} \ui \omega_k(\beta_{k;l}-\alpha_{k;l})}
\prod_{k=1}^d x_{k;l}^{\alpha_{k;l}} \xi_{k;l}^{\beta_{k;l}}\,.
\label{sc_n_order_gen_func}
\end{equation}
As mentioned above this solution of the homological equation is unique if we
require $W_n$ to be in $\text{Im } \opD  \big|_{\ccW^n}$.

\subsubsection{Integrals of the classical motion from the $N^\text{th}$ order  classical normal
form}
\label{sec:classicalintegrals}

The $N^\text{th}$ order  classical normal form $\Hcnf^{(N)}$ is a polynomial
in the  $N^\text{th}$ order phase space coordinates \eqref{eq:Nthordernormalformcoordinates} which, in order to 
keep
the notation in this section simple,
we will denote by $(q,p)$, i.e., we will omit superscripts $(N)$ on the phase
space coordinates.
As discussed in  the previous section, it follows that if we perform the
symplectic complex linear change of coordinates $x_1=q_1$, $\xi_1=p_1$ and
\begin{equation}
x_k   :=\frac{1}{\sqrt{2}}(q_k-\ui p_k) \,\, ,\quad
\xi_k :=\frac{1}{\sqrt{2}}(p_k- \ui q_k)  \,\, ,\qquad k=2,\ldots ,d\,,
\end{equation}
then the coordinate pairs $x_k$ and $\xi_k$ will have equal integer exponents
for all $k=1,\ldots,d$ in each monomial of $\Hcnf^{(N)}$.
As a consequence the functions

\begin{equation} \label{eq:integrals_classical}
I =p_1 q_1 = \xi_1 x_1 \, , \quad
J_k=\frac{1}{2} \big( p_k^2+ q_k^2 \big) = \ui \xi_k x_k \,,\quad k=2,\dots,d \,,
\end{equation}

\noindent
are integrals of the motion generated by $\Hcnf^{(N)}$.
This assertion is simple to verify with the following computations:

\begin{equation}
\frac{\ud}{\ud t} I  =\{I,\Hcnf^{(N)}\}=0\,,\quad
\frac{\ud}{\ud t}J_k=\{J_k,\Hcnf^{(N)}\}=0\,,\qquad k=2,\dots,d\,.
\end{equation}

The integrals of the motion $I$ and $J_k$ can be used to define
action angle variables. We therefore define the conjugate angles

\begin{equation}
\begin{split}
\varphi_1 &=
\left\{
\begin{array}{ccc}
\tanh^{-1}\big( \frac{q_1+p_1}{q_1-p_1} \big) &,& p_1
q_1 <0\\
\tanh^{-1}\big( \frac{q_1-p_1}{q_1+p_1} \big) &,& p_1
q_1 >0
\end{array}
\right.
\,,\\
\varphi_k &= \text{arg}( p_k+\ui \, q_k )\,,\qquad k=2,\dots,d\, .
\end{split}
\end{equation}
It is not difficult to see that the map $(q,p) \mapsto
(\varphi_1,\dots,\varphi_d,I,J_2,\dots,J_d)$ is symplectic.

For $k=2,\dots,d$, the ranges of the  $\varphi_k$ are $[0,2\pi)$ and the
ranges of the $J_k$ are $[0,\infty)$.
The maps $(q_k,p_k)\mapsto (\varphi_k,J_k)$ are singular at $q_k=p_k=0$
where the angles $\varphi_k$ are not defined.
Away from the singularities the maps are one to one.
In contrast, the range of  both $\varphi_1$ and $I$ is
$\R$ ($\varphi_1$ thus is not an angle in the usual sense). The map
$(q_1,p_1)\mapsto (\varphi_1,I_1)$ is singular on the lines $p_1=0$ and
$q_1=0$ which map to $I=0$ with $\varphi_1=\infty$ and
$\varphi_1=-\infty$, respectively. Even away from the singularities each $(\varphi_1,I)$ has
still two preimages $(q_1,p_1)$ which correspond to the two branches of the
hyperbola $I=p_1q_1$. The coordinate lines of the action angle variables are
shown in Fig.~\ref{fig:actionangle}.

We note that in terms of the coordinates $(Q,P)$ with $(Q_k,P_k)=(q_k,p_k)$,
$k=2,\ldots,d$, and
\begin{equation} \label{eq:def_Q_P}
Q_1 = \frac{1}{\sqrt{2}}(q_1-p_1)\,,
\quad P_1 = \frac{1}{\sqrt{2}}(q_1+p_1)\,,
\end{equation}
the integrals $J_k$, $k=2,\ldots,d$, are of the same form while $I$ changes to
\begin{equation}
  I = \frac12 \big(P_1^2 - Q_1^2 \big)\,.
\end{equation}

The angles $\varphi_k$, $k=1,\ldots,d$, are cyclic, i.e., the Hamilton
function $\Hcnf^{(N)}$ effectively depends only on the integrals $I$ and
$J_k$, $k=2,\ldots,d$.
To indicate this and for later reference we introduce the function
$K_{\text{CNF}}^{(N)}$
defined via
\begin{equation}\label{eq:def_K}
\begin{split}
\Hcnf^{(N)} &= K_{\text{CNF}}^{(N)}(I,J_2,\ldots,J_d)\\
&= E_0 + \lambda I + \omega_2
J_2 + \ldots + \omega_d J_d + \text{ higher order terms }\,.
\end{split}
\end{equation}
Here the higher order terms are of order greater than 1
and less than or equal to $[N/2]$ in the integrals, where $[N/2]$
denotes the integer part of $N/2$. Note
that since the Hamilonian in normal form does not have any odd order terms,
only the case of even $N$ is of interest.

\def\figactionangle{%
The left figure shows
contourlines of the action angle variables $I$ and $\varphi_1$  (hyperbolae
and straight lines, respectively) in the saddle plane with coordinates
$(q_1,p_1)$ and $(Q_1,P_1)$ which are rotated versus each other by 45$^\circ$ degrees.
The right figure shows contourlines of the action angle variables
$J_k$ and  $\varphi_k$, $k=2,\dots,d$, (circles and straight lines,
respectively)
in the centre planes with coordinates $(q_k,p_k)=(Q_k,P_k)$.
}
\def\FIGactionangle{
\centerline{
\includegraphics[angle=0,width=12cm]{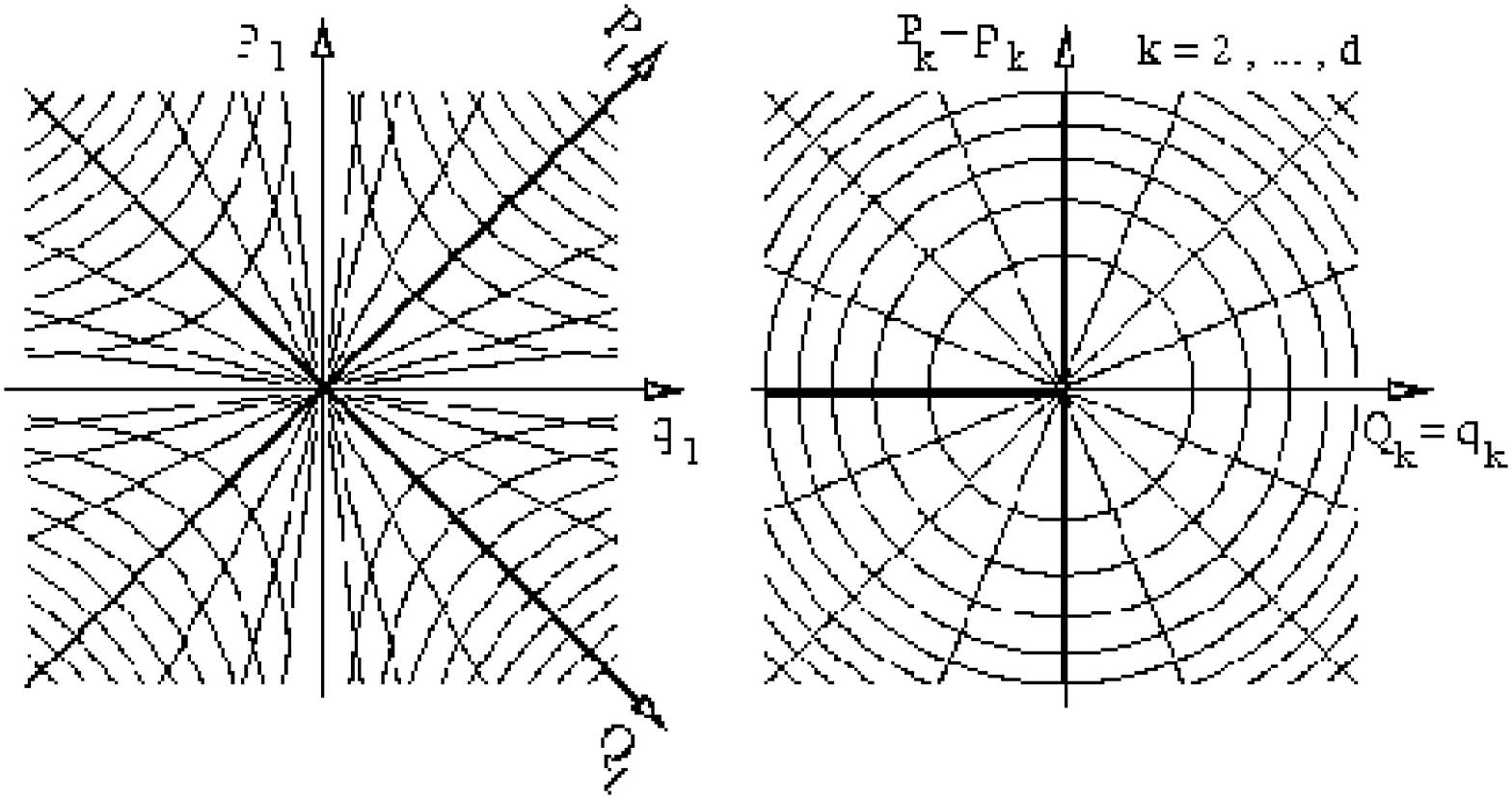}
}
}
\FIGo{fig:actionangle}{\figactionangle}{\FIGactionangle}

As we will see, the classical integrals of motion are extremely
useful for characterising, and realising, classical phase space
structures. However,  the obvious question arises, and must be
answered. These are constants of the motion  for the $N^\text{th}$
order  classical normal form $\Hcnf^{(N)}$.  How close to being
constant are they on trajectories of the full Hamiltonian? Also,
we will use them to construct certain invariant manifolds for the
$N^\text{th}$ order   classical normal form $\Hcnf^{(N)}$. How
close to being invariant will these manifolds be for the full
Hamiltonian? These questions must be asked, and answered, on a
problem-by-problem basis. A number of studies have recently shown
that for moderate $N$ (e.g. 10-14), these integrals are ``very
close'' to constant for the full Hamiltonian dynamics for most
practical purposes and that the invariant manifolds constructed
from them are ``almost invariant'' for the full Hamiltonian
dynamics.

We emphasise again that in this section we omitted superscripts $(N)$ on the
coordinates in order to keep the notation simple and that the integrals of the
motion of the $N^{\text{th}}$ order normal form  only
assume the simple form in \eqref{eq:integrals_classical} if they are expressed  in terms of the
$N^{\text{th}}$ order normal form coordinates \eqref{eq:Nthordernormalformcoordinates}.

%% file: semiclassical_quantum_NF.tex
\section{Quantum Normal Form Theory}
\label{sec:SNF}

In this section we develop a normal form theory for quantum
mechanics that is algorithmically the same as  the one presented
for classical mechanics in the previous section,
Sect.~\ref{sec:CNF}. However, the objects manipulated by the
algorithm in the quantum mechanical case are different, and this
is what we now describe.

In quantum mechanics the role of a Hamilton function in classical
mechanics is played by a self-adjoint operator, the Hamilton
operator. \New{While the Hamilton function in classical mechanics acts
on a phase space, which was $\R^{2d}$ in
Sect.~\ref{sec:CNF}, a Hamilton operator acts on a Hilbert space,
which will be $L^2(\R^d)$ in our case.}

The quantum mechanical analogue of a symplectic transformation in classical mechanics is
a unitary transformation.
The conjugation of a Hamilton operator $\widehat{H}$ by a unitary operator $\widehat{U}$
gives the new operator
\begin{equation}
\widehat{H}'= \widehat{U}^* \widehat{H} \widehat{U}\,,
\end{equation}
where $\widehat{U}^*$ denotes the adjoint of
$\widehat{U}$. The operator $\widehat{H}'$ is again self-adjoint
and has the same spectral properties as the original Hamilton
operator $\widehat{H}$. We will use unitary transformations to
simplify the Hamilton operator in the same way that we used
symplectic transformations to simplify the classical Hamilton
function. In the classical setting the symplectic transformations
were obtained as the time-one maps of a Hamiltonian flow, where
the Hamiltonian, $W$, was referred to as the generating function.
In the quantum mechanical setting we will analogously consider a
self-adjoint operator $\widehat{W}$ which gives the unitary
operator 
\begin{equation}
\widehat{U}=\ue^{-\frac{\ui}{\hbar}\widehat{W}}\,.
\end{equation}
The operator $\widehat{W}$ is called the generator of
$\widehat{U}$. Analogous to the development of
\eqref{eq:dAdepsilon_classical1} and the results that follow, we
now consider the one parameter family of self-adjoint operators
defined by 
\begin{equation}\label{eq:H-family}
\widehat{H}(\epsilon):=\ue^{\frac{\ui}{\hbar}\epsilon\widehat{W}}\widehat{H}\ue^{-\frac{\ui}{\hbar}\epsilon\widehat{W}}\,,
\end{equation}
where the parameter $\epsilon$ is real. Note that
$\widehat{H}'=\widehat{U}^*\widehat{H} \widehat{U}=\widehat{H}(\epsilon=1)$,
  and
$\widehat{H}=\widehat{H}(\epsilon =0)$. If we differentiate
\eqref{eq:H-family} with respect to $\epsilon$ we obtain the
Heisenberg equation 
\begin{equation}\label{eq:Heisenberg}
\frac{\ud}{\ud \epsilon}\widehat{H}(\epsilon)=\frac{\ui}{\hbar}[\widehat{W},\widehat{H}(\epsilon)]\,,
\end{equation}
where $[\cdot,\cdot]$ denotes the commutator which, for two operators $\widehat{A},\widehat{B}$, is defined as
$[\widehat{A},\widehat{B}]=\widehat{A}\widehat{B}-\widehat{B}\widehat{A}$.
Therefore $\widehat{H}'$ can be obtained from  the solution of \eqref{eq:Heisenberg}
with initial condition $\widehat{H}(\epsilon=0)=\widehat{H}$. Equation \eqref{eq:Heisenberg}
will play the same role for the development of the quantum normal form  as
Equation~\eqref{eq:dAdepsilon_classical} played for the classical normal form. This is consistent with the usual
quantum-classical correspondence where
the commutator $\frac{\ui}{\hbar}[\cdot,\cdot]$ is related to the Poisson bracket $\{\cdot,\cdot\}$.
In the next section we will make this correspondence  more precise.

One of the key properties of the classical normal form in the
neighbourhood of a nonresonant saddle-centre-...-centre
equilibrium point is that the Hamilton function in normal form is
a function of the classical integrals, see \eqref{eq:def_K}. We
will see in Sec.~\ref{sec:classical}  that this feature will help
us to understand the local classical dynamics and identify the phase space
structures that control the dynamics near a non-resonant
saddle-centre-...-centre equilibrium point. In the quantum
mechanical case the classical integrals will correspond to
``elementary'' operators with well known spectral properties.
Analogously to symplectic transformations in the classical case,
we will use unitary transformations in the quantum mechanical case
to bring the Hamilton operator into a simpler form in which it
will be a function of these elementary operators only. In the same
manner as in the classical case, this simplification will be
obtained ``order by order''. To give notions like `order'  and
`equilibrium point' a meaning  for quantum operators and also to
derive an explicit algorithm to achieve the desired simplification
we will have to relate quantum operators to classical phase space
functions and vice versa. This is the subject of the following
section, Sec.~\ref{sec:correspondence}.
The formalism developed in Sec.~\ref{sec:correspondence} is then used in
Sec.~\ref{sec:conjunitary} to transform Hamilton operators through
conjugation  by unitary operators.
In Sec.~\ref{sec:compscnf} we will define when a
Hamilton operator is in quantum normal form,
and show how a given  Hamilton operator can be
transformed to quantum normal form to any desired order.
In Sec.~\ref{sec:examp_comp_scnf}  we study the nature of the
quantum normal form for our case of interest, which is in a
neighborhood of a non-resonant saddle-centre-$\cdots$-centre
equilibrium point of a corresponding classical Hamiltonian system.
As a first explicit example,  we show how the quantum
normal form can be computed for one-dimensional potential barriers in  Sec.~\ref{sec:1Dpotentialbarriers}.


\subsection{The Classical-Quantum Correspondence}
\label{sec:correspondence}

The basis for our quantisation of  the classical normal form
described in Section~\ref{sec:defcompcnf} is the Weyl quantisation
and the associated Weyl calculus. Before we use the Weyl calculus
to define the quantum normal form in Sec.~\ref{sec:compscnf} we
want to  give some background on the general theory, which
provides a systematic way of formulating the quantum-classical
correspondence. General references for the material in this
section that contain much more detail and background are
\cite{Fol89,DimSjo99,Mar02}.


\subsubsection{Weyl quantisation}

A quantisation is a rule which associates operators on a Hilbert space
 with functions on a phase space. We will use here the Weyl quantisation, which is the one
most commonly used.
Let $q_k$ and $p_k$, $k=1, \ldots , d$, be the components of the position and
momentum vectors $q$ and $p$, respectively.  These are quantised in such a way that
they act on a wavefunction $\psi(q)$ according to
\begin{equation}
\widehat{q}_k\psi(q)=q_k\psi(q)\,\, ,\quad \widehat{p}_k\psi(q)=\frac{\hbar}{\ui}\frac{\pa\psi(q)}{\pa q_k}\,\, .
\end{equation}
The Weyl quantisation extends these prescriptions to general functions of
$(q,p)$ by requiring that, for  $\xi_q,\xi_p\in\R^d$, the quantisation of the exponential function
\begin{equation}
\ue^{\frac{\ui}{\hbar}(\la\xi_p,q\ra+\la \xi_q,p\ra)}
\end{equation}
is the phase space translation operator
\begin{equation}
\widehat{T}_{\xi_q,\xi_p}=\ue^{\frac{\ui}{\hbar}(\la\xi_p
,\hat{q}\ra+\la\xi_q ,\hat{p}\ra)}\,\, .
\end{equation}
 Using Fourier inversion we can represent
a function on phase space as
\begin{equation}\label{eq:Afourier}
A(q,p)=\frac{1}{(2\pi\hbar)^{2d}}\int_{\R^d} \int_{\R^d} \overline{A}(\xi_q ,\xi_p)
\ue^{\frac{\ui}{\hbar}(\la \xi_p, q\ra+\la\xi_q, p\ra)} \,\, \ud \xi_q\ud \xi_p\,\, ,
\end{equation}
where
\begin{equation}
\overline{A}(\xi_q,\xi_p)=\int_{\R^d} \int_{\R^d} A(q,p) \ue^{-\frac{\ui}{\hbar}(\la \xi_p, q\ra+\la\xi_q, p\ra)} \,\, \ud q\ud p\,\,
\end{equation}
is the Fourier transform of $A$. The \emph{Weyl quantisation}
$\Op[A]$ of $A$ is then defined by replacing the factor
$\ue^{\frac{\ui}{\hbar}(\la \xi_p, q\ra+\la\xi_q, p\ra)}$ in the
integral \eqref{eq:Afourier} by the operator
$\widehat{T}_{\xi_q,\xi_p}$, i.e. 
\begin{equation}\label{def:Weyl-quant}
\Op[A]=\frac{1}{(2\pi\hbar)^{2d}} \int_{\R^d}\int_{\R^d}  \overline{A}(\xi_q
,\xi_p)\widehat{T}_{\xi_q,\xi_p}\,\, \ud \xi_q\ud \xi_p\,\, .
\end{equation}
In order to manipulate these operators and understand their
mathematical properties we will need the appropriate definitions
and notation. We will say that $A\in \cS_{\hbar}(\R^d\times \R^d)$
if $A$ depends smoothly on $(\hbar,q,p)$ and  if for all
$\alpha,\beta\in \N^d$ and $k\in \N$ there exists a constant
$C_{\alpha,\beta,k}$ such that 
\begin{equation}\label{eq:def-S-hbar}
(1+\abs{q}+\abs{p})^k \abs{\pa_q^{\alpha}\pa_p^{\beta}A(\hbar,q,p)}\leq C_{\alpha,\beta,k}\,\, .
\end{equation}
The space $ \cS_{\hbar}(\R^d\times \R^d)$ is similar to the usual Schwartz
space. The only difference is that we allow the functions to depend additionally on the parameter
$\hbar$ in a smooth way. For $A\in \cS_{\hbar}(\R^d\times \R^d)$ the Fourier transform is again a Schwartz function
and so the Weyl quantisation \eqref{def:Weyl-quant} gives a well defined  bounded operator.
But the quantisation can be extended to larger
classes of functions. One such larger standard
class of functions for which the Weyl quantisation is well behaved
is $S^m(\R^d\times\R^d)$ for some $m\in \R$. Here $A\in
S^m(\R^d\times\R^d)$ if $A$ satisfies the estimates
\begin{equation}\label{eq:symbol-est}
\abs{\pa_{q}^{\alpha}\pa_p^{\beta}A(\hbar, q,p)}\leq C_{\alpha,\beta}
(1+\abs{q}+\abs{p})^m\,\, \quad
 \text{for all} \,\,\, \alpha,\beta\in\N^d\,\, .
\end{equation}
If $A\in S^m(\R^d\times\R^d)$ then $\Op[A]:\cS_{\hbar}(\R^d)\to\cS_{\hbar}(\R^d)$ (see, e.g., \cite{DimSjo99}). Here 
$\cS_{\hbar}(\R^d)$ is defined analogously to $\cS_{\hbar}(\R^d\times \R^d)$ in \eqref{eq:def-S-hbar}. 
The function $A$ is called the (Weyl) {\em symbol} of the operator
$\Op[A]$. If the symbol $A$  also depends on the parameter $\hbar$
we will usually  assume that, for small $\hbar$,  $A$ has an
asymptotic expansion in integer powers of $\hbar$,
\begin{equation}
A(\hbar,q,p)\sim A_0(q,p)+\hbar A_1(q,p)+\hbar^2 A_2(q,p)+\ldots \,\,.
\end{equation}
Here the leading order term $A_0(q,p)$ is then called the {\em principal symbol} and
it is interpreted as the classical phase space function corresponding to $\Op[A]$.

The quantisation \eqref{def:Weyl-quant} can also be inverted. Let $\widehat{A}$ be an operator, then
\begin{equation}
A(\hbar,q,p):=\Tr \big(\widehat{T}^*(q,p) \widehat{A}\big)\,,
\end{equation}
is the Weyl symbol of $\widehat{A}$, i.e. we have
$\Op[A]=\widehat{A}$, with $\Tr$ denoting the trace, and $\widehat{T}^*$ denoting the adjoint
 of $ \widehat{T}$.

The advantage of this representation of  operators is
that many properties of the operators are nicely reflected in their
symbols. For later reference we collect two such relations:
\begin{enumerate}
\item For the adjoint operator one has $\Op[A]^{*}=\Op[A^*]$, \New{where $A^*$ denotes the 
complex conjugate symbol of $A$}. Hence, a real valued symbol gives a symmetric operator.
\item If $ A\in S^0(\R^d\times\R^d)$, i.e., the symbol and all its derivatives
  are bounded, then the corresponding operator is bounded  as an operator
on $L^2(\R^d)$. This is known as the Calderon-Vaillancourt
  theorem \cite{DimSjo99}. This implies in particular that
a real valued symbol $A\in S^0(\R^d\times\R^d)$ gives a self-adjoint operator $\Op[A]$.
\end{enumerate}

For example, the symbol $J=(p^2+q^2)/2$ on $\R\times\R$ is in
$S^2(\R\times\R)$. Its principal symbol is
$(p^2+q^2)/2$ and the Weyl quantisation gives
\begin{equation} \label{eq:opJ}
\Op[J]=-\frac{\hbar^2}{2}\frac{\ud^2}{\ud q^2}+\frac{1}{2}q^2\,.
\end{equation}
Similarly, the symbol $I=p\, q$ is in $S^2(\R\times\R)$ with principal symbol $p\, q$ and is quantised as
\begin{equation} \label{eq:opI}
\Op[I]=\frac{\hbar}{\ui}\bigg(q\frac{\ud}{\ud q}
+\frac{1}{2}\bigg)\,\, .
\end{equation}

\noindent These are  the quantisations of the classical integrals
obtained in Section \ref{sec:classicalintegrals}, and they will form the
building blocks of the quantum normal form associated with a
saddle-centre-$\cdots$-centre equilibrium point
in Section~\ref{sec:examp_comp_scnf}.


\subsubsection{The Moyal bracket}

The main idea behind the introduction of symbols of operators is that one can use 
the symbols to study properties of the operators, as we already indicated in 
the last subsection. Since the symbols are functions they are in general much easier to study 
than operators. One can probably say that the single most useful fact about 
pseudodifferential operators, i.e., operators whose symbols satisfy estimates like 
\eqref{eq:symbol-est}, is that they form an algebra, i.e., the product of two such 
operators is again of this type, and that one can compute the symbol of a product from
 the symbols of the operators which are multiplied. 

The quantum normal form algorithm we will develop will rely essentially on this product 
formula for symbols. Given two    
functions $A,B$, one can find a function $A*B$ such
that $\Op[A]\Op[B]=\Op[A*B]$, see \cite{DimSjo99}. This so called {\em star  product} of
$A$ and $B$ is
given by
\begin{equation}\label{eq:star-product}
A*B(q,p)= A(q,p) \exp\bigg(\ui\frac{\hbar}{2}[ \la
\stackrel{\leftharpoonup}{\pa}_{q},\stackrel{\rightharpoonup}{\pa}_{p}\ra-
\la
\stackrel{\rightharpoonup}{\pa}_{q},\stackrel{\leftharpoonup}{\pa}_{p}\ra
]\bigg) B(q,p) \,,
\end{equation}
where the arrows indicate whether the partial differentiation acts
to the left (on $A$) or to the right (on $B$). For the precise
meaning of the expression on the right hand side of this equation
we refer the reader to \cite{Fol89,DimSjo99,Mar02}. However, by
expanding the exponential we obtain the more explicit asymptotic
expansion in powers of $\hbar$ that will suffice for our purposes
\begin{equation}
\begin{split}
A*B(q,p)&\sim \sum_{k=0}^{\infty}\frac{1}{k!}\bigg(\frac{\ui\hbar}{2}\bigg)^k A(q,p)[ \la
\stackrel{\leftharpoonup}{\pa}_{q},\stackrel{\rightharpoonup}{\pa}_{p}\ra-
\la
\stackrel{\rightharpoonup}{\pa}_{q},\stackrel{\leftharpoonup}{\pa}_{p}\ra
]^kB(q,p)\\
&=A(q,p)B(q,p)+\frac{\ui\hbar}{2}\{A,B\}(q,p)+\cdots \,,
\end{split}
\end{equation}
where $\{\cdot,\cdot \}$ again denotes the Poisson bracket defined
in \eqref{pb_def}. In particular, if $A\in S^m(\R^d\times\R^d)$
and $B\in S^{m'}(\R^d\times\R^d)$ then $A*B\in
S^{m+m'}(\R^d\times\R^d)$ (\cite{Fol89,DimSjo99,Mar02}). It is
worth mentioning that even if $A$ and $B$ are independent of
$\hbar$, the product $A*B$ will in general depend on $\hbar$ with
the principal symbol being given by $A\, B$, i.e., the usual
product of the functions $A$ and $B$.

From the Heisenberg equation~\eqref{eq:Heisenberg} we see that the commutator plays an important role when one
wants to conjugate an operator with a one-parameter family of unitary operators.
Applying the product formula \eqref{eq:star-product} to the expression for the
commutator of $\Op[A]$ and $\Op[B]$,
\begin{equation}
\Op[A]\Op[B]-\Op[B]\Op[A]=\Op[A*B]-\Op[B*A]=\Op[A*B-B*A]\,,
\end{equation}
we obtain the formula for the symbol of a commutator
\begin{equation}
\label{eq:commut_moyal} \left( A*B-B*A \right) (q,p) =
\frac{\hbar}{\ui} \left\{ A, B \right\}_{M} (q,p),
\end{equation}

\noindent where $\{\cdot ,\cdot \}_M$ is the \emph{Moyal bracket} which
is defined as
\begin{equation}
\{A,B\}_M(q,p)=\frac{2}{\hbar} A(q,p) \sin
\bigg(\frac{\hbar}{2}[\la\stackrel{\leftharpoonup}{\pa}_{p},\stackrel{\rightharpoonup}{\pa}_{q}\ra
-\la\stackrel{\rightharpoonup}{\pa}_{p},\stackrel{\leftharpoonup}{\pa}_{q}\ra]\bigg)
B(q,p)\,\, .
\end{equation}
For the precise interpretation of the right hand side of this
equation  we  again refer the reader to
\cite{Fol89,DimSjo99,Mar02}. However, as above, by  expanding the
sine we can obtain an explicit asymptotic expansion for small
$\hbar$ that will suffice for our purposes, 
\begin{equation}\label{eq:Moyal}
\{A,B\}_M(q,p)\sim\sum_{k=0}^{\infty}
\bigg(\frac{\hbar}{2}\bigg)^{2k}\frac{(-1)^k}{(2k+1)!} A(q,p) [\la
\stackrel{\leftharpoonup}{\pa}_{p},\stackrel{\rightharpoonup}{\pa}_{q}
\ra - \la \stackrel{\rightharpoonup}{\pa}_{p},
\stackrel{\leftharpoonup}{\pa}_{q} \ra]^{(2k+1)} B(q,p)\,\, .
\end{equation}
Note that in the case where one of the functions $A,B$ is a polynomial the
sum terminates at some  finite $k$ and gives the exact expression
for the Moyal product. In what follows, all our explicit calculations will use from the
Weyl quantisation only the asymptotic formula \eqref{eq:Moyal} for the Moyal
product.
Since we will only work  with finite Taylor series,
the asymptotic expansion will always terminate and give the exact result.

From \eqref{eq:Moyal} we see that
\begin{equation}\label{eq:moyal-to-poisson}
\{A,B\}_M(q,p)=\{A,B\}(q,p)+O(\hbar^2)\,\, ,
\end{equation}
i.e., in leading order the Moyal bracket is equal to the Poisson
bracket, and moreover, if at least one of the functions $A,B$ is a second
order polynomial then 
\begin{equation}\label{eq:moyal=poisson}
\{A,B\}_M(q,p)=\{A,B\}(q,p)\,\, .
\end{equation}


\subsubsection{Localising in phase space} \label{sec:localising}

One important application of the product formula \eqref{eq:star-product} is that it allows to localise operators 
in phase space, a technique often called {\em micro-localisation}, which in fact gave the whole 
field of microlocal analysis its name. 
Let $\rho\in  S^0(\R^d\times\R^d)$ be a cutoff function, i.e., there is a set 
$U\subset \R^d\times\R^d$ such that 
\begin{equation}
\rho|_{U}=1
\end{equation}
 and 
$\rho$ has support in a small neighbourhood of $U$. 
Then we will call $\Op[\rho]$ a cutoff operator (associated with $U$), and  we can use it to 
split any operator 
$\Op[H]$ into two parts 
\begin{equation}
\Op[H]=\Op[\rho]\Op[H]+(1-\Op[\rho])\Op[H]=\Op[\Hloc]+\Op[H_{\rm rem}]
\end{equation}
where $\Hloc=\rho*H$ and $H_{\rm rem}=H-\rho*H$. By the product formula \eqref{eq:star-product} the symbol 
$\Hloc$ is concentrated near the support of $\rho$ and $H_{\rm rem}$ is concentrated 
on the complement of the support of $\rho$. In this sense the usual procedure to 
localise the study of functions and dynamical systems by multiplication with 
cutoff functions can be quantised. In particular we have 
$\Hloc=\rho H+O(\hbar)$, so the leading order is actually the classical localisation. 
If $\Op[\rho]$ is a cutoff operator associated with some phase space region $U$ we will call 
$\Op[\Hloc]=\Op[\rho]\Op[H]$ the localisation of $H$ to $U$. 
 
The localisation appears to be a very natural object to consider with regard to 
the application we are interested in, namely   
the study the dynamics of a chemical reaction which is described by  a
Hamilton operator $\Op[H]$ whose principal symbol has a 
saddle-centre-$\cdots$-centre equilibrium point. The neighbourhood of the equilibrium point is 
the most important region for the chemical reactions. This is 
where the reactants combine to form the activated complex 
which then decays into the products. So it is natural to use the above 
procedure to localise the Hamiltonian to a neighbourhood of  the equilibrium point in phase space. In fact,  we will derive the quantisation of the classical normal form 
procedure for a Hamiltonian which is localised. 

The localisation has another advantage which is of a more technical nature. 
The Hamilton operators we 
will encounter have symbols with polynomial growth  in $p$ and $q$ 
for large $p$ and $q$, and this leads to some technical complications 
concerning questions like self-adjointness and unitarity. If we localise 
our Hamiltonians by multiplication with a cut-off operator we end up working with 
operators with bounded symbols only, for which self-adjointness 
is easy to show. This will make many proofs technically much easier.




\subsection{Transformation of  Operators through Conjugation with
  Unitary Operators Using the Weyl Calculus}
\label{sec:conjunitary}

We will now apply the Weyl calculus to the problem outlined in the
beginning of this section. For an operator $\widehat{A}=\Op[A]$
with symbol $A$ we consider its conjugation by  a unitary operator
$\widehat{U}=\ue^{\frac{\ui}{\hbar}\widehat{W}}$, where
$\widehat{W}=\Op[W]$ has symbol $W$. Our aim is to find the symbol
$A'$ such that 
\begin{equation}\label{eq:Aprime_as_conjugation_of_A}
\Op[A']=\ue^{\frac{\ui}{\hbar}
\Op[W]}\Op[A]\ue^{-\frac{\ui}{\hbar} \Op[W]}\,\, .
\end{equation}
If we introduce the one parameter family of operators
\begin{equation}\label{eq:symb_schroed1}
\widehat{A}(\flowparam)=\Op[A(\flowparam)]=\ue^{\frac{\ui}{\hbar}\flowparam\Op[W]}
\Op[A]\ue^{-\frac{\ui}{\hbar} \flowparam \Op[W]}
\end{equation}
then $A'=A(\flowparam=1)$ and the Heisenberg equation
\eqref{eq:Heisenberg} can be written in terms of the Moyal bracket
as an equation for the symbol $A(\flowparam)$, 
\begin{equation}\label{eq:symb_schroed}
\frac{\ud}{\ud \flowparam} A(\flowparam)=\{W,A(\flowparam)\}_M\,\,
.
\end{equation}
In order to obtain $A'$ we thus have to solve
\eqref{eq:symb_schroed} with initial condition $A(0)=A$. Note the
similarity between \eqref{eq:symb_schroed} and
\eqref{eq:dAdepsilon_classical} in Section~\ref{sec:conj_symp}
which expresses the correspondence between the Heisenberg equation
\eqref{eq:Heisenberg} and the classical equation
\eqref{eq:dAdepsilon_classical} in the framework of the Weyl
calculus.

We will now discuss methods of how to solve equation
\eqref{eq:symb_schroed} for certain choices of $W$. Recall that if
$W$ is a polynomial of order less than or equal to two, then the
Moyal bracket reduces to the Poisson bracket (see
\eqref{eq:moyal=poisson}) and hence
Equation~\eqref{eq:symb_schroed} reduces to
\eqref{eq:dAdepsilon_classical}, and we recalled earlier in our
development of the classical normal form theory that polynomials
of order less than or equal to two generate affine linear
symplectic transformations (see Sec.~\ref{sec:defcompcnf} and
reference~\cite{Fol89}). The following Lemma tells us that the
symbols of operators transform in the same way as classical phase
space functions under such transformations.


\begin{lem}[Exact Egorov]\label{lem:ex-eg}
Assume $W(q,p)$ is a polynomial of order less than or equal to $2$ with real valued coefficients, and let
$\flow^1_W$ be the time one map of the Hamiltonian flow generated by $W$ (see \eqref{eq:class_flow}). Then
\begin{equation}
\widehat{U}=\ue^{-\frac{\ui}{\hbar} \Op[W]}
\end{equation}
is unitary, and for every $A\in S^m(\R^d\times \R^d)$, we have 
\begin{equation}
\ue^{\frac{\ui}{\hbar} \Op[W]}\Op[A]\ue^{-\frac{\ui}{\hbar}
\Op[W]}=\Op[A']
\end{equation}
with $A'\in S^m(\R^d\times \R^d)$ given by 
\begin{equation}\label{eq:exact-Egorov}
A'=A\circ \Phi_W^{-1}\,\, .
\end{equation}
\end{lem}


\begin{proof} For the full proof we refer the reader to the appendix to chapter 7 in \cite{DimSjo99}. The main ideas are as follows.
It is well known that $\Op[W]$ is essentially self-adjoint (see
e.g. \cite{DimSjo99}), and therefore $\widehat{U}$ is unitary. In
order to find $H'$ we have to solve \eqref{eq:symb_schroed}. Since
$W$ is a polynomial of order two or less than two
Equation~\eqref{eq:symb_schroed} reduces to
\eqref{eq:dAdepsilon_classical}. From
Equation~\eqref{eq:class_flow} we see that $A(\flowparam)=A\circ
\Phi_W^{-\flowparam}$, and at $\flowparam=1$ we obtain
\eqref{eq:exact-Egorov}. Now if $W$ is a polynomial of order less
than or equal to two, then $\Phi_W^{-1}$ is an affine linear
transformation. Hence if $A\in S^m(\R^d\times \R^d)$, then $A'\in
S^m(\R^d\times \R^d)$.
\end{proof}

This result is called `exact Egorov' because there is a more
general theorem due to Egorov~\cite{Ego69} which states that, for
a large class of $W$, a similar result holds asymptotically for
$\hbar\to 0$. However, only for polynomials of degree equal to or less than two,
the higher order terms in $\hbar$ vanish.

For later reference we consider the following example. For $(q,p)\in\R^2$, let
\begin{equation}\label{eq:symbol-45}
W(q,p)=-\frac{\pi}{4}\frac12 (p^2+q^2)\,,
\end{equation}
which is the Hamilton function of an harmonic oscillator. The
factor $-\pi/4$ is introduced for convenience.
The function $W$ generates the vector field
\begin{equation}
X_W (q,p) =  (\partial_p W(q,p),-\partial_q W(q,p)) = \frac{\pi}{4}(-p,q)\,.
\end{equation}
The corresponding flow is given by
\begin{equation}
   \big(q(\flowparam),p(\flowparam)\big)=\flow^\flowparam_W (q,p)
= \big(\cos(\flowparam\frac{\pi}{4}) \,q - \sin (\flowparam\frac{\pi}{4}) \,p, \sin
(\flowparam\frac{\pi}{4}) \,q + \cos (\flowparam\frac{\pi}{4}) \,p\big)\,.
\end{equation}
The harmonic oscillator thus generates rotations in the $(q,p)$-plane. In particular,
the time one map of the flow generated by $W$ gives the map from the coordinates $(q,p)$
to the new coordinates
\begin{equation} \label{eq:def45rotation}
(Q,P) = \flow_W^1(q,p) = \frac{1}{\sqrt{2}} (q-p,q+p)\,,
\end{equation}
which we already considered in Sec.~\ref{sec:examp_comp_cnf}.
Transforming $I(q,p)=pq$ under this flow we get
\begin{equation}\label{eq:I45rotation}
I'(Q,P) = I \circ \flow^{-1}_W (Q,P) = \frac12(P^2-Q^2)\,,
\end{equation}
which gives the operator
\begin{equation}\label{eq:opI45rotation}
\Op[I'] = - \frac{\hbar^2}{2} \frac{\ud^2}{\ud Q^2} - \frac12
Q^2\,.
\end{equation}
We will refer to $\Op[I']$ as the $Q$ representation of $\Op[I]$,
and for later reference we denote the unitary transformation which classically
generates the 45$^\circ$ rotation \eqref{eq:def45rotation} as
\begin{equation} \label{eq:defUr}
\widehat{U}_{\text{r}} = \ue^{-\frac{\ui}{\hbar} \Op[W]}\, ,
\end{equation}
where $W$ is given by \eqref{eq:star-product}.

Having discussed this particular example of an application of
Lemma~\ref{lem:ex-eg} (exact Egorov) we now
turn to the case of higher order polynomials in $W$.
To this end we will develop a power series approach analogous to what
we described in Sec.~\ref{sec:conj_symp}. This will provide
higher order approximations in $\hbar$ as well as give an
explicit expression for the symbol 
of the transformed in \eqref{eq:Aprime_as_conjugation_of_A} 
in the case where $W$ or $A$
are polynomials.

We begin by simplifying the notation and define the
Moyal-adjoint action. For two smooth
functions $W$ and $A$, we define analogously to the adjoint action in
\eqref{eq:ad_op_class} the  Moyal-adjoint action as
\begin{equation}
\label{eq:ad_op_quant}
 \adM_WA:=\{W,A\}_M\,\, .
\end{equation}
Using the Moyal adjoint  Equation~\eqref{eq:symb_schroed} becomes
\begin{equation}
\frac{\ud A(\flowparam)}{\ud \flowparam}=\adM_WA(\flowparam).
\end{equation}
We compute higher order derivatives of $A(\flowparam)$
with respect to $\flowparam$ in a manner analogous to
\eqref{eq:2deriv_class} and \eqref{eq:ho_deriv_class}. We
successively differentiate \eqref{eq:symb_schroed} and apply the
notation \eqref{eq:ad_op_quant} to obtain
\begin{equation}\label{eq:n-der=adn}
\frac{\ud^n}{\ud \flowparam^n} A(\flowparam)=\big[\adM_W\big]^nA(\flowparam)\,\, .
\end{equation}
Hence, the (formal) Taylor series in $\flowparam$ around
$\flowparam=0$ is given by

\begin{equation}\label{eq:Egorov_Lie1}
A(\flowparam)=\sum_{n=0}^{\infty}
\frac{\flowparam^n}{n!}\big[\adM_W\big]^nA\,\, ,
\end{equation}
and setting $\flowparam=1$ we obtain the formal sum
\begin{equation}\label{eq:Egorov_Lie}
A'=\sum_{n=0}^{\infty} \frac{1}{n!}\big[\adM_W\big]^nA\,\, .
\end{equation}
This expression is completely analogous to
\eqref{eq:Atransf_classical}, and as we will see in more detail,
can be used in a similar fashion to compute the symbol $A'$ up to
any desired order in $\hbar$ and $(q,p)$. In particular, analogous
to equation~\eqref{eq:Atransf_classical}, it gives the Taylor
expansion with respect to $\epsilon$, evaluated at $\epsilon=1$,
for the symbol $A'$  of the operator obtained after conjugation of
the operator defined by the symbol $A$  by the unitary
transformation generated by $W$.  This formula forms the basis of
the quantum normal form method where the idea is to ``simplify''
(or ``normalise'') the symbol  whose quantisation will then
correspond to the normal form of the Hamilton operator.  As in the
classical case, the computation of the Taylor expansion is carried
out ``order by order'' using power series expansions of the symbol
in $(q,p)$ and $\hbar$.  The series is expanded about an
equilibrium point of the principal symbol, and therefore the
quantum normal form will be valid in a neighbourhood of this
point. Hence, as in the classical case, the quantum normal
form is a ``local object'' whose operator nature requires more
technical details for a rigorous characterization of its
properties (cf. Section \ref{sec:localising} and Definition
\ref{def:remainder}), and we will describe these in more detail in
the following.

Therefore similar to the mathematical formalism required for
computing the classical normal form, normalising the symbol of the
operator that will correspond to the quantum normal form will
require us to manipulate monomials which in addition to $(q,p)$
now also have factors of $\hbar$. In order to describe this we
adopt a notation introduced by Crehan \cite{Cre90} and define the
spaces 
\begin{equation}\label{eq:order-s}
\cW^s =
\text{span }\bigg\{
\hbar^j q^{\alpha}p^{\beta}:= \hbar^j \prod_{k=1}^d
q_k^{\alpha_k}p_k^{\beta_k}\,:\, \abs{\alpha}+\abs{\beta}+2j =s\bigg\}\, .
\end{equation}
These spaces $\cW^s$ are closely related to the spaces $\ccW^s$
spanned by the polynomials \eqref{eq:classicalmonomials} in the
classical case.
In fact we have
\begin{equation}\label{eq:quant-sum-class}
\cW^s=\bigoplus_{k=0}^{[s/2]} \hbar^k \ccW^{s-2k}\,,
\end{equation}
where $[s/2]$ denotes the integer part of $s/2$.

Below we want to use functions $W\in \cW^s$ in order to construct
unitary operators of the form $\ue^{-\frac{\ui}{\hbar}\Op[W]}$.
However, the quantisation of a function $W\in \cW^s$   will give
an unbounded operator and this makes the discussion of
self-adjointness of $\Op[W]$, and hence the unitarity of
$\ue^{-\frac{\ui}{\hbar}\Op[W]}$, more  complicated. But since we
are interested in the local quantum dynamics generated by a
Hamilton operator in the neighbourhood of an equilibrium point of
its principal symbol it will be sufficient to have a local version of the
spaces $ \cW^s$. We thus apply the localisation procedure from Section \ref{sec:localising}. 
We say that $W\in \cWl^s$ if $W\in
\cS_{\hbar}(\R^d\times\R^d)$ and there is an open neighbourhood
$U$ of $z_0=0\in \R^d\times\R^d$ such that 
\begin{equation}
W|_{U}\in \cW^s\,\, .
\end{equation}
The quantisation of elements of $\cWl^s$ will then give bounded
operators. Therefore, if $W\in \cWl^s$ is real valued then $\Op[W]$ will be
self-adjoint and thus $\widehat{U}=\ue^{-\frac{\ui}{\hbar} \Op[W]}$ will be unitary.

We will frequently use Taylor expansions and want to modify them
in such a way that the terms in the expansion  are in $\cWl^s$. In
order to make our discussion of this property precise we will need
the following definition.


\begin{Def}
\label{def:remainder}
We will say a function $O_N\in \cS_{\hbar}(\R^d\times \R^d)$ is
\emph{a remainder of order}
$N$ (around $(q,p)=(0,0)$) if there is an open neighbourhood $U$ of
$(q,p)=(0,0)$  and $c>0$ such that
\begin{equation}\label{eq:def-oreder-N}
|O_N(\varepsilon^2\hbar, \varepsilon q,\varepsilon p)| < c \varepsilon^N
\end{equation}
for $\hbar<1$, $(q,p)\in U$ and $\varepsilon < 1$.
\end{Def}



We then can formulate
\begin{lem}\label{lem:S-Taylor}

Let $A\in \cS_{\hbar}(\R^d\times\R^d)$, then there exist $A_s\in \cWl^s$ such that
for any $N\in\N$ there is a remainder $O_N\in \cS_{\hbar}(\R^d\times\R^d)$ of order $N$  such that
\begin{equation}
A=\sum_{s=0}^{N-1}A_s +O_N\,\, .
\end{equation}
\end{lem}


\begin{proof}

Let us take the ordinary Taylor expansion of $A(\hbar,q,p)$ around $(\hbar,q,p)=(0,0,0)$
and order the terms according to the definition of order in \eqref{eq:order-s}. This
gives us an expansion $A=\sum_{s=0}^{N-1} \tilde{A}_s + R_N$ with
\begin{equation}
\tilde{A}_s=\sum_{\abs{\alpha}+\abs{\beta}+2j=s}\frac{1}{j!\alpha!\beta!}
\pa_{\hbar}^k\pa_q^{\alpha}\pa_p^{\beta}A(0,q_0,p_0)q^{\alpha}p^{\beta}\hbar^j\in \cW^s
\end{equation}
 and
$R_N(\varepsilon^2\hbar, \varepsilon q,\varepsilon p)=O(\varepsilon^N)$. 
Now choose a function
$\rho\in \cS_{\hbar}(\R^d\times\R^d)$ with $\rho|_U\equiv 1$ for some 
open neighbourhood $U$ of $0$, and set
$A_s:=\rho \tilde{A}_s$. Then it follows directly that \New{$A_s\in \cWl^s$} and
$O_N:=A-\sum_{s=0}^{N-1}A_s\in\cS_{\hbar}(\R^d\times\R^d)$
is a remainder of order $N$.
\end{proof}

The main reason for defining the order $s$ according to
\eqref{eq:order-s}, i.e., the reason for double counting the powers of $\hbar$,  
is that it behaves nicely with respect to the
Moyal product. This is reflected in the following lemmata. The first one is  the 
analogue of Lemma~\ref{lem:classical} in the classical case.


\begin{lem}\label{lem:Mad-order}
Let $W\in \cWl^{s'}$, $A\in \cWl^{s}$, $s,s'\ge1$,
then 
\begin{equation}
\{W,A\}_M\in \cWl^{s+s'-2}\,,
\end{equation}
and for $n\ge0$, 
\begin{equation}
\big[\adM_W\big]^nA\in \cWl^{n(s'-2)+s}\,\, ,
\end{equation}
if $n(s'-2)+s\ge 0$ and $\big[\adM_W\big]^nA=0$ otherwise. 
\end{lem}


\begin{proof}
We can write the  Moyal bracket \eqref{eq:Moyal} as

\begin{equation}\label{eq:Moyal2}
\{W,A\}_M=\sum_{k}\bigg(\frac{\hbar}{2}\bigg)^{2k}\frac{(-1)^k}{(2k+1)!}
D^{(2k+1)}(W,A)(q,p)
\end{equation}
with the  bi-differential operators

\begin{equation}
D^{(2k+1)}(W,A)(q,p):= W(q,p) [\la
\stackrel{\leftharpoonup}{\pa}_{p},\stackrel{\rightharpoonup}{\pa}_{q}
\ra - \la \stackrel{\rightharpoonup}{\pa}_{p},
\stackrel{\leftharpoonup}{\pa}_{q} \ra]^{(2k+1)} A(q,p)\,\, ,
\end{equation}
Now the bi-differential operator $D^{(2k+1)}$ is of order $2k+1$ 
in the arguments involving $A$ and $W$
individually, and therefore 
\begin{equation}
D^{(2k+1)}:\cWl^s\times \cWl^{s'}\to
\cWl^{s-(2k+1)+s'-(2k+1)}\,\, .
\end{equation}

On the other hand, multiplication by $\hbar^{2k}$ maps
$\cWl^{s-(2k+1)+s'-(2k+1)}$ to \\ $\cWl^{s-(2k+1)+s'-(2k+1)+4k}=
\cWl^{s+s'-2}$, and therefore every term in the series
\eqref{eq:Moyal2} is in $\cWl^{s+s'-2}$. But the order of $W$ and
$A$ as polynomials in $(q,p)$ near $(q,p)=(0,0)$ is at most $s$ and $s'$,
respectively, and therefore the terms in the series \eqref{eq:Moyal2} vanish
near $(q,p)=(0,0)$ for $2k+1>\min(s,s')$. Hence
\begin{equation}
\{W,A\}_M\in \cWl^{s+s'-2}\,\, .
\end{equation}
The second result follows then by induction.
\end{proof}

We can now turn our attention to the computation of the symbol of
a conjugated operator when the generator of the unitary operator
has order larger than 2. The computation will proceed in two steps,
in the first lemma we show that conjugation respects the class of
symbols we are working with.



\begin{lem}\label{lem:Beals-conjug}

Let $W\in \cWl^s$ and $A\in \cS_{\hbar}(\R^d\times\R^d)$, then there exists
an $A'\in \cS_{\hbar}(\R^d\times\R^d)$ such that $\Op[A']=\ue^{\frac{\ui}{\hbar}
\Op[W]}\Op[A]\ue^{-\frac{\ui}{\hbar} \Op[W]}$.
\end{lem}


The techniques for proving this lemma are different from the ones we use
in the rest of the paper. In order not to interrupt the flow of the paper, we therefore
present the proof in  Appendix~\ref{sec:proof-Beals-conjug}.

By Lemma~\ref{lem:Beals-conjug} we know that the symbol of $\ue^{\frac{\ui}{\hbar}
\Op[W]}\Op[A]\ue^{-\frac{\ui}{\hbar} \Op[W]}$ is a function in
$\cS_{\hbar}(\R^d\times\R^d)$.  With the help of Lemma~\ref{lem:Mad-order} we can reorder the
terms in the formal expansion \eqref{eq:Egorov_Lie} to turn it into a well defined Taylor
expansion in the sense of Lemma \ref{lem:S-Taylor}. This is the content of the
following Lemma which can be considered to be the analogue of Lemma~\ref{lem:classical2} in
the classical case.
%
%
%
%


\begin{lem}\label{lem:conjug}

Let $W\in \cWl^{s'}$, $s'\geq 3$, and $A\in \cS_{\hbar}(\R^d\times\R^d)$ with
Taylor expansion $A=\sum_{s=0}^{\infty}A_s$, $A_s\in \cWl^s$. Then
the symbol $A'$  of $\ue^{\frac{\ui}{\hbar}
\Op[W]}\Op[A]\ue^{-\frac{\ui}{\hbar} \Op[W]}$ has the Taylor
expansion 
\begin{equation}
A'=\sum_{s=0}^{\infty}A_s'
\end{equation}
with 
\begin{equation}\label{eq:H-k}
A_s'=\sum_{n=0}^{[\frac{s}{s'-2}]} \frac{1}{n!}
[\adM_{W}]^nA_{s-n(s'-2)}\in \cWl^s\,\, ,
\end{equation}
i.e., for every $N\in\N$ there exists a remainder $O_N\in \cS_{\hbar}(\R^d\times\R^d)$ of order $N$ such
that
\begin{equation}
A'=\sum_{s=0}^{N-1}A_s'+O_N\,\, .
\end{equation}
\end{lem}



\begin{proof}

By Lemma \ref{lem:Beals-conjug} we know that $A'\in \cS_{\hbar}(\R^d\times\R^d)$, and we have to compute
its Taylor series. With \eqref{eq:n-der=adn}
we can use the Taylor expansion of $A'(\flowparam)$ to write 
\begin{equation}
A'=\sum_{n=0}^{N-1} \frac{1}{n!}\big[\adM_W\big]^nA +O_N'
\end{equation}
with 
\begin{equation}
O_N'=\frac{1}{(N-1)!}\int_0^1(1-\flowparam)^{N-1}
\big[\adM_W\big]^NA'(\flowparam)\,\, \ud \flowparam\,\, ,
\end{equation}
being just the standard remainder formula for Taylor expansions.
Since $A'(\flowparam)\in \cS_{\hbar}(\R^d\times\R^d)=\cWl^0$ we
have by Lemma~\ref{lem:Mad-order} that $O_N'\in
\cS_{\hbar}(\R^d\times\R^d)$ is a remainder of order $N$. If we
next insert the Taylor expansion for $A$ we get 
\begin{equation}
A'=\sum_{l=0}^{N-1}\sum_{n=0}^{N-1}
\frac{1}{n!}\big[\adM_W\big]^nA_l +O_N\,,
\end{equation}
where $O_N\in \cS_{\hbar}(\R^d\times\R^d)$ denotes the collection of all the
remainder terms of order $N$. Using  Lemma~\ref{lem:Mad-order} we can collect
all the terms of order $k$ in the sum which gives \eqref{eq:H-k}. To this end
one can proceed completely analogously to the  proof of
Lemma~\ref{lem:classical2} and we therefore omit the details.

\end{proof}


\subsection{Definition and Computation of the Quantum Normal Form}
\label{sec:compscnf}

We will now define when a Hamilton operator is in quantum normal form.
Similar to the case of the classical normal form, in general a Hamilton operator is not in quantum
normal form. However, as we will show, the formalism based on the Weyl calculus developed in
the previous two sections can be used to construct an explicit algorithm
which will allow us
to transform  a Hamilton operator to normal form to any desired order of
its  symbol.
The algorithm will consist of two parts. The first part operates on the
level of the symbols of operators, and this part of the algorithm will be very
similar to the normalisation algorithm in the classical case.  In the second part the symbols are
quantised, i.e., the operators corresponding to the symbols will be determined.

The starting point is a Hamilton operator $\Op[H]$ which is the Weyl
quantisation of a symbol $H(\hbar,q,p)$.
Assume that the Hamiltonian dynamical system defined by the \emph{principal symbol} 
has an \emph{equilibrium point}  at
$z_0=(q_0,p_0)$, i.e., the gradient of the principal symbol vanishes at $z_0$.
%
%
%
%
Let $H_2(z) \in \cW^2$ 
denote the second order term of the Taylor
expansion of the symbol $H$ about $z_0$ and $\Op[H_2]$ its Weyl quantisiation.
We now make the

\begin{Def}[Quantum Normal Form]\label{def:qnf}
We say that $\Op[H]$ is in {\em quantum normal form} with respect to the
equilibrium point $z_0$ of its principal symbol
if
\begin{equation}
\big[\Op[H_2],\Op[H]\big]=0\,,
\end{equation}
or equivalently in terms of the symbol,
\begin{equation} \label{eq:def_qnf_symbol}
\ad_{H_2}  H \equiv \{ H_2, H \} =0\,.
\end{equation}
 \end{Def}

The equivalence of the two equations in Definition~\ref{def:qnf}
derives from the fact that the Moyal bracket reduces to the
Poisson bracket if one of its argument is quadratic. Moreover, we
remark that $H_2$ and the quadratic part of the principal symbol
differ at most by a term that consists of  $\hbar$ with a constant
prefactor. Since the Poisson bracket vanishes if one of its two
arguments is a constant it does not make a difference in
Definition~\ref{def:qnf} if $H_2$ in \eqref{eq:def_qnf_symbol}
would be replaced by the second order term of the Taylor expansion
of the principal symbol.

Like in the case of a Hamilton function being in classical normal form the
property of a Hamilton operator to be in quantum normal form has strong
implications which in the quantum case lead to a considerable simplification of the
study of the spectral properties of the operator.
To this end recall that two commuting operators have a joint set of
eigenfunctions. Hence, if an operator is in quantum normal form
the study of its spectral properties will be simplified
considerably, since the spectrum and eigenfunctions of an operator
$\Op[H_2]$ with  a symbol of order 2 are well known.

Similar to the classical case a Hamilton operator is in general not in quantum
normal form. However, we will now show how the formalism developed in the previous two
sections can be used  to transform a Hamilton operator to quantum normal form
to any desired order of its symbol. Similar to  the classical case we will
truncate the symbol at a certain order and show that the corresponding Hamilton
operator will lead to a very good approximation of many interesting spectral properties of
the original Hamilton operator.

We develop the following procedure.
Let $H=H^{(0)}$ denote the symbol of our original Hamilton operator.
We will construct a consecutive sequence of transformations of the symbol
according to
\begin{equation}\label{eq:transformation-sequence_quantum}
H=:H^{(0)}\to H^{(1)}\to H^{(2)}\to H^{(3)}\to\cdots \to H^{(N)}
\end{equation}
by requiring the symbol $H^{(n)}$, for $n\ge 1$, to derive from the symbol $H^{(n-1)}$ by
conjugating $\Op[H^{(n-1)}]$ with a unitary transformation according to
\begin{equation}\label{eq:Hn-1toHn}
\Op[H^{(n)}]=\ue^{\frac{\ui}{\hbar}\Op[W_n]}\Op[H^{(n-1)}]\ue^{-\frac{\ui}{\hbar}\Op[W_n]}\,,
\end{equation}
where the symbol $W_n$ of the generator  of the unitary
transformation is in $\cWl^n$. 
Like in the series of symplectic transformations in the classical case in \eqref{eq:transformation-sequence}, 
$N$ in
\eqref{eq:transformation-sequence_quantum} is again a sufficiently
large integer at which we will truncate the quantum normal form
computation. The algorithm for normalising the symbol will be
identical to the classical case. The key difference is that the
Poisson bracket of the classical case is replaced by the Moyal
bracket in the quantum case. With this replacement, the
mathematical manipulations leading to normalisation of the symbol
are virtually identical.

Towards this end, using \eqref{eq:Egorov_Lie} we see that,
analogously to \eqref{n_trans_Ham} in the classical case, we have
\begin{equation}
H^{(n)} = \sum_{k=0}^\infty \frac{1}{k!} \big[ \adM_{W_n} \big]^k H^{(n-1)}\,.
\end{equation}

Like in the classical case the first two steps, $n=1,2$, in \eqref{eq:transformation-sequence_quantum} differ somewhat in nature from the steps for $n\ge 3$. The first step serves to shift the equilibrium
point to the origin and the second step serves to simplify the quadratic part of the symbol.
It follows from Lemma~\ref{lem:ex-eg} (exact Egorov) that we
achieve these affine linear transformations
by choosing the symbols $W_1$ and $W_2$
identical to the generators of the corresponding symplectic transformations in
the classical case. We thus have
\begin{equation} \label{eq:def_H11_z1_quantum}
H^{(1)}(\hbar,z) = H^{(0)}(\hbar,z+z_0)
\end{equation}
and
\begin{equation} \label{eq:def_H22_z2_quantum}
H^{(2)}(\hbar,z) = H^{(1)}(\hbar, M^{-1} z)\,,
\end{equation}
where $M$ is a suitable  symplectic $2d\times2d$ matrix which
achieves the simplification of the quadratic part of the symbol
analogously to the classical case. It is important to note that we
do not explicitly need the generators $W_1$ and $W_2$ which, as
mentioned in Sec.~\ref{sec:defcompcnf}, might be difficult to
compute.

Before we proceed with the normalisation of the higher order terms, $n\ge3$,
we will assume that we
localise around the equilibrium point which is now at the origin, see Section
\ref{sec:localising}, i.e., by multiplying
$H^{(2)}$ by a suitable cutoff function concentrated about the origin we can assume
$H^{(2)}\in \cS_{\hbar}(\R^d\times\R^d)$ and the terms $H_s^{(2)}$ of the
Taylor expansion of $H^{(2)}$ to be in $\cWl^s$.

For the higher order terms, $n\ge 3$, we find by
\eqref{eq:H-k} in Lemma~\ref{lem:conjug} that the terms $H^{(n)}_s$ can
be computed from the terms of the power series of $H^{(n-1)}$
according to 
\begin{equation}\label{eq:higher-ordertransform}
H^{(n)}_s=\sum_{k=0}^{[\frac{s}{n-2}]} \frac{1}{k!}
[\adM_{W_n}]^kH^{(n-1)}_{s-k(n-2)}\,\, .
\end{equation}

The normalisation procedure for the terms of order $n\ge 3$ of the
symbol has very similar properties as the corresponding procedure
in the classical case. In particular a transformation at a given
order does not affect lower order terms.  This is made more
precise in the following lemmata that are the analogues of
Lemma~\ref{lem:order2} and Lemma~\ref{lem:lower_order} for the
classical case from Section~\ref{sec:defcompcnf}.

\begin{lem}
\label{lem:order2_scnf}
 $H_2^{(n)} = H_2^{(2)}$, $n \ge 3$.
\end{lem}

\begin{proof}
The proof is
completely analogous to the proof of  Lemma~\ref{lem:H2unchanged} and
is therefore omitted.
\end{proof}

Like in the classical case, Lemma~\ref{lem:order2_scnf} motivates
the adoption of the following notation for the operator
\begin{equation} \label{eq:def_D_quantum}
  \opD := \ad_{H_2^{(2)}} = \{ H_2^{(2)}, \cdot \}.
\end{equation}

\begin{lem}
\label{lem:lower_order_scnf} For $n\ge 3$  and $0 \le s < n$, $H^{(n)}_s=H^{(n-1)}_s$.
\end{lem}

\begin{proof}
The proof is
completely analogous to the proof of  Lemma~\ref{lem:lower_order} and
is therefore omitted.
\end{proof}

Like in the classical case the $n^{\text{th}}$ order term in
$H_n^{(n)}$ indicates how to choose $W_n$ for $n \ge 3$.

\begin{lem}[Quantum Homological Equation]
\label{lem:hom_eq_scnf} For $s=n \ge 3$,
\begin{equation}
H^{(n)}_n = H^{(n-1)}_n -  \opD  W_n,
 \label{eq:homological}
\end{equation}
\end{lem}

\begin{proof}
The proof is completely analogous to the proof of
Lemma~\ref{lem:hom_eq} and is therefore omitted. \rem{ {\bf
Details..} We have used Lemma \ref{lem:order2_scnf}
($H^{(n-1)}_2=H^{(2)}_2$) and, since $H^{(2)}_2$  is of order two,
$\{W_n,H^{(n-1)}_2\}_M=\{W_n,H^{(2)}_2\}$
} 

\end{proof}

The homological equation \eqref{eq:homological} is solved in
exactly the same way as the homological equation in the classical
normal form computation described in Sec.~\ref{sec:solvhomequ}.
The only difference is that we now deal with a symbol that in
contrast to the classical Hamilton function in general depends on
$\hbar$. But due to the splitting $\cW^s=\bigoplus_{k=0}^{[s/2]}
\hbar^k \ccW^{s-2k}$, see \eqref{eq:quant-sum-class}, the results
on the solution of the classical homological equation can be
transferred directly. In particular the notion of solvability 
introduced in Definition \ref{def:solvable} carries over 
verbatim. 

We note that so far we have only shown how to transform the
Hamilton operator to quantum normal form on the level of its
symbol. We have not yet discussed the implications for the
corresponding transformed operator. As we will see, similar to the
question of how to explicitly solve the homological equation, the
nature of the transformed Hamilton operator depends on the type of
the equilibrium point of the principal symbol. In the next
section, Sec.~\ref{sec:examp_comp_scnf}, we will discuss this in
detail for the case of a non-resonant
saddle-centre-$\cdots$-centre equilibrium point.


We summarise our findings in the following

\begin{thm}
\label{theorem_qm_NF}
Assume the principal symbol of $\Op[H]$ has an equilibrium point at $z_0\in
\R^d\times\R^d$, and that the homological equation is solvable in the sense of Def. \ref{def:solvable}.  
Then for every $N\in\N$
there is a unitary transformation $\widehat{U}_N$ such that
\begin{equation} \label{eq:def_H_QNF_N}
\widehat{U}_N^*\Op[H]\widehat{U}_N=\Op[H_{\text{QNF}}^{(N)} ]+\Op[O_{N+1}]
\end{equation}
where $\Op[H_{\text{QNF}}^{(N)}]$
is in quantum normal form (with respect to 0) and $O_{N+1}$ is of order $N+1$.

\end{thm}


\begin{proof}

As we have seen in this section the conjugations of a Hamilton
operator by unitary transformations to transform it to quantum
normal form can be carried out  on the level of the
symbols of the operators involved. This makes the proof of
Theorem~\ref{theorem_qm_NF} very similar to the proof of
Theorem~\ref{theorem_classicalNF} in the classical case.
In fact, the proof of Theorem~\ref{theorem_classicalNF} 
carries over verbatim when one replaces the Poisson bracket by 
the Moyal bracket. Then Lemma~\ref{lem:H2unchanged} is replaced by 
Lemma~\ref{lem:order2_scnf} and Lemma~\ref{lem:lower_order} by 
 Lemma~\ref{lem:lower_order_scnf}.

Using the scheme \eqref{eq:transformation-sequence_quantum} with 
\eqref{eq:Hn-1toHn} gives then
the unitary transformation $\widehat{U}_N$ in \eqref{eq:def_H_QNF_N} 
as
\begin{equation} \label{eq:defUN_in_proof}
\widehat{U}_N=\ue^{-\frac{\ui}{\hbar}\Op[W_1]}
\ue^{-\frac{\ui}{\hbar}\Op[W_2]}\ue^{-\frac{\ui}{\hbar}\Op[W_3]}\cdots 
\ue^{-\frac{\ui}{\hbar}\Op[W_{N}]}\,\, .
\end{equation}
The first two generators, $W_1$ and $W_2$ are chosen exactly as in the 
classical normal form algorithm, see \eqref{eq:def_H11_z1_quantum} and the following paragraph, by 
Lemma~\ref{lem:ex-eg} (exact Egorov) this induces the same transformation 
of the symbols as in the classical case. The other generators 
$W_n$, $n\ge 3$, are then chosen recursively as solutions of the 
homological equation, see Lemma \ref{lem:hom_eq_scnf}, where after 
each step we have to determine $H^{(n)}$ up to order $N$ from \eqref{eq:higher-ordertransform}.  


\end{proof}

Similar to  the classical case the definition of the quantum normal form in
Defintion~\ref{def:qnf} is of little value for practial purposes since we cannot
expect the quantum normal computation to converge if we carry it out for
$N\rightarrow \infty$ as required by  Defintion~\ref{def:qnf}.
For applications it is more useful to consider the \emph{truncated quantum normal form}.

 \begin{Def}[$N^{th}$ Order Quantum Normal Form]
Consider a Hamilton operator $\Op[H]$ whose principal symbol has
an equilibrium point at $z_0\in\R^d\times\R^d$ which, for
$N\in\N$, we normalise according to Theorem~\ref{theorem_qm_NF}.
Then we refer to the operator $\Op[H^{(N)}_{\text{QNF}}]$ in
Equation~\eqref{eq:def_H_QNF_N} as the \emph{$N^{th}$ order
quantum normal form}
 (QNF) of $\Op[H]$.

\rem{
Consider a symbol
 \begin{equation}
H^{(N)}=\sum_{s=0}^\infty H^{(N)}_s=\sum_{s=0}^N
H^{(N)}_s+O_{N+1}\, \,,
\end{equation}
where $H^{(N)}_s$, $3 \le s \le N$ have all been
normalised, i.e., $ \opD
  H^{(N)}_s = \{ H_2^{(2)}, H^{(N)}_s \} =0, \, 3 \le s \le
 N$, and $O_{N+1}\in \cS_{\hbar}(\R^d\times\R^d)$ is a remainder of order $N+1$.
Then we refer to the symbol
 \begin{equation} \label{def:NthorderQNF}
 H^{(N)}_{QNF} = E_0 + H_2^{(2)} + H_3^{(N)} + \cdots +
 H_{N-1}^{(N)} + H_N^{(N)},
 \end{equation}
 as the \emph{$N^{th}$ order quantum normal form}
 (QNF) of the symbol $H$. Similarly we call the corresponding operator
 $\Op[H^{(N)}_{QNF}]$ the  quantum normal form of $\Op[H]$.
} 

 \end{Def}

We have seen that the procedure to construct the quantum normal
form is very similar to the procedure to compute the classical
normal form. In particular the homological equations
\eqref{eq:homological_classical} and \eqref{eq:homological} which
determine the choice of the successive transformations
\eqref{eq:transformation-sequence} and
\eqref{eq:transformation-sequence_quantum}, respectively, look
identical since the Poisson bracket reduces to the Moyal bracket
if one of its argument is a polynomial of order less than or equal
to 2. However, it is important to point out that this does not
mean that the Moyal bracket completely disappears from the
procedure in the quantum case. In fact, while the normalisation
transformation at a given order does not modify lower order terms,
it does modify all higher order terms, and the Moyal bracket plays
an important role in this, see \eqref{eq:higher-ordertransform}.
Consequently the terms in the Taylor expansions of $H^{(n)}$ and
the generators $W_n$ will in general depend on $\hbar$.

Since  the Moyal bracket tends to the Poisson bracket in the limit $\hbar\to 0$ we expect that the
symbol of the quantum normal form should tend to the classical normal form, too. This is
indeed the case.


\begin{prop}

The principal symbol of the $N^{\text{th}}$ order quantum normal
$\Op[H_{\text{QNF}}^{(N)}]$ is  the classical
normal form of order $N$, i.e.,
\begin{equation}
H_{\text{QNF}}^{(N)}(\hbar,q,p)=H_{\text{CNF}}^{(N)}(q,p)+O(\hbar)\,\, .
\end{equation}
\end{prop}


\begin{proof}
This follows from an inspection of the construction of the classical and quantum normal forms. The first two steps are identical by
Lemma  \ref{lem:ex-eg} (exact Egorov). The homological equation determining the choices of the $W_n$ is as well identical. What
is different however is the transformation of the higher order terms, $k>n$.
Here we have
Equation~\eqref{n_trans_Ham_s}
in the classical case and Equation~\eqref{eq:higher-ordertransform} in the
quantum case, and these equations differ by
the use of the adjoint versus the Moyal adjoint. But since $\adM_WA=\ad_WA+O(\hbar)$ and therefore
\begin{equation}
\adM_W^kA=\ad_W^kA+O(\hbar)
\end{equation}
the differences in the higher order terms between the classical and the quantum transformation schemes are
always of order $\hbar$. This implies that the difference between the
symbol of the quantum normal form and the classical normal form are
of order $\hbar$.
\end{proof}


\subsection{Nature and Computation of the Quantum Normal Form in a
  Neighbourhood of an Equilibrium  Point
of the Principal Symbol of  Saddle-Centre-$\cdots$-Centre Type}
\label{sec:examp_comp_scnf}

We now describe how the quantum normal form of a Hamilton operator  can be
computed in the case where the principal symbol has an equilibrium point
of  saddle-centre-$\cdots$-centre type, i.e., the matrix associated with the
linearisation of the Hamilonian vector field generated by the
principal symbol  has  two real eigenvalues,
$\pm \lambda$, and $d-1$ complex conjugate pairs of imaginary eigenvalues $\pm
\ui\, \omega_k, \, k=2, \ldots d$. We will assume that the
$\omega_k, \, k=2, \ldots d$, are nonresonant in the sense that
they are linearly independent over the integers, i.e.,  $k_2 \omega_2 + \ldots
+ k_d \omega_d \ne 0$ for all
$(k_2, \ldots k_d) \in \Z^{d-1} - \{0 \}$.

As mentioned in the previous section it follows from Lemma~\ref{lem:ex-eg}
(exact Egorov) that we can use the same affine linear symplectic
transformations that we used in the classical case in Sec.~\ref{sec:CNF} to shift the
equilibrium point to the orgin of the coordinate system and to simplify the
second order term  of the symbol.
We thus have
\begin{equation}
H^{(2)} =  E_0 + H_2^{(2)} + \sum_{s=3}^\infty H_s^{(2)}\,,
\end{equation}
where
\begin{equation}
H_2^{(2)} (\hbar,q,p) = \lambda q_1 p_1 + \sum_{k=2}^d \frac{\omega_k}{2}
(p_k^2 + q_k^2) + c \hbar \,,
\end{equation}
where $c$ is some real constant.

We note that in terms of the coordinates $(Q,P)$ we defined in
Sec.~\ref{sec:examp_comp_cnf} $H^{(2)}_2$ is given by
\begin{equation}
H_2^{(2)} (\hbar,q,p) = \frac{\lambda}{2} \big( P_1^2 - Q_1^2 \big) + \sum_{k=2}^d \frac{\omega_k}{2}
\big( P_k^2 + Q_k^2 \big) + c \hbar \,,
\end{equation}
which is the analogue of Equation~\eqref{eq:H22_QP_classical} in the classical case.

\subsubsection{Solution of the homological equation}

We will solve the homological equation in the spaces $\cW^n$. The solution will then be localised 
by multiplication with a cutoff function afterwards to obtain elements in $\cWl^n$.
This will ensure that the quantizations of these symbols  
are bounded and generate unitary operators. 

In order to solve the homological equation in Lemma~\ref{lem:hom_eq_scnf} we perform the
linear symplectic complex change of coordinates $(q,p)\mapsto(x,\xi)$ given by
$x_1=q_1$, $\xi_1=p_1$ and
\begin{equation}
x_k   :=\frac{1}{\sqrt{2}}(q_k-\ui p_k) \,,\quad
\xi_k :=\frac{1}{\sqrt{2}}(p_k- \ui q_k) \,,\qquad k=2,\ldots,d\, \,.
\end{equation}
In terms of these coordinates the operator $\opD$ defined in \eqref{eq:def_D_quantum} assumes
the simple form
\begin{equation}
\opD =    \lambda
(\xi_1\pa_{\xi_1}-x_1\pa_{x_1})  + \sum_{k=2}^{d}\ui
\omega_k(\xi_k\pa_{\xi_k}-x_k\pa_{x_k})\,.
\end{equation}
In terms of these coordinates the spaces $\cW^n$ defined in \eqref{eq:order-s}
are given by
\begin{equation}
\cW^n =
\text{span }\bigg\{
\hbar^j x^{\alpha}\xi^{\beta}:= \hbar^j \prod_{k=1}^d
x_k^{\alpha_k}\xi_k^{\beta_k}\,:\, \abs{\alpha}+\abs{\beta}+2j =n\bigg\}\,,
\end{equation}
and the operator $\opD$ acts on an element $\hbar^j
x^{\alpha}\xi^{\beta}\in \cW^n$
according to
\begin{equation} \label{eq:Dmonomial_quantum}
  \opD \, \hbar^j \prod_{k=1}^d
x_k^{\alpha_k} \xi_k^{\beta_k} = \bigg(\lambda ( \beta_1 -
\alpha_1) + \sum_{k=2}^{d}\ui \omega_k(\beta_k-\alpha_k)\bigg)\,
\hbar^j \prod_{k=1}^d x_k^{\alpha_k} \xi_k^{\beta_k}\,.
\end{equation}
This means that the map $\opD$ can again be diagonalised and similar to the classical
case we have that $\cW^n$ is given by the direct sum of
the kernel of $\opD$ acting on $\cW^n$,
 $\text{Ker } \opD  \big|_{\cW^n}$ and the image of $\opD$ acting on
 $\cW^n$, $\text{Im } \opD  \big|_{\cW^n}$, i.e.,

\begin{equation}
\cW^n =\text{Ker }  \opD \big|_{\cW^n}
\oplus \text{Im }
\opD \big|_{\cW^n}\,.
\label{ds_decomp_quantum}
\end{equation}

Now we can express $H_n^{(n-1)}$ as
\begin{equation}
H_n^{(n-1)} = H_{n; \text{Ker}}^{(n-1)} + H_{n; \text{Im}}^{(n-1)},
\label{order_decomp_quantum}
\end{equation}
where $H_{n; \text{Ker}}^{(n-1)}\in \text{Ker }  \opD
\big|_{\cW^n}$ and  $H_{n; \text{Im}}^{(n-1)}\in \text{Im }
\opD \big|_{\cW^n}$. We can therefore choose $W_n$ such
that
\begin{equation}
 \opD W_n = H_{n; \text{Im}}^{(n-1)}\,\,,
\label{trans_sol_quantum}
 \end{equation}
and therefore

\begin{equation}
 H_n^{(n)} =H_{n; \text{Ker}}^{(n-1)}.
\label{n_order_nform_quantum}
 \end{equation}
Similar to the classical case the choice of $W_n$ is not unique since one can always add terms from the
kernel of $\opD\big|_{\cW^n}$. However, we will require $W_n \in \text{Im }
\opD  \big|_{\cW^n}$, i.e., we will invert  $\opD$ on its
image $\text{Im } \opD  \big|_{\cW^n}$.

Using our assumption  that the frequencies $\omega_2,\dots,\omega_d$ are nonresonant, i.e.,
linearly independent over $\Z$, we see from \eqref{eq:Dmonomial_quantum}
that a monomial $\hbar^j x^\alpha \xi^\beta$ is mapped to zero if and only
if $\alpha_k=\beta_k$, $k=1,\dots,d$.
In particular $ \text{Ker }
\opD  \big|_{\cW^s} = \{0\} $ if $s$  is odd. This implies that unitary transformations can
be constructed such that all odd order terms in the symbol of the conjugated
Hamilton operator are eliminated. Moreover, for $s$ even, the terms that {\em cannot} be
eliminated are those which are sums of monomials for which $x_k^{(s)}$ and
$\xi_k^{(s)}$ have equal integer exponents for all
$k=1, \ldots, d$.

Concretely, we can compute  $W_{n}$  from \eqref{trans_sol_quantum} as follows. We
assume that $H^{(n-1)}_{n;\text{ Im}}$ is the linear combination of
$L$ monomials of order $n$,

\begin{equation}
H^{(n-1)}_{n;\text{ Im}} = \sum_{l=1}^L h_l \, \hbar^{j_l} \prod_{k=1}^d
x_{k}^{\alpha_{k;l}} \xi_{k}^{\beta_{k;l}},
\label{sc_n_order_image_quantum}
\end{equation}

\noindent with $2j_l + \sum_{k=1}^d \alpha_{k;l} + \beta_{k;l} = n$ for all
$l=1,\dots,L$, and for all $l=1,\dots,L$, there is at least one $k=1,\dots,d$
for which  $\alpha_{k;l} \ne \beta_{k;l}$ (i.e., the vectors
$(\alpha_{1;l},\ldots,\alpha_{d;l})$ and $(\beta_{1;l},\ldots,\beta_{d;l})$
are different for all $l=1,\dots,L$).  Upon
inspecting  \eqref{eq:Dmonomial_quantum},  and using \eqref{trans_sol_quantum}, we
see that  a suitable generating function is given by

\begin{equation}
W_{n} = \sum_{l=1}^L \frac{h_l}{\lambda(\beta_{1;l}-\alpha_{1;l}) +
\sum_{k=2}^{d} \ui \omega_k(\beta_{k;l}-\alpha_{k;l})}
\hbar^{j_l} \prod_{k=1}^d x_{k;l}^{\alpha_{k;l}} \xi_{k;l}^{\beta_{k;l}}\,.
\end{equation}

As mentioned above this solution of the homological equation is unique if we
require $W_n$ to be in $\text{Im } \opD \big|_{\cW^n}$.



\subsubsection{Structure of the Hamilton operator in $N^\text{th}$ order quantum normal form}
\label{sec:structure_Hamilton_operator_saddle}

In the previous section we have seen how to obtain the quantum normal form to
order $N$ in the case where the equilibrium point is of
saddle-centre-$\cdots$-centre type. So far these computations were carried out
on the level of the symbols of the Hamilton operators.
We now discuss the implications for the
structure of the corresponding Hamilton operator in quantum normal form itself.

In the classical case in Sections~\ref{sec:solvhomequ}
and~\ref{sec:classicalintegrals}
we have shown that in each monomial of the Hamilton function in $N^\text{th}$
order classical normal form the coordinate pairs $(x_k,\xi_k)$ (or
equivalently $(q_k,p_k)$), $k=1,\ldots,d$, occur with equal integer exponents and that
this implies that the Hamilton function in $N^\text{th}$
order classical normal form is effectively a function of $d$ integrals,
see Equation~\eqref{eq:def_K}.

In the previous section we saw that in the monomials that form the symbol of a Hamilton operator in $N^\text{th}$
order quantum  normal form the coordinate pairs $(x_k,\xi_k)$ (or
equivalently $(q_k,p_k)$), $k=1,\ldots,d$, again have equal integer exponents. Hence, the
symbol is effectively a function of $I=p_1q_1$, $J_k=\frac12(p_k^2+q_k^2)$,
$k=2,\dots,d$.
We will now show that analogously to \eqref{eq:def_K} the Hamilton operator in $N^\text{th}$
order quantum  normal form is a function of the $d$ operators
\begin{equation} \label{eq:defIhatJhat}
\quad \hat{I}:=\Op[I] \,\, , \qquad  \hat{J_k}:=\Op[J_k]\,\, ,\quad k=2,\ldots,d\,,
\end{equation}
see Equations~\eqref{eq:opJ} and \eqref{eq:opI}. \New{To this end recall that the Hamilton operator in quantum normal form is localised 
near the equilibrium point. We will say that two operators $\Op[A]$ and $\Op[B]$ are equal near a point $z=(q,p)$ in phase space if their symbols $A$ and $B$ are equal in a neighbourhood of $z$. To indicate this we write 
\begin{equation}
\Op[A]\equiv_z\Op[B]\,\, .
\end{equation}
}

\begin{thm}\label{thm:QNF}

Let $\Op[H_{\text{QNF}}^{(N)}]$ be a Hamilton operator in $N^{\text{th}}$
order quantum normal form with respect to an equilibrium point of its principal symbol of
saddle-centre-$\cdots$-centre type, and assume
furthermore that the frequencies $\omega_2,\ldots, \omega_d$
associated with the $d-1$ centres are linearly independent
over $\Z$. Then there
exists a polynomial $\Hqnf^{(N)}:\R^d\to \R$ of order $[N/2]$ such that 
\begin{equation}\label{eq:defQNF}
\Op[H_{\text{QNF}}]\equiv_0\Hqnf^{(N)}(\hat{I},\hat{J_2}, \dots , \hat{J_d})\,.
\end{equation}
\end{thm}

In this theorem $[N/2]$ denotes the integer part of $N/2$. 
The proof of this Theorem is based on the following

\begin{lem}\label{lem:int-n-exp}
Let $I=pq$, $J=\frac{1}{2}(p^2+q^2)$, and $\hat{I}=\Op[I]$,
$\hat{J}=\Op[J]$, respectively, then there are integers
$\Gamma_{n,k}$ such that for any $n\in\N$, 
\begin{equation}\label{eq:In-exp}
\Op[I^n]=\sum_{k=0}^{[n/2]}(-1)^k\Gamma_{n,k}\bigg(\frac{\hbar}{2}\bigg)^{2k}\hat{I}^{n-2k}\,\,,
\end{equation}
and

\begin{equation}\label{eq:Jn-exp}
\Op[J^n]=\sum_{k=0}^{[n/2]}\Gamma_{n,k}\bigg(\frac{\hbar}{2}\bigg)^{2k}\hat{J}^{n-2k}\,\,.
\end{equation}
Here $[n/2]$ denotes the integer part of $n/2$, and the coefficients $\Gamma_{n,k}$ are determined by the recursion relation
\begin{equation}
\Gamma_{n+1,k}=\Gamma_{n,k}+n^2\Gamma_{n-1,k-1}\,\,\quad
\text{for}\,\, k\geq 1 \label{eq:recursion-a}
\end{equation}
and $\Gamma_{n,0}=1$.
\end{lem}


\begin{proof}

We start by considering the case of $I=p q$. The strategy will be
to use the Weyl calculus to determine the symbol of $\Op[I^n]$ as
a function of $I$, and then to invert this relation. The symbol of
$\Op[I^n]$ is $I^{*n}:=I*I*\cdots*I$, the $n$-fold star product of
$I$. Using $I=p q$ and the definition of the star product in \eqref{eq:star-product} we find the
recursion relation 
\begin{equation} \label{eq:Ipowerrecursion}
I*I^n=I^{n+1}+\bigg(\frac{\hbar}{2}\bigg)^2 n^2 I^{n-1}\,\, .
\end{equation}
This can be rewritten as 
\begin{equation}\label{eq:recursion}
\Op[I^{n+1}]=\hat{I}\Op[I^n] -\bigg(\frac{\hbar}{2}\bigg)^2 n^2
\Op[I^{n-1}]\,,
\end{equation}
which can be used to determine the $ \widehat{I^n}:=\Op[I^n] $
recursively. If we insert the ansatz \eqref{eq:In-exp} into the recursion
relation \eqref{eq:recursion} we find the recursion for the coefficients \eqref{eq:recursion-a}.

In order to show the validity of Equation~\eqref{eq:Jn-exp} we apply the same strategy and find instead of
\eqref{eq:recursion}
\begin{equation}\label{eq:recursionJ}
\Op[J^{n+1}]=\hat{J}\Op[J^n] +\bigg(\frac{\hbar}{2}\bigg)^2 n^2
\Op[J^{n-1}]\,,
\end{equation}
and inserting now \eqref{eq:Jn-exp} as an ansatz into this equation leads again to the
relation \eqref{eq:recursion-a} for the coefficients.
\end{proof}

We note that the closed formulae for $\hat{I}^n$ and $\hat{J}_k^n$ given
in \cite{Cre90} are not correct. We now prove Theorem~\ref{thm:QNF}.

\begin{proof}[Proof of Theorem \ref{thm:QNF}].
It follows from our construction that the
symbol of a Hamilton operator in quantum normal form is \New{near $(q,p)=(0,0)$} a
polynomial in $I$, $J_k$, $k=2,\ldots,d$ that can be written in
the following form:

\begin{equation}
  H_{\text{QNF}}^{(N)} = \sum_{l=1}^L h_l \hbar^{j_l} I^{\alpha_{1;l}}
  J_2^{\alpha_{2;l}}\cdots J_d^{\alpha_{d;l}}\,,
\end{equation}
where $2j_l + 2 \sum_{k=1}^d {\alpha_{k;l}} \le N$, or equivalently $j_l +  \sum_{k=1}^d {\alpha_{k;l}} \le N/2$, for all $l=1,\ldots,L$.
For $\Op[ H_{\text{QNF}}^{(N)} ]$ we thus find
\begin{equation}\label{eq:hqn234}
\Op[  H_{\text{QNF}}^{(N)} ]= \sum_{l=1}^L h_l \hbar^{j_l} \Op[I^{\alpha_{1;l}}]
  \Op[J_2^{\alpha_{2;l}}] \cdots \Op[J_d^{\alpha_{d;l}}]\,.
\end{equation}
If we insert the  expansions from Lemma \ref{lem:int-n-exp} into
\eqref{eq:hqn234} we obtain
\begin{equation}\label{eq:horriblelongexpression}
\begin{split}
\Op[  H_{\text{QNF}}^{(N)} ]= \sum_{l=1}^L h_l \hbar^{j_l}
\sum_{k_1=0}^{ [\alpha_{1;l}/2] }
\sum_{k_2=0}^{ [\alpha_{2;l}/2] }
\cdots
\sum_{k_d=0}^{ [\alpha_{d;l}/2] } &
(-1)^{\alpha_{1;l}}
\Gamma_{\alpha_{1;l},k_1}\cdots \Gamma_{\alpha_{d;l},k_d} \times \\
&\times
\bigg(\frac{\hbar}{2}\bigg)^{2(k_1+\ldots+k_d)}
\hat{I}^{\alpha_{1;l}-2 k_1}  \hat{J}_2^{\alpha_{2;l}-2 k_2} \cdots
\hat{J}_d^{\alpha_{d;l}-2 k_d} \,.
\end{split}
\end{equation}
Since $j_l +  \sum_{k=1}^d {\alpha_{k;l}} \le N/2$ for all $l=1,\ldots,L$
it follows that the RHS of \eqref{eq:horriblelongexpression}
is a polynomial  of
order $[N/2]$ in $\hat{I},\hat{J}_2,\ldots ,\hat{J}_d$. This polynomial
defines the function  $\Hqnf^{(N)}$.
\end{proof}

We note that for $\hbar\to 0$ the polynomial $\Hqnf^{(N)}$ tends to the
polynomial $K_{\text{CNF}}^{(N)}$ defined in \eqref{eq:def_K} that gives the
$N^{\text{th}}$ order  classical normal form as a function of the integrals
$I$ and $J_k$, $k=2,\dots,d$. Though this is obvious from the proof of
Theorem~\ref{thm:QNF} it is worth mentioning that
in general the coefficients in the polynomial $\Hqnf^{(N)}$ differ from the
polynomial that is obtained from writing $H_{\text{QNF}}^{(N)}$ as a
function of $I$ and $J_k$, $k=2,\dots,d$. We will see this in the example
presented in Sec.~\ref{sec:1Dpotentialbarriers}.

\rem{
Since the remainder term $O_{N+1}$ in \eqref{eq:SCNF&QNF} is small
({\bf In what norm?}) in a neighbourhood of the equilibrium point
$z_0$, the local dynamics of the system around $z_0$ can be
expected to be well described by $\Hqnf^{(N)}(\hat{I},\hat{J}_2,
\dots , \hat{J}_d)$. We will verify this later on.
}

Theorem~\ref{thm:QNF} is a crucial result. It tells us that the truncated quantum
normal form simply is a polynomial in the operators $\hat{I}$ and $\hat{J}_k$,
$k=2,\ldots,d$,  whose spectral properties are well known. As we will see in
more detail in Sections~\ref{sec:smatrix} and \ref{sec:resonances} this will allow us to compute quantum reaction rates and quantum
resonances with high efficiency.


\subsection{Quantum normal form for one-dimensional potential barriers}
\label{sec:1Dpotentialbarriers}

In the following we present the explicit computation of the
quantum normal form for Hamilton operators of
one-dimensional systems of type ``kinetic plus potential'' where the potential has a  maximum.
It is important to point out that
the applicability of the normal form algorithms -- both classical and quantum -- are not restricted
to systems  of the form ``kinetic plus potential''
(i.e., for example Coriolis terms in the Hamiltonian function or Hamilton operator due to a magnetic field
or a rotating coordinate frame are allowed).
Since even for this simple one-dimensional problem the expressions for the symbols
and operators involved  soon become very lengthy we will
carry out the quantum normal form algorithm only to order 4.
We note that we implemented the normalisation algorithm in the programming language {\sf C}$++$.
In our object oriented implementation the number of dimensions  and the order of truncation of the normal form can be chosen
arbitrarily. This {\sf C}$++$ program will be used to compute the high order quantum
normal forms for the more complicated examples given in Section~\ref{sec:examples}.


For now let us consider a Hamilton operator of the form
\begin{equation} \label{eq:Hamilton_op_def_1D_barrier}
\widehat{H} = -\frac{\hbar^2}{2m} \frac{\ud^2}{\ud q^2}+V(q)\,,
\end{equation}
where the potential $V$ is assumed to have a (non-degenerate) maximum at $q=q_0$.
The Weyl symbol of $\widehat{H}$ is given by
\begin{equation}
H(\hbar,q,p) = \frac{1}{2m}p^2+V(q)\,,
\end{equation}
i.e., $\Op[H]=\widehat{H}$.
Since the symbol $H$ does not depend on $\hbar$, the symbol agrees with the
principal symbol.
Hamilton's equations for the Hamiltonian function given
by $H$ then have an equilibrum point at $(q,p)=(q_0,0)$ which is of
saddle stability type,  i.e., the matrix associated with the linearisation of
the Hamiltonian vector field about the equilibrium point has a pair of  real
eigenvalues $\pm \lambda$. Here $\lambda$ is given by
\begin{equation}
\lambda = \sqrt{-\frac{1}{m} V''(q_0)}\,.
\end{equation}

\rem{
The Taylor expansion of
$H$ to second order about $(q,p)=(q_0,0)$ gives
\begin{equation}
\label{eq:eckart_quadratic}
H = V_0 + \frac{1}{2m}p^2 - \frac12 m \lambda^2 q^2
\end{equation}
with
\begin{equation}
V_0 = V(0) \, \,, \quad\lambda = \sqrt{-\frac{1}{m} V''(0)} \,.
\end{equation}
} 

The first two steps in the sequence of transformations
\eqref{eq:transformation-sequence_quantum}
serve to shift the equilibrium point of the (principal) symbol to the origin and to simplify the
quadratic part of the symbol. As mentioned in Sec.~\ref{sec:compscnf}, it
follows from Lemma~\ref{lem:ex-eg} (exact Egorov) that the transformations of
the symbol $H$ to achieve these goals agree with the corresponding classical transformations.

Classically, we  shift the equilibrium point to the origin of the coordinate
system by transforming the coordinates according to
\begin{equation}
(q,p) \mapsto (q-q_0,p)\,.
\end{equation}
For completeness, we note that this transformation can be obtained from the
time one map of the flow generated by the first order polynomial
\begin{equation}
W_1(q,p)  
=  - q_0 p \,,
\end{equation}
i.e., $\flow_{W_1}^1(q,p)=(q-q_0,p)$.
The Weyl quantisation of $W_1$ is given by
\begin{equation}
\Op[W_1] = q_0 \,\ui \hbar \frac{\ud }{\ud q} \,.
\end{equation}
It follows from Lemma~\ref{lem:ex-eg} that
\begin{equation}
\ue^{\frac{\ui}{\hbar}\Op[W_1]} \Op[H]  \ue^{- \frac{\ui}{\hbar}\Op[W_1]}
\end{equation}
has the symbol
\begin{equation}
   H^{(1)} (\hbar,q,p) = H \circ \flow_{W_1}^{-1} (\hbar,q,p) = H(\hbar,q+q_0,p)
   = \frac{1}{2m} p^2 + V(q+q_0) \,.
\end{equation}

We now want to find a unitary transformation such that the quadratic part of
the symbol $H^{(2)}$ of the transformed Hamilton operator assumes the form
\begin{equation}
H_2^{(2)}(\hbar,q,p) = \lambda \,p\, q\,.
\end{equation}
Classically, this is achieved by the transformation
\begin{equation} \label{eq:quadratic_trafo_first_example}
(q,p) \mapsto \bigg(\sqrt{m\lambda}\, q,\frac{1}{\sqrt{m\lambda}} p\bigg)
\end{equation}
followed by the 45$^\circ$ rotation
\begin{equation} \label{eq:quadratic_trafo_second_example}
(q,p) \mapsto  \bigg( \frac{1}{\sqrt{2}}(p+q) ,\frac{1}{\sqrt{2}}(p-q)\bigg) \,.
\end{equation}
Both these transformations are symplectic.

Again for completeness, we note that
the transformation
\eqref{eq:quadratic_trafo_first_example} can be
obtained from the time one map of the
flow generated by
\begin{equation}
W_2(q,p) = \ln\big( \sqrt{m \lambda}\big) \, p\,q\,,
\end{equation}
i.e.,
\begin{equation}
\flow^1_{W_2} (q,p) = \bigg(\sqrt{m\lambda}\, q,\frac{1}{\sqrt{m\lambda}} p\bigg)\,.
\end{equation}
The Weyl quantisation of $W_2$ is given by
\begin{equation}
\Op[W_2] = \ln\big( \sqrt{m\lambda}\big)\,\frac{\hbar}{\ui}\bigg( q \frac{\ud}{\ud
  q} + \frac12 \bigg)\,,
\end{equation}
see Equation~\eqref{eq:opI}.
The transformation
\eqref{eq:quadratic_trafo_second_example} can be obtained from the time one map of the
flow generated by 
\begin{equation}
W'_2(q,p) = \frac{\pi}{4}\frac12  ( q^2 + p^2) \,,
\end{equation}
which gives
\begin{equation}
\flow^1_{W'_2} (q,p) = \bigg( \frac{1}{\sqrt{2}}(p+q) ,\frac{1}{\sqrt{2}}(p-q)   \bigg)\,, 
\end{equation}
see the example after Lemma~\ref{lem:ex-eg} (exact Egorov), \eqref{eq:symbol-45}. The Weyl quantisation of $W'_2$ is given by
\begin{equation}
\Op[W'_2] = \frac{\pi}{2} \bigg( -\frac{\hbar^2}{2}\frac{\ud^2}{\ud q^2} +
\frac12 q^2\bigg)\,,
\end{equation}
see Equation~\eqref{eq:opJ}.

Using Lemma~\ref{lem:ex-eg} it follows that the symbol of
\begin{equation}
\Op[H^{(2)}]=
\ue^{\frac{\ui}{\hbar} \Op[W'_2] }\, \ue^{\frac{\ui}{\hbar} \Op[W_2]}\,
\Op[H^{(1)}] \,
\ue^{-\frac{\ui}{\hbar} \Op[W_2]} \, \ue^{-\frac{\ui}{\hbar} \Op[W'_2]}
\end{equation}
is given by
\begin{equation} \label{eq:normal_general}
\begin{split}
H^{(2)}(\hbar,q,p) &= H^{(1)} \circ \flow_{W'_2}^{-1} \circ \flow_{W_2}^{-1}
(\hbar,q,p) \\
& = V_0 + \lambda \, q\, p  + \sum_{k=3}^\infty \sum_{n=0}^k V_{n;k-n} p^{n} q^{k-n}
=: \sum_{k=0}^\infty H_k^{(2)} (\hbar,q,p)\,,
\end{split}
\end{equation}
where
\begin{equation}
H_0^{(2)}(\hbar,q,p) = V_0:= V(q_0)\,,\quad
H_1^{(2)} (\hbar,q,p) =0\,,\quad  H_2^{(2)}(\hbar,q,p) = \lambda \, p \,q \,.
\end{equation}
The coefficients of the monomials in (\ref{eq:normal_general}) of cubic or higher degree  are
\begin{equation}
V_{n;j} = (-1)^n \frac{1}{n!j!} \frac{1}{(2m\lambda)^{(n+j)/2}}
 \frac{ \mbox{d}^{n+j} V(q_0)}{\mbox{d} q^{n+j}} \,, \qquad n+j\ge3\,.
\end{equation}

So far, i.e., up to order 2, the transformations involved in the quantum normal form
algorithm agree with their counterparts in the classical normal form
algorithm. We now want
to study the next steps in the sequence \eqref{eq:transformation-sequence_quantum}  which give the quantum normal form of
order three and four.
To make these transformations well defined we from now on assume that we use the scheme
outlined in Sec.~\ref{sec:localising} to localise the Hamilton operator $H^{(2)}$ and the
operators which will generate the required unitary transformations about the origin.
The monomials in the third and fourth order polynomials $H_3^{(2)} $ and $H_4^{(2)} $ have coefficients
\begin{eqnarray}
\label{eq:V30}
V_{3;0} &=& - V_{0;3} = -\frac13 V_{2;1} = \frac13 V_{1;2}
= -\frac16 \frac{1}{(2m\lambda)^{3/2}} V^{'''}(q_0) \,, \\
V_{4;0} &=& V_{0;4} = -\frac14 V_{3;1} = - \frac14 V_{1;3}  =\frac16 V_{2;2}
= \frac{1}{24} \frac{1}{(2m\lambda)^{2}} V^{''''}(q_0) \,,
\end{eqnarray}
respectively, where the primes denote derivatives.

It follows from Equation~\eqref{eq:higher-ordertransform} that for $W_3\in
\cWl^{3}$, the symbol of the transformed operator
\begin{equation}
\Op[H^{(3)}] = \ue^{\frac{\ui}{\hbar} \Op[W_3]} \Op[H^{(2)}] \ue^{-\frac{\ui}{\hbar} \Op[W_3]}
\end{equation}
is given by
\begin{equation}
H^{(3)} = H_0^{(3)} + H_1^{(3)} + H_2^{(3)} + H_3^{(3)} + H_4^{(3)} + \dots \,,
\end{equation}
where following Lemma~\ref{lem:lower_order_scnf}, the terms $H_k^{(3)}$ and $H_k^{(2)}$ agree for $k \le 2$, and
\begin{eqnarray}
H_3^{(3)} &=& H_3^{(2)} + \adM_{W_3} H_2^{(2)} = H_3^{(2)} + \{W_3, H_2^{(2)} \}\,,  \label{eq:hom_example} \\
H_4^{(3)} &=& H_4^{(2)} + \adM_{W_3} H_3^{(2)} + \frac12 \left[\adM_{W_3} \right]^2 H_2^{(2)}\,. \label{eq:H43}
\label{eq:exampleH43}
\end{eqnarray}
Equation~\eqref{eq:hom_example} is the homological equation.
Introducing the operator
\begin{equation}
\opD = \{H_2^{(2)},\cdot \}
\end{equation}
the homological equation takes the form
\begin{equation}
  H_3^{(3)} = H_3^{(2)} - \opD W_3 \,,
\end{equation}
which agrees with the form of the homological equation in
Lemma~\ref{lem:hom_eq_scnf}.
Following Sec.~\eqref{sec:examp_comp_scnf} we need to solve the homological
equation, i.e., choose $W_3$, such that $\opD H_3^{(3)}=0$.
Since $\cW^3 = \text{Im } \opD  \big|_{\cW^3}$, or equivalently
$\text{Ker } \opD  \big|_{\cW^3}=\{0\}$, we have to choose $W_3$ such that $H_3^{(3)}=0$.
From  \eqref{eq:hom_example}
we see that this is achieved by setting
\begin{eqnarray}
W_3(\hbar,q,p) &=& - \sum_{n=0}^3 \frac{1}{\lambda(2n - 3)} V_{n;3-n} p^{n} q^{3-n}\,,\\
&=& -\frac{V_{3;0}}{3\lambda} \left(  p^3 -9 p^2q -9 p q^2 +  q^3 \right) \,.
\end{eqnarray}
Inserting this $W_3$ into (\ref{eq:exampleH43}) gives
\begin{equation}
\label{eq:H43new}
\begin{split}
H_4^{(3)}(\hbar,q,p) = &
V_{4;0}\left( p^4 -4 p^3 q + 6 p^2 q^2 - 4 p q^3 + q^4 \right)-\\
& \frac{V_{3;0}^2}{\lambda}\left( 3p^4+12p^3q-30p^2q^2+12p q^3 +3q^4-4\hbar^2 \right)  \,.
\end{split}
\end{equation}
Note the occurrence of the term inolving $\hbar^2$. It is a consequence of the second term on the right hand side of \eqref{eq:H43}
which involves the Moyal bracket of two polynomials which are of degree higher than
two for which the Moyal bracket no longer coincides with the Poisson bracket.

Using Equation~\eqref{eq:higher-ordertransform} again we see that for $W_4\in \cWl^{4}$,
the symbol of the transformed operator
\begin{equation}
\Op[H^{(4)}] = \ue^{\frac{\ui}{\hbar} \Op[W_4]} \Op[H^{(3)}] \ue^{-\frac{\ui}{\hbar} \Op[W_4]}
\end{equation}
is given by
\begin{equation}
H^{(4)} = H_0^{(4)} + H_1^{(4)} + H_2^{(4)} + H_3^{(4)} + H_4^{(4)} + \dots\,,
\end{equation}
where it again follows from Lemma~\ref{lem:lower_order_scnf} that
$H_k^{(4)}=H_k^{(3)}$ for $k \le 3$. For $k=4$ we obtain the homological equation
\begin{equation}
H_4^{(4)} = H_4^{(3)} + \adM_{W_4} H_2^{(3)} = H_4^{(3)} -  \opD  W_4 \,.
\end{equation}
We need to choose $W_4$ such that $\opD H_4^{(4)} = 0$. We therefore decompose
$H_4^{(3)}$ according to
\begin{equation}
 H_4^{(3)}   = H^{(3)}_{4;\text{Ker}} + H^{(3)}_{4;\text{Im}}\,,
\end{equation}
where $ H^{(3)}_{4;\text{Ker}} \in \text{Ker } \opD\big|_{\cW^4} $ and  $ H^{(3)}_{4;\text{Im}} \in \text{Im } \opD\big|_{\cW^4} $.
It follows from Sec.~\ref{sec:examp_comp_scnf} that $H^{(3)}_{4;\text{Ker}}$
consists of all monomials  of $ H_4^{(3)}$ in which $p$ and $q$ have the same
integer exponent and $H^{(3)}_{4;\text{Im}}$
consists of all monomials  of $ H_4^{(3)}$ in which $p$ and $q$ have different
integer exponents. We thus have
\begin{equation}
 H^{(3)}_{4;\text{Ker}} = 6 V_{4;0}  p^2 q^2  +
\frac{V_{3;0}^2}{\lambda}\left( 30p^2q^2 + 4\hbar^2 \right)
\end{equation}
and
\begin{equation}
 H^{(3)}_{4;\text{Im}} = V_{4;0}\left( p^4 -4 p^3 q  - 4 p q^3 + q^4 \right)-
 \frac{V_{3;0}^2}{\lambda}\left( 3p^4+12p^3q + 12p q^3 +3q^4 \right) \,.
\end{equation}
To achieve $\opD H_4^{(4)} = 0$ we choose
\begin{equation}
W_4(\hbar,q,p) =
\frac{V_{3;0}}{4\lambda^2}
\left(
3 p^4 + 24 p^3 q -24 p q^3 -3 q^4
\right) -
\frac{V_{4;0}}{4\lambda}\left( p^4- 8p^3q +8 p q^3 - q^4  \right)\,.
\end{equation}

We thus get
\begin{equation}
H^{(4)}(\hbar,q,p) = V_0 + \lambda p q + 6\,V_{4;0}  p^2 q^2  +
\frac{V_{3;0}^2}{\lambda}\left(  30p^2q^2 + 4\hbar^2 \right)  + O_5\,,
\end{equation}
where the remainder $O_5$ is defined according to
Definition~\ref{def:remainder}.
Neglecting $O_5$ gives the symbol of the 4$^{\text{th}}$ order quantum normal form.
In order to get the corresponding operator we have to replace the factors $I=p q$ by the operator
$\hat{I}=\Op[I]$. To this end we use the recurrence~\eqref{eq:In-exp} in
Lemma~\ref{lem:int-n-exp} to get
\begin{equation}\label{eq:I2_quantisation}
\Op[I^2] = \hat{I}^2 - \frac{\hbar^2}{4}\,.
\end{equation}
The 4$^{\text{th}}$ order quantum normal form of the operator $\widehat{H}$ in \eqref{eq:Hamilton_op_def_1D_barrier} is thus given by
\begin{eqnarray} \label{eq:opH4}
K_{\text{QNF}}^{(4)} (\hat{I})&=&
V_0 + \lambda \hat{I} + \left( 30 \frac{V_{3;0}^2}{\lambda} + 6 V_{4;0}   \right) \hat{I}^2 - \frac{\hbar^2}{2} \left( 7 \frac{V_{3;0}^2}{\lambda}
 + 3 V_{4;0} \right)
\\
&=&
V_0  + \lambda \hat{I}  + \frac{1}{16 m^2 \lambda^2}
 \left( \frac{5}{3 m \lambda^2} \big( V'''(q_0)\big)^2 +  V''''(q_0)  \right) \hat{I}^2  \nonumber \\
& & \qquad -  \frac{1}{64 m^2\lambda^2}
\left( \frac{7}{ 9 m \lambda^2} \big(V'''(q_0)\big)^2 +
 V''''(q_0)  \right)\hbar^2 \,\, .
\end{eqnarray}
This gives the first correction term to the well known quadratic approximation which consists of
approximating the potential barrier by an inverted parabola.
The corresponding classical normal form is given by
\begin{equation}
K_{\text{CNF}}^{(4)} (I)
=
V_0  + \lambda I  + \frac{1}{16 m^2 \lambda^2}
 \left( \frac{5}{3 m \lambda^2} \big( V'''(q_0)\big)^2 +  V''''(q_0)  \right) I^2\,.
\end{equation}
We see that the polynomials  $K_{\text{QNF}}^{(4)}$ (in $\hat{I}$) has two
more terms than the polynomial $K_{\text{CNF}}^{(4)}$ (in $I$).  These are the
terms involving $\hbar$, and their occurrence is due to the Moyal bracket (see the comment
following Equation~\eqref{eq:H43new}) which enters the quantum normal form
computation on the level of the symbols and the Weyl quantisation of powers of
the classical integral $I=p\,q$ (see \eqref{eq:I2_quantisation}) which is
required to obtain the Hamilton operator from its symbol.


%% file: class_phase_space.tex
\section{Classical Reaction Dynamics and Reaction Probabilities}
\label{sec:classical}

In this section we give an overview of the theory of reaction
dynamics that is firmly rooted in the dynamical arena of phase
space and has recently been developed in
\cite{WWJU01,UJPYW01,WaalkensBurbanksWigginsb04,WaalkensWiggins04,WaalkensBurbanksWiggins05,
WaalkensBurbanksWiggins05b, WaalkensBurbanksWiggins05c}. This
section is organised as follows. In
Sec.~\ref{sec:phasespacestruct} we describe the geometric
structures in \emph{phase space} near an equilibrium point of
saddle-centre-$\cdots$-centre stability type (see
Sec.~\ref{sec:examp_comp_cnf}) that control the classical dynamics
of reactions.  These phase space structures are ``realised''
through the classical normal form, and details of this are given
in Sec.~\ref{sec:nfcpss} where we also provide a detailed
discussion of how these phase space structures constrain
trajectories of Hamilton's equations. In Sec.~\ref{sec:fls} we
describe how the integrability of the truncated normal form gives
rise to the foliation of the phase space near the saddle by Lagrangian manifolds.
This will be of central importance for the quantum mechanics of reactions as
we will see in Sec.~\ref{sec:smatrix}.
In Sec.~\ref{sec:classicalrate} we will show how the normal form can
be used to compute the directional flux through the dividing
surface. As we will see the normal form obtained from truncating
the normal form algorithm at a suitable order gives a very
accurate description of the local dynamics. Means to verify the
accuracy are discussed in Sec.~\ref{sec:accuracy_NF}. While the
normal form technique is ``locally applicable'' in a neighbourhood
of the  reaction region, in Sec.~\ref{sec:classicglobal} we
discuss how the local structures mentioned above can be globalised
in a way that their influence on reactions outside this ``local''
region can be determined.  Finally, in
Sec.~\ref{sec:fluxfluxMiller} we comment on the flux-flux
autocorrelation function formalism to compute classical reaction
probabilities that is frequently utilised in the chemistry
literature, its relation to our phase space theory,  and the
computational benefits of our approach over the flux-flux
autocorrelation function formalism.
\rem{
Finally, in Section~\ref{sec:EMM_3dof_ex} we will show how these
ideas can be implemented for the example of a three-degree-of-freedom
system -- a Hamiltonian system consisting of an Eckart barrier that is coupled
to two Morse oscillators. \todo{move this stuff to the section Examples?}
} 


\subsection{Phase Space Structures that Control Classical Reaction Dynamics: An Overview of the Geometry}
\label{sec:phasespacestruct}

Our starting point is an  equilibrium point of Hamilton's
equations of saddle-centre-$\cdots$-centre stability type. Near
(and we will discuss what we mean by ``near'' in
Section~\ref{sec:accuracy_NF})
such equilibrium points there exist lower
dimensional manifolds that completely dictate the dynamics of the
evolution of trajectories from reactants to products (or
vice-versa). The normal form theory developed in
Sec.~\ref{sec:CNF} provides a transformation to a new set of
coordinates, referred to as the {\em normal form coordinates}, in
which these manifolds can be identified and explicitly computed,
and then mapped back into the original, ``physical'' coordinates
via the normal form transformation. In this section we give a
brief description of  these phase space structures, and in
Sec.~\ref{sec:nfcpss} we will describe how they constrain
trajectories.

We let $E_0$ denote the energy of the saddle, and we consider a
fixed energy $E>E_0$ (and ``sufficiently close'' to $E_0$). We
will also restrict our attention to a certain neighbourhood $U$,
local to the equilibrium point.  We will defer  a discussion
exactly how this region is chosen to  
Sec.~\ref{sec:accuracy_NF}; suffice it to say for now that the region
is chosen so that an integrable nonlinear approximation to the
dynamics yields structures to within a given desired accuracy.

Near this equilibrium point the $(2d-1)$-dimensional energy
surface in the $2d$-dimensional phase space $\R^{2d}$ has the
structure of a ``spherical cylinder'' $S^{2d-2}\times\mathbb{R}$,
i.e., the Cartesian product of a $(2d-2)$-dimensional sphere
$S^{2d-2}$ and a line $\R$. The dividing surface that we construct
locally separates the energy surface into two components;
``reactants'' and ``products''. This dividing surface which we
denote by $\ts$ has the structure of  a $(2d-2)$-dimensional
sphere $S^{2d-2}$. It can be shown to have the following
properties:

\begin{itemize}

\item The only way that trajectories can evolve from the reactants
component to the  products component
(and vice-versa), without leaving the local region $U$, is by
crossing $\ts$. We refer to this property of $\ts$ as the
``bottleneck property''.\footnote{ Here we inserted the
restriction `without leaving the local region $U$' to exclude the
case where the dividing surface does not divide the full (global)
energy surface into two disjoint components. For example, two
regions in an energy surface might be connected by channels
associated with two different saddle-centre-$\cdots$-centre
equilibrium points. }

\item The dividing surface that we construct is free of local
recrossings; any trajectory which crosses $\ts$ must leave the
neighbourhood $U$ before it might possibly cross $\ts$ again.

\item A consequence of the previous property of the dividing
surface is that it minimizes the (directional) flux. It is thus
the optimal dividing surface sought for in variational transition
state theory \cite{WaalkensWiggins04}.

\end{itemize}

\rem{
In the normal form coordinates on the energy surface $H_{CNF}=E$ the
dividing surface $\ts$ is given by the equation $p_1=q_1$, which
defines a $(2d-2)$-sphere. It is of codimension one in the energy
surface. It divides the energy surface into two components: the
reactant region, given by $p_1-q_1>0$ in the normal form
coordinates, and the product region, given by $p_1-q_1<0$ in the
normal form coordinates.
} 

The dividing surface $\ts$ itself is
divided into two hemispheres: the {\em forward reactive
hemisphere} $\tsf$,
and the {\em backward reactive hemisphere}
$\tsb$.
The hemispheres $\tsf$ and $\tsb$ are topological $(2d-2)$-balls.
These two hemispheres are separated by the equator of $\ts$, which
itself is a sphere of dimension $(2d-3)$. On $\tsf$ and $\tsb$ the
Hamiltonian vector field is transversal to each of these surfaces.
This transversality is the mathematical manifestation of ``no
recrossing''. Heuristically, ``transversal'' means that the
Hamiltonian vector field ``pierces'' the surfaces, i.e., there is
not point where it is tangential to the surface. Now the
Hamiltonian vector field pierces the surfaces $\tsf$ and $\tsb$ in
opposite directions. Since the vector field varies smoothly from
point to point, it must be tangential to the  equator on which
$\tsf$ and $\tsb$ are joined.  More mathematically, the fact that
the Hamiltonian vector field is tangential to the equator means
that the equator is an {\em invariant manifold}. In fact, it is a
so-called {\em normally hyperbolic invariant manifold}
 (NHIM) \cite{Wiggins94}, denoted by $\nhim$, where normal
hyperbolicity means that the expansion and contraction rates
transverse to the manifold dominate those tangent to the manifold,
and there are an equal number of independent expanding and
contracting directions transverse to the manifold at each point on
the manifold. This implies that it is ``saddle like'' in terms of
stability (in our set-up there is one expanding direction and one
contracting direction normal to the NHIM at each point on the
NHIM).
Heuristically, one can think of it as a ``big saddle like
surface''. In fact, the
$(2d-3)$-dimensional NHIM is the energy surface of an
invariant subsystem which has $d-1$ degrees of freedom, i.e., one degree of
freedom less than the full system.
In chemistry terminology this subsystem is the ``activated complex'',
which may be thought of as representing an oscillating (unstable)
``supermolecule'' poised between reactants and products
\cite{Eyring35,Pechukas81,Miller98}.

Normally hyperbolic invariant manifolds have stable and unstable
manifolds, which themselves are invariant manifolds. In
particular, the  NHIM, $\nhim$,
has $(2d-2)$-dimensional stable and unstable manifolds $W^s(E)$
and $W^u(E)$ which are isoenergetic, i.e., contained in the energy
surface.
These invariant manifolds have the topology of spherical cylinders
$S^{2d-3}\times \R$. Since they are of codimension one in the energy surface, i.e.,
they are of one dimension less than the energy surface, they act
as impenetrable barriers. The importance of these particular
geometrical structures is that all reactive trajectories (both
forward and backward) must lie inside regions of the energy surface that are
enclosed by the NHIM's stable and unstable manifolds.
This can be described more precisely by first noting that $W^s(E)$
and $W^u(E)$ each have two branches that ``join' at the NHIM. We call
these branches of the {\em forward} and {\em backward} branches of $W^s(E)$
and  $W^u(E)$,
and denote them by $W^s_f(E)$, $W^s_b(E)$, $W^u_f(E)$ and $W^u_b(E)$, respectively.
We call the union of the forward branches,
$W_f (E) := W^s_f (E) \cup W^u_f (E)$, the {\em forward reactive spherical
cylinder}. Trajectories with initial conditions enclosed by $W_f (E)$ in the
reactants component of the energy
surface evolve towards the forward hemisphere of the dividing surface $\tsf$, cross
$\tsf$, and evolve into a region of the products component of the energy
surface that is enclosed by  $W_f (E)$.
Similarly, we call the union of the backward branches,
$W_b(E):=W^s_b(E)\cup W^u_b(E)$, the {\em backward reactive
spherical cylinder}. Trajectories with initial conditions
enclosed by $W_b(E)$ in the products component of the energy surface
evolve towards $\tsb$, cross $\tsb$, and evolve into
a region of the reactants component of the energy
surface that is enclosed by  $W_b(E)$.
\emph{All} forward  reactive trajectories are enclosed by $W_f(E)$ and all
\emph{all} backward reactive trajectories are enclosed by $W_b(E)$.
As we will see in the next section these structures can be computed from the
normal form developed in Section~\ref{sec:CNF}.


\subsection{The Normal Form Coordinates: Phase Space
  Structures and Trajectories of Hamilton's Equations}
\label{sec:nfcpss}

We now  describe how the phase space structures mentioned in the
previous section can be identified and computed from the normal
form algorithm, and how they influence trajectories of Hamilton's
equations. From the discussion in Section
\ref{sec:examp_comp_cnf}, after $N$ steps of the normal form
algorithm, we have constructed a coordinate transformation from
the original, ``physical'' coordinates to new, ``normal form''
coordinates
$(q^{(N)}_1,\ldots,q^{(N)}_d,p^{(N)}_1,\ldots,p^{(N)}_d)$, and in
these new coordinates the Hamiltonian truncated at order $N$ takes
the form

\begin{equation}  \label{ham_cnf_int}
\begin{split}
\Hcnf^{(N)} &= K_{\text{CNF}}^{(N)}(I^{(N)},J^{(N)}_2,\ldots,J^{(N)}_d) \\
&= E_0 + \lambda I^{(N)} + \omega_2
J^{(N)}_2 + \ldots + \omega_d J^{(N)}_d + \text{ higher order terms }\,,
\end{split}
\end{equation}
 where
\begin{equation} \label{nf_int}
I^{(N)} = q^{(N)}_1 p^{(N)}_1\,, \quad J^{(N)}_k =
\frac{1}{2} \left( \left(q^{(N)}_k \right)^2 + \left(p^{(N)}_k
\right)^2 \right)\,, \quad k=2, \ldots, d\,,
\end{equation}
and the higher order terms are of order greater than 1
and less than or equal to $[N/2]$ (in the integrals), see Equation~\eqref{eq:def_K} in Sec.~\ref{sec:examp_comp_cnf}.

\noindent The quantities (\ref{nf_int}) are integrals of the
motion (``conserved quantities''), i.e, they are constant on
trajectories of the Hamiltonian vector field given by the $N^{\rm
th}$ order classical normal form Hamiltonian \footnote{The fact
that there are $d$ constants of motion is a consequence of the
non-resonance assumption on the linear frequencies $\omega_k$,
$k=2,\ldots,d$. If there are resonances amongst the $\omega_k$, then
there will be fewer integrals.}. Henceforth we will drop the
superscripts $(N)$ for the sake of a less cumbersome notation, but
it should be understood that the normal form procedure is
truncated at some fixed order $N$.
\rem{
Special attention should be
given to the integral $I=q_1 p_1$ in the following discussion. One
could view it as the {\em reaction integral}. We will see in the
discussion that a trajectory must have $I >0$ for reaction to be
possible.
\todo{Why is/was this mentioned here??}
}

In the normal form coordinates, using \eqref{ham_cnf_int} and
\eqref{nf_int}, Hamilton's equations take the form

\begin{equation} \label{hameq_nf}
\begin{array}{crrr}
\dot{q}_1 = & \phantom{-} \frac{\partial K_{\text{CNF}}}{\partial I}(I, J_2, \ldots, J_d)\, q_1
\equiv & \phantom{-} \Lambda(I, J_2, \ldots, J_d)\, q_1\,,& \\[1.1ex]
\dot{p}_1 = & -\frac{\partial K_{\text{CNF}}}{\partial I}(I, J_2, \ldots, J_d) \, p_1
\equiv & -\Lambda(I, J_2, \ldots, J_d) \, p_1\,, & \\[1.1ex]
\dot{q}_k = & \phantom{-} \frac{\partial K_{\text{CNF}}}{\partial J_k}(I,J_2, \ldots,
J_d) \,p_k
\equiv & \phantom{-} \Omega_k(I,J_2, \ldots, J_d) \,p_k\,, & \\[1.1ex]
\dot{p}_k = & -\frac{\partial
K_{\text{CNF}}}{\partial J_i}(I, J_2, \ldots, J_d)  \, q_k
\equiv & -\Omega_k(I, J_2, \ldots, J_d ) \, q_k\,, & k=2, \ldots, d\,.
\end{array}
\end{equation}
These equations appear ``decoupled''. It is important to understand this
statement in quotations since  the equations are not ``decoupled'' in the usual
fashion. Nevertheless, effectively, this is the case since the
coefficient $\Lambda$ and the nonlinear frequencies $\Omega_k$,
$k=2,\ldots,d$,  are constant  on a given trajectory. This follows
from the fact that they are functions of the integrals $I$ and
$J_k$, $k=2,\ldots,d$. Hence, once the initial conditions for a
trajectory are chosen, then the coefficients of (\ref{hameq_nf})
are constant (in time), and in this sense the equations are
decoupled and can be easily integrated.
The reason we have this property is a result of the $d$
independent integrals given in (\ref{nf_int}).
However,  $\Lambda$ and the nonlinear frequencies $\Omega_k$,
$k=2,\ldots,d$, will generally vary from trajectory to trajectory and
the equations are hence not decoupled in the classical sense.
We could view them as being  ``decoupled on trajectories'' as a result of the $d$
integrals being constant on trajectories. In  mathematical terms
this means that the equations of motion are \emph{ integrable}.
The notion `integrability' can be viewed as a generalisation of the notion `separability'.
The latter refers to the property of the equations of motion
that allows to achieve a decoupling of the form
\eqref{hameq_nf} from a transformation that involves the
configuration space variables $q$ only (which then entails a transformation of the
momenta $p$ to give a symplectic transformation of the full phase space coordinates).
Historically, separability has played an important role in developing
approximate transition state theory and analyzing tunnelling
effects, see, e.g., \cite{jr,em,Miller76}. Indeed, if the full
dynamics is separable near the saddle point (in phase
space) then the construction of a dividing surface with no
recrossing is trivial and the choice of reaction coordinate is
``obvious''. However, it is important to point in the neighbourhood of a
saddle-centre-$\cdots$-centre equilibrium point the equations of
motions are in general \emph{not} separable but the normal form transformation
leading to the decoupling in \eqref{hameq_nf} in general involves a symplectic
transformation which mixes configuration and momentum variables.

The general solution of \eqref{hameq_nf} is given by

\begin{equation} \label{hameq_nf_solutions}
\begin{split}
q_1(t) & = A_1 \exp \big( \phantom{-}\Lambda(I, J_2, \ldots, J_d)\,t \big)\,,\\
p_1(t) & = B_1 \exp \big( -\Lambda(I, J_2, \ldots, J_d)\,t \big)\,,\\
q_k(t) & = A_k \sin \big( \Omega_k(I, J_2, \ldots, J_d)\,t+ \varphi_k \big)\,,\\
p_k(t) & = A_k \cos \big( \Omega_k(I, J_2, \ldots, J_d)\,t+ \varphi_k \big)\,,\quad
 k=2, \ldots, d\,,
\end{split}
\end{equation}
where the $A_1,\ldots,A_d$, $\varphi_2,\ldots,\varphi_d$ and $B_1$ are $2d$
constants determined by the initial conditions
$(q_1(0),\ldots,q_d(0),p_1(0),\ldots,p_d(0))$. The constants in \eqref{hameq_nf_solutions}
determine the integrals according to
\begin{equation}
I = A_1 \,B_1\,,\quad J_k = \frac12 A_k^2 \,,\quad k=2,\ldots,d\,.
\end{equation}
From the general solution \eqref{hameq_nf_solutions} we see that
the motion is generally hyperbolic (i.e., ``saddle like'') in the
plane of the coordinates $(q_1,p_1)$ associated with the saddle
and rotational in the planes of the coordinate pairs $(q_k,p_k)$,
$k=2,\ldots,d$, associated with the centre directions.

In the following, we show how the normal form, which is valid in
the neighbourhood of the saddle-centre-$\cdots$-centre equilibrium
point, gives {\em explicit formulae} for the various manifolds
described in Sec.~\ref{sec:phasespacestruct}. At the same time, we
show how trajectories of Hamilton's equations expressed in the
normal form coordinates, are constrained by these manifolds. Many
more details can be found  in
\cite{UJPYW01,WaalkensBurbanksWigginsb04}. The geometrical
illustrations that we give  are for three degrees of freedom.  In
fact, conceptually, the step from two to three degrees of freedom
is the big step; once the case of three degrees of freedom is well
understood, it is not difficult to incorporate more degrees of
freedom.  We begin by describing the local structure of the energy
surfaces.


\vspace*{0.3cm} \noindent {\bf The structure of an energy surface
near a saddle point: } For $E<E_0$, the energy surface consists of
two disjoint components.  The two components correspond to
``reactants'' and ``products.''  The top panel of
Fig.~\ref{fig:esurfs} shows how the two components project to the
various planes of the normal form coordinates.  The projection to
the plane of the saddle coordinates $(q_1,p_1)$ is bounded away
from the origin by the two branches of the hyperbola, $q_1p_1=I<0$, where $I$ is
given implicitly by the energy equation with the centre actions
$J_k$, $k=2,\dots,d$, set equal to zero:
$K_{\text{CNF}}(I,0,\ldots,0)=E<E_0$.  The projections to the
planes of the centre coordinates, $(q_k,p_k)$, $k=2,\dots,d$, are
unbounded.

\def\figesurfs{%
Projection of energy surfaces (turquoise regions) to the planes of
the normal form coordinates.  The energy surface in the top panel
has $E<E_0$; the energy surface in the bottom panel has $E>E_0$. }
\def\FIGesurfs{
\centerline{\includegraphics[width=12.0cm]{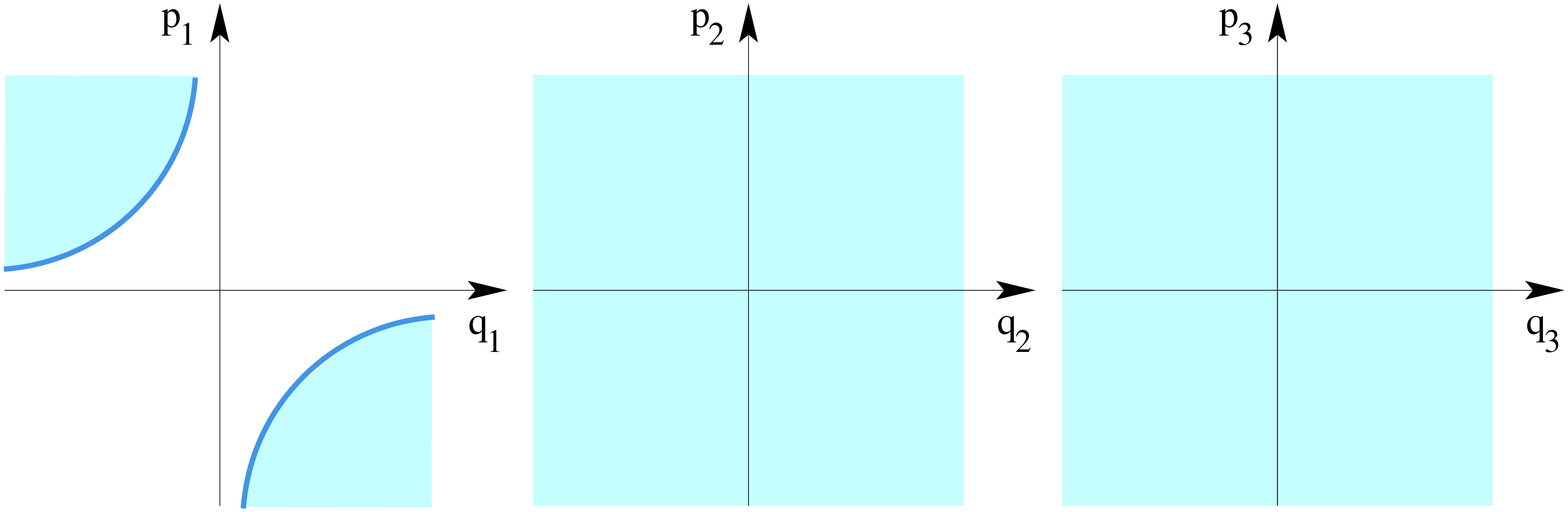}}
\centerline{\includegraphics[width=12.0cm]{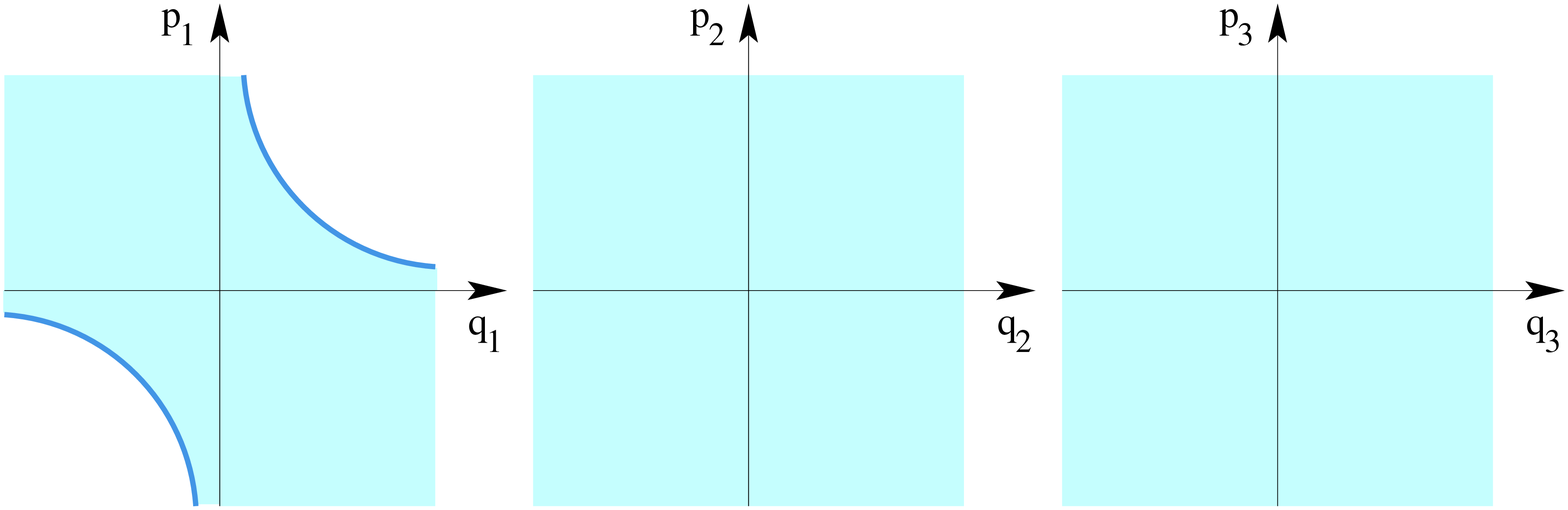}}
}
\FIGo{fig:esurfs}{\figesurfs}{\FIGesurfs}

At $E=E_0$, the formerly disconnected components merge (the energy
surface bifurcates), and for $E>E_0$ the energy surface has
locally the structure of a spherical cylinder, $S^{2d-2}\times
\R$.  Its projection to the plane of the saddle coordinates now
includes the origin.  In the first and third quadrants it is
bounded by the two branches of the  hyperbola, $q_1p_1=I>0$, where $I$ is again given
implicitly by the energy equation with all centre actions equal to
zero, but now with an energy greater than $E_0$:
$K_{\text{CNF}}(I,0,\dots,0)=E>E_0$. The projections to the planes
of the centre coordinates are again unbounded. This is illustrated
in the bottom panel of Fig.~\ref{fig:esurfs}.


%
\vspace*{0.3cm}
\noindent
{\bf The dividing surface and reacting and nonreacting trajectories:}
On an energy surface with $E>E_0$, we define the dividing surface by
$q_1=p_1$.  This gives a $(2d-2)$-sphere which we denote by $\ts$.
Its projection to the saddle coordinates simply gives a line segment
through the origin which joins the boundaries of the projection of the
energy surface, as shown in Fig.~\ref{fig:tst}.  The projections of the
dividing surface to the planes of the centre coordinates are bounded by
circles $(p_k^2+q_k^2)/2=J_k$, $k=2,\dots,d$, where $J_k$ is
determined by the energy equation with the other centre actions, $J_l$,
$l\ne k$, and the saddle integral, $I$, set equal to zero.  The
dividing surface divides the energy surface into two halves,
$p_1-q_1>0$ and $p_1-q_1<0$, corresponding to reactants and products.

As mentioned above, trajectories project to
hyperbolae in the plane of the saddle coordinates, and to circles
in the planes of the centre coordinates.  The sign of $I$
determines whether a trajectory is nonreacting or reacting, see
Fig.~\ref{fig:tst}.  Trajectories which have $I<0$ are nonreactive and for one branch of the hyperbola $q_1p_1=I$ they stay on the
reactants side and for the other branch they stay on the products side; trajectories with $I>0$ are reactive, and for one  branch of the hyperbola $q_1p_1=I$ they react
in the forward direction, i.e., from reactants to products, and for the other branch they react in the
backward direction, i.e., from products to reactants.  The
projections of reactive trajectories to the planes of the centre
coordinates are always contained in the projections of the dividing
surface. In this, and other ways, the geometry of the reaction is
highly constrained.  There is no analogous restriction on the
projections of nonreactive trajectories to the centre coordinates.

\def\figtst{%
Projection of the dividing surface and reacting and nonreacting
trajectories to the planes of the normal form coordinates. In the
plane of the saddle coordinates, the projection of the dividing
surface is the  dark red diagonal line segment, which has
$q_1=p_1$.  In the planes of the centre coordinates, the
projections of the dividing surface are the dark red discs.
Forward and backward reactive trajectories (yellow and blue)
project to the first and third quadrant in the plane of the saddle
coordinates, respectively, and pass through the dividing surface.
The red and green curves mark nonreactive trajectories on the
reactant side ($p_1-q_1>0$), and on the product side
($p_1-q_1<0$), of the dividing surface, respectively.  The
turquoise regions indicate the projections of the energy surface.
}
\def\FIGtst{
\begin{center}
\includegraphics[width=12.0cm]{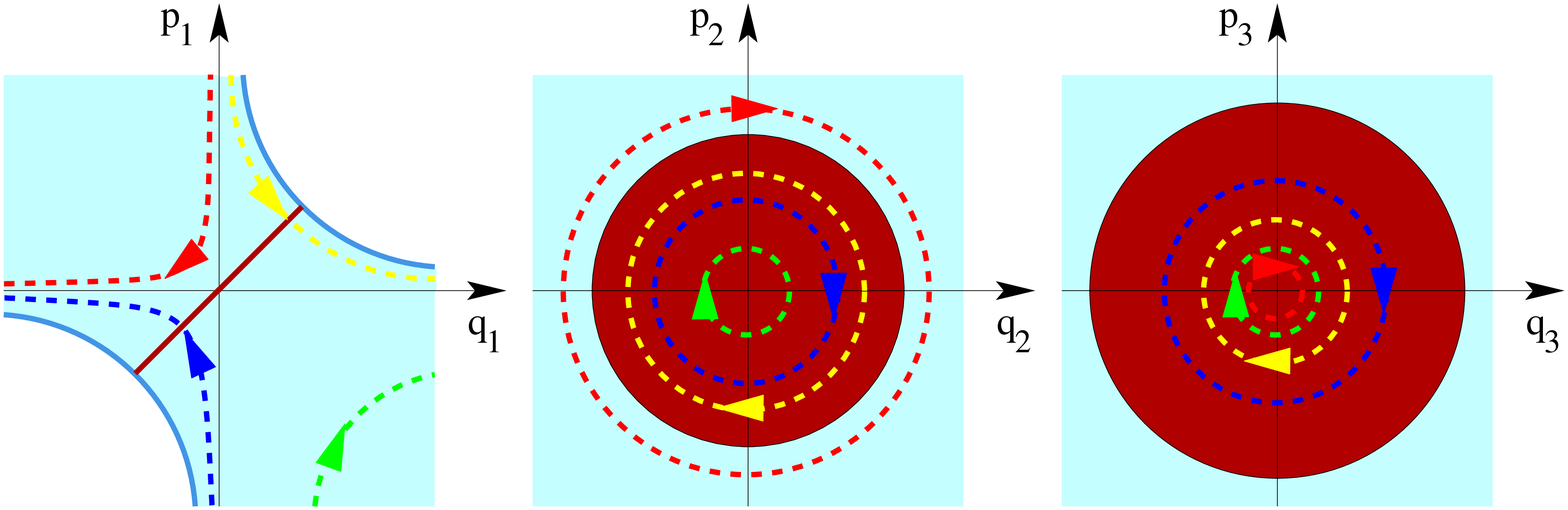}
\end{center}
}
\FIGo{fig:tst}{\figtst}{\FIGtst}

\rem{
\begin{figure}[htb!]
\begin{center}
\includegraphics[width=12.0cm]{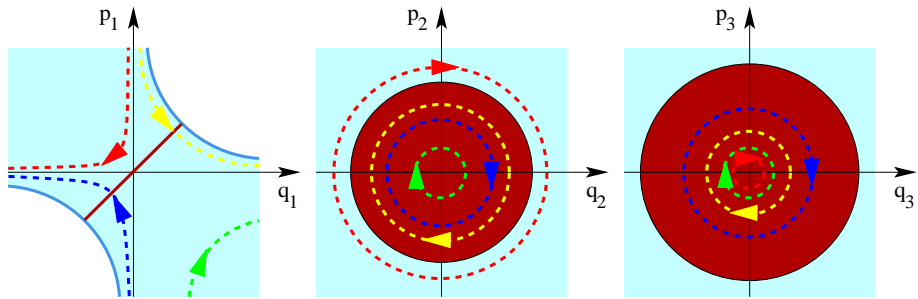}
\end{center}
\caption{Projection of the dividing surface and reacting and
nonreacting trajectories to the planes of the normal form coordinates.
In the plane of the saddle coordinates, the projection of the
dividing surface is the bold diagonal line segment, which has
$q_1=p_1$.  In the planes of the centre coordinates, the projections of
the dividing surface are the discs bounded by the thin circles.
Forward and backward reactive trajectories (bold curves) project to
the first and third quadrant in the plane of the saddle coordinates,
respectively, and pass through the dividing surface.  The dashed
curves mark nonreactive trajectories on the reactant side
($p_1-q_1>0$), and on the product side ($p_1-q_1<0$), of the
dividing surface.  The shaded regions indicate the projections of the
energy surface.  }
\label{fig:tst}
\end{figure}
}


\vspace*{0.3cm} \noindent {\bf The normally hyperbolic invariant
manifold (NHIM) and its relation to the `activated complex':} On
an energy surface with $E>E_0$, the NHIM is given by $q_1=p_1=0$.
The NHIM has the structure of a $(2d-3)$-sphere, which we denote
by $\nhim$.  The NHIM is the equator of the dividing surface; it
divides it into two ``hemispheres'': the \emph{forward dividing
surface}, which has $q_1=p_1>0$, and the \emph{backward dividing
surface}, which has $q_1=p_1<0$.  The forward and backward
dividing surfaces have the structure of $(2d-2)$-dimensional
balls, which we denote by $\tsf$ and $\tsb$, respectively.  All
forward reactive trajectories cross $\tsf$; all backward reactive
trajectories cross $\tsb$.  Since $q_1=p_1=0$ in the equations of
motion \eqref{hameq_nf} implies that $\dot{q}_1=\dot{p}_1=0$, the
NHIM is an invariant manifold, i.e., trajectories started in the
NHIM stay in the NHIM for all time.  The system resulting from
$q_1=p_1=0$ is an invariant subsystem with one degree of freedom
less than the full system.  
In fact, $q_1=p_1=0$ 
defines the centre manifold associated with the saddle-centre-$\cdots$-centre equilibrium point, and 
the NHIM at an energy $E$ greater than the energy of the quilibrium point is given by the intersection of the centre manifold with the energy surface of this energy $E$ \cite{UJPYW01,WaalkensWiggins04}.

This subsystem is the ``activated complex'' (in
phase space), located between reactants and products (see
Sec.~\ref{sec:phasespacestruct}).  
The NHIM can be considered to
be the energy surface of the activated complex.  In particular,
all trajectories in the NHIM have $I=0$.

The equations of motion \eqref{hameq_nf} also show that
$\dot{p}_1-\dot{q}_1<0$ on the forward dividing surface $\tsf$, and
$\dot{p}_1-\dot{q}_1>0$ on the backward dividing surface $\tsb$.
Hence, except for the NHIM, which is is an invariant manifold, the
dividing surface is everywhere transverse to the Hamiltonian flow.
This means that a trajectory, after having crossed the forward or
backward dividing surface, $\tsf$ or $\tsb$, respectively, must leave
the neighbourhood of the dividing surface before it can possibly cross
it again. Indeed, such a trajectory must leave the local region in
which the normal form is valid before it can possibly cross the
dividing surface again.

The NHIM has a special structure: due to the conservation of the centre
actions, it is filled, or {\em foliated}, by invariant
$(d-1)$-dimensional tori, $\T^{d-1}$.  More precisely, for $d=3$ degrees of freedom, each
value of $J_2$ implicitly defines a value of $J_3$ by the energy
equation $K_{\text{CNF}}(0,J_2,J_3)=E$.  For three degrees of freedom, the NHIM is thus
foliated by a one-parameter family of invariant 2-tori.  The end
points of the parameterization interval correspond to $J_2=0$ (implying
$q_2=p_2=0$) and $J_3=0$ (implying $q_3=p_3=0$), respectively.  At the
end points, the 2-tori thus degenerate to periodic orbits, the
so-called {\em Lyapunov periodic orbits}.   As we will
discuss in more detail in Sections~\ref{sec:smatrix} and \ref{sec:resonances},
the fact that the NHIM is foliated by invariant tori has important
consequences for the corresponding quantum system.

\def\fignhim{%
The projection of the NHIM and the local parts of its stable
and unstable manifolds, $W^s(E)$ and $W^u(E)$, to the planes of the
normal form coordinates.  In the plane of the saddle coordinates,
the projection of the NHIM is the origin marked by the blue bold point, and the projection of
$W^s(E)$ and $W^u(E)$ are the $p_1$-axis and $q_1$-axis, respectively.
$W^s(E)$ consists of the forward and backward branches $W_f^s(E)$ and
$W_b^s(E)$, which have $p_1>0$ and $p_1<0$, respectively; $W^u(E)$
consists of $W_f^u(E)$ and $W_b^u(E)$, which have $q_1>0$ and $q_1<0$,
respectively.  In the plane of the centre coordinates, the projections of
the NHIM, $W^s(E)$, and $W^u(E)$ (the blue circular discs)
coincide with the projection of the dividing surface in
Fig.~\ref{fig:tst}.  The turquoise regions mark the projections
of the energy surface.
}
\def\FIGnhim{
\begin{center}
\includegraphics[width=12.0cm]{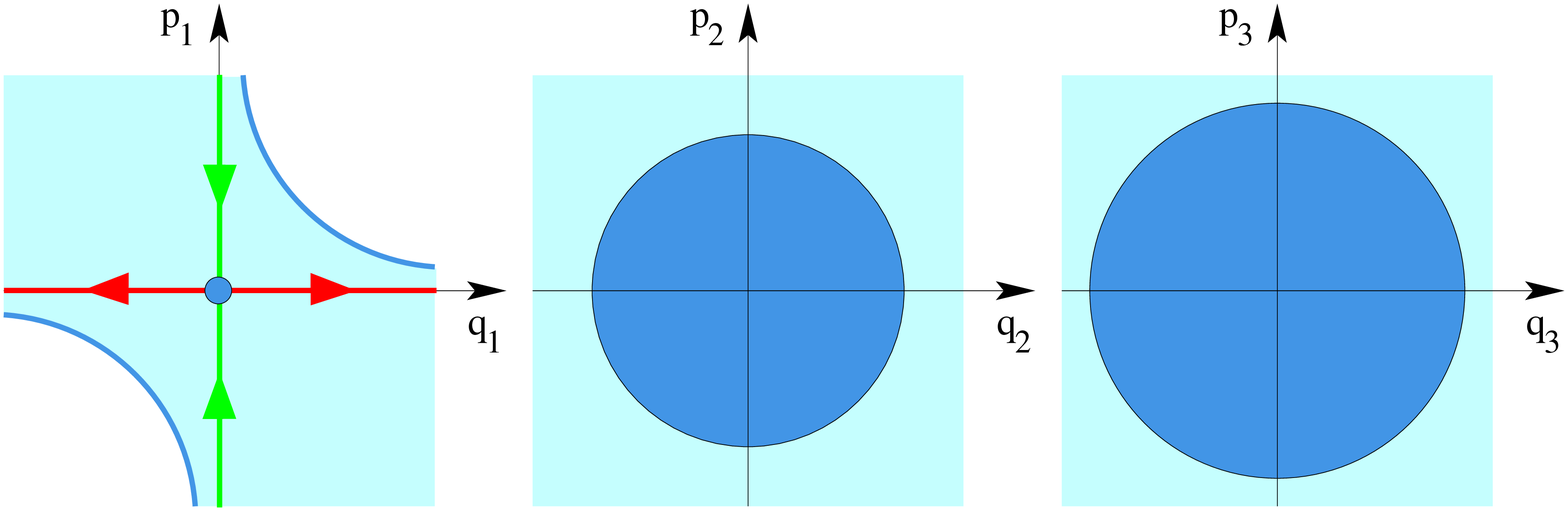}
\end{center}
}
\FIGo{fig:nhim}{\fignhim}{\FIGnhim}

\rem{
\begin{figure}[htb!]
\begin{center}
\includegraphics[width=12.0cm]{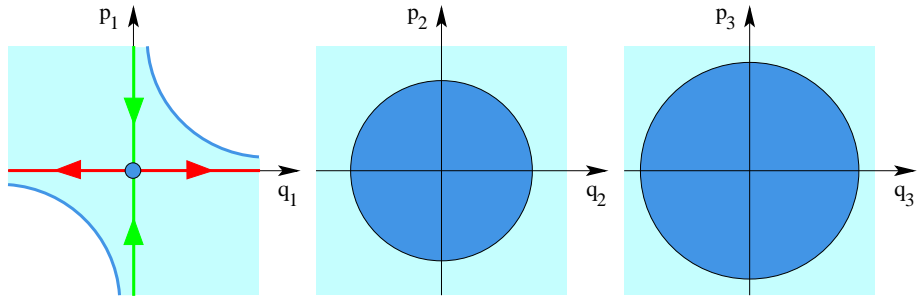}
\end{center}
\caption{The projection of the NHIM and the local parts of its stable
and unstable manifolds, $W^s(E)$ and $W^u(E)$, to the planes of the
normal form coordinates.  In the plane of the saddle coordinates,
the projection of the NHIM is the origin, and the projection of
$W^s(E)$ and $W^u(E)$ are the $p_1$-axis and $q_1$-axis, respectively.
$W^s(E)$ consists of the forward and backward branches $W_f^s(E)$ and
$W_b^s(E)$, which have $p_1>0$ and $p_1<0$, respectively; $W^u(E)$
consists of $W_f^u(E)$ and $W_b^u(E)$, which have $q_1>0$ and $q_1<0$,
respectively.  In the plane of the centre coordinates, the projections of
the NHIM, $W^s(E)$, and $W^u(E)$ (the darkly shaded circular discs)
coincide with the projection of the dividing surface in
Fig.~\ref{fig:tst}.  The lightly shaded regions mark the projections
of the energy surface.  }
\label{fig:nhim}
\end{figure}
}

\vspace*{0.3cm} \noindent {\bf The stable and unstable manifolds
of the NHIM forming the phase space conduits for reactions:} Since
the NHIM is of saddle stability type, it has stable and unstable
manifolds, $W^s(E)$ and $W^u(E)$.  The stable and unstable
manifolds have the structure of spherical cylinders,
$S^{2d-3}\times\R$.  Each of them consists of two branches: the
``forward branches'', which we denote by $W_f^s(E)$ and
$W_f^u(E)$, and the ``backward branches'', which we denote by
$W_b^s(E)$ and $W_b^u(E)$.  In terms of the normal form
coordinates, $W_f^s(E)$ is given by $q_1=0$ with $p_1>0$,
$W_f^u(E)$ is given by $p_1=0$ with $q_1>0$, $W_b^s(E)$ is given
by $q_1=0$ with $p_1<0$, and $W_b^u(E)$ is given by $p_1=0$ with
$q_1<0$, see Fig.~\ref{fig:nhim}.  Trajectories on these manifolds
have $I=0$.

Since the stable and unstable manifolds of the NHIM are of one less
dimension than the energy surface, they enclose volumes of the energy
surface.  We call the union of the forward branches, $W_f^s(E)$ and
$W_f^u(E)$, the {\em forward reactive spherical cylinder} and denote it by
$W_f(E)$.  Similarly, we define the {\em backward reactive spherical
cylinder}, $W_b(E)$, as the union of the backward branches, $W_b^s(E)$ and
$W_b^u(E)$.

\def\figvolumes{%
Projections of the reactive volumes enclosed by the forward
and backward reactive spherical cylinders, $W_f(E)$ and $W_b(E)$, and
the forward and backward reactions paths, to the planes of the normal
form coordinates.  The volumes enclosed by $W_f(E)$ and $W_b(E)$
project to the dark pink and green regions in the first and third quadrant
in the plane of the saddle coordinates, respectively.
These volumes project to the dark green/dark pink brindled disks in the planes of the centre
coordinates, where their
projections coincide with the
projection of the NHIM and the dividing surface in
Figs.~\ref{fig:tst}~and~\ref{fig:nhim}.  The forward and backward
reaction paths project to the two branches of a hyperbola marked blue in the first and
third quadrant in the plane of the saddle coordinates, respectively,
and to the origins (bold blue points) in the planes of the centre
coordinates.  The turquoise regions mark the projections of the
energy surface.
}
\def\FIGvolumes{
\begin{center}
\includegraphics[width=12.0cm]{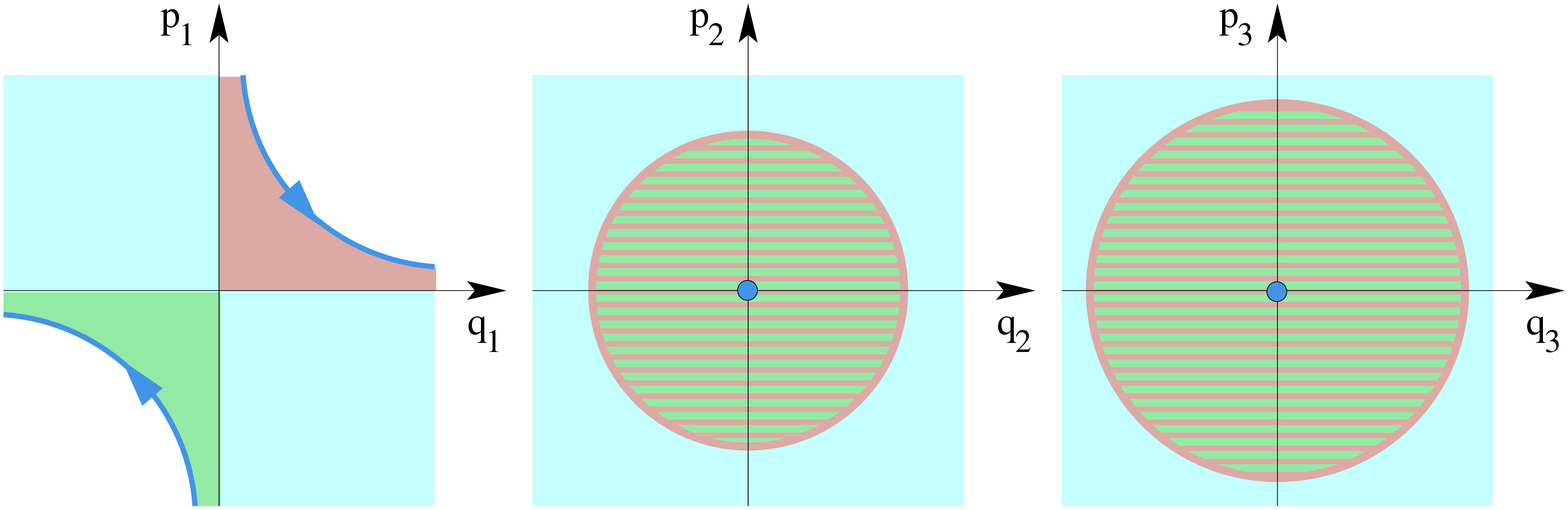}
\end{center}
}
\FIGo{fig:volumes}{\figvolumes}{\FIGvolumes}

The reactive volumes enclosed by $W_f(E)$ and $W_b(E)$ are shown
in Fig.~\ref{fig:volumes} as their projections to the normal form
coordinate planes.  In the plane of the saddle coordinates, the
reactive volume enclosed by $W_f(E)$ projects to the first
quadrant.  This projection is bounded by the corresponding
hyperbola $q_1 p_1= I$, with $I$ obtained from
$K_{\text{CNF}}(I,0,\dots,0)=E$.  Likewise, $W_b(E)$ projects to
the third quadrant in the $(q_1,p_1)$-plane.  $W_f(E)$ encloses
{\em all} forward reactive trajectories; $W_b(E)$ encloses {\em
all} backward reactive trajectories.  {\em All} nonreactive
trajectories are contained in the complement.


%
\vspace*{0.3cm}
\noindent
{\bf Forward and backward reaction paths:}
The local geometry of $W_f(E)$ and $W_b(E)$ suggests a natural definition of
{\em dynamical} forward and backward reaction paths as the unique
paths in {\em phase space} obtained by putting all of the energy of a
reacting trajectory into the reacting mode, i.e., setting
$q_2=\dots=q_d=p_2=\dots=p_d=0$.  This gives the two branches of the  hyperbola $q_1
p_1 = I$, with $I$ obtained from $K_{\text{CNF}}(I,0,\dots,0)=E$, which in phase space are contained in the plane of
the saddle coordinates, see Fig.~\ref{fig:volumes}.  This way, the
forward (respectively, backward) reaction path can be thought of as
the ``centre curve'' of the relevant volume enclosed by the forward
(resp., backward) reactive spherical cylinder $W_f(E)$ (resp.,
$W_b(E)$).  These reaction paths are the special reactive trajectories
which intersect the dividing surface at the ``poles'' (in the sense of
North and South poles, where $q_1=p_1$ assumes its maximum and minimum
value on the dividing surface).

\vspace*{0.3cm} \noindent {\bf The Transmission Time through the
Transition State Region:} The normal form coordinates provide a way
of computing the time for {\em all} trajectories to cross the
transition region. We illustrate this with a forward reacting
trajectory (a similar argument and calculation can be applied to
backward reacting trajectories). We choose the boundary for the
entrance to the reaction region to be $p_1-q_1=c$ for some
constant $c>0$, i.e., initial conditions which lie on the reactant
side of the transition state, and the boundary for exiting the
reaction region to be $p_1-q_1=-c$ on the product side. We now
compute the {\em time of flight} for a forward reacting trajectory
with initial condition on $p_1-q_1=c$ to reach $p_1-q_1=-c$ on the
product side.  The solutions are
$q_1(t)=q_1(0) \exp (
\Lambda (I,J_2,\dots,J_d) t )$ and
$p_1(t)=p_1(0) \exp ( - \Lambda (I,J_2,\dots,J_d) t)$ (see
\eqref{hameq_nf_solutions}),
where $\Lambda (I,J_2,\dots,J_d)$ is determined by the
initial conditions.  This gives the time of flight as

\begin{equation}
T = \left( \Lambda(I,J_2,\dots,J_d) \right)^{-1}
 \ln \left( \frac{p_1(0)}{q_1(0)} \right).
\end{equation}

\noindent
 The time diverges logarithmically as $q_1(0)\rightarrow
0$, i.e., the closer the trajectory starts to the boundary
$W_f(E)$. It is not difficult to see that the time of flight is
shortest for the centre curve of the volume enclosed by $W_f(E)$,
i.e., {\em the trajectory which traverses the transition state
region fastest is precisely our forward reaction path}.  A similar
construction applies to backward reactive trajectories.

In fact, the normal
form can be used to map trajectories through the transition state
region, i.e. the phase space point at which a trajectory enters
the transition state region can be mapped analytically to the
phase space point at which the trajectories exits the transition
state region.



\subsection{The Normal Form Coordinates: The Foliation of the
Reaction Region by Lagrangian Submanifolds} \label{sec:fls}

In Section~\ref{sec:nfcpss} we have indicated that the different types
of possible motion near a saddle-centre$\cdots$-centre equilibrium
point can be described in terms of the integrals.
In fact, the existence of the $d$ integrals  \eqref{nf_int}
leads to even further constraints on the classical motions and hence to even
more detailed structuring of the phase space near a
saddle-centre-$\cdots$-centre equilibrium point  than we already
described in Sec.~\ref{sec:nfcpss}. As we will see in Sect.~\ref{sec:smatrix}
this structure will have important consequences for the quantum mechanics of reactions.
In order to describe this structure we  introduce the so called \emph{momentum map} ${\cal
M}$ \cite{Guillemin94,MarsdenRatiu99} which maps a point
$(q_1,\ldots,q_d,p_1,\ldots,p_d)$ in the phase space $\R^{2d}$ to
the integrals evaluated at this point:
\begin{equation} \label{eq:def_momentum_map}
{\cal M}(q_1,\ldots,q_d,p_1,\ldots,p_d) \mapsto (I,J_2,\ldots,J_d)\,.
\end{equation}
The preimage of a value for the constants of motion $(I,J_2,\ldots,J_d)$ under ${\cal M}$  is called a \emph{fibre}. A fibre thus
corresponds to the common level set of the integrals in
phase space.
\rem{
\begin{equation}
  \{(q,p)\in \R^{2d}\,:\,
  p_1q_1=I,\,\frac12\big(p_2^2+q_2^2\big)=J_2\,,\ldots\,,\frac12\big(p_d^2+q_d^2\big)=J_d \}\,.
\end{equation}
} 

A point $(q_1,\ldots,q_d,p_1,\ldots,p_d)$ is called a \emph{regular
point} of the momentum map if the linearisation of the momentum map, $D\M$, has rank $d$
at this point, i.e., if the gradients of the $d$ integrals $I$, $J_k$,
$k=2,\ldots,d$, with respect to the
phase space coordinates $(q,p)$ are linearly independent at this point.
If the rank of $D\M$ is less than $d$ then the
point is called an irregular point. A regular fibre is a fibre
which consists of regular points only. The regular fibres of the
momentum map in \eqref{eq:def_momentum_map} are $d$-dimensional
manifolds given by the Cartesian product of an hyperbola $q_1 p_1 =
I$ in the saddle plane $(q_1,p_1)$ and $d-1$ circles
$S^1$ in the centre planes $(q_k,p_k)$, $k=2,\ldots,d$.
Since hyperbola $q_1 p_1 =
I$ consists of two branches
each of which have the topology of a line $\R$,  the
regular fibres consist of two disjoint \emph{toroidal cylinders}, $\T^{d-1}
\times \R $,
which are the Cartesian products of a
$(d-1)$-dimensional torus and a line.
We denote these toroidal cylinders by
\begin{equation} \label{eq:def_Lambda_pm}
\Lambda^+_{I,J_2,\ldots,J_d} =
\{(q,p)\in \R^{2d}\,:\,
  p_1q_1=I,\,\frac12\big(p_2^2+q_2^2\big)=J_2\,,\ldots\,,\frac12\big(p_d^2+q_d^2\big)=J_d\,,q_1>0 \}
\end{equation}
and
\begin{equation}
\Lambda^-_{I,J_2,\ldots,J_d} =
\{(q,p)\in \R^{2d}\,:\,
  p_1q_1=I,\,\frac12\big(p_2^2+q_2^2\big)=J_2\,,\ldots\,,\frac12\big(p_d^2+q_d^2\big)=J_d\,,q_1<0 \}\,.
\end{equation}
$\Lambda^+_{I,J_2,\ldots,J_d}$ and $\Lambda^-_{I,J_2,\ldots,J_d}$ are
\emph{Lagrangian manifolds} \cite{Arnold78}. Prominent examples
of Lagrangian manifolds are tori which foliate the
neighbourhood of a centre-$\ldots$-centre equilibrium point and whose
semiclassical quantisation
often lead to a very good approximation of part of the energy spectra of
the corresponding bounded system \cite{Ozorio88}. In our case the Lagrangian
manifolds are unbounded. They are the products of
$(d-1)$-dimensional tori $\T^{d-1}$ and unbounded lines $\R$. The
toroidal base of these cylinders will again lead to semiclassical quantisation
conditions and as we will see in Sections~\ref{sec:smatrix} and
\ref{sec:resonances} this will have important consequences for the
computation of quantum reaction rates and resonances.

\def\figmomentummap{%
Sketch of the image of the energy surface  of energy  $E>E_0$ under
the momentum map ${\cal M}$ in Equation~\eqref{eq:def_momentum_map}
in the space of the integrals $I$ and $J_k$,
$k=2,\ldots,d$, for the case of $d=3$ degrees of freedom.
The green/dark pink brindled piece of the image of the energy surface has
$I>0$; the turquoise piece has $I<0$.
The intersections with the
planes $I=0$, $J_2=0$ and $J_3=0$ (pieces of which are visualised by semitransparent planes
for clarity) form the bifurcation diagram of
the energy surface. The image of the energy surface is not bounded in the
direction of negative $I$ as indicated by the dashed line  at the bottom.
The topology of the fibres ${\cal M}^{-1}(I,J_2,J_3)$ is
indicated for the various points  $(I,J_2,J_3)$ marked by dots.
The fibre of a point  $(I,J_2,J_3)$ with $I\ne 0$ consists of two
disconnected manifolds as indicated by the factor of 2. The fibre of
a point  $(I,J_2,J_3)$ with $I = 0$ consists of a single
connected manifold.
}
\def\FIGmomentummap{
\centerline{
\includegraphics[angle=0,width=8cm]{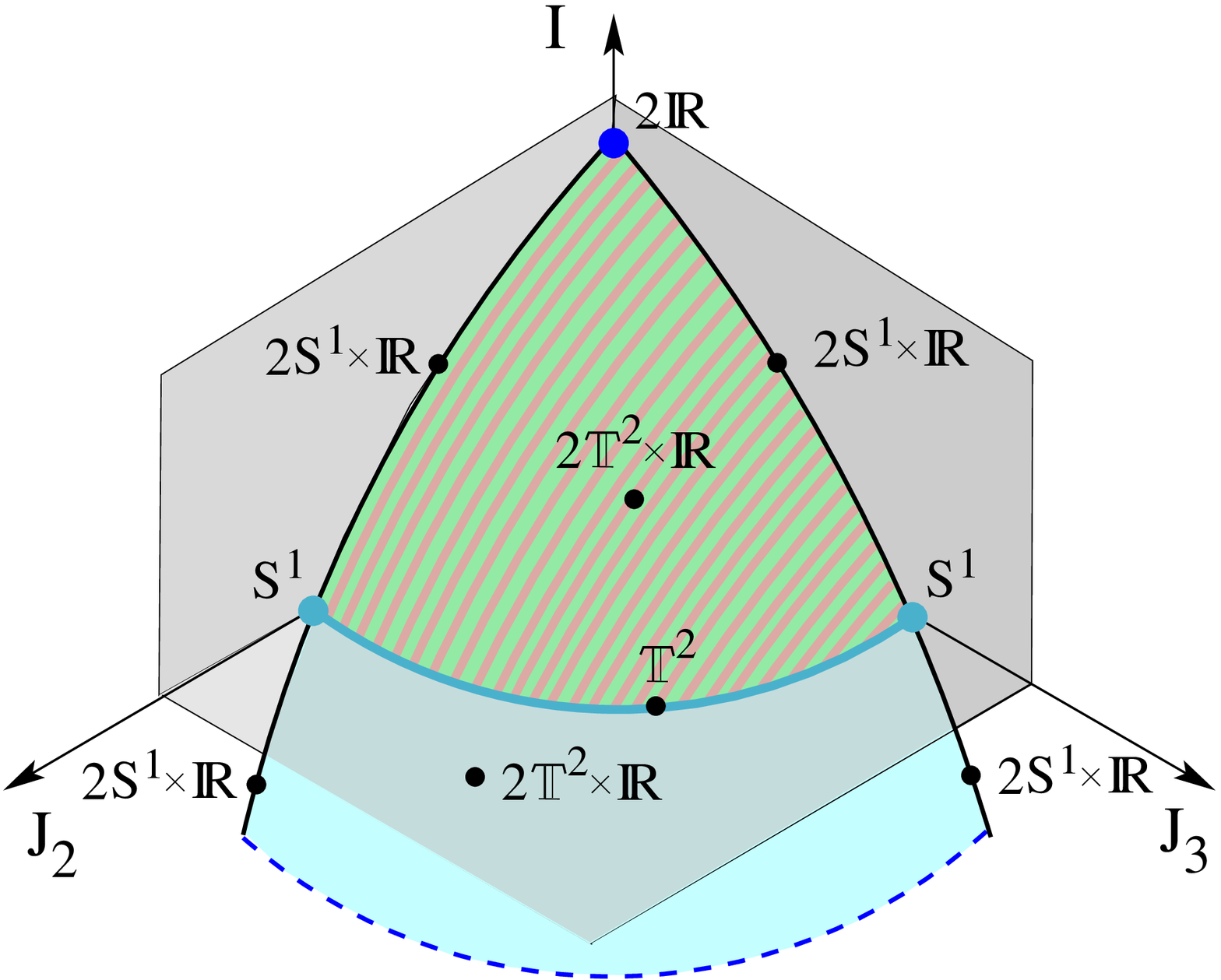}
}
}
\FIGo{fig:momentummap}{\figmomentummap}{\FIGmomentummap}

If the fibre contains an irregular point then the fibre is called singular.
The image of the singular fibres under the momentum map is called the
\emph{bifurcation diagram}. It is easy to see that the bifurcation diagram
consists of the set of $(I,J_2,\ldots,J_d)$ where one or more of the integrals vanish.
In Fig.~\ref{fig:momentummap} we show the image of the energy surface with
energy $E>E_0$ under the momentum map ${\cal M}$  in the space of the
integrals for $d=3$ degrees of freedom. The bifurcation diagram (of the energy
surface) consists of the intersections of the image of the energy surface (the
turquoise and green/dark ping brindled surface in Fig.~\ref{fig:momentummap})
with one of the planes $I=0$, $J_2=0$ or $J_3=0$.
Upon approaching one of the edges that have $J_2=0$ or $J_3=0$ the
circle in the plane $(q_2,p_2)$ or $(q_3,p_3)$, respectively, shrinks to a
point, and accordingly
the regular fibres $\T^{2}\times \R$ reduce to cylinders or `tubes'
$S^{1}\times \R$. At the top corner in  Fig.~\ref{fig:momentummap} both
$J_2$ and $J_3$ are zero.
Here both circles in the centre planes $(q_2,p_2)$ and $(q_3,p_3)$ have
shrinked to points.
The corresponding singular fibre consists of two
lines, $\R$, which are the forward and backward reaction paths, respectively (see
also Fig.~\ref{fig:volumes}).

The  fibres mentioned so far all have $I\ne 0$ and each consist of a pair of two
disconnected components.
For $I<0$,  one member of each pair is located on the reactants
side and the other on the products side of the dividing surface.
For $I>0$, one member of each pair consists of trajectories
evolving from reactants to products and the other member consists of
trajectories that evolve from products to reactants.
In fact the two members of a fibre which has $I>0$ are contained in the energy
surface volume enclosed by the forward and backward reactive spherical
cylinders $W_f(E)$ and $W_b(E)$, see Fig.~\ref{fig:volumes}.
For this reason we marked the piece of the image of the energy surface under
the momentum map which has $I>0$ by the same green/dark pink colour in
Fig.~\ref{fig:momentummap} that we used Fig.~\ref{fig:volumes}.
Green corresponds to forward reactive trajectories and dark pink corresponds
to backward reactive trajectories. Under the momentum map these trajectories
have the same image.

The light blue line in Fig.~\ref{fig:momentummap} which has $I=0$
is  the image of the NHIM under the momentum map. For three
degrees of freedom the NHIM is a 3-dimensional sphere, and as
mentioned in Sec.~\ref{sec:nfcpss} and indicated in
Fig.~\ref{fig:momentummap} it is foliated by a one-parameter
family of invariant 2-tori which shrink to periodic orbits, i.e.
circles $S^1$, at the end points of the parameterisation interval.
As we already indicated in Sec.~\ref{sec:nfcpss} this foliation of the NHIM
has important consequences for the quantum mechanics of reactions which we
will dicuss in Sections~\ref{sec:smatrix} and \ref{sec:resonances}.
Moreover, we will see in Sec.~\ref{sec:classicalrate} that, for $d=3$ degrees
of freedom, the area enclosed by the image of the NHIM in the
plane $(J_2,J_3)$ gives, up to a prefactor, the directional
flux through the dividing surface.


\subsection{The Directional Flux Through the Dividing Surface}
\label{sec:classicalrate}

A key ingredient of transition state theory and the classical
reaction rate is the directional flux through the dividing surface
defined in Sec.~\ref{sec:phasespacestruct}. Given the Hamiltonian
function in normal form expressed as a function of the integrals
\eqref{nf_int}, and a fixed energy $E$ above the energy of the
saddle-centre-$\cdots$-centre,  $E_0$, it is shown in
\cite{WaalkensWiggins04} that the directional flux through the
dividing surface is given by

\begin{equation} \label{eq:fluxactions_classical}
\flux(E) = (2\pi)^{d-1} {\cal V}(E)\,,
\end{equation}

\noindent
 where ${\cal V}(E)$ is the volume in the space of the
actions $(J_2,\dots,J_d)$ enclosed by the contour
$\Hcnf(0,J_2,\dots,J_d)=E$. 
\def\figflux{%
Contour $K_{\text{CNF}}(0,J_2,\ldots,J_d)=E$ (blue line) in the
space of the centre integrals $(J_2,\ldots,J_d)$ for $d=3$ degrees
of freedom. Up to the prefactor $(2\pi)^{d-1}$, the area ${\cal
V}(E)$ of the enclosed region (marked green) gives the directional
flux through the dividing surface, see
Equation~\eqref{eq:fluxactions_classical}. The green region agrees
with the projection of the piece of the image of the energy
surface under the momentum map which has $I>0$ in
Fig.~\ref{fig:momentummap} to the $(J_2,J_3)$-plane. }
\def\FIGflux{
\centerline{\includegraphics[width=5.0cm]{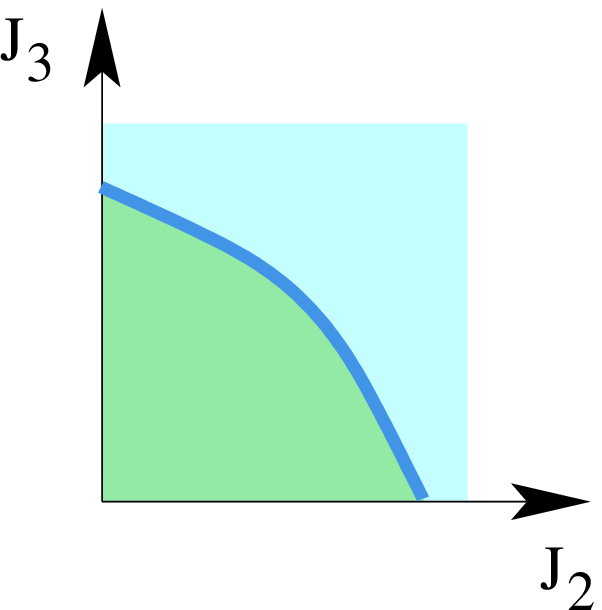}}
}
\FIGo{fig:flux}{\figflux}{\FIGflux}
For $E<E_0$, the directional flux is zero. For the case of a
system with three degrees of freedom for which we sketched the
image of the energy surface in the space of the integrals in
Fig.~\ref{fig:momentummap}, the volume ${\cal V}(E)$ is given by
the area in the $(J_2,J_3)$ plane enclosed by the light blue line
corresponding to the NHIM in Fig.~\ref{fig:momentummap}. For
clarity we illustrate this area again in Fig.~\ref{fig:flux}. As
we mentioned in Sec.~\ref{sec:nfcpss} the NHIM can be considered
as the energy surface of an invariant subsystem with one degree of
freedom less than the full system which is referred to as the
activated complex in the chemistry literature. Therefore the flux
can be interpreted as the volume  enclosed by the  energy surface
(given by the NHIM) in the phase space of this invariant
subsystem. This gives a direct connection between the directional
flux through the dividing surface and the activated complex.
\rem{
({\bf
It would be good if we could find a chemistry reference where this
connection explains some speculation or phenomena.})

({\bf This paragraph is a bit ``disjoint'' from the previous
paragraph. Do we want to say something like ``the connection
between the classical directional flux through the dividing
surface and quantum mechanics made through Weyl's
approximation...? Again relating this to the chemistry literature,
e.g. approximate quantum transition state theories, would be
good.})
} 
In fact, the dimensionless quantity
\begin{equation} \label{eq:NWeyl_def}
\Nweyl(E) = \frac{\flux(E)}{(2\pi \hbar)^{d-1}} \,,
\end{equation}
where $2\pi \hbar$ is Planck's constant, is Weyl's approximation
of the integrated density of states, or equivalently the mean
number of quantum states of the activated complex with energies
less than or equal to $E$ (see, e.g. \cite{Gutz90}). As we will see in
Sec.~\ref{sec:smatrix} $\Nweyl(E)$ can be interpreted as the mean
number of open quantum ``transition channels''  at energy $E$.

In the case where we only take into account the quadratic part of
the normal form, or equivalently, if we linearise Hamilton's equations,
we have $\Hcnf(I,J_2,\dots,J_d)=\lambda I+ \sum_{k=2}^d \omega_k J_k$ and
the energy surface $\Hcnf(0,J_2,\dots,J_d)=E$ encloses a simplex
in $(J_2,\dots,J_d)$ whose volume leads to the well-known result
\cite{MacKay1990}
\begin{equation} \label{eq:linear_flux}
\flux(E) = \frac{ E^{d-1}}{(d-1)!} \prod_{k=2}^{d}
\frac{2\pi}{\omega_k}\,.
\end{equation}
This shows, e.g, that the flux scales with $E^{d-1}$ for energies close to
the saddle energy. The key advantage of the normal form algorithm that we
presented in Sec.~\ref{sec:CNF} is that it allows one to include the
non-linear corrections to \eqref{eq:linear_flux} to any desired order.

Here we give a brief outline of the essential elements of the
derivation of  the expression for the flux in
\eqref{eq:fluxactions_classical} following the discussion
\cite{WaalkensWiggins04}. It is important to note that our work is
firmly rooted in phase space. In particular, we are considering
the (directional) flux of a vector field on phase space
(Hamilton's equations) through a dividing surface in phase space
(which has been proven to have the ``no recrossing'' property as
discussed earlier). For this reason the modern notation of
differential forms, especially in light of its importance in the
modern formulation of Hamiltonian mechanics, proves to be most
convenient and notationally economical.

Therefore we begin by considering the phase space volume form
$\Omega=\mbox{d}p_1\wedge\mbox{d}q_1\wedge\cdots\wedge\mbox{d}p_d\wedge\mbox{d}q_d$,
which in terms of the symplectic 2-from  $\omega=\sum_{k=1}^d
\mbox{d} p_k \wedge \mbox{d} q_k$ can be written as
$\Omega=\omega^d/d!$. Note that in our case the phase space
coordinates $(q,p)$ used here will be $N^\text{th}$ order normal
form coordinates and we do not use superscripts $(N)$ to indicate
this. However the quantities introduced in the following do not
depend on the chosen coordinate system. They are invariant under
symplectic coordinate transformations. Let $\eta$ be an energy
surface volume form defined via the property $\mbox{d}H\wedge \eta
= \Omega$. Then the flux through a codimension one submanifold of
the $(2d-1)$-dimensional energy surface $H=E$ is obtained  from
integrating over it the ``flux'' form $\Omega'$ given by the
interior product of the Hamiltonian vector field $X_H$ with $\eta$
\cite{MacKay1990}, i.e.
\begin{equation}
\label{eq:omegaprime}
\Omega' =i_{X_H}\eta = \frac{1}{(d-1)!} \omega^{d-1}\,,
\end{equation}
where $i_{X_H}\eta(\xi_1,\dots,\xi_{2d-2})=\eta(\xi_1,\dots,\xi_{2d-2},X_H)$ for any $2d-2$ vectors $\xi_k$. The second equality
in (\ref{eq:omegaprime}) is easily established on a non-critical energy surface, i.e. on an energy surface
which contains no equilibria.
The flux form $\Omega'$ is exact. In fact the generalised ``action'' form
\[
\phi = \sum_{k=1}^d p_k\mbox{d} q_k \wedge \frac{1}{(d-1)!} \omega^{d-2}
\]
has the property $\mbox{d} \phi = \Omega'$ and facilitates the use
of Stokes' theorem to compute the flux. In the case of two
degrees of freedom we simply have $ \Omega' = \omega=\mbox{d}p_1
\wedge \mbox{d}q_1 + \mbox{d}p_2 \wedge \mbox{d}q_2 $ and $\phi$
becomes the usual action form $ \phi = p_1 \mbox{d}q_1 + p_2
\mbox{d}q_2\,. $ Since the dividing surface $\ts$ is a sphere,
that is, a manifold without boundary, it follows from Stokes'
theorem that the integral of $\Omega'$ over  $\ts$ is zero. In
order to compute reaction rates one has to distinguish between the
directions in which the Hamiltonian flow crosses the dividing
surface (i.e., distinguish between forward and backward reactive
trajectories). Given a normal bundle \footnote{Roughly speaking,
at each point of the dividing surface we  consider the normal
vector in the energy surface. The {\em normal bundle} is the union
of all vectors taken over all points on the dividing surface.}
over $\ts$ the direction can be specified by the sign of the
scalar product between the normal vectors and the Hamiltonian
vector field. This scalar product is strictly positive on one
hemispheres of $\ts$, strictly negative on the other hemisphere
and zero only at the equator of $\ts$, i.e. at the normally
hyperbolic invariant manifold $\nhim$, where the Hamiltonian
vector field is tangent to $\ts$. Likewise, the flux form
$\Omega'$ on $\ts$ vanishes nowhere on $\tsf$ and $\tsb$ and is
identically zero on $\nhim$. It is natural to take as the
orientation of $\tsf$ and $\tsb$ the orientation they inherit from
the dividing surface. Without restriction we may assume that the
orientation of $\ts$ is such that $\Omega'$ is positive on the
forward hemisphere $\tsf$ and negative on the backward hemisphere
$\tsb$, i.e. $\Omega'$ and $-\Omega'$ can be considered as {\em
volume forms} on $\tsf$ and $\tsb$, respectively. It follows from
Stokes' theorem that the flux through the forward and backward
hemispheres, $\int_\tsf \Omega'$ and $\int_\tsb \Omega'$, have the
same magnitude but opposite sign and can be computed from
integrating the action form $\phi$ over the NHIM:
\begin{equation}
\flux(E) = \int_\tsf \Omega' =-\int_\tsb \Omega' =
\bigg|\int_\nhim \phi \bigg|\,.
\end{equation}
We call the positive quantity $\int_\tsf \Omega'$ the {\em forward flux} and
the negative quantity $\int_\tsf \Omega'$ the {\em backward flux} through $\ts$ .

Writing the flux form $\Omega'$ in terms of ``angle-action
variables'' $(\varphi_1,\dots,\varphi_d,I,J_2,\dots,J_d)$ (these
were derived in terms of the integrals of the normal form in
Section \ref{sec:solvhomequ}) we obtain the result that the
forward flux through the dividing surface is given by the expression in Equation~\eqref{eq:fluxactions_classical}.


\subsection{The Normal Form Coordinates: Issues Associated with
Truncation} \label{sec:accuracy_NF}

The final question to address concerns the ``validity'' of the
normal form transformation. More precisely, this means how large
can the neighbourhood (in phase space) $U$ of the
saddle-centre-$\cdots$-centre equilibrium point be taken so that
the geometric structures given by the normal form are accurate for
the ``full equations''. Actually, there are a number of questions
to be answered related to ``validity''.

\begin{itemize}

\item In truncating the Taylor expansion of the Hamiltonian at
degree $N$, how do you determine $N$?

\item What is the region of validity of the normal form
transformation for the Taylor expanded Hamiltonian truncated at
degree $N$?

\item How ``accurate'' are the phase space structures (e.g., the
dividing surface, the NHIM) for the normal form of the Hamiltonian
truncated at degree $N$?

\item How accurate are trajectories of the normal form of the
Hamiltonian at order $N$?

\end{itemize}

First, general theory  ensures us that the phase
space structures {\em exist}, and have the properties described
above (e.g., normal hyperbolicity, the bottleneck property, etc.),
{\em for energies sufficiently close to that of the
saddle-centre-$\cdots$-centre equilibrium point} \cite{Wiggins94}. The normal form
computation is merely an approach for realizing the geometrical
structures that the theory tells us must exist.

In practice, one Taylor expands the Hamiltonian and then truncates
it at a degree that one {\em thinks} will provide sufficient
accuracy for the range of energies of interest. Experience will
generally provide some good ``rules of thumb'', e.g. for the HCN
isomerization work described in \cite{WaalkensBurbanksWigginsb04},
an expansion up to degree 10 was found to provide sufficient
accuracy in the range of energies studied (up to $0.2$ eV above the
saddle-centre-$\cdots$-centre equilibrium point).

There is still the question of accuracy. Once the normal form is
computed to the desired degree (and, most importantly, the
transformation and its inverse between the original coordinates
and the ``normal form coordinates''), and the energy is fixed, we
have explicit formulae for the dividing surface, the NHIM (the
``equator'' of the dividing surface), and the (local) stable and
unstable manifolds of the NHIM\footnote{Here ``local'' means that
we only have realizations of the stable and unstable manifolds in
a neighbourhood of the saddle-centre-$\cdots$-centre equilibrium
point where the normal form transformation has the desired
accuracy}. We next need to check their ``accuracy''. There are
several tests that we employ, and these tests are carried out at
{\em fixed energy}.

\begin{itemize}

\item Numerically verify that the dividing surface satisfies the
``bottleneck property'', i.e. it (locally) separates the energy
surface into two components, and the only way a trajectory can
pass between components (while remaining in this region)  is by
passing through the dividing surface.

\item Using the inverse of the normal form transformation map the
NHIM and its (local) stable and unstable manifolds back into the
original coordinates and check that the full (i.e., not a
truncated Taylor expansion) Hamiltonian vector field is tangent to
these surfaces. This is a requirement for these surface to be
``invariant manifolds''. The tests are carried out pointwise on a
grid of points covering the surfaces.

\item The integrals (\ref{nf_int}) are constant in time on
trajectories of the normal form of the truncated Taylor expansion.
We check how they vary in time on trajectories of the full
Hamiltonian.

\end{itemize}

If the desired accuracy is obtained for this energy, then the
energy may be increased and the accuracy tests are repeated at the
higher energy. If accuracy is inadequate, then a higher degree
Taylor expansion can be computed. As energy is increased,
ultimately two factors may lead to break down of this approach for
realizing these phase space structures. One is that the energy
surface may deform in such a way that the bottleneck property does
not hold. Another is that the approach will require such a high
degree Taylor expansion that it becomes computationally
intractable.


\subsection{The Global Dynamics Associated with the Manifolds Constructed in the Reaction Region}
\label{sec:classicglobal}

As we have shown, the  normal form transformation to normal form
coordinates provides a method for providing a complete
understanding of the geometry of reaction dynamics in a
neighbourhood $U$ (in phase space)  of the
saddle-centre-$\cdots$-centre equilibrium point of  Hamilton's
equations. By this, we mean that in the normal form coordinates we
can give an explicit equation for the surfaces and, as a result of
the ``simple''  structure of Hamilton's equations in the normal
form coordinates, we can describe precisely the influence of these
geometrical structures on trajectories of Hamilton's equations. In
Tab.~\ref{tab_surf} we summarize the results obtained this far by
providing a list of the different surfaces that control the
evolution of trajectories from reactants to products in the
neighbourhood $U$ in Fig. \ref{tab_surf}.

\begin{table}[!htb]
\begin{center}
\begin{tabular}{|c|c|} \hline
{\bf Geometrical Structure} & {\bf Equation in Normal Form Coordinates} \\
\hline \hline dividing surface, $\ts$&  $q_1 = p_1$ \\
\hline forward reactive hemisphere, $\tsf$ & $q_1 = p_1 >0$ \\
\hline backward reactive hemisphere, $\tsb$ & $q_1 = p_1 <0$  \\
\hline NHIM, $\nhim$  & $q_1 = p_1 =0$\\
\hline stable manifold of the NHIM, $W^s(E)$& $q_1 =0$, $p_1\ne 0$  \\
\hline unstable manifold of the NHIM, $W^u(E)$& $p_1 =0$, $q_1\ne 0$  \\
\hline forward branch of $W^s(E)$, $W^s_f(E)$& $q_1 =0$, $p_1>0$  \\
\hline backward branch of $W^s(E)$, $W^s_b(E)$& $q_1 =0$, $p_1<0$  \\
\hline forward branch of $W^u(E)$, $W^u_f(E)$& $p_1 =0$, $q_1>0$  \\
\hline backward branch of $W^u(E)$, $W^u_b(E)$& $p_1 =0$, $q_1<0$  \\
\hline forward reactive spherical cylinder & $p_1 q_1 = 0$, $p_1,q_1\ge 0$, $q_1\ne p_1$ \\
 $W_f(E) \equiv W^s_f(E) \cup W^u_f(E)$ & \\
\hline backward reactive spherical cylinder & $p_1 q_1 = 0$, $p_1,q_1\le 0$, $q_1\ne p_1$ \\
 $W_b (E) \equiv W^s_b(E) \cup W^u_b(E)$ & \\
\hline forward reaction path & $q_2=\dots=q_d=p_2=\dots=p_d=0$, $p_1>0$
 \\
\hline
backward reaction path & $q_2=\dots=q_d=p_2=\dots=p_d=0$,  $p_1<0$
\\
\hline
\end{tabular}
\end{center}
\caption{Table of phase space surfaces influencing reaction
dynamics and their representations in normal form coordinates on an energy
surface of energy greater than the energy of the saddle equilibrium point.}
\label{tab_surf}
\end{table}

However, all of these surfaces, and associated dynamical
phenomena, are only ``locally valid'' in the neighbourhood $U$. The
next step is to understand their influence on the dynamics outside
of $U$, i.e., their influence on the dynamics of reaction
throughout phase space in the original coordinates (as opposed to
the normal form coordinates). In order to do this we will need the
normal form transformation constructed in Section \ref{sec:CNF}
and given in (\ref{eq:Nthordernormalformcoordinates}), to order
$N$ (where $N$ is determined according to the desired accuracy
following the discussion in Section \ref{sec:accuracy_NF}). We
rewrite (\ref{eq:Nthordernormalformcoordinates}) below:

\begin{eqnarray}
&& z^{(1)} = z-z_0, \nonumber \\
&& z^{(2)} = M z^{(1)}, \nonumber \\
&& (q_1^{(N)}, \ldots, q_d^{(N)}, p_1^{(N)}, \ldots, p_d^{(N)})
\equiv z^{(N)} =\flow_{W_N}^{1} \circ \cdots \circ \flow_{W_3}^{1}
(z^{(2)}).
 \label{eq:Nthordernormalformcoordinates_2}
\end{eqnarray}

\noindent We refer to the original coordinates as the ``physical
coordinates'' where reading from top to bottom,
\eqref{eq:Nthordernormalformcoordinates_2} describes the sequence
of transformations from physical coordinates to normal form
coordinates as follows. We translate the
saddle-centre-$\cdots$-centre equilibrium point to the origin, we
``simplify'' the linear part of Hamilton's equations, then we
iteratively construct a sequence of nonlinear coordinate
transformations that successively ``simplify'' the order 3, 4,
$\ldots$, N terms of the Hamiltonian according to the algorithm
described in Section \ref{sec:CNF}. We can invert each of these
transformations to return from the normal form coordinates to the
physical coordinates.

\paragraph{Computation of $W^u_b(E)$ and $W^u_f(E)$:}
Our approach to computing the stable and unstable manifolds of a
NHIM is, in principle, the same as for computing the stable and
unstable manifolds of a hyperbolic trajectory (however, the
practical implementation of the algorithm in higher dimensions is
a different matter and one that deserves much more investigation).

We describe the computation of $W^u_f(E)$ as follows.

\begin{itemize}

\item In the normal form coordinates, choose a distribution of
initial conditions on the NHIM and displace these initial
conditions ``slightly'' in the direction of the forward branch of
$W^u(E)$ ($p_1=0, \, q_1 = \varepsilon >0$, $\varepsilon$
``small'').

\item Map these  initial conditions back into the physical
coordinates using the inverse of the normal form  transformation.

\item Integrate the initial conditions forward in time using
Hamilton's equations in  the physical coordinates, for the desired
length of time (typically determined by accuracy considerations)
that will give the manifold of the desired ``size''. Since the
initial conditions are in the unstable manifold they will leave
the neighbourhood $U$ in which the normal form transformation is
valid (which is why we integrate them in the original
coordinates with respect to the original equations of motion).

\end{itemize}

\noindent The backward branch of $W^u(E)$ can be computed in an
analogous manner by displacing the initial conditions on the NHIM
in the direction of the backward branch of $W^u(E)$ ($p_1=0, \,
q_1 = \varepsilon < 0$, $\varepsilon$ ``small'').

\paragraph{Computation of $W^s_b(E)$ and $W^s_f(E)$:} The forward
and backward branches of $W^s(E)$ can be computed in an analogous
fashion, except the initial conditions are integrated {\em
backward} in time.

\paragraph{Computation of the forward and backward reaction
paths:} Here the situation is, numerically, much simpler since we
only have to integrate a trajectory. We consider the case of the
forward reaction path. The backward reaction path is treated in
the same way, after the obvious changes of sign for the
appropriate quantities.

Recalling that the dividing surface in normal form coordinates is
given by $q_1= p_1$, the intersection of the forward reaction path
with the dividing surface is given by

\begin{eqnarray}
&& q_2=\dots=q_d=p_2=\dots=p_d=0, \nonumber \\
&& q_1^2 = I, q_1 = p_1 >0, \, \mbox{with} \,
K_{\text{CNF}}(I,0,\dots,0)=E\,. \label{int_frp_ds}
\end{eqnarray}

\noindent
We transform this point in normal form coordinates into
physical coordinates using the inverse of the transformations
given in (\ref{eq:Nthordernormalformcoordinates_2}). Integrating
this point forward in time  using Hamilton's equations in  the
physical coordinates gives the forward reaction path immediately
{\em after} passage through the dividing surface. Integrating the
point backward in time gives the forward reaction path immediately
{\em before} passage through the dividing surface.

\paragraph{Computation of reactive volumes:}

\New{
Consider a region of the energy surface of some fixed energy $E$ whose entrance and exit channels are associated 
with saddle-centre-$\cdots$-centre equilibrium points. 
Near each such equilibrium point we can construct a dividing surface that a trajectory of energy $E$ must cross in order to enter the region.
Suppose that the region is compact and simply connected. An example is the phase space region associated with the potential well that 
corresponds to an isomer in an isomerization reaction \cite{WaalkensBurbanksWigginsb04}.
It is then  possible to give a formula for
the energy surface volume corresponding to trajectories of the energy 
$E$ that
will leave that region of the energy surface. 

This formula is expressed
in terms of the phase space flux across the dividing surfaces
controlling access to this region of the energy surface and the
corresponding ``mean first passage times'' of trajectories
entering the region through the dividing surfaces. This theory is
described in detail in
\cite{WaalkensBurbanksWiggins05,WaalkensBurbanksWiggins05c} and
here we just outline the results and show how the phase space
structures discussed above in a region of the transition state are
``globalized'' to give this result.

 We consider an energy surface region 
 to which entrance  is only possible through a
number of dividing surfaces, $B_{\text{ds,\,f};i}^{2d-2}(E)$ ($i$ is the index for the
number of forward dividing surfaces that control access to the
region under consideration
in the sense that trajectories initialized on this
surface and integrated in {\em forward} time enter the region), and we compute the {\em
energy surface volume} of {\em reactive initial conditions}, i.e.,
the initial conditions of trajectories that can leave the region under consideration
through one of the dividing surfaces. The phase space transport
theory described above is crucial for this computation as it
allows us to define entrance and exit channels {\em uniquely} in
terms of dividing surfaces that have the property of ``no
recrossing of trajectories'' and minimal directional flux.

If the region under consideration is compact and connected it is a simple
consequence of the Poincar\'e recurrence theorem
\cite{Arnold78} that reactive initial conditions in the region lie
(up to a set of measure zero, or ``zero volume'') on trajectories
which in the future escape from the region \emph{and} in the past
entered the region. Hence, for each point on a particular dividing
surface hemisphere $B_{\text{ds,\,f};i}^{2d-2}(E)$, there exists a time $t$ (which
depends on the point) for the trajectory starting at this point to
spend in the region before it
escapes through the same, or another, dividing surface. We define
the {\em mean passage time associated with} $B_{\text{ds,\,f};i}^{2d-2}(E)$ as,

\begin{equation}
\label{eq:mpt} \langle t\rangle_{\text{enter};i} (E)  = \left(
\int_{B_{\text{ds,\,f};i}^{2d-2}(E)} t\, \Omega \right) / \left( \int_{B_{\text{ds,\,f};i}^{2d-2}(E)}
\Omega\, \right).
\end{equation}

\noindent
 Here we use the more concise language of differential forms also used in Section \ref{sec:classicalrate} to express the measure
 on the dividing surface over which we
integrate the passage time. This measure is give by
$\Omega=\omega^{d-1}/(d-1)!$, where $\omega$ denotes the canonical
symplectic two-form $\sum_{k=1}^d \mbox{d} p_k \wedge \mbox{d}
q_k$. It then follows from arguments analogous to those that lead
to the so called classical spectral theorem proven by Pollak
 in the context of bimolecular collisions \cite{Pollak81}, that the
energy surface volume of reactive initial conditions in an
energy surface region is given by

\begin{equation}
\label{eq:volume} {\cal V}_{\text{react}} (E) = \sum_i \langle
t\rangle_{\text{enter};i}(E) \,\, \flux_{\text{enter};i}(E)\,,
\end{equation}

\noindent where the summation runs over all dividing surfaces
$B_{\text{ds,\,f};i}^{2d-2}(E)$ controlling access to the region under consideration, 
and each entrance/exit
channel contributes to the total reactive volume by the product of
the associated mean passage time and the  (directional) flux,

\begin{equation}
\label{eq:flux} \flux_{\text{enter};i} (E) = \int_{B_{\text{ds,\,f};i}^{2d-2}(E)} \Omega.
\end{equation}

\noindent The mean passage time for a given dividing surface
hemisphere can be computed from a Monte Carlo sampling of that
hemisphere. Performing such a sampling, uniformly with respect to
the measure $\Omega$, is straightforward in the normal form
coordinates. The flux through a dividing surface hemisphere is
also computed easily from the normal form as described in 
Sec.~\ref{sec:classicalrate}. The efficiency of this procedure has been demonstrated for concrete examples  in 
\cite{WaalkensBurbanksWiggins05,WaalkensBurbanksWiggins05c}

} 

\paragraph{Practical considerations:} By their very definition,
invariant manifolds consist of  trajectories, and the common way
of computing them, and visualizing them, that works well in low
dimensions is to integrate a distribution of initial conditions
located on the invariant manifold (hence, this illustrates the
value of the normal form coordinates and transformation for
locating appropriate initial conditions). In high dimensions there
are numerical and algorithmic issues that have yet to be fully
addressed. How does one choose a mesh on a $2d-3$ dimensional
sphere? As this mesh evolves in time, how does one ``refine'' the
mesh in such a way that the evolved mesh maintains the structure
of the invariant manifold?


\subsection{The flux-flux autocorrelation function formalism for
  computing classical reaction probabilities}
\label{sec:fluxfluxMiller}

In the chemistry literature (see
\cite{Yamamoto60,MillerSchwartzTromp83,Miller98}) the accepted
expression for the flux that goes in to the expression for the
classical reaction rate is given by

\begin{equation} \label{eq:flux_miller}
  \flux(E) = \int_{\R^d}  \int_{\R^d} \delta(E - H(q,p))
  F(q,p) P_{\text{r}}(q,p) \,\, \ud q\,  \ud p\,.
\end{equation}
We want to explain the relation of this expression for the flux
to the one derived in Section \ref{sec:classicalrate}.  We begin
by explaining the dynamical significance of each function in
(\ref{eq:flux_miller}). The function $\delta(E-H)$ restricts the
integration to the energy surface of energy $E$ under
consideration. The remaining functions in the integral are defined
on the basis of a dividing surface which is defined as the zero
level set of a function $s$, i.e. the dividing surface is given by
\begin{equation} \label{eq:ds_miller}
\{(q,p)\in \R^{2d} \,:\,  s(q,p)=0\}\,.
\end{equation}
It is assumed that this surface divides the phase space into two
components: a reactants component which has  $s(q,p)<0$ and a products
component which has
$s(q,p)>0$.
In the chemistry literature $s$
is usually a function of $q$ only, i.e. ``it is a dividing surface defined in
configuration space.'' However, it is crucial to note that this restriction is
not important.

If we let $\Theta$ denote the Heaviside function (which is zero if
its argument is negative and one if its argument is positive) then
the composition $\Theta \circ s$ can be viewed as a characteristic
function on phase space which  vanishes on the reactants
components and is identically one on the products component. The
function $F$ occurring in \eqref{eq:flux_miller} at a point
$(q,p)$ is then defined as the time derivative of  $\Theta \circ s
(\flow_H^t (q,p))$ at time $t=0$, i.e.,
\begin{equation}
  F(q,p) = \frac{\ud }{\ud t} \left.\Theta \circ s (\flow_H^t(q,p))\right|_{t=0}
=
\delta(s(q,p))  \{s,H\}(q,p) \,,
\end{equation}
where $\{\cdot,\cdot\}$ again denotes the Poisson bracket.
This means that $F$ is a $\delta$ function in $s$ that is weighted by the scalar product
between the gradient of the surface $s$ and the Hamiltonian vector field $X_H$,
\begin{equation} \label{eq:rewrite_F(q,p)_classical}
 F(q,p) = 
 \delta(s(q,p)) \langle  \nabla s(q,p), X_H(q,p) \rangle\,.
\end{equation}
Due to the function $\delta(s)$ in $F$ the integral \eqref{eq:flux_miller}
is effectively restricted to the dividing surface  \eqref{eq:ds_miller}, or if
we also
take into acount the function $\delta(E-H)$, the integral
\eqref{eq:flux_miller} is effectively a $(2d-2)$-dimensional integral over the
intersection of the dividing surface \eqref{eq:ds_miller} with the energy
surface of energy $E$.
It is not difficult to see that if we disregard the factor $P_{\text{r}}$ in \eqref{eq:flux_miller},
then the restriction of the resulting measure $\{s,H\}\,\ud q \ud p$ to the
intersection of the dividing surface with the energy surface agrees with the
measure $\Omega'$ that we defined in \eqref{eq:omegaprime} in
Sec.~\ref{sec:classicalrate}. This implies that the expression for the flux
\eqref{eq:flux_miller} is invariant under symplectic coordinate transformations.

The function  $P_{\text{r}}$ in \eqref{eq:flux_miller} is defined as
\begin{equation} \label{eq:defPr}
   P_{\text{r}}(q,p)= \lim_{t\rightarrow \infty} \Theta(s(\flow_H^t(q,p))\,,
\end{equation}
which evaluates to one if the trajectory with initial conditions
$(q,p)$  has  $s(q(t),p(t))>0$ and hence proceeds to products for
$t\rightarrow \infty$ and to zero otherwise. In this way  the
function $P_{\text{r}}$ in \eqref{eq:flux_miller} acts as a
characteristic function on the intersection of the dividing
surface with the energy surface.

Equation~\eqref{eq:flux_miller}
can be rewritten as
\begin{equation} \label{eq:fluxfluxintegral}
 \flux(E) =  \int_0^\infty C_{F}(t)\,\ud t\,,
\end{equation}
where
\begin{equation}
  C_{F}(t) = \int_{\R^d} \int_{\R^d} \delta(E-H(q,p)) F(q,p) F(q(t),p(t)) \,\, \ud q\,  \ud p \,,
\end{equation}
which is referred to as the flux-flux autocorrelation function.
This result is obtained from using the identity
\begin{equation} \label{eq:modifyPr}
\begin{split}
   P_{\text{r}}(q,p) &= \int_0^\infty  \frac{\ud}{\ud t} \Theta \circ s
  (\flow_H^t (q,p))
  \,\ud t\\
                     &= \int_0^\infty F(\flow_H^t(q,p)) \,\ud t\,,
\end{split}
\end{equation}
and changing the order of the time and phase space integrals.
In \eqref{eq:modifyPr} it is tacitly assumed that
$\Theta(s(q,p))=0$, which means that if we want to use the form of
$P_{\text{r}}$ given in \eqref{eq:modifyPr} in the integral \eqref{eq:flux_miller}
then it is assumed that $\Theta(s(q,p))$ evaluates to zero on the dividing
surface. This means that  one assumes that a trajectory with initial condition at  a point $(q,p)$ on the dividing surface
\eqref{eq:ds_miller} still
requires an infinitesimal time to actually cross the dividing
surface \eqref{eq:ds_miller}, i.e. more correctly \eqref{eq:fluxfluxintegral}
should be
\begin{equation} \label{eq:fluxfluxintegral_correct}
 \flux(E) =  \lim_{\epsilon\rightarrow 0+}\int_{-\epsilon}^\infty C_{F}(t)\,\ud t\,.
\end{equation}
We emphasised this point since it is important for understanding the time dependence of the
function $ C_{F}$ to which we come back below.

As stated in the chemistry literature (see e.g. \cite{Miller98})
the equivalent expressions for the flux in  \eqref{eq:flux_miller} and \eqref{eq:fluxfluxintegral}
do not depend on the particular choice of
the dividing surface.
To see this recall that an arbitrarily chosen dividing surface will in general have the
recrossing problem that we mentioned in Sec.~\ref{sec:phasespacestruct}. This
means that there are either
\begin{itemize}
\item ``nonreactive recrossings'': nonreactive trajectories that cross the dividing surface, or

\item ``reactive recrossings'': reactive trajectories that cross the dividing surface more
  than once,
\end{itemize}

or both.

In fact reactive and nonreactive recrossings are independent, i.e.
one can construct a dividing surface that only has nonreactive
recrossings, or only has reactive recrossings or has both (or no
recrossings at all like the dividing surface that we construct).
From the definition of the function $P_{\text{r}}$  in
\eqref{eq:defPr} it is clear that those nonreactive recrossings
that result from trajectories that approach the dividing surface
from the side of reactants, cross the dividing surface
\eqref{eq:ds_miller} (two or an even number larger than two times)
and return to the side of reactants do not contribute to the
integral \eqref{eq:flux_miller}. In order to see that the factor
$P_{\text{r}}$  in the expression for the flux in
\eqref{eq:flux_miller} also takes care of nonreactive trajectories
that approach the dividing surface from the products side  and
also of reactive recrossings one needs to note that $F(q,p)$ takes
into account in which direction a trajectory crosses the dividing
surface: The sign of the scalar product between the Hamiltonian
vector field and the gradient of the function $s$ that defines the
dividing surface depends on the direction in which the Hamiltonian
vector field pierces the dividing surface (see
\eqref{eq:rewrite_F(q,p)_classical}). In this way a family of
nonreactive trajectories that  approach the dividing surface from
the products side, crosses the dividing surface
\eqref{eq:ds_miller} (two or an even number larger than two times)
and returns to the side of products will have a vanishing net
contribution  to the integral \eqref{eq:flux_miller}. Similarly,
if a family of reactive trajectories crosses the dividing surface
on its  way from reactants to products $n$ times (where $n$ must
be odd for the trajectories to be reactive) then the net
contribution of the first $n-1$ intersections of this family of
trajectories to the integral \eqref{eq:flux_miller}  is zero. This
can be rigorously proven using the methods described in
\cite{WaalkensWiggins04} but we omit the details here.

The benefits that result from \eqref{eq:flux_miller} formally not
depending on the particular choice of the dividing surface are
diminished by the fact that  the implementation of the
characteristic function $P_{\text{r}}$  is computationally very
expensive. In practice (i.e., in numerical computations) one
cannot carry out the integration of Hamilton's equations to
$t=\infty$  in order to evaluate $P_{\text{r}}$ according to
\eqref{eq:defPr}. Instead one attempts to truncate the integration
after a finite time $t_0$ after which trajectories are
\emph{assumed} not to come back to the dividing surface. This is
equivalent to assuming that the flux-flux autocorrelation function
$ C_{F}(t)$ is essentially zero for times $t>t_0$ such that
the integral in \eqref{eq:fluxfluxintegral} can be truncated at
time $t_0$. A smaller time $t_0$  required for this
assumption to hold means that the amount of  numerical
computations required is reduced. This implies that some dividing surfaces
are better suited for numerical computations than others
\cite{PredescuMiller05}, but this is generally not known a priori.

We note that our dividing surface is free of recrossings. In order
to use expression \eqref{eq:flux_miller} to get our result for the
flux in \eqref{eq:fluxactions_classical} we define the function
$s$ according to $s(q,p)=q_1-p_1$ where $(q,p)$ are the normal
form coordinates that we used in Sec.~\ref{sec:nfcpss}. The delta
function $\delta(E-H(q,p))$ in the integral \eqref{eq:ds_miller}
then restricts the integration to the isoenergetic dividing
surface that we constructed in Sec.~\ref{sec:nfcpss}. In our case
$P_{\text{r}}$ simply needs to effectively restrict the integral
\eqref{eq:ds_miller} to the forward reactive hemisphere of our
dividing surface. We therefore set
\begin{equation} \label{eq:ourPr}
  P_{\text{r}}(q,p) = \Theta( q_1 -p_1 )\,.
\end{equation}
In this way we recover the expression for the flux that we have
given in \eqref{eq:fluxactions_classical}.  It is crucial to note
that in our case the evaluation of $ P_{\text{r}}$ does not
require the integration of Hamilton's equations and is therefore
computationally much cheaper than using \eqref{eq:flux_miller}
with $P_{\text{r}}$ defined according to \eqref{eq:defPr}  for an
arbitrarily chosen dividing surface. Equivalently, using the fact
that in our case we have $F = \{H,P_{\text{r}}\}$ it is easy to
see that the flux-flux autocorrelation function  $C_{F}(t)$
becomes the function $\delta(t) $ times our result for the flux
given in \eqref{eq:fluxactions_classical}. The time integration in
\eqref{eq:fluxfluxintegral_correct} (or in its  corrected version
\eqref{eq:fluxfluxintegral_correct}) becomes trivial in our case.
For an arbitrarily chosen dividing surface $C_{F}$ will as
a function of time gradually approach zero -- in a monotonic or an
oscillatory manner depending on the portions of reactive and
nonreactive recrossings of the dividing surface (see e.g.
\cite{PredescuMiller05}).

%% file: smatrix.tex
\section{Quantum Reaction Dynamics and Cumulative Reaction Probabilities}
\label{sec:smatrix}

As described in the introduction, in this section we develop the
quantum version of the classical reaction rate theory developed in
Section \ref{sec:classical}. We especially emphasize the roles of
the classical and quantum normal forms. In particular, the
classical coordinates in this section are the normal form
coordinates. Moreover, we will see that the classical phase space
structures that are realized through the classical normal form the
``skeleton'' on which the quantum dynamics evolves.



\subsection{Quantum normal form}
\label{sec:qnf_eigenfunctions}

We consider a Hamilton operator whose principal symbol has an equilibrium point of
saddle-centre-$\cdots$-centre stability type.
In Sec.~\ref{sec:examp_comp_scnf} we have shown how such a Hamilton operator
can be transformed to quantum
normal form to any desired order $N$ of its symbol by conjugating it with
suitable unitary transformations.
The resulting $N^{\text{th}}$ order quantum normal form $\widehat{H}^{(N)}_{\text{QNF}}$  is
a polynomial of order $[N/2]$  in the operators

\begin{equation}\label{eq:elementary_operators_smatrix}
\hat{I} = \frac{\hbar}{\ui}\bigg(q_1\frac{\ud}{\ud q_1}
+\frac{1}{2}\bigg) \quad \text{and}\quad
\hat{J}_k = -\frac{\hbar^2}{2}\frac{\ud^2}{\ud q_k^2}+\frac{1}{2}q_k^2\,,\quad k=2,\ldots,d\,,
\end{equation}

\noindent i.e., $ \widehat{H}^{(N)}_{\text{QNF}}$ is of the form

\begin{equation} \label{eq:qnf_operator_smatrix}
\begin{split}
   \widehat{H}^{(N)}_{\text{QNF}} &=
   K^{(N)}_{\text{QNF}}(\hat{I},\hat{J}_2,\ldots,\hat{J}_d) \\
                            &= E_0 + \lambda \hat{I} + \omega_2 \hat{J}_2 +
   \ldots +
   \omega_d \hat{J}_d+ c\hbar + \text{ higher order terms}\,,
\end{split}
\end{equation}
where $c\in\R$ is a constant and the higher order terms are of order greater than
one and  less than $[N/2]$
in the operators $\hat{I}$ and $\hat{J}_k$, $k=2,\ldots,d$.

\New{
From the structure of $\widehat{H}^{(N)}_{\text{QNF}}$ in
\eqref{eq:qnf_operator_smatrix} it follows that its eigenfunctions
are products of the eigenfunctions of the individual operators in
\eqref{eq:elementary_operators_smatrix}. 
This structure is the quantum
manifestation of the integrability of the classical normal form
described in Section \ref{sec:classicalintegrals}. In the
classical case integrability leads to a particular simple form of
Hamilton's equations which provides  a complete understanding of the
phase space structure and dynamics in a neighborhood of the
 saddle-center-$\cdots$-center equilibrium point. Similarly, we
 see that the quantum manifestation of classical integrability
 will lead to a simple structure for the corresponding quantum
 Hamilton operators in such a way that multidimensional problems
 are rendered ``solvable''.
}

The operators $\hat{J}_k$
are the Hamilton operators of one-dimensional harmonic oscillators
(with unit frequency). Their eigenvalues are $\hbar(n_k+1/2)$,
$n_k\in\N_0$, and the corresponding eigenfunctions are given by
\begin{equation}\label{eq:harm_osc_eigenfunctions}
\psi_{n_k}(q_k) = \frac{1}{(\pi\hbar)^{1/4}\sqrt{2^{n_k}n_k!}}   H_{n_k}\bigg(\frac{x}{\sqrt{\hbar}}\bigg) \ue^{-\frac{q_k^2}{2\hbar}}\,,
\end{equation}
where $H_{n_k}$ is the $n_k^{\text{th}}$ Hermite polynomial
\cite{AbraSteg65,LandauLifschitz01}. 

We will choose the eigenfunctions of $\hat{I}$ in such a way that their
product with the harmonic oscillator eigenfunctions \eqref{eq:harm_osc_eigenfunctions}
give incoming and outgoing scattering wavefunctions of the system
described by the Hamilton operator in \eqref{eq:qnf_operator_smatrix}.
For clarity, we start with the one-dimensional case.


\subsection{Scattering states for one-dimensional systems}
\label{sec:scatt_states}

The scattering states and S-matrix associated with a saddle equilibrium point
in a one-dimensional system have been studied in
\cite{ColindeVerdiereParisse94,ColindeVerdiereParisse94b,ColindeVerdiereParisse99}
and in the following we mainly follow their presentation.

For one-dimensional systems a Hamilton operator in quantum normal form  is a
polynomial function of the operator $\Ihat=-\ui \hbar\big(q \ud/\ud q + 1/2\big)$.
The scattering states $\psiI$ are the eigenfunctions of $\Ihat$, i.e., solutions of
\begin{equation}\label{eq:scat-states}
\Ihat\psiI(q) \equiv -\ui \hbar\bigg(q \frac{\ud}{\ud q} + \frac12\bigg) \psiI(q) = I\psiI(q)
\end{equation}
with eigenvalues $I\in \R$. Two solutions of this equation are given by
\begin{equation}\label{eq:out-states}
\begin{split}
\psi_{I\text{o;r}}(q)  &= \Theta(-q)\abs{q}^{-1/2+\ui  I/\hbar}\,,\\
\psi_{I\text{o;p}}(q)  &= \Theta(\phantom{-}q)\abs{q}^{-1/2+\ui  I/\hbar}\,,
\end{split}
\end{equation}
\rem{
\begin{equation}\label{eq:out-states}
\psi^{\text{out}}_{\text{react/prod}}(q)  = \Theta(\mp q)\abs{q}^{-1/2+\ui I/\hbar}\,,
\end{equation}
} 
where $\Theta$ is the Heaviside function, and
the index `o' is for `outgoing to' and `r' and `p' are for `reactants' and
`products', respectively.
The motivation for this notation becomes
clear from viewing the solutions \eqref{eq:out-states} as Lagrangian states,
i.e., we rewrite
them as
\begin{equation}
\begin{split}
\psi_{I\text{o;r/p}}(q) &= A_{I\text{o;r/p}}(q) \ue^{\ui \varphi_{I\text{o;r/p}}(q)/\hbar}\,,
\end{split}
\end{equation}
where the amplitude and phase functions are given by
\begin{equation}
A_{I\text{o;r/p}}(q)=\Theta(\mp q)\abs{q}^{-1/2}\,,\qquad
\varphi_{I\text{o;r/p}}(q)=  I\ln \abs{q}\,,
\end{equation}
respectively.
This way we can associate the one-dimensional Lagrangian manifolds
\begin{equation} \label{eq:lagrman_scatt}
\begin{split}
\Lambda_{{I\text{o;r}}} &= \bigg\{ (q,p)=\bigg(q,\frac{\ud}{\ud
q} \varphi_{I\text{o;r}}(q)\bigg)=\bigg(q,\frac{I}{q}\bigg)\,:\, q<0 \bigg\}\,,\\
\Lambda_{{I\text{o;p}}}  &=\bigg\{ (q,p)=\bigg(q,\frac{\ud }{\ud
q} \varphi_{I\text{o;p}}(q)\bigg) =\bigg(q,\frac{I}{q}\bigg)\,:\, q>0 \bigg\}
\end{split}
\end{equation}
with the states $\psi_{I\text{o;r}}$ and $\psi_{I\text{o;p}}$.
\rem{
The momentum functions associated with these solutions are
\begin{equation}
p_{\text{react/prod}}(q)=\Theta(\mp q) \varphi'_{\text{react/prod}}(q)= \mp \Theta(\mp q) \frac{I}{q}\,\,.
\end{equation}
} 
From the presentation of
$\Lambda_{{I\text{o;r}}}$ and
$\Lambda_{{I\text{o;p}}}$ in  Fig.~\ref{fig:lagrange} we see that
for $q\rightarrow -\infty$,
$\psi_{I\text{o;r}}$ is the outgoing state to reactants,
and for $q\rightarrow +\infty$,
$\psi_{I\text{o;p}}$ is the outgoing state to products.

\def\figlagrange{%
Lagrangian manifolds $\Lambda_{{I\text{o/i;r/p}}}$ associated with the wavefunctions
$\psi_{I\text{o/i;r/p}}$. The arrows indicate the Hamiltonian
vector field
generated by $I=p q$.
}
\def\FIGlagrange{
\centerline{
\includegraphics[angle=0,height=5cm]{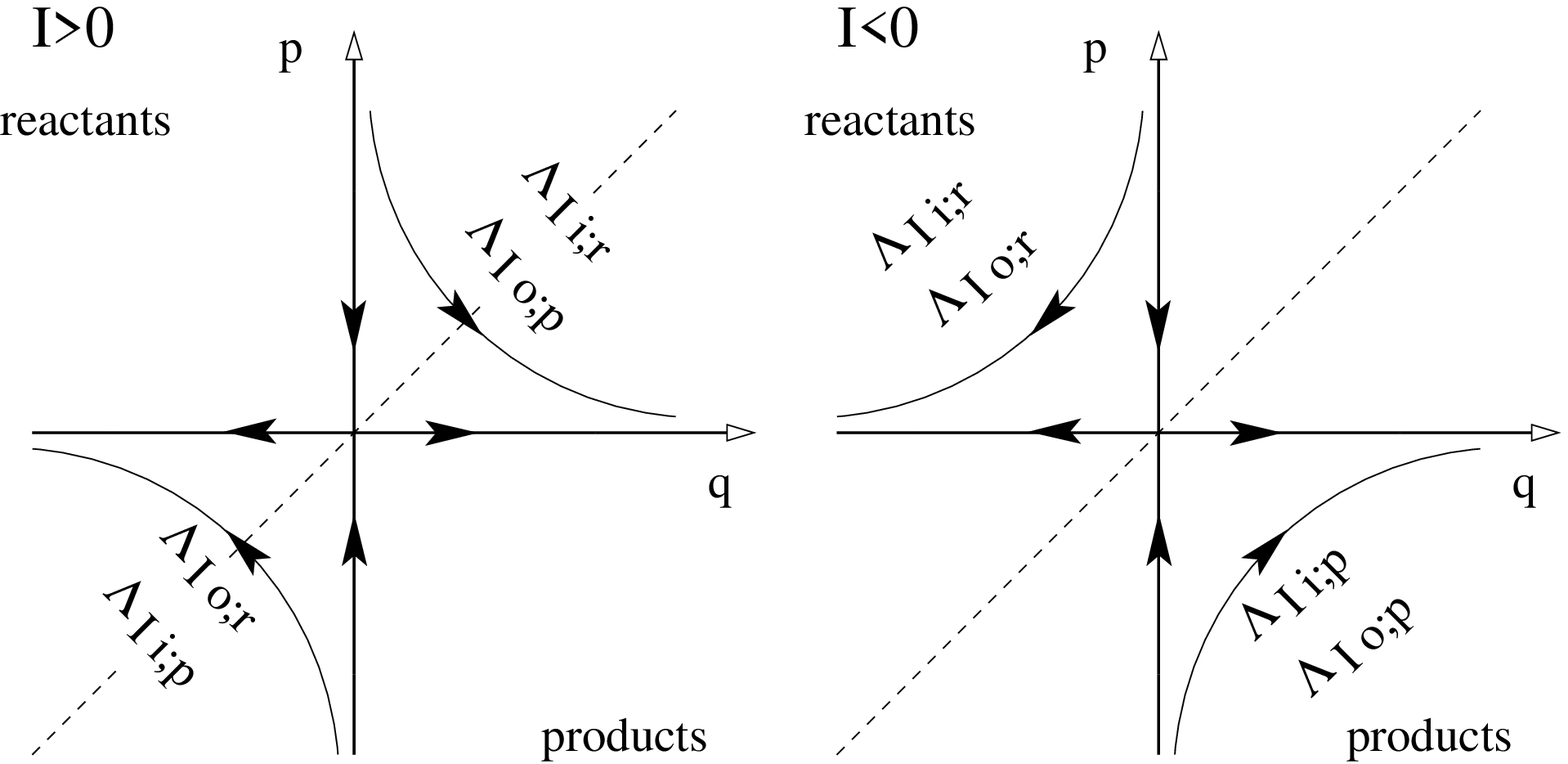}
}
}
\FIGo{fig:lagrange}{\figlagrange}{\FIGlagrange}

We define another set of eigenfunctions of $\hat{I}$ which will correspond to
incoming states by requiring their momentum representations to be given by
\begin{equation} \label{eq:def_psi_in}
\overline{\psi}_{I\text{i;r}}(p) = \psi^*_{I\text{o;p}}(p)\,,\qquad
\overline{\psi}_{I\text{i;p}}(p) = \psi^*_{I\text{o;r}}(p)\,.
\end{equation}
\rem{
If we observe that conjugation with the
Fourier transform maps $\Ihat$ to $-\Ihat$, we find a second
pair of solutions as Fourier transform of  $\psi^{\text{out}}_{\text{react/prod}}$
with $I$ replaced by $-I$,
}
Here `$*$' denotes complex conjugation.
The corresponding position representations are obtained from the Fourier
transforms of \eqref{eq:def_psi_in} giving
\begin{equation}\label{eq:psi_in_pm}
\begin{split}
\psi_{I\text{i;r}}(q)&=
\frac{1}{\sqrt{2\pi\hbar}} \int
\overline{\psi}_{I\text{i;r}}(p)
\ue^{\frac{\ui}{\hbar} qp}\,\, \ud p =
\frac{1}{\sqrt{2\pi\hbar}} \int_0^{\infty} p^{-1/2-\ui I/\hbar}
\ue^{\frac{\ui}{\hbar} qp}\,\, \ud p\,\, ,\\
\psi_{I\text{i;p}}(q)&=
\frac{1}{\sqrt{2\pi\hbar}} \int
\overline{\psi}_{I\text{i;p}}(p)
\ue^{\frac{\ui}{\hbar} qp}\,\, \ud p =
\frac{1}{\sqrt{2\pi\hbar}} \int_{-\infty}^0 (-p)^{-1/2-\ui I/\hbar}
\ue^{\frac{\ui}{\hbar} qp}\,\, \ud p\,\, .
\end{split}
\end{equation}
\rem{
\begin{equation}\label{eq:psi_in_pm}
\begin{split}
\psi^{\text{in}}_{\text{react}}(q)&=
\frac{1}{\sqrt{2\pi\hbar}} \int
\hat{\psi}^{\text{in}}_{\text{react}}(p)
\ue^{\frac{\ui}{\hbar} qp}\,\, \ud p =
\frac{1}{\sqrt{2\pi\hbar}} \int_0^{\infty} p^{-1/2-\ui I/\hbar}
\ue^{\frac{\ui}{\hbar} qp}\,\, \ud p\,\, ,\\
\psi^{\text{in}}_{\text{prod}}(q)&=
\frac{1}{\sqrt{2\pi\hbar}} \int
\hat{\psi}^{\text{in}}_{\text{prod}}(p)
\ue^{\frac{\ui}{\hbar} qp}\,\, \ud p =
\frac{1}{\sqrt{2\pi\hbar}} \int_{-\infty}^0 (-p)^{-1/2-\ui I/\hbar}
\ue^{\frac{\ui}{\hbar} qp}\,\, \ud p\,\, .
\end{split}
\end{equation}
} 
\rem{
\begin{equation}
\begin{split}
\psi^{\text{in}}_{\text{react/prod}}(q)&=\frac{1}{\sqrt{2\pi\hbar}} \int {\psi^{\text{out}}_{\text{prod/react}}(p)}^*
\ue^{\frac{\ui}{\hbar} qp}\,\, \ud p\\
&=\frac{1}{\sqrt{2\pi\hbar}} \int_0^{\infty} p^{-1/2-\ui I/\hbar}
\ue^{\pm\frac{\ui}{\hbar} qp}\,\, \ud p\,\, .
\end{split}
\end{equation}
} 
The integrals in \eqref{eq:psi_in_pm}  are not absolutely convergent, but can be defined as
oscillatory integrals.  The motivation for defining incoming states according
to  Equation~\eqref{eq:def_psi_in} becomes clear from
considering the stationary phase contributions to the integrals \eqref{eq:psi_in_pm}. These come from the
$p$ satisfying
\begin{equation}
\frac{\ud}{\ud p}(-I \ln \abs{p} + q p)=  0 \,,
\end{equation}
i.e., $p=I/q$, where $p>0$ for $\psi_{I\text{i;r}}$ and
$p<0$ for $\psi_{I\text{i;p}}$. This way we can associate with the incoming states the
Lagrangian manifolds
\begin{equation} \label{eq:defLambdairp}
\begin{split}
\Lambda_{{I\text{i;r}}} &= \bigg\{ (q,p) = \bigg(q,\frac{I}{q}\bigg)\,:\, p>0 \bigg\}\,,\\
\Lambda_{{I\text{i;p}}}  &= \bigg\{ (q,p) = \bigg(q,\frac{I}{q}\bigg)\,:\, p<0 \bigg\}\,.
\end{split}
\end{equation}
These manifolds are also shown in Fig.~\ref{fig:lagrange} and we see that for
$p\rightarrow + \infty$,
$\psi_{I\text{i;r}}$ is an incoming state from reactants
and for $p\rightarrow - \infty$,
$\psi_{I\text{i;p}}$ is an incoming state from products.

In order to evaluate the integrals \eqref{eq:psi_in_pm} we use the well known formula
\begin{equation}
\int_0^{\infty} y^{z-1}\ue^{-ky}\,\, \ud y =\ue^{-z\ln k}\Gamma(z)\,\, .
\end{equation}
This is valid for $\Re k >0$, and we will use the analytic continuation
to $\Re k=0$, in which case the left hand side is defined as an oscillatory
integral. We then obtain
\begin{equation} \label{eq:psi_in_react}
\psi_{I\text{i;r}}(q)=\begin{cases} \frac{\ue^{\ui\frac{\pi}{4}}}{\sqrt{2\pi}}
\ue^{-\ui\frac{I}{\hbar}\ln\hbar} \ue^{\frac{\pi}{2}\frac{I}{\hbar}}
\Gamma\big(\frac{1}{2}-\ui\frac{I}{\hbar}\big)\, q^{-1/2+\ui I/\hbar}\,,
& q>0\\
\frac{\ue^{-\ui\frac{\pi}{4}}}{\sqrt{2\pi}}
\ue^{-\ui\frac{I}{\hbar}\ln\hbar} \ue^{-\frac{\pi}{2}\frac{I}{\hbar}}
\Gamma\big(\frac{1}{2}-\ui\frac{I}{\hbar}\big)\, (-q)^{-1/2+\ui I/\hbar}\,,
& q<0\end{cases}\,\,.
\end{equation}
This can be rewritten as
\begin{equation} \label{eq:psi_in_react_dec}
\psi_{I\text{i;r}}=\frac{\ue^{\ui\frac{\pi}{4}}}{\sqrt{2\pi}}
\ue^{-\ui\frac{I}{\hbar}\ln\hbar}
\Gamma\bigg(\frac{1}{2}-\ui\frac{I}{\hbar}\bigg)
\big(\ue^{\frac{\pi}{2}\frac{I}{\hbar}}\psi_{I\text{o;p}}
-\ui\ue^{-\frac{\pi}{2}\frac{I}{\hbar}} \psi_{I\text{o;r}}\big)\,\,.
\end{equation}
In the same way we obtain
\begin{equation} \label{eq:psi_in_prod_dec}
\psi_{I\text{i;p}}=\frac{\ue^{\ui\frac{\pi}{4}}}{\sqrt{2\pi}}
\ue^{-\ui\frac{I}{\hbar}\ln\hbar}
\Gamma\bigg(\frac{1}{2}-\ui\frac{I}{\hbar}\bigg)
\big(
\ue^{\frac{\pi}{2}\frac{I}{\hbar}} \psi_{I\text{o;r}}
-\ui\ue^{-\frac{\pi}{2}\frac{I}{\hbar}}\psi_{I\text{o;p}}
\big)\,\, .
\end{equation}

For what follows in Sec.~\ref{sec:flux-flux_quantum} it is useful
to discuss how the eigenfunctions $\psi_{I\text{o;r/p}}$ and
$\psi_{I\text{i;r/p}}$ are related to the more standard
eigenfunctions of the operator $\Ihat$ in the $Q$-representation
that we introduced in Sec.~\ref{sec:conjunitary} (see
\eqref{eq:def45rotation}-\eqref{eq:opI45rotation}).

The eigenvalue equation \eqref{eq:scat-states} then becomes
\begin{equation}
\Ihat\chi_I(Q) = \bigg(-\frac{\hbar^2}{2}\frac{\ud^2}{\ud Q^2}- \frac12 Q^2 \bigg) \chi_I(Q) = I\chi_I(Q)\,.
\end{equation}
Two solutions of this equation are given by
\begin{equation} \label{eq:def_psi_I_pm}
  \chi_{I\pm} (Q)= \frac{1}{\sqrt{2\pi^2\hbar}} \left(\frac{1}{2\hbar
  }\right)^{1/4} \ue^{\frac{I}{\hbar}\frac{\pi}{4}}\Gamma\bigg(\frac12 -\ui \frac{I}{\hbar}\bigg)
  D_{-\frac12 +\ui \frac{I}{\hbar} }\left(\pm \ue^{-\ui\frac{\pi}{4}}\sqrt{ \frac{2 }{\hbar}} Q\right)\,,
\end{equation}
where $D_{\nu}$ is the parabolic cylinder function \cite{AbraSteg65,LandauLifschitz01}.
In fact, the eigenfunctions $\psi_{I\text{i;r/p}}$
are the images of  $\chi_{I+/-}$
under  the unitary transformation $\widehat{U}_{\text{r}}$ that we defined in \eqref{eq:defUr}, or equivalently
\begin{equation}
\begin{split}
  \chi_{I+} = \widehat{U}^*_{\text{r}} \, \psi_{I\text{i;r}}\,,\quad
  \chi_{I-} = \widehat{U}^*_{\text{r}} \, \psi_{I\text{i;p}}\,.
\end{split}
\end{equation}
This relationship is discussed in great detail in
\cite{Chruscinski03,Chruscinski03b} where  it is also shown
that the pairs of
eigenfunctions $\psi_{I\text{i;r/p}}$,
$\psi_{I\text{o;r/p}}$ and $\chi_{I+/-}$ are orthogonal and fulfill the
completeness relations
\begin{equation}\label{eq:completeness_rel_general}
\begin{split}
  \int_\R \big(
  \psi^{*}_{I\text{i;r}}(q)\psi_{I\text{i;r}}(q') +
  \psi^{*}_{I\text{i;p}}(q)\psi_{I\text{i;p}}(q')
  \big) \,\ud I&= \delta(q-q')\,,\\
  \int_\R \big(
  \psi^{*}_{I\text{o;r}}(q)\psi_{I\text{o;r}}(q') +
   \psi^{*}_{I\text{o;p}}(q)\psi_{I\text{o;p}}(q')
  \big) \,\ud I&= \delta(q-q')\,,\\
  \int_\R \big(
   \chi_{I+}^{*} (Q) \chi_{I+} (Q') + \chi_{I-}^{*} (Q) \chi_{I-} (Q')
   \big) \,\ud I&= \delta(Q-Q')\,.
\end{split}
\end{equation}

\subsection{S-matrix and transmission probability for one-dimensional systems}
\label{sec:oneDSmatrix}

The incoming and outgoing wavefunctions defined in Sec.~\ref{sec:scatt_states} are not independent.
Each solution $\psiI$
of \eqref{eq:scat-states} can be written as a linear combination
of $\psi_{I\text{o;r/p}}$ or $\psi_{I\text{i;r/p}}$,
\begin{align}
\psiI&= \alpha_p \psi_{I\text{o;p}}+\alpha_r \psi_{I\text{o;r}}\,,\\
\psiI&= \beta_p \psi_{I\text{i;p}}+\beta_r \psi_{I\text{i;r}} \,.
\end{align}
These representations are connected by the S-matrix,
\begin{equation}
\begin{pmatrix}\alpha_p \\ \alpha_r\end{pmatrix}
=\mathcal{S}(I)\begin{pmatrix}\beta_p\\ \beta_r \end{pmatrix}\,\, .
\end{equation}
We can  read off the entries of the S-matrix from
\eqref{eq:psi_in_react_dec} and \eqref{eq:psi_in_prod_dec} and obtain
\begin{equation}\label{eq:local-S}
\mathcal{S}(I)=\frac{\ue^{\ui\frac{\pi}{4}}}{\sqrt{2\pi}}
\ue^{-\ui\frac{I}{\hbar}\ln\hbar}
\Gamma\bigg(\frac{1}{2}-\ui\frac{I}{\hbar}\bigg)
\begin{pmatrix}
-\ui \ue^{-\frac{\pi}{2}\frac{I}{\hbar}} & \ue^{\frac{\pi}{2}\frac{I}{\hbar}}  \\
\ue^{\frac{\pi}{2}\frac{I}{\hbar}} & -\ui \ue^{-\frac{\pi}{2}\frac{I}{\hbar}}
\end{pmatrix}\,\, .
\end{equation}
\rem{
and hence
\begin{equation}\label{eq:local-S}
S=\frac{\ue^{-\ui\frac{\pi}{4}}}{\sqrt{2\pi}}
\ue^{\ui\frac{I}{\hbar}\ln\hbar}
\Gamma\bigg(\frac{1}{2}+\ui\frac{I}{\hbar}\bigg)
\begin{pmatrix} \ue^{\frac{\pi}{2}\frac{I}{\hbar}} & -\ui \ue^{-\frac{\pi}{2}\frac{I}{\hbar}} \\ -\ui \ue^{-\frac{\pi}{2}\frac{I}{\hbar}} &\ue^{\frac{\pi}{2}\frac{I}{\hbar}} \end{pmatrix}\,\, .
\end{equation}
}
Using the relation $\Gamma(1/2+\ui y)\Gamma(1/2-\ui y)=\pi/\cosh(\pi y)$
it is easy to see that $\mathcal{S}(I)^*\mathcal{S}(I)=1$, i.e., $\mathcal{S}(I)$ is unitary.

From the S-matrix we can determine the transmission coefficient
\begin{equation}
\mathcal{T}(I) =\abs{\mathcal{S}_{12}(I)}^2=\frac{\ue^{\pi\frac{I}{\hbar}}}{\ue^{\pi\frac{I}{\hbar}}+\ue^{-\pi\frac{I}{\hbar}}} =\frac{1}{1+\ue^{-2\pi \frac{I}{\hbar}}}\,\, ,
\end{equation}
and the reflection coefficient
\begin{equation}
\mathcal{R}(I) =\abs{\mathcal{S}_{11}(I)}^2=\frac{\ue^{-\pi\frac{I}{\hbar}}}{\ue^{\pi\frac{I}{\hbar}}+\ue^{-\pi\frac{I}{\hbar}}} =\frac{1}{1+\ue^{2\pi \frac{I}{\hbar}}}\,\, .
\end{equation}
As required we have $\mathcal{T}(I) + \mathcal{R}(I) = 1$.
We see that the relevant scale is $I/\hbar$. $\mathcal{T}$ tends to $1$ if $I\gg \hbar$ and to $0$ if $I\ll -\hbar$.

We can generalise this now easily to operators $\widehat{H}_{\text{QNF}}=K_{\text{QNF}}(\Ihat)$, where
$K_{\text{QNF}}$ is a polynomial function of $\Ihat$. In this case the incoming and
outgoing states defined in Sec.~\ref{sec:scatt_states} are also eigenfunctions of $\widehat{H}_{\text{QNF}}$. We have
\begin{equation}
\widehat{H}_{\text{QNF}}\psi_{I\text{i/o;r/p}} = E \psi_{I\text{i/o;r/p}}\,,
\end{equation}
where $E=K_{\text{QNF}}(I)$ with $I$ being the corresponding eigenvalue of $\Ihat$.  The expression for the
S-matrix in \eqref{eq:local-S} remains valid with $I$ replaced by
$I(E):=K_{\text{QNF}}^{-1}(E)$, where we have to assume that the energy is close enough to the equilibrium energy so that 
$K_{\text{QNF}}(E)$ is invertible.  We thus obtain the S-matrix for the scattering problem described by the Hamilton operator
$\widehat{H}_{\text{QNF}}=K_{\text{QNF}}(\Ihat)$\,,
\begin{equation}
S(E)=\mathcal{S}(I(E))\,.
\end{equation}
The corresponding transmission coefficient is given by
\begin{equation}
T(E)=\mathcal{T}(I(E))=\frac{1}{1+\exp\big(-2\pi \frac{I(E)}{\hbar}\big)}\,,
\end{equation}
and similarly the reflection coefficient is given by
$R(E)=\mathcal{R}(I(E))$.
This is a simple generalisation of the previous example.
However, it is a very important result because we see that we can
use the quantum normal form  to compute the local
S-matrix and the transmission and reflection coefficients to any desired order
of the symbol of the Hamilton operator that describes the scattering problem.

\subsection{S-matrix and cumulative reaction probability for multi-dimensional systems}
\label{sec:smatrixmultiD}

We now consider the multi-dimensional case. In this case the Hamilton operator
in quantum normal form is given by
$\widehat{H}_{\text{QNF}}=K_{\text{QNF}}(\Ihat, \Jhat_2,\dots ,\Jhat_{d})$,
where $K_{\text{QNF}}$ is a polynomial function, and
$\hat{J}_k=(-\hbar^2\pa_{q_k}^2 +q_k^2)/2$, $k=2, \dots ,d$, are
one-dimensional harmonic oscillators. Let $\psi_{n_k}$, $n_k\in\N_0$, be the $n_k^\text{th}$
harmonic oscillator eigenfunction \eqref{eq:harm_osc_eigenfunctions}, i.e.,
\begin{equation}\label{eq:harm-osc-E}
\hat{J}_k\psi_{n_k}=\hbar({n_k}+1/2)\psi_{n_k}\,\, .
\end{equation}
Then the
incoming and outgoing scattering states are given by
\begin{equation}\label{eq:multscatteringwaves}
\begin{split}
\psi_{(I,\nscatt)\,\text{i;r/p}}(q_1,\ldots ,q_d)&=
\psi_{I\text{i;r/p}}(q_1) \psi_{n_2}(q_2)\cdots\psi_{n_{d}}(q_{d})\,, \\
\psi_{(I,\nscatt)\,\text{o;r/p}}(q_1,\ldots ,q_d)&=
 \psi_{I\text{o;r/p}}(q_1) \psi_{n_2}(q_2)\cdots\psi_{n_{d}}(q_{d})\,,
\end{split}
\end{equation}
where ${\nscatt}=(n_2,\ldots ,n_{d})\in\N_0^{d-1}$
is a $(d-1)$-dimensional vector of scattering quantum numbers.

The S-matrix connecting incoming to outgoing states is then
block-diagonal with
\begin{equation}\label{eq:Smatrixfull}
S_{\nscatt,\mscatt}(E)
=\delta_{\nscatt,\mscatt} \mathcal{S}(I_{\nscatt}(E)) \,\, ,
\end{equation}
where $\delta_{\nscatt,\mscatt}$ is the multi-dimensional Kronecker symbol,
$\cal{S}(I)$ is given by \eqref{eq:local-S} and $I_{\nscatt}(E)$ is
determined by
\begin{equation}\label{eq:I-of-E}
K_{\text{QNF}}\big(I_{\nscatt}(E) ,\hbar(n_2+1/2), \dots ,\hbar(n_{d}+1/2)\big)= E\,\, .
\end{equation}
We will assume that this equation has a unique solution $I_{\nscatt}(E)$, which is guaranteed if the energy is close enough 
to the equilibrium energy since $K_{\text{QNF}}$ starts linearly in the actions, see \eqref{eq:qnf_operator_smatrix}. 


\New{We can now define the transition matrix $T$ as the diagonal sub-block of the S-matrix which has
the $(1,2)$-components of the matrices in \eqref{eq:Smatrixfull} on the diagonal, i.e.,
\begin{equation} \label{eq:transmission1D}
\begin{split}
T_{\nscatt,\mscatt}(E)&= \delta_{\nscatt,\mscatt} \mathcal{S}_{1,2}(I_{\nscatt}(E))\\
&=\delta_{\nscatt,\mscatt}
\bigg[1+\exp\bigg(-2\pi\frac{I_{\nscatt}(E)}{\hbar}\bigg) \bigg]^{-1}\,.
\end{split}
\end{equation}}
The \emph{cumulative reaction probability} $N(E)$ is then defined  as (see,
e.g., \cite{Miller98})
\begin{equation}
N(E) = \text{Tr } T(E)T(E)^\dagger\,.
\end{equation}
Using \eqref{eq:transmission1D} we thus get
\begin{equation}\label{eq:cum_react}
N(E) = \sum_{{\nscatt}} T_{\nscatt,\nscatt}(E)=
\sum_{{\nscatt}\in\N_0^{d-1}} \bigg[1+\exp\bigg(-2\pi\frac{I_{\nscatt}(E)}{\hbar}\bigg) \bigg]^{-1}\,.
\end{equation}

The cumulative reaction probability $N(E)$ is the quantum analogue of the
classical flux $\flux(E)$ or, more precisely, of the dimensionless quantity
$\Nweyl(E)=\flux(E)/(2\pi \hbar)^{d-1}$ that we defined in
Equation~\eqref{eq:NWeyl_def} in Sec.~\ref{sec:classicalrate}.
To see this let us consider $N(E)$ in the semiclassical limit
$\hbar\rightarrow 0$.
To this end first note that
\begin{equation}
\bigg[1+\exp\bigg(-2\pi I/\hbar \bigg) \bigg]^{-1} \rightarrow \Theta(I) \text{ as } \hbar \rightarrow 0\,,
\end{equation}
where $\Theta$ is the Heaviside function. This means that
the transmission coefficients $ T_{\nscatt,\nscatt}(E)$ in \eqref{eq:cum_react}
are essentially characteristic functions, i.e.,  in the semiclassical limit,
$ T_{\nscatt,\nscatt}(E)$ is 0 or 1 if the solution of
$K(I_{\nscatt},\hbar(n_2+1/2),\ldots,\hbar(n_d+1/2))=E$ for $I_{\nscatt}$
is negative or positive, respectively.
This way the cumulative reaction probability can be considered to be  a counting function. For a given
energy $E$, it counts how
many of the solutions $I_\nscatt$ of the  equations
$K_{\text{QNF}}(I_{\nscatt},\hbar(n_2+1/2),\ldots,\hbar(n_d+1/2))=E$ with scattering quantum numbers
$\nscatt = (n_2,\ldots,n_d) \in\N_0^{d-1}$ are positive:
\begin{equation}
N(E) \rightarrow   \# \{ I_{\nscatt}>0 \,:\,
K_{\text{QNF}}(I_{\nscatt},\hbar(n_2+\frac12),\ldots,\hbar(n_d+\frac12))=E,\,
\nscatt\in\N_0^{d-1} \}\,,
\end{equation}
as $\hbar \rightarrow 0$. In other words, $N(E)$ can be considered to count the
number of open `transmission channels', where a transmission channel with quantum numbers $\nscatt$
is open if the corresponding transmission coefficient
$T_{\nscatt,\nscatt}(E)$ is close to 1.

\def\fighbargrid{%
(a) Lines $(I,\hbar(n_2+1/2),\ldots,\hbar(n_d+1/2))$, $I\in\R$,
$n_k\in\N_0$, $k=2,\ldots,d$, in the space
$(I,J_2,\ldots,J_d)\in\R\times[0,\infty)^{d-1}$ for $d=3$ and their
intersections with the surface $K_{\text{QNF}(I,J_2,J_3)}=E$.
(b) Grid points  $(\hbar(n_2+1/2),\ldots,\hbar(n_d+1/2))$ in the space
$(J_2,\ldots,J_d)$ for $d=3$. The blue line marks the  contour
$K_{\text{QNF}}(0,J_2,\ldots,J_d)=E$. In this plot only the scattering states for
which the quantum numbers $(n_2,n_3)$ have the values $(0,0)$,
$(0,1)$, $(1,0)$ or $(1,1)$ correspond to ``open transmission channels'', see text.
}
\def\FIGhbargrid{
\centerline{
  \raisebox{6cm}{a)}\includegraphics[width=8.0cm]{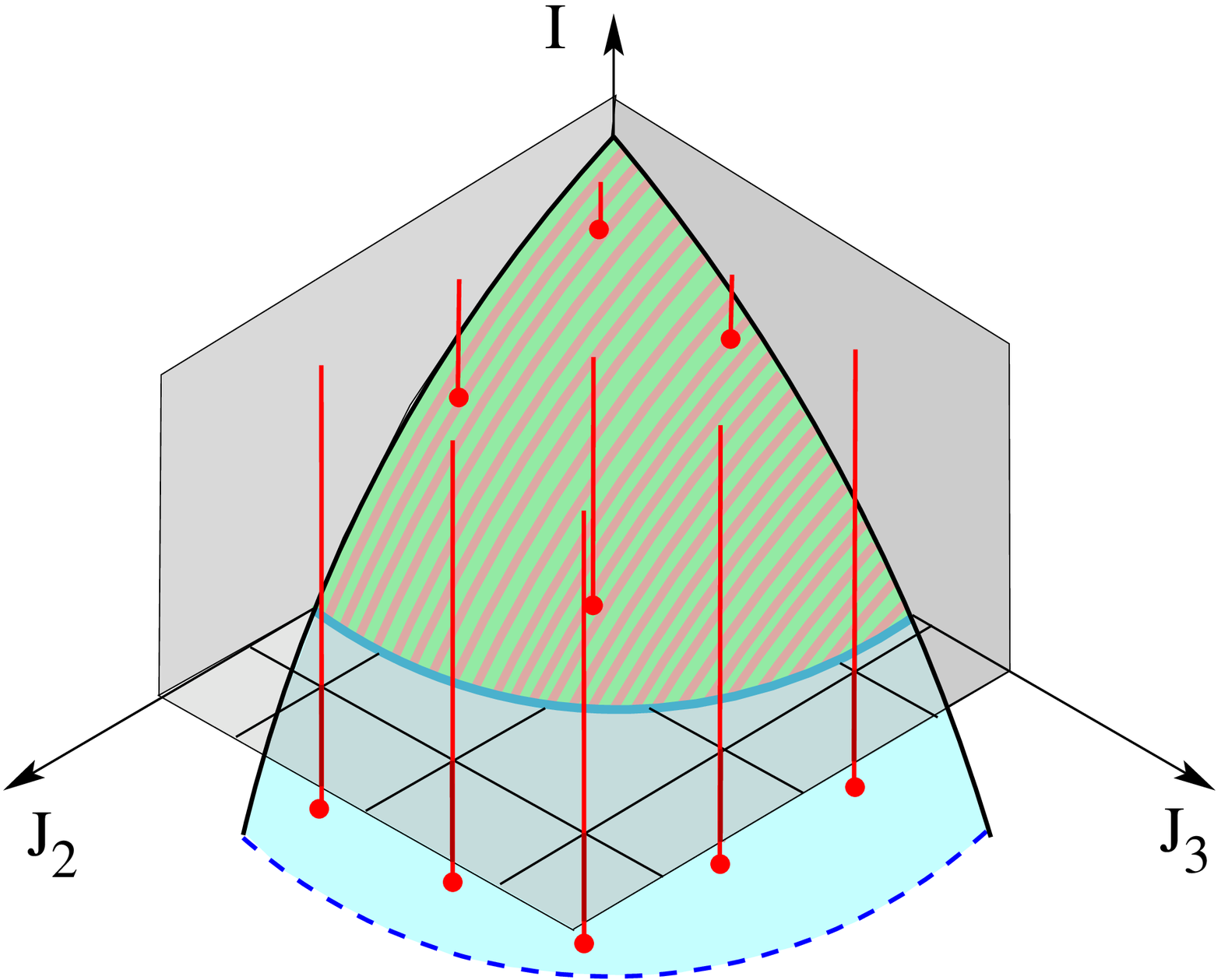}
  \raisebox{6cm}{b)}\includegraphics[width=5.0cm]{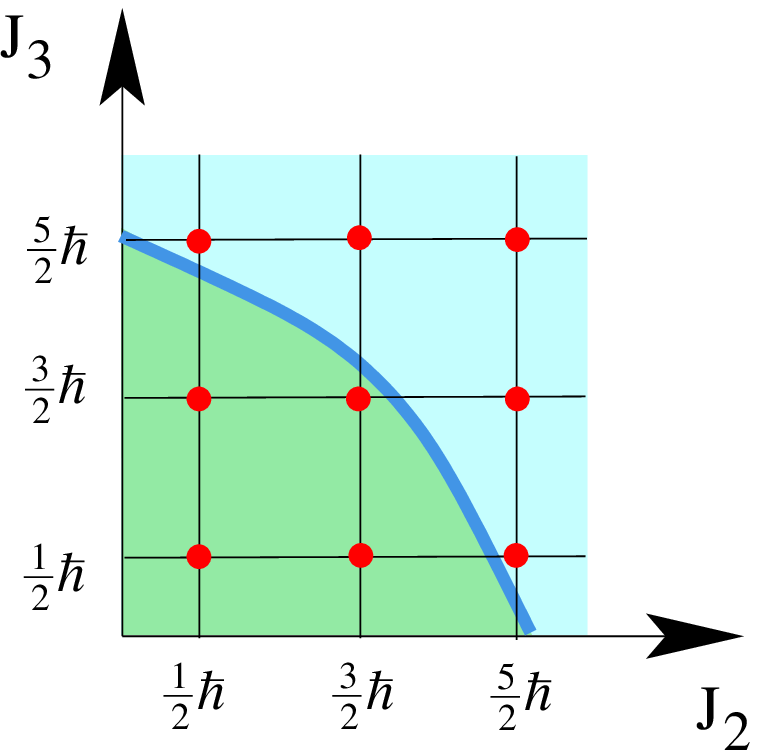}
}
}
\FIGo{fig:hbargrid}{\fighbargrid}{\FIGhbargrid}

We can interpret $N(E)$ graphically as the number of grid points
$(\hbar(n_2+1/2),\ldots,\hbar(n_d+1/2))$ in the space of $(J_2,\ldots,J_d)\in[0,\infty)^{d-1}$
that are enclosed by the contour
$K_{\text{QNF}}(0,J_2,\ldots,J_d)=E$, see Fig.~\ref{fig:hbargrid}.
The number of grid points is approximately given by the volume in the space of
$(J_2,\ldots,J_d)\in[0,\infty)^{d-1}$ enclosed by $K_{\text{QNF}}(0,J_2,\ldots,J_d)=E$ divided by
$\hbar^{d-1}$. Using the fact that for $\hbar\rightarrow 0$,  $K_{\text{QNF}}$ becomes the function
$K_{\text{CNF}}$  which gives the classical energy as a function of the
classical integrals $(I,J_2,\ldots,J_d)$ we find that
the  volume in the space of
$(J_2,\ldots,J_d)$ enclosed by $K_{\text{CNF}}(0,J_2,\ldots,J_d)=E$ is given by the
classical flux $\flux(E)$ divided by $(2\pi)^{d-1}$, see \eqref{eq:fluxactions_classical} in
Sec.~\ref{sec:classicalrate},
and the cumulative reaction probability
$N(E)$ is thus approximately given by $\Nweyl(E)=\flux(E)/(2\pi\hbar)^{d-1}$ defined in  \eqref{eq:NWeyl_def} in
Sec.~\ref{sec:classicalrate}. This way we verified our statement in
Sec.~\ref{sec:classicalrate} that  $\Nweyl(E)$ gives the mean number of
open transmission channels. In fact,
as mentioned in Sec.~\ref{sec:classicalrate}, the
classical flux $\flux(E)$ can be considered to be the phase space
volume enclosed by the energy contour of energy $E$ of the invariant
subsystem which has one degree of freedom less than the full scattering
system and which as the so called activated complex is located between
reactants and products. $\Nweyl(E)$ counts how many elementary quantum
cells of volume $(2\pi \hbar)^{d-1}$ fit into this phase space volume and this way gives the Weyl approximation of the cumulative reaction probability $N(E)$.

It is important to note here that like the flux in the classical case
 the cumulative reaction probability is
determined by  local properties of the Hamilton operator embodied in 
its symbol in the neighbourhood of the equilibrium point only. 
All
one needs to know is the quantum normal form, which enters through
the relation \eqref{eq:I-of-E} and which determines $I_{\nscatt}(E)$.

\subsection{Distribution  of the scattering states in phase space}
\label{sec:husimi_scatt}

At the end of the  previous section we have seen how the cumulative reaction probability is related to the classical flux.
In this section we want to further investigate the quantum classical
correspondence by studying the distribution of the scattering states in phase
space and relating these distributions to the classical phase space structures that
control classical reaction dynamics as discussed in
Sections~\ref{sec:phasespacestruct} and \ref{sec:nfcpss}.

The standard tool to describe the phase space distribution of a wavefunction is the 
Wigner function, but since the scattering wavefunctions are not square integrable the Wigner functions 
will be distributions. Therefore it is more 
convenient to study the phase space distribution in terms of  their
\emph{Husimi representation} which is obtained from projecting the
scattering states onto a coherent state basis (see \cite{Har88,Bal98}) and this way leads to smooth functions.  
For a point $(q_0,p_0)\in \R^d\times \R^d$
we define a coherent state with wavefunction
\begin{equation}\label{def:coherent_state}
\psi_{q_0,p_0}(q)=\frac{1}{(\pi\hbar)^{d/4}} \ue^{\frac{\ui}{\hbar}( \langle p_0,q \rangle -\langle q_0 , p_0 \rangle /2)} \ue^{-\frac{1}{2\hbar}\langle q-q_0,q-q_0\rangle}\,\, .
\end{equation}
This wavefunction is concentrated around $q=q_0$ and its Fourier transform, i.e.
its momentum representation, is concentrated around
$p=p_0$.  In phase space the coherent state \eqref{def:coherent_state} is thus concentrated around $(q_0,p_0)$. The
Husimi function of a state $\psi$ is now defined by the  modulus square of the projection
onto a coherent state,
\begin{equation}
H_{\psi}(q,p):=\frac{1}{(2\pi\hbar)^d}\abs{ \la \psi_{p,q},\psi \ra}^2\,\, .
\end{equation}
It has the important property that the expectation value of an operator
$\Op[A]$ with respect to a state $\psi$ is given by
\begin{equation}
\la \psi, \Op[A]\psi\ra =\iint_{\R^d\times \R^d} A(q,p)H_{\psi}(q,p)\,\, \ud q\ud p +O(\hbar)\,\, .
\end{equation}
Furthermore, we have $H_{\psi}(q,p)\geq 0$, i.e., the Husimi function can be considered
to be a probability density on phase space and describes how a quantum state is distributed 
in phase space.


The Husimi functions of the  scattering states
$\psi_{(I,\nscatt)\,\text{i/o;r/p}}$
inherit the product structure \eqref{eq:multscatteringwaves}, i.e.
we have
\begin{equation}\label{eq:hus_multscatteringwaves}
\begin{split}
H_{\psi_{(I,\nscatt)\,\text{i/o;r/p}}}(q_1,\ldots ,q_d,p_1,\ldots,p_d)&=
H_{\psi_{I\text{i/o;r/p}}}(q_1,p_1)
H_{\psi_{n_2}}(q_2,p_2)  \cdots
H_{\psi_{n_{d}}}(q_{d},p_d) \,\, .
\end{split}
\end{equation}

The Husimi functions of the eigenfunctions $\psi_{n_k}$ of the one-dimensional harmonic oscillators $\hat{J}_k$
are well known (see, e.g.,\cite{KorschMuellerWiescher97}),
\begin{equation}  \label{eq:harmHussimi}
H_{\psi_{n_k}}(q_k,p_k)=  \frac{1}{2\pi\hbar 2^{n_k} n_k!}
\frac{(p_k^2+q_k^2)^{n_k}}{\hbar^{n_k}} \ue^{-\frac{p_k^2+q_k^2}{2\hbar}}\,\, .
\end{equation}
The first three of these Husimi functions are shown in Fig.~\ref{fig:husscatt}.
They are concentrated on the circles $p_k^2+q_k^2=2n_k\hbar$ and have an $n_k$-fold zero at the origin.

The computation of the Husimi functions for the one-dimensional scattering states
$\psi_{I\text{o;r/p}}$ in \eqref{eq:out-states}
can be found in \cite{NonVor97} where it is shown that for the linear combination
\begin{equation} \label{eq:alphabetacombination}
\psi^{\alpha,\beta}=\alpha \psi_{I\text{o;p}}+\beta\psi_{I\text{o;r}}\,\,  ,\quad  \alpha,\beta\in \C\,\, ,
\end{equation}
one gets
\begin{equation}
\begin{split}
H_{\psi^{\alpha,\beta}}(q,p)&=\frac{\sqrt{\pi}}{2\pi\hbar\cosh(\pi I/\hbar)}\ue^{-\frac{1}{2\hbar} (p^2+q^2)}\\
&\qquad\biggabs{\alpha D_{-\frac{1}{2}-\frac{\ui I}{\hbar}}\bigg(- \frac{q-\ui p}{\sqrt{\hbar}} \bigg)
+\beta D_{-\frac{1}{2}-\frac{\ui I}{\hbar}}\bigg( \frac{q-\ui p}{\sqrt{\hbar}} \bigg)}^2
\,\, ,
\end{split}
\end{equation}
\rem{
For $I=0$ the functions simplify a bit. Here we have
\begin{equation}
D_{-1/2}(y) = \bigg(\frac{y}{2\pi}\bigg)^{1/2} K_{1/4}(y^2/4)
\end{equation}
and
\begin{equation}
\frac{1}{2}(D_{-1/2}(y)\pm D_{-1/2}(-y)) =\ue^{\mp\ui\pi/8} (\pi y)^{1/2}J_{\mp 1/4}(\ui y^2/4)\,.
\end{equation}
} 
where $D_\nu$ again denotes the parabolic cylinder function \cite{AbraSteg65}.
Fig.~\ref{fig:husscatt} shows contour plots of the Husimi representation of
the state $\psi_{I\text{i;r}}$ for
 different values of the eigenvalue $I$. Here $\alpha$ and $\beta$ in \eqref{eq:alphabetacombination}
 are determined from \eqref{eq:psi_in_react_dec}.
In accordance with the classical dynamics where trajectories with $I<0$ are non-reactive and trajectories with $I>0$ are reactive,
most of the state
$\psi_{I\text{i;r}}$ is reflected to the reactants side for $I<0$
while it is mainly transmitted to the products side for $I>0$.
The borderline case between these two situations is given by $I=0$. Here the state is
localised in phase space at the hyperbolic equilibrium point with ridges
along the reactants branches of the
stable and unstable manifold and the products branch of the unstable manifold.

\def\fighusscatt{%
Contour plots of the harmonic oscillator Husimi functions $H_{\psi_{n_k}}$ in the $(q_k,p_k)$-plane for
$n_k=0$ (a)
$n_k=1$ (b) and
$n_k=2$ (c),
and contour plots of the Husimi functions $H_{\psi_{I\text{i;r}}}$ in the $(q_1,p_1)$-plane for
$I=-1$ (d) $I=0$ (e) and $I=1$ (f).
Red corresponds to low values; blue corresponds to high values. In (a)-(c) the spacing between the values of the contourlines
is decreasing exponentially. ($\hbar=0.1\,\,.$)
}
\def\FIGhusscatt{
\centerline{
\includegraphics[angle=0,width=10cm]{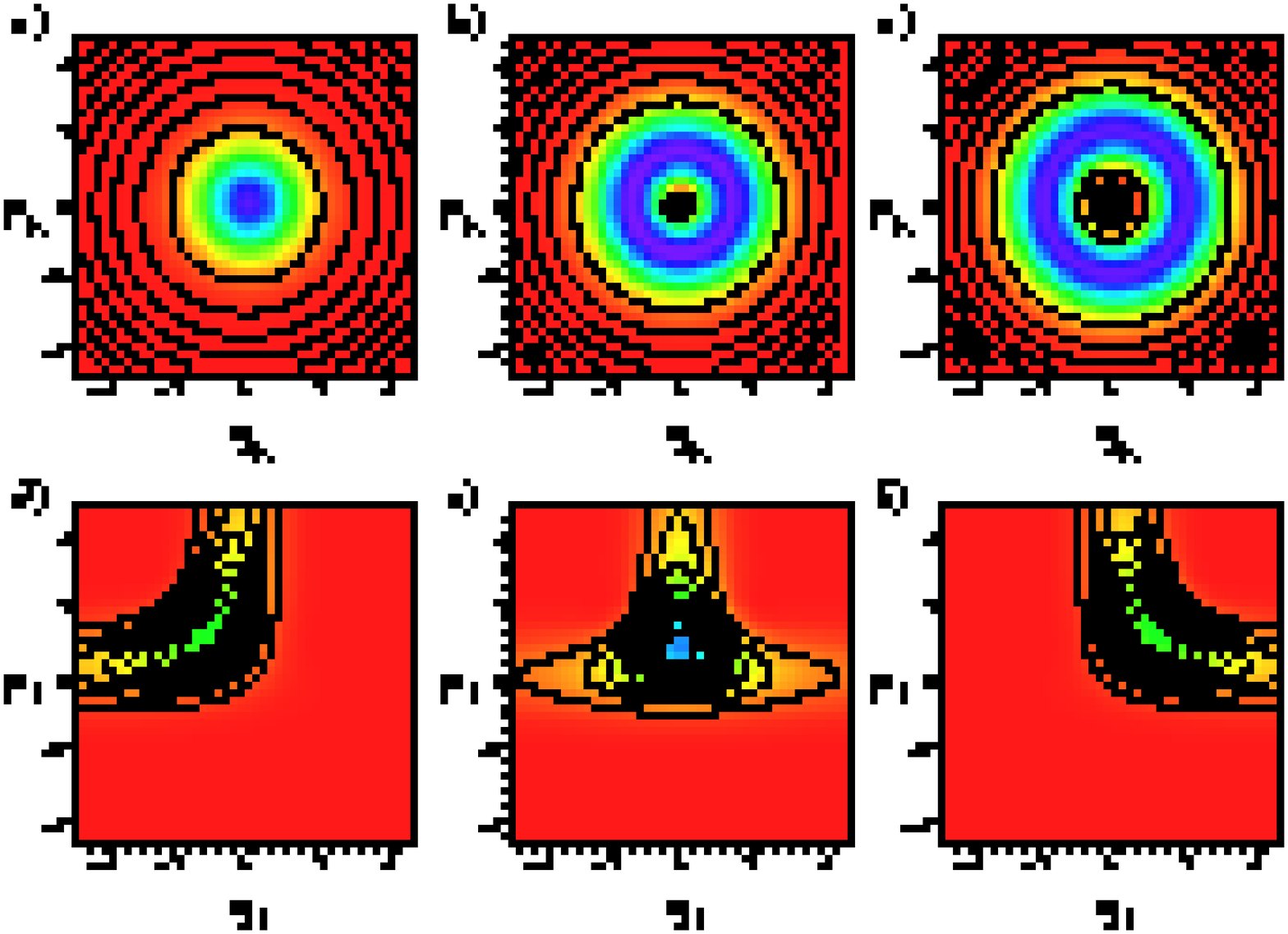}
}
}
\FIGo{fig:husscatt}{\fighusscatt}{\FIGhusscatt}

Fig.~\ref{fig:husscatt} indicates that the Husimi functions of the scattering
states $\psi_{I\text{i;r}}$ are localised on
the Lagrangian manifolds
\begin{equation} \label{eq:lagr_manf_scatt_multi}
\Lambda_{(I,{\nscatt)\,\text{i/o;r/p}}} =
\Lambda_{{I\text{i/o;r/p}}} \times \Lambda_{{n_2}}\times \cdots \times \Lambda_{{n_d}}\,,
\end{equation}
where the $\Lambda_{I\text{i/o;r/p}}$ are defined in \eqref{eq:lagrman_scatt}
and \eqref{eq:defLambdairp}, and
\begin{equation}
 \Lambda_{{n_k}} = \{ (q_k,p_k)\in \R^2 \,:\, q_k^2+p_k^2=2\hbar n_k\}\,,\quad k=2,\dots,d,
\end{equation}
are the Lagrangian manifolds associated with one-dimensional harmonic
oscillator eigenfunctions.
Quantum mechanics thus picks out those Lagrangian manifolds
$\Lambda^{\pm}_{I,J_2,\ldots,J_d}$
foliating the classical phase space (see Sec.~\ref{sec:fls}) for which the
actions, $J_2,\ldots,J_d$, fulfill Bohr-Sommerfeld quantisation conditions.
More precisely we find that  the outgoing scattering states
$\psi_{I;\text{o;r/p}}$
are localised on the the Lagrangian manifolds
\begin{equation}
\begin{split}
\Lambda_{(I,{\nscatt)\,\text{o;r}}} &= \Lambda^-_{I,\hbar n_2,\ldots,\hbar
  n_d}\,,\\
\Lambda_{(I,{\nscatt)\,\text{o;p}}} &= \Lambda^+_{I,\hbar n_2,\ldots,\hbar
  n_d}\,,
\end{split}
\end{equation}
and the incoming scattering states $\psi_{I;\text{i;r/p}}$ are localised
on the Lagrangian manifolds
\begin{equation}
\begin{split}
\Lambda_{(I,{\nscatt)\,\text{i;r}}} &=
\left\{
\begin{array}{cc}
\Lambda^+_{I,\hbar n_2,\ldots,\hbar
  n_d}\,, & I>0 \\
\Lambda^-_{I,\hbar n_2,\ldots,\hbar
  n_d}\,, & I<0
\end{array}
\right.\,, \\
\Lambda_{(I,{\nscatt)\,\text{i;p}}} &=
\left\{
\begin{array}{cc}
\Lambda^-_{I,\hbar n_2,\ldots,\hbar
  n_d}\,, & I>0 \\
\Lambda^+_{I,\hbar n_2,\ldots,\hbar
  n_d}\,, & I<0
\end{array}
\right.\,.
\end{split}
\end{equation}

The projection of the Lagrangian manifolds
$\Lambda_{(I,{\nscatt)\,\text{i/o;r/p}}}$  to
the centre planes $(q_k,p_k)$, $k=2,\ldots,d$,
is thus restricted to the discrete circles $p_k^2+q_k^2=2 n_k \hbar$,
$n_k\in\N_0$. If we fix the total energy $E$ then this also entails a
discretisation of the projection of the manifolds
\eqref{eq:lagr_manf_scatt_multi}
to the saddle plane $(q_1,p_1)$ since the eigenvalue $I$ needs to satisfy the
energy equation $K_{\text{QNF}}(I,\hbar(n_2+1/2),\ldots,\hbar(n_d+1/2))=E$.
For the Lagrangian manifold $\Lambda_{(I,\nscatt)\,\text{i;r}}$ this is
depicted in Fig.~\ref{fig:nfplanesquantum}. Depending on whether $I$ is
positive or negative
the Lagrangian manifold $\Lambda_{(I,\nscatt)\,\text{i;r}}$ is either located
inside or outside of the energy surface volume enclosed by the forward reactive spherical cylinder
$W_f(E)$ defined in Sec.~\ref{sec:classical}, and  hence is either composed
of reactive or nonreactive trajectories of the classical dynamics.
From our discussion at the end of  Sec.~\ref{sec:smatrixmultiD} it then follows that the cumulative
reaction probability $N(E)$ is approximately given by
the total number
of Lagrangian manifolds $\Lambda_{(I,\nscatt)\,\text{i;r}}$ which, for
scattering quantum numbers $\nscatt=(n_2,\ldots,n_d)\in\N_0^{d-1}$, are
located inside of the energy surface volume enclosed by $W_f(E)$.

\def\fignfplanesquantum{%
Projections of the Lagrangian manifolds
$\Lambda_{(I,\nscatt)\,\text{i;r}}$ defined in
Equation~\eqref{eq:lagr_manf_scatt_multi} to the normal form coordinate
planes for the same setup as in Fig.~\ref{fig:hbargrid}.
The scattering quantum numbers are $\nscatt=(n_2,n_3)$ with $0\le n_2,n_3 \le
3$. For the values $(0,0)$, $(0,1)$, $(1,0)$ and $(1,1)$ of the quantum
numbers $(n_2,n_3)$, the Lagrangian manifolds
$\Lambda_{\psi_{\nscatt\,\text{react}}^{\text{in}}}$ are contained in the
energy surface volume (green region) enclosed by the forward reactive spherical cylinder
$W_f(E)$ defined in Sec.~\ref{sec:classical}. For the other values of the quantum numbers the
Lagrangian manifolds $\Lambda_{(I,\nscatt)\,\text{i;r}}$ are
located in the reactants component of the energy surface.
}
\def\FIGnfplanesquantum{
\centerline{\includegraphics[width=12.0cm]{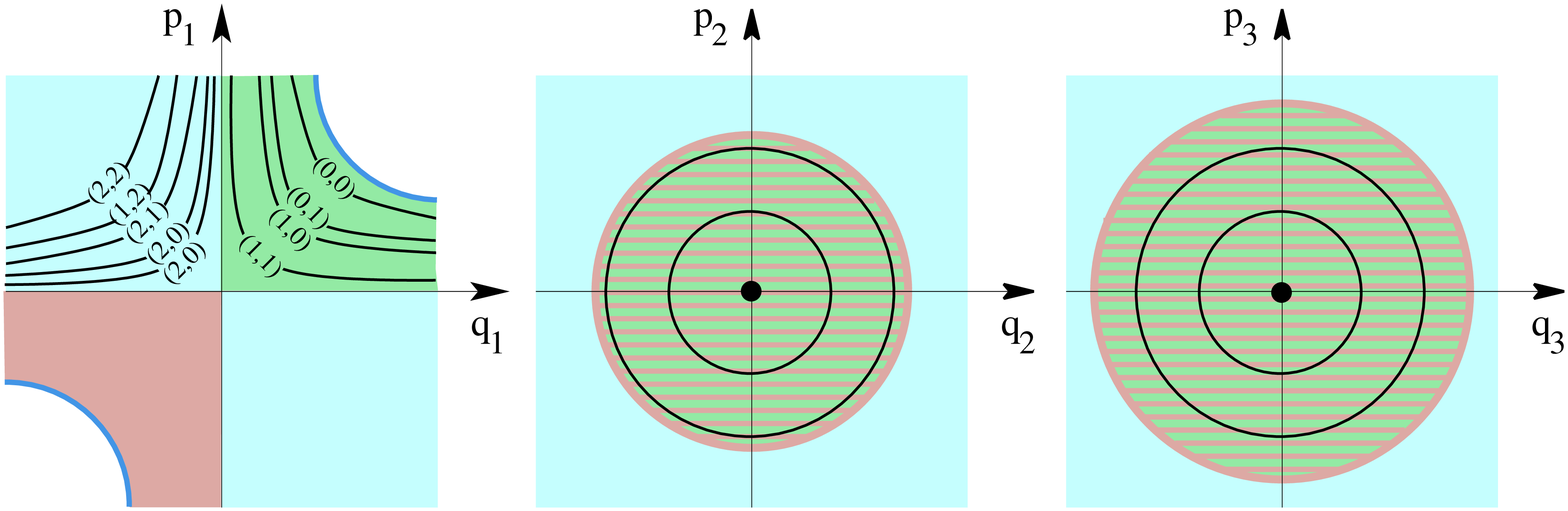}}
}
\FIGo{fig:nfplanesquantum}{\fignfplanesquantum}{\FIGnfplanesquantum}

\subsection{The global S-matrix}
\label{sec:global-S-matrix}

\New{
It is important to emphasize again that, so far, our approach to
quantum reaction dynamics has been local, i.e., it is derived
completely from the properties of the quantum normal form that is
valid in the neighborhood of the saddle-centre-$\cdots$-centre equilibrium point. 
The property of the resulting S-matrix in \eqref{eq:Smatrixfull} being block-diagonal reflects the fact that the  quantum normal form is integrable
in the sense that the
basis of scattering states can be chosen in the product form \eqref{eq:multscatteringwaves}. In a different basis the
matrix will lose this feature, and phenomena like mode mixing are related to how other incoming
and outgoing scattering states are related to this special basis. It is natural to embed  
the study of this phenomenon in a study of the global dynamics which we will describe in this section.
The global formalism is in particular required in order to compute general state-to-state
reaction rates.
}

Let us start by
describing the scattering or reaction process in classical
mechanics by using Poincar{\'e} sections. Recall that a
Poincar{\'e} section at energy $E$ is given by a smooth
hypersurface $\Sigma(E)$ of the energy surface with energy $E$
which is transversal to the flow ($\Sigma(E)$ is allowed to have
several components). If we have two such Poincar{\'e} sections
$\Sigma_1(E)$ and $\Sigma_2(E)$ such that all the flow lines
intersecting $\Sigma_1(E)$ intersect at a later time
$\Sigma_2(E)$, too, then moving along the flow from $\Sigma_1(E)$
to $\Sigma_2(E)$ defines a Poincar{\'e} map 
\begin{equation}
P^{(2,1)}(E):\Sigma_1(E)\to \Sigma_2(E)\,\, .
\end{equation}
Such Poincar{\'e} maps can be composed. If $\Sigma_3(E)$ is another Poincar{\'e} section which lies behind
$\Sigma_2(E)$ in the sense that the flow lines that intersect $\Sigma_2(E)$ also intersect $\Sigma_3(E)$ at a later time,
and if $P^{(3,2)}(E):\Sigma_2(E)\to \Sigma_3(E)$ is the corresponding Poincar{\'e} map, then the Poincar{\'e} map
\begin{equation}
P^{(3,1)}(E):\Sigma_1(E)\to \Sigma_3(E)\,\,
\end{equation}
is given by
\begin{equation}
P^{(3,1)}(E)=P^{(3,2)}(E) \circ P^{(2,1)}(E)\,\, .
\end{equation}

Using this construction we can describe transport through phase space regions by a sequence of maps.
Given some Poincar{\'e} section  $\Sigma_{\text{initial}}(E)$ located in the area of initial points
in the reactants region
where we prepare the system and a Poincar{\'e} section $\Sigma_{\text{final}}(E)$ in the products region
where we measure the outcome, a succession
of Poincar{\'e} maps
\begin{equation}\label{eq:Poincare_sect_succession}
\Sigma_{\text{initial}}(E)\to \Sigma_{1}(E)\to \Sigma_{2}(E)\to \cdots \to \Sigma_{\text{final}}(E)
\end{equation}
tells us how the initial points are transported through the
system.\footnote{ We here ignore the difficulties involved in
constructing global Poincar{\'e} sections (see, e.g.,
\cite{DullinWittek95}); we assume that the sequence of
Poincar{\'e} sections \eqref{eq:Poincare_sect_succession} is
intersected transversally by the trajectories with initial points
from a suitable open subset in the reactants region.}

The advantage of subdividing the flow into a sequence of maps lies
in the fact that different regions in phase space might need
different techniques to compute the flow. In our case of interest
Poincar{\'e} sections can be constructed to the products and
reactants side of a saddle-centre-$\cdots$-centre equilibrium point. The dynamics `across'
this equilibrium point can then be described by the normal form
while the dynamics between neighbourhoods of different saddle
points can be obtained from integrating the original equations of
motions \cite{Creagh04,Creagh05,WaalkensBurbanksWiggins05b}.
Moreover, the phase space structures obtained from the local
normal form can be ``globalized'' following the discussion in
Section \ref{sec:classicglobal}.

A similar procedure can be developed in the quantum case. The Poincar{\'e} maps
\begin{equation}
P^{(j,i)}(E):\Sigma_i(E)\to \Sigma_j(E)\,\,
\end{equation}
are symplectic maps, and as such can be quantised using the theory of Fourier integral operators.
The quantisations will be unitary operators which we interpret as local S-matrices,
\begin{equation}
S^{(j,i)}(E):L^2_{\Sigma_i(E)}\to L^2_{\Sigma_j(E)}\,\, ,
\end{equation}
where $L^2_{\Sigma(E)}$ is a Hilbert space obtained by geometric quantisation of $\Sigma(E)$, see, e.g., \cite{Kir01}.
This is similar to the quantisation developed in \cite{Bogomolny92}.
As in classical dynamics we can  compose these matrices to obtain a global S-matrix
\begin{equation}
S^{(\text{final}, \text{initial})}(E)=S^{(\text{final},n)}(E)S^{(n,n-1)}(E)\cdots S^{(1, \text{initial})}(E)
\end{equation}
which tells us how initial states in
$L^2_{\Sigma_{\text{initial}}(E)}$ are transformed into final
states in $L^2_{\Sigma_{\text{final}}(E)}$. The reasons for
introducing this splitting of the S-matrix are the same as in the
classical case. We can employ different techniques for computing
the S-matrices according to different  local properties of the system.
Near equilibrium  points the dynamics can be described by the quantum
normal form we developed in this paper. 
Notice that the neighbourhoods of the saddle-centre-$\cdots$-centre
equilibriuml points are the regions where we expect quantum effects to
be of most importance due to partial reflection at and tunnelling
through the barriers associated with saddle points. 
The quantum transport between neighbourhoods of
different equilibrium points can be described by a standard van Vleck
type formalisms, using, e.g, {\em initial value representations} (IVRs) 
which are very common in theoretical chemistry
(see, e.g., \cite{Miller98,Miller1} for references).

\subsection{The flux-flux autocorrelation function formalism to compute quantum reaction probabilities}
\label{sec:flux-flux_quantum}

The main approach to compute quantum mechanical reaction rates that is most heavily
pursued in the chemistry literature
is the quantum version of the flux-flux autocorrelation function formalism
that we reviewed in Sec~\ref{sec:fluxfluxMiller}. This approach was developed
by Miller and others (see
\cite{Yamamoto60,MillerSchwartzTromp83,Miller98}) and in the following we will
mainly follow their presentation.
We will see that the cumulative reaction
probability $N(E)$ is the quantum mechanical flux through a dividing surface
and hence is the analogue of the classical flux.
The goals of this section are twofold.
Firstly, we will show that we recover our result for the cumulative reaction probability
in \eqref{eq:cum_react}
when we evaluate the quantum flux-flux
autocorrelation function expression for the cumulative reaction probability $N(E)$ in terms of
the quantum normal form and for our choice of the dividing surface that we discussed
in Sec.~\ref{sec:classical}. This way will see that the flux-flux autocorrelation function
formalism and our result for the cumulative reaction probability are formally
equivalent and hence, our result for $N(E)$ can be viewed as a quantum
mechanical flux through a dividing surface.
Secondly, we will argue that, like in the classical case,
the application of the flux-flux autocorrelation formalism in its original form,
which does not depend on the
specific choice of a dividing surface, is computationally much more expensive than
our quantum normal form approach.

Following \cite{Yamamoto60,MillerSchwartzTromp83,Miller98},
a quantisation of the flux-flux autocorrelation function formalism in
Sec.~\ref{sec:fluxfluxMiller}, or more precisely of the
dimensionless quantity
\begin{equation}\label{eq:Nweyl_integral}
\Nweyl(E)=\flux(E)/(2\pi \hbar)^{d-1} = 2\pi \hbar \int_{\R^d}\int_{\R^d}
\delta(E-H) F P_{\text{r}} \,
\frac{\ud q \,\ud p}{(2\pi \hbar)^d}\,
\end{equation}
is obtained by
replacing the classical phase space integral in \eqref{eq:Nweyl_integral}
by the trace of the associated operators in
the form
\begin{equation} \label{eq:Millers_quantum_N(E)}
N(E) = 2\pi \hbar \,\text{Tr}\, \delta(E-\widehat{H} ) \widehat{F} \widehat{P}_{\text{r}}  \,.
\end{equation}
Following the quantum classical correspondence principle the operator
$\widehat{F}$ is obtained from its classical counterpart $F$ by replacing
the Poisson bracket in the classical expression $F=\{\Theta(s),H\}$ by the
corresponding commutator to give
\begin{equation}
\widehat{F} = -\frac{\ui}{\hbar} [\widehat{\Theta(s)},\widehat{H}]\,.
\end{equation}
Here $\widehat{\Theta(s)}$ is a quantisation  (to
which we will come back below) of the composition of the
Heaviside function with a function $s$ that defines the dividing surface
according to $s(q,p)=0$ as discussed in Sec.~\ref{sec:fluxfluxMiller}.
Similarly, the quantisation of the projection function $P_{\text{r}}=\lim_{t
  \rightarrow\infty} \Theta\big(s (\flow^t ) \big)$ in \eqref{eq:defPr} is given by the operator
\begin{equation} \label{eq:defPr_quantum}
\widehat{P}_{\text{r}} = \lim_{t\rightarrow \infty}
\ue^{\frac{\ui}{\hbar}\widehat{H}t}\widehat{\Theta(s)} \ue^{-\frac{\ui}{\hbar}\widehat{H}t}\,.
\end{equation}
The application of $\widehat{P}_{\text{r}}$ to a state $\psi$ is thus
obtained from taking the limit $t\rightarrow \infty$ 
in the process 
of
letting the time evolution operator, $\exp (-\frac{\ui}{\hbar}\widehat{H}t)$,
act on $\psi$ for the time $t$, then apply $\widehat{\Theta(s)}$ to determine whether $\psi$
has evolved to products after time $t$ (see below for the details),
and then evolve the state $\psi$
backward in time by applying the inverse of the time evolution operator,
$\exp (\frac{\ui}{\hbar}\widehat{H}t)$. In fact, the operator
$\widehat{P}_{\text{r}}$  is given by the limit $t\rightarrow \infty$ of the
Heisenberg picture of the operator $\widehat{\Theta(s)}$.

\New{Using
\begin{equation} \label{eq:Pr_integral_quantum}
\begin{split}
   \widehat{P}_{\text{r}} &= \int_0^\infty \frac{\ud}{\ud t}
   \left( \ue^{\frac{\ui}{\hbar}\widehat{H}t}
\widehat{\Theta(s)} \ue^{-\frac{\ui}{\hbar}\widehat{H}t} \right)  \,\ud t \\ 
  &=\int_0^\infty  \ue^{\frac{\ui}{\hbar}\widehat{H}t}\widehat{F}
  \ue^{-\frac{\ui}{\hbar}\widehat{H}t}\,\ud t\,, 
\end{split}
\end{equation}
we find that analogously to \eqref{eq:fluxfluxintegral} the cumulative
reaction probability can be rewritten as an autocorrelation function\footnote{Formally Eq.~\ref{eq:Pr_integral_quantum} still contains a term $\widehat{\Theta(s)}$. 
But this term  will give no contribution to $N(E)$  for the same reason as in the classical flux-flux autocorrelation formalism (see the 
discussion after  \eqref{eq:modifyPr}). In the examples below this can be seen explicitly since 
we define the operator $\widehat{\Theta(s)}$ in normal form coordinates as a multiplication 
operator by a characteristic function. Then the same reasoning as in the classical case applies.}:
\begin{equation}
  N(E) = 2\pi \hbar \, \int_0^\infty C_{\widehat{F}}(t)\,\ud t\,,
\end{equation}
where
\begin{equation}
 C_{\widehat{F}}(t)  = \text{Tr}\,  \delta(E-\widehat{H})\widehat{F}\ue^{\frac{\ui}{\hbar}\widehat{H}t}\widehat{F} \ue^{-\frac{\ui}{\hbar}\widehat{H}t}\,.
\end{equation}
}

We illustrate the application of the flux-flux autocorrelation function formalism in the following sections.

\subsubsection{Example: 1D parabolic barrier}

As a first example we consider a one-dimensional system and a surface defined
according to $s(q,p)=q-q_0=0$.
In the position representation the quantisation of the function $\Theta(s)$ is then defined by its action on a wavefunction
$\psi(q)$ according to
\begin{equation}\label{eq:def_theta_s_operator}
  \widehat{\Theta(s)} \psi(q) = \Theta(q-q_0)  \psi(q)\,.
\end{equation}
A state $\psi$ thus is an eigenfunction with eigenvalue 1 of the
operator $ \widehat{P}_{\text{r}}$ if its wavefunction $\psi(q)$
is concentrated in $q>q_0$ if evolved forward in time to time
$t=\infty$. Likewise, $\psi$ is an eigenfunction with eigenvalue 0
of the operator $ \widehat{P}_{\text{r}}$ if its wavefunction
$\psi(q)$ is concentrated in $q<q_0$ if evolved forward in time to
time $t=\infty$. For a Hamilton operator of type `kinetic plus
potential',
$\widehat{H}=\frac{1}{2m}\widehat{p}^2+V(\widehat{q})$, with the
quantisation of the operators $\widehat{q}$ and  $\widehat{p}$
given in \eqref{eq:opI45rotation}, the operator $\widehat{F}$
becomes
\begin{equation}  \label{eq:operator_F_gen}
\begin{split}
\widehat{F}&=
  -\frac{\ui}{\hbar}[ \widehat{\Theta(s)} , \widehat{H} ] =
   -\frac{\ui}{\hbar}[ \widehat{\Theta(s)} , \frac{1}{2m} \widehat{p}^2 ]
=- \frac{\ui}{\hbar}  \frac{1}{2m} \big( \widehat{p}[
  \widehat{\Theta(s)},\widehat{p}] + [ \widehat{\Theta(s)},\widehat{p}]
  \widehat{p}   \big) \\
&=
\frac{1}{2m} \big( \widehat{p} \delta(q_0) + \delta(q_0) \widehat{p}  \big)\,.
\end{split}
\end{equation}
For the expectation value of $ \widehat{F}$ with respect to a state
$\psi$ we thus get\footnote{In the following it will be notationally more convenient 
to use the Dirac notation for scalar products. Here $\langle\psi | A | \psi\rangle$ is the same 
as $\langle \psi, A\psi\rangle$ for any operator $A$ and state $\psi$.}
\begin{equation} \label{eq:expect_F_gen}
  \langle \psi | \widehat{F} | \psi \rangle = -\ui \frac{\hbar}{2m} \big(
  \psi^*(q_0)\psi'(q_0) - \psi'^*(q_0)\psi(q_0) \big) \,,
\end{equation}
where the primes denote the derivatives.
This agrees with the standard definition of the quantum probability current density
that can be found in any quantum mechanics textbook (see, e.g., \cite{LandauLifschitz01}).

To make the  example more concrete we consider a parabolic barrier described by the Hamilton operator
\begin{equation}\label{eq:parab_barrier_original}
\widehat{H} = -\frac{\hbar^2}{2m}\frac{\ud^2 }{\ud q^2} - \frac12 m \lambda^2 q^2\,.
\end{equation}
The spectrum of $\widehat{H}$ is $\R$.
We choose energy eigenfunctions $\psi_{E\,\pm}$  such that they correspond to wavefunctions
moving  in positive and negative $q$ direction, respectively, i.e., besides
\begin{equation}
  \widehat{H} \psi_{E\,\pm} = E \psi_{E\,\pm}
\end{equation}
we have
\begin{equation}\label{eq:Ppsi+-}
 \widehat{P}_{\text{r}} \psi_{E\,+} = \psi_{E\,+}\,,\quad
 \widehat{P}_{\text{r}} \psi_{E\,-} = 0\,.
\end{equation}
For the trace \eqref{eq:Millers_quantum_N(E)} to be well defined we need to
require that the states $\psi_{E\,\pm}$ are normalised in such a way that they
satisfy the completeness relation
\begin{equation} \label{eq:psi+-complete}
\int_\R \big(
\psi^{*}_{E\,+} (q) \psi_{E\,+} (q') +
\psi^{*}_{E\,-} (q) \psi_{E\,-} (q')
\big)\,\ud E = \delta(q-q')\,.
\end{equation}
The eigenfunctions $\psi_{E\,\pm}$ having the
properities \eqref{eq:Ppsi+-} and \eqref{eq:psi+-complete}
are given by
\begin{equation}\label{eq:harm_osc_eigen_standard}
  \psi_{E\,\pm}(q) = \frac{1}{\sqrt{2\pi^2\hbar}} \left(\frac{m}{2\hbar
  \lambda}\right)^{1/4} \ue^{\frac{E}{\lambda \hbar}\frac{\pi}{4}}\Gamma \bigg(
   \frac12 -\ui \frac{E}{\hbar\lambda}  \bigg)
  D_{-\frac12-\ui\frac{E}{\hbar\lambda}}\left(\pm \ue^{-\ui\frac{\pi}{4}}\sqrt{ \frac{2m\lambda}{\hbar}} q\right)\,,
\end{equation}
where $D_\nu$ again denotes the parabolic cylinder function \cite{AbraSteg65}.
In fact, the wavefunctions $\psi_{E\,\pm}$ can be obtained from a suitable scaling of the wavefunctions
$\chi_{I\pm}$ that we defined in \eqref{eq:def_psi_I_pm} and which satisfy the
completeness relations \eqref{eq:completeness_rel_general}.
For $\psi_{E\,\pm}$, we have
\begin{equation}
  -\ui \frac{\hbar}{2m} (  \psi_{E\,\pm}^* \psi_{E\,\pm}' - \psi_{E\,\pm}'^{*} \psi_{E\,\pm} ) =
 \pm \frac{1}{2\pi \hbar} \frac{1}{1+\ue^{-2\pi E/(\lambda\hbar)}}\,,
\end{equation}
and hence using \eqref{eq:expect_F_gen} and \eqref{eq:Ppsi+-} we get for the
cumulative reaction probability,
\begin{equation} \label{eq:N(E)_parabolic_barrier_original}
\begin{split}
  N(E) &= 2\pi \hbar \text{Tr } \delta(E-\widehat{H} ) \widehat{F} \widehat{P}_{\text{r}}\\
&= 2\pi \hbar \, \int_\R \big(
\langle \psi_{E'\,+} | \delta(E-\widehat{H} ) \widehat{F} \widehat{P}_{\text{r}}|
\psi_{E'\,+} \rangle +
\langle \psi_{E'\,-} |\delta(E-\widehat{H} ) \widehat{F} \widehat{P}_{\text{r}} |
\psi_{E'\,-} \rangle \big) \,\ud E'\\
&=
2\pi \hbar \int_\R \delta(E-E') \langle \psi_{E'\,+} |\widehat{F} |\psi_{E'\,+}
\rangle \,\ud E' \\
&=
\frac{1}{1+\ue^{-2\pi E/(\lambda\hbar)}}\,,
\end{split}
\end{equation}
which is the exact quantum mechanical reflection coefficient for a parabolic
barrier \cite{LandauLifschitz01}.

We now want to repeat the calculation above by inserting for $\widehat{H}$
the quantum normal form of the parabolic barrier in \eqref{eq:Millers_quantum_N(E)}.
This will show two things. Firstly, this will lead to our result for
the cumulative reaction probability $N(E)$
that we  have given in \eqref{eq:cum_react}
(which for the one-dimensional case reduces the reflection coefficient derived in Sec.~\ref{sec:oneDSmatrix}).
Secondly, we will see that our result agrees with $N(E)$ in
\eqref{eq:N(E)_parabolic_barrier_original}, i.e., our result for $N(E)$ in
terms of the quantum normal form is exact for parabolic barriers.

From our discussion in Sec.~\ref{sec:1Dpotentialbarriers} it follows that the
quantum normal form of \eqref{eq:parab_barrier_original} is given by
\begin{equation}
  \widehat{H}_{\text{QNF}} = K_{\text{QNF}}(\hat{I}) = \lambda \Ihat\,.
\end{equation}
In order to evaluate \eqref{eq:Millers_quantum_N(E)} for our dividing
surface which in terms  of the normal form coordinates is given by
$s(q,p)=q-p=0$  (see Sec.~\ref{sec:nfcpss}) it is convenient to work
with the rotated coordinates
\begin{equation}
(Q,P) = \frac{1}{\sqrt{2}} (q-p,q+p)\,.
\end{equation}
The $Q$ representation of the operator $\widehat{\Theta(s)}$ is then defined
analogously to \eqref{eq:def_theta_s_operator}, i.e.,
\begin{equation} \label{eq:def_Theta(Q)}
  \widehat{\Theta(s)} \psi(Q) = \Theta(Q)  \psi(Q)\,.
\end{equation}
As we have seen in the example of the application of
Lemma~\ref{lem:ex-eg} (exact Egorov) in Sec.~\ref{sec:conjunitary}
the $Q$ representation of the operator $\Ihat$ reads
\begin{equation}\label{eq:Ihat_Q_1Dexample}
\Ihat = -\frac{\hbar^2}{2}\frac{\ud^2}{\ud Q^2} - \frac12 Q^2
\end{equation}
(see Equation~\eqref{eq:opI45rotation}). In
Sec.~\ref{sec:scatt_states} we showed that the eigenfunction of
\eqref{eq:Ihat_Q_1Dexample} are given by $\chi_{I\pm}$ defined in
\eqref{eq:def_psi_I_pm}.
In fact, the eigenfunctions $\chi_{I\pm}$
formally agree with the eigenfunctions
$\psi_{E\,\pm}$
in \eqref{eq:harm_osc_eigen_standard} if $m$ and $\lambda$ are replaced by 1, 
 and $E$ is replaced by $I$.
Analogously to \eqref{eq:operator_F_gen} we have
\begin{equation}
- \frac{\ui}{\hbar}[\widehat{\Theta(s)},\Ihat] = \frac12( \widehat{P} \delta(Q) +
\delta(Q) \widehat{P} )\,,
\end{equation}
and for an arbitrary state $\psi$,
\begin{equation}\label{eq:theta_Ihat_exp_1D}
\langle \psi |-\frac{\ui}{\hbar}[\widehat{\Theta(s)},\Ihat] | \psi\rangle
= -\ui  \frac{\hbar}{2}  (\psi^*(0) \psi'(0)  - \psi'^*(0) \psi(0))\,.
\end{equation}
Evaluating this expression for the eigenfunctions $\chi_{I\,\pm}$ we get
\begin{equation} \label{eq:chi_current_1D}
-\ui \frac{\hbar}{2} (  \chi^*_{I\,\pm} \chi'_{I\,\pm}  - \chi'^*_{I\,\pm}  \chi_{I\,\pm} ) =
 \pm \frac{1}{2\pi \hbar} \frac{1}{1+\ue^{-2\pi I /\hbar}}\,.
\end{equation}
Using this result and the fact
that $\chi_{I\,+}$ and $\chi_{I\,-}$  are moving in positive and
negative $Q$ direction and hence
are eigenfunction of
$\widehat{P}_{\text{r}}$ with eigenvalues 1 and 0, respectively,
we get
\begin{equation}
\begin{split}
   N(E) &= 2\pi \hbar \text{Tr } \delta(E-K_{\text{QNF}}(\Ihat) ) \widehat{F}
   \widehat{P}_{\text{r}} \\
&= 2\pi \hbar\int_\R \big(
\langle \chi_{I\,+} | \delta(E-K_{\text{QNF}}(\Ihat) ) \widehat{F} \widehat{P}_{\text{r}}|
\chi_{I\,+} \rangle +
\langle \chi_{I\,-} |\delta(E-K_{\text{QNF}}(\Ihat) ) \widehat{F} \widehat{P}_{\text{r}} |
\chi_{I\,-} \rangle \big)\, \ud I \\
&=
2\pi \hbar \int_\R \delta(E-K_{\text{QNF}}(I) ) \lambda \langle \chi_{I\,+} |-\frac{\ui}{\hbar}[\widehat{\Theta(s)},\Ihat] |\chi_{I\,+}
\rangle \,\ud I \\
&=
\frac{1}{1+\ue^{-2\pi E/(\lambda\hbar)}}\,.
\end{split}
\end{equation}
This formally agrees with the expression for $N(E)$ that we have given in
\eqref{eq:cum_react} and also with the exact result in
\eqref{eq:N(E)_parabolic_barrier_original}, i.e., our
quantum normal form computation of $N(E)$  is exact for parabolic barriers.

\subsubsection{Example: General barriers in 1D}

Let us now use the quantum normal form in the flux-flux autocorrelation
formalism in
 the more general case of a one-dimensional system with a Hamilton
operator whose principal symbol has a saddle equilibrium point but  is not necessarily
quadratic.
Like in the previous section we again work in the $Q$ representation,
i.e., our dividing surface is defined by $s(Q,P)=Q=0$, and the operators
$\widehat{\Theta(s)}$ and $\Ihat$ are defined by \eqref{eq:def_Theta(Q)}
and \eqref{eq:Ihat_Q_1Dexample}, respectively.
In order to evaluate \eqref{eq:Millers_quantum_N(E)} for a general Hamilton
operator in quantum normal form,
$\widehat{H}_{\text{QNF}}=K_{\text{QNF}}(\Ihat)$, where
$K_{\text{QNF}}(\Ihat)$ is a polynomial in $\Ihat$, we use that for $n\in \N$,
we have
\begin{equation}
[\widehat{\Theta(s)}, \Ihat^n] =
\sum_{k=0}^{n-1} \Ihat^{n-k-1} [\widehat{\Theta(s)}, \Ihat]  \Ihat^{k} \,.
\end{equation}
This can be shown by direct calculation.
For the eigenfunction $\chi_{I\,\pm}$ of $\Ihat$ we thus have
\begin{equation}
\langle \chi_{I\,\pm}  | [\widehat{\Theta(s)},\Ihat^n]  | \chi_{I\,\pm}    \rangle
=  \langle \chi_{I\,\pm}  | [\widehat{\Theta(s)},\Ihat]  | \chi_{I\,\pm}    \rangle n I^{n-1}\,,
\end{equation}
and hence
\begin{equation}\label{eq:psiIFpsiI_1D}
\langle \chi_{I\,\pm}  | [\widehat{\Theta(s)},K_{\text{QNF}}(\Ihat)]  | \chi_{I\,\pm}    \rangle
= \langle \chi_{I\,\pm}  | [\widehat{\Theta(s)},\Ihat]  | \chi_{I\,\pm}
\rangle
\frac{\ud K_{\text{QNF}}(I)}{\ud I}\,.
\end{equation}
Using this together with \eqref{eq:theta_Ihat_exp_1D} and \eqref{eq:chi_current_1D} we find that
the cumulative reaction probability is given by
\begin{equation}
\begin{split}
 N(E) &= 2\pi \hbar\, \big(
\langle \chi_{I\,+} | \delta(E-K_{\text{QNF}}(\Ihat) ) \widehat{F} \widehat{P}_{\text{r}}|
\chi_{I\,+} \rangle +
\langle \chi_{I\,-} |\delta(E-K_{\text{QNF}}(\Ihat) ) \widehat{F} \widehat{P}_{\text{r}} |
\chi_{I\,-} \rangle \big) \\
&=
2\pi \hbar \int_\R \delta(E-K_{\text{QNF}}(I) ) \langle \chi_{I\,+} |-\frac{\ui}{\hbar}[\widehat{\Theta(s)},\Ihat] |\chi_{I\,+}
\rangle \frac{\ud K_{\text{QNF}}(I)}{\ud I} \,\ud I \\
&=
\frac{1}{1+\ue^{-2\pi I(E)/\hbar}}\,,
\end{split}
\end{equation}
where $I(E)$ is the solution of $E=K_{\text{QNF}}(I(E))$, and we have assumed that there is only 
one such solution (compare with the remark after \eqref{eq:I-of-E}).
We thus recover our result for $N(E)$ that we have given in
\eqref{eq:cum_react}.

\subsubsection{Example: General barriers in arbitrary dimensions}

We now consider the $d$-dimensional case with a Hamilton operator in quantum
normal form given by
$\widehat{H}_{\text{QNF}}=K_{\text{QNF}}(\Ihat,\Jhat_2,\ldots,\Jhat_d)$.
Again we work in the $Q$ representation in terms of which our
dividing surface is defined as
$s(Q_1,\ldots,Q_d,P_1,\ldots,P_d)=Q_1=0$.
The quantisation of $\Theta(s)$ is then defined by its action on a wavefunction
$\psi(Q_1,\ldots,Q_d)$ according to
\begin{equation}
 \widehat{\Theta(s)} \psi(Q_1,\ldots,Q_d) = \Theta(Q_1) \psi(Q_1,\ldots,Q_d)\,.
\end{equation}
The $Q$ representation of the incoming eigenfunctions \eqref{eq:multscatteringwaves}
is given
by
\begin{equation}
\begin{split}
\chi_{(I,\nscatt)\,\text{i;r}}(Q_1,\ldots ,Q_d) &:=  \chi_{I\,+}(Q_1)
\psi_{n_2}(Q_2)\cdots\psi_{n_{d}}(Q_{d})\,,\\
\chi_{(I,\nscatt)\,\text{i;p}}(Q_1,\ldots ,Q_d) &:=  \chi_{I\,-}(Q_1)
\psi_{n_2}(Q_2)\cdots\psi_{n_{d}}(Q_{d})
\end{split}
\end{equation}
with $I\in \R$ and scattering quantum numbers $\nscatt =(n_2,\ldots,n_d)\in \N_0^{d-1}$.
It then follows from the one-dimensional case discussed in the previous
section that
\begin{equation}
\begin{split}
&\langle \chi_{(I,\nscatt)\,\text{i;r}} | [\widehat{\Theta(s)},K_{\text{QNF}}(\Ihat,\Jhat_2,\ldots,\Jhat_d)]  |  \chi_{(I,\nscatt)\,\text{i;r}}    \rangle \\
= & \langle  \chi_{(I,\nscatt)\,\text{i;r}} | [\widehat{\Theta(s)},\Ihat]  |
\chi_{(I,\nscatt)\,\text{i;r}} \rangle
\frac{\partial }{\partial I}
K_{\text{QNF}}(I,\hbar(n_2+\frac12),\ldots,\hbar(n_d+\frac12))
\end{split}
\end{equation}
(see Equation~\eqref{eq:psiIFpsiI_1D}). Using the completeness of the states
$\chi_{(I,\nscatt)\,\text{i;r/p}}$ we find for the cumulative reaction probability,
\begin{equation}
\begin{split}
N(E)  = 2\pi \hbar \sum_{\nscatt\in \N_0^{d-1}} \int_\R
  & \big(
\langle  \chi_{(I,\nscatt)\,\text{i;r}}  |
\delta(E-K_{\text{QNF}}(\Ihat,\Jhat_2,\ldots,\Jhat_d))
\widehat{F} \widehat{P}_{\text{r}}
|  \chi_{(I,\nscatt)\,\text{i;r}}  \rangle \\
& +
\langle  \chi_{(I,\nscatt)\,\text{i;p}}  |
\delta(E-K_{\text{QNF}}(\Ihat,\Jhat_2,\ldots,\Jhat_d))
\widehat{F} \widehat{P}_{\text{r}}
|  \chi_{(I,\nscatt)\,\text{i;p}}  \rangle \big)\, \ud I  \\
\\
=  2\pi \hbar \sum_{\nscatt\in \N_0^{d-1}} \int_\R &
\delta(E-K_{\text{QNF}}(I,\hbar(n_2+\frac12),\ldots,\hbar(n_d+\frac12)))
  \times \\
&  \langle
\chi_{(I,\nscatt)\,\text{i;r}}  | -\frac{\ui}{\hbar}
[ \widehat{\Theta(s)} , K_\text{QNF}(\Ihat,\Jhat_2,\ldots,\Jhat_d)]
|  \chi_{(I,\nscatt)\,\text{i;r}}  \rangle \big)\, \ud I  \\
=
\sum_{\nscatt \in\N_0^{d-1}}
\bigg[1+  & \exp  \bigg(-2\pi\frac{I_{\nscatt}(E)}{\hbar}\bigg)  \bigg]^{-1}\,,
\end{split}
\end{equation}
where $I_{\nscatt}(E)$ solves
$K_{\text{QNF}}(I,\hbar(n_2+1/2),\ldots,\hbar(n_d+1/2))=E$ for $\nscatt=(n_2,\ldots,n_d)\in\N_0^{d-1}$, and we assume there is 
only one such solution (compare, again,  with the remark after \eqref{eq:I-of-E}).  
We thus recover our  result in \eqref{eq:cum_react}.

Though we showed that if the flux-flux
autocorrelation function formalism is evaluated in terms of the quantum normal form
then it reproduces our results for the cumulative reaction
probability that we developed in Sec.~\ref{sec:smatrixmultiD}
it is important to point out the computational differences between the flux-flux
autocorrelation function formalism in its original form and the quantum normal
form approach to compute cumulative reaction
probabilities. The main problem with the implementation of the
flux-flux autocorrelation function formalism is the occurrence of the
projection operator $\widehat{P}_{\text{r}}$ in the trace in
\eqref{eq:Millers_quantum_N(E)}. The presence of the operator
$\widehat{P}_{\text{r}}$
is crucial in order to ensure that only
states that evolve from reactants to products contribute to the trace in
\eqref{eq:Millers_quantum_N(E)}.
The extraction of this information
for an arbitrarily chosen dividing surface and without any insight
into the quantum dynamics requires one to look at the full time evolution of
states as embodied in the definition of
the operator $\widehat{P}_{\text{r}}$  in
\eqref{eq:defPr_quantum}.
Though various techniques like Monte Carlo path integration and
{\em initial value representation} (IVR)  
\cite{Miller98,Miller1}  have been developed in
order to solve this time evolution problem that is involved in the
evaluation of the trace in \eqref{eq:Millers_quantum_N(E)} due to
the presence of $\widehat{P}_{\text{r}}$ it remains a formidable
numerical task to apply \eqref{eq:Millers_quantum_N(E)} to
specific systems.
In contrast to this, the computation of the cumulative
reaction probability from the quantum normal form does not involve the
solution of a time evolution problem.
The reason for this is that the quantum normal form yields an unfolding
of the quantum dynamics in the reaction region.
As a result the
S-matrix expressed in terms of the corresponding scattering states
is diagonal, i.e., the scattering states can be immediately classified and the
reaction probabilities can be immediatedly determined without explicitly
looking at the time evolution.
The numerical effort to implement and evaluate the quantum normal form is
comparable to the classical normal form computation described in
Sections~\ref{sec:CNF} and \ref{sec:classical}. In
Sec.~\ref{sec:examples} we will illustrate the efficiency of the quantum
normal form computation of the cumulative reaction probability for several concrete examples.

%% file: resonances.tex
\section{Quantum Resonances}
\label{sec:resonances}

\New{
In this section we consider quantum resonances and the corresponding resonance states. 
The role of quantum resonances in the context of chemical reactions has been 
studied for the first time explicitly in the chemistry literature by
Friedman and Truhlar \cite{FriedmanTruhlar91} and Miller \cite{SeidemanMiller91}.
The  Quantum resonances 
are viewed as another  imprint of the activated complex in addition to the quantisation of the cumulative reaction probability discussed in the previous section, Sec.~\ref{sec:smatrix}.
Recent developments in high resolution spectroscopic techniques allow one to probe the dynmaics of quantum mechanical reactions with unprecedented accuracy. There is therefore an immense interest in quantum resonances both in experimental and computational chemistry 
\cite{Zare06,SkodjeYang04,Skodjeetal00}.

We will show that the quantum normal form provides us with a very efficient algorithm for computing quantum resonances and also 
the corresponding resonance states.  In our discussion of the classical reaction dynamics we could identify the activated complex with the centre manifold of the saddle-centre-$\cdots$-centre equilibrium point, i.e. with an invariant subsystem 
with one degree of freedom less than the full system located between reactants and products
(see Sec.~\ref{sec:phasespacestruct}). 
As will discuss in detail in  Sec.~\ref{sec:lifetime-act-complex}, the Heisenberg uncertainty relation excludes the existence of an invariant quantum subsystem. In fact, the
quantum resonances will describe how a wavepacket initialised near the classically 
invariant subsystem will decay in time.
}

Quantum resonance can be introduced in several ways. A common definition is based on  the S-matrix. 
If one can  
extend the S-matrix analytically to complex energies, then the  resonances are defined as  
its poles in the complex energy plane. 
We therefore could  use the results of the previous section to 
determine the  resonances from the quantum normal form. However, 
we will choose a different approach to introduce resonances which will make 
their dynamical meaning much more clear. 


\subsection{Definition of quantum resonances}

We will define resonances as  
the poles of the resolvent operator.
This is in line with the the convention in the  mathematical 
literature (see, e.g., \cite{Zworski99}). 
Let us recall the necessary notions. 

For an operator  $\cH: L^2(\R^d)\to L^2(\R^d)$,  the 
resolvent set $r(\cH)$ of $\cH$ is defined as the set of $E\in \C$ such that 
$\cH-E$ is invertible. The spectrum of 
$\cH$ is the complement of the  resolvent set. 
For $E\in r(\cH)$, the resolvent of $\cH$ is defined as
\begin{equation}
\widehat{R}(E)=(\cH-E)^{-1}:   L^2(\R^d)\to L^2(\R^d)\,\, .
\end{equation}
If $\cH$ is selfadjoint, then the spectrum 
of $\cH$ is contained in $\R$.  The resolvent is thus defined 
at least for all $E\in \C\backslash\R$. The resolvent is related to the 
time evolution operator $\widehat{U}(t)=\exp(-\frac{\ui}{\hbar}t \cH)$  by 
Laplace transformation. For $\Im E \geq 0$, 
\begin{equation}\label{eq:lapl-tr}
\widehat{R}(E)=\frac{\ui}{\hbar}
\int_0^{\infty}\ue^{\frac{\ui}{\hbar}Et}\widehat{U}(t)\,\, \ud t\,\, ,
\end{equation}
and by Mellin transform 
\begin{equation}\label{eq:mell-tr}
\widehat{U}(t)=\frac{1}{2\pi\ui}\int_{\Im E=c}\widehat{R}(E)\ue^{\frac{\ui}{\hbar}tE}\,\, \ud E \,,
\end{equation}
where $c>0$. The path of integration in the Mellin integral should 
be thought of as encircling the spectrum of $\cH$. Hence, if 
$\cH$ has only isolated eigenvalues $E_n$ then Cauchy's theorem gives 
\begin{equation}\label{eq:time-evo-exp-unitary}
\widehat{U}(t)=\sum \ue^{-\frac{\ui}{\hbar}tE_n} \widehat{P}_n
\end{equation}
with the projectors 
\begin{equation}
\widehat{P}_n:=\frac{1}{2\pi\ui}\int_{C_n}\widehat{R}(E)\,\, \ud E\,,
\end{equation}
where  the $C_n$ are a closed paths encircling only $E_n$. 
This is the usual spectral theorem which shows how eigenvalues and 
eigenfunctions (contained in the projectors $\widehat{P}_n$) determine the 
time evolution of a system with discrete spectrum. 

In the case where the spectrum of $\cH$ is not discrete  the sum over 
eigenvalues is replaced by an integral, and it becomes harder 
to read off properties of the time-evolution directly. Physically, 
a continuous spectrum corresponds to an open system 
like a scattering system
where  wavepackets can 
decay by spreading out to infinity. This will be  described by 
resonances. 

Let us assume $\cH$ has continuous spectrum. The resolvent $\widehat{R}(E)$ 
is an analytic function of $E$ for $\Im E>0$, and the resonances 
are defined as the poles of the meromorphic continuation of 
$\widehat{R}(E)$ to the region $\Im E \leq 0$. Since the operator $\cH$ 
is selfadjoint on $L^{2}(\R^d)$ and has continuous spectrum, 
there is no meromorphic continuation of $\widehat{R}(E)$ as an operator 
from $L^2(\R^d)\to L^2(\R^d)$. Instead one looks for 
a continuation of $\widehat{R}(E)$ as an operator 
\begin{equation}
\widehat{R}(E):L^2_{\rm{comp}}(\R^d)\to L^2_{\rm{\text{loc}}}(\R^d) \,\, ,
\end{equation} 
where $L^2_{\rm{comp}}(\R^d)$ and $L^2_{\rm{\text{loc}}}(\R^d)$ denote 
the spaces of functions that are in $L^2(\R^d)$ and have compact support, or that locally are in $L^2(\R^d)$, respectively.
More directly, let $\varphi,\psi\in L^2_{\rm{comp}}(\R^d)$, then
\emph{quantum resonances} are the poles of the meromorphic continuation of the 
matrix elements 
\begin{equation}
\la \varphi, \widehat{R}(E)\psi\ra
\end{equation}
from the region $\Im E> 0$ to $\Im E\leq 0$. 
Assuming we have found such a meromorphic continuation with 
poles at $E_n\in\C$, $n\in\N$, $\Im E_n<0$, then we can use 
\eqref{eq:mell-tr} to get
\begin{equation}
\la \varphi, \widehat{U}(t)\psi\ra=
\frac{1}{2\pi\ui}\int_{\Im E=c}\la \varphi, \widehat{R}(E)\psi\ra\ue^{\frac{\ui}{\hbar}tE}\,\, \ud E \,\, .
\end{equation}
Shifting the contour of integration and picking up the contribution 
from the poles gives us an expansion in terms of the  resonances $E_n$
\begin{equation}\label{eq:decay-time}
\la \varphi, \widehat{U}(t)\psi\ra \sim \sum \ue^{-\frac{\ui}{\hbar}tE_n} 
\la \varphi,\widehat{P}_n\psi\ra 
\end{equation}
with the projectors 
\begin{equation}
\widehat{P}_n:=\frac{1}{2\pi\ui}\int_{C_n}\widehat{R}(E)\,\, \ud E\,,
\end{equation}
where $C_n$ is a closed path encircling only the resonance $E_n$. 
This looks formally  like \eqref{eq:time-evo-exp-unitary}, 
but there are two important differences. Firstly, $\Im E_n <0$ 
which means that $\abs{\ue^{-\frac{\ui}{\hbar}tE_n}}=\ue^{t\Im E_n}$, and
hence the terms in the sum are exponentially decreasing for $t\to \infty$ (since $\Im E_N <0$). 
Secondly, the projectors $\widehat{P}_n$ are no longer orthogonal projectors 
in $L^2(\R^d)$. Futhermore, we can take the expansion only as far as the 
meromorphic continuation allows us to, and even if it extends to $\C$, 
the resulting sum could be divergent. The range of the 
meromorphic continuation and  the convergence properties of the sum 
can depend on $\varphi$ and $\psi$ (see \cite{Zworski99} for a more detailed 
description). 

The relation \eqref{eq:decay-time} reveals the dynamical meaning of the 
resonances. 
Resonance states are not stationary, and the reciprocal value of the imaginary 
part of the resonance energies determines their lifetime.

\subsection{Computation of resonances of the quantum normal form}

We now turn to explicit calculations and show how one can compute quantum resonances of the 
quantum normal form. 

\subsubsection{Resonances of one-dimensional systems}
\label{sec:oneDrescompâ}

We start with the simplest one-dimensional example ( $d=1$)
and consider 
the operator 
\begin{equation}\label{eq:basic-op}
\cH=\lambda \hat{I}=\lambda \frac{\hbar}{\ui}\bigg(q\pa_q+\frac{1}{2}\bigg)\,,
\end{equation}
where $\lambda >0$. 
For this operator the Schr{\"o}dinger equation can be 
solved explicitly and the time evolution operator is given by 
\begin{equation}
\widehat{U}(t)\psi(q)=\ue^{-\frac{\lambda t}{2}} \psi(\ue^{-\lambda t}q)\,\, .
\end{equation}
This operator is of course unitary, i.e. it preserves the $L^2$-norm. In time, 
the state $\widehat{U}(t)\psi(q)$ spreads out at an exponential rate. If we look at the overlap of  $\widehat{U}(t)\psi(q)$  with another localised state we expect an 
exponential decay, and this is exactly what the resonances describe. 
Let $\varphi, \psi\in C_0^{\infty}(\R)$,  then 
\begin{equation}
\la\varphi, \widehat{U}(t)\psi\ra 
=\ue^{-\frac{\lambda t}{2}} \int \varphi^*(q) \psi(\ue^{-\lambda t}q)\,\, \ud q
\end{equation}
and if we insert for $\psi$ its Taylor series 
\begin{equation}
\psi(q)=\sum_{n=0}^{N}\frac{1}{n!}\psi^{(n)}(0) q^n +R_{N+1}(q) \,\, ,
\end{equation}
with $\abs{R_{N+1}(q)}\leq C_{N+1}\abs{q}^{N+1}$, 
then we obtain 
\begin{equation}
\la\varphi, \widehat{U}(t)\psi\ra 
= \sum_{n=0}^{N}\ue^{-\lambda (n+1/2)t} \frac{1}{n!}\psi^{(n)}(0)
\int \varphi^*(q) q^n\,\, \ud q+O\big(\ue^{-\lambda(N+1+1/2) t}\big)
\end{equation}
for $t\geq 0$. Inserting this equation into \eqref{eq:lapl-tr} 
leads to the meromorphic continuation of $\widehat{R}(E)$ to the domain 
$\Im E> -\hbar \lambda (N+1+1/2)$ with poles at 
\begin{equation}
E_n =-\ui \hbar \lambda (n+1/2)\,\, ,\quad  n=0,\ldots, N\,\, .
\end{equation}
These are the resonances of the operator $\cH$ given in \eqref{eq:basic-op}.

We can furthermore read off the projection operators 
\begin{equation}
\widehat{P}_n\psi(q):=\frac{1}{n!}\psi^{(n)}(0)q^n\,\, , 
\end{equation}
and a direct calculation shows that $q^n$ is an eigenfunction with complex 
eigenvalue $E_n=-\ui\hbar\lambda (n+1/2)$, 
\begin{equation}
\cH q^n=-\ui\hbar\lambda (n+1/2)q^n\,\, .
\end{equation}

We now extend this analysis to the case  of a Hamilton operator in  quantum normal form for $d=1$, 
i.e., $\cH=K(\hat{I})$, where $K$ is a polynomial or an analytic function in 
$I$. We will require furthermore the condition 
\begin{equation}\label{eq:H-cond}
\Im K(-\ui x)<0\,\, ,\quad \text{for}\quad x>0\,\, .
\end{equation}
By expanding $K$ in a power series we find 
\begin{equation}
\cH q^n=K\big(-\ui\hbar (n+1/2)\big)q^n
\end{equation}
and  solving the Schr{\"o}dinger equation yields 
$\widehat{U}(t)q^n=\exp[-\frac{\ui}{\hbar}t K\big(-\ui\hbar (n+1/2)\big)]q^n$. 
Hence, if $\psi(q)$ is analytic, we have 
\begin{equation}
\widehat{U}(t)\psi(q)=\sum_{n=0}^{\infty} \frac{1}{n!} \psi^{(n)}(0)\ue^{-\frac{\ui}{\hbar} t K(-\ui\hbar (n+1/2))}q^n \,,
\end{equation}
and by condition \eqref{eq:H-cond} we can use \eqref{eq:lapl-tr} to see that 
the resonances are given by 
\begin{equation}
E_n = K\big(-\ui \hbar  (n+1/2)\big)\,\, ,\quad  n=0,1,2,\ldots \,\, .
\end{equation}
This can be regarded as  a kind of imaginary Bohr-Sommerfeld quantisation condition for the resonances. 
Moreover, we can formally write the resonance states $\phi_n(q)=q^n$  as ``complex'' Lagrangian states,
\begin{equation}\label{eq:complex_lagr_states}
\phi_n(q) = q^n = (\text{sgn } q)^n \abs{q}^{-1/2+\ui I_n /\hbar}
\end{equation}
with $I_n=-\ui\hbar(n+1/2)$. This reveals the formal similarity of the resonance states to 
the scattering states \eqref{eq:out-states} with the main difference being that in the case of resonances 
$I$ fulfils an imaginary Bohr-Sommerfeld quantisation condition while in the case of scattering the spectrum of $\Ihat$ is continuous and real.
With the states \eqref{eq:complex_lagr_states} we can associate the complex Lagrangian manifolds
\begin{equation}\label{eq:comp-lag1}
\Lambda_{\phi_n} = \{ (q,p)=(q,I_n/p) \,:\, q \in \R\} \subset \R \times \ui\R\,.
\end{equation}

\subsubsection{Resonances of multi-dimensional quantum normal form}

Finally,  we consider the case of a $d$-dimensional system in
quantum normal form, 
i.e. let $\cH=K(\hat{I} ,\hat{J}_2,\dots ,\hat{J}_{d})$ and 
$\varphi_{n_k}$ denote the $n_k^{\text{th}}$ harmonic oscillator 
eigenfunction (see \eqref{eq:harm-osc-E}).
For $n=(n_1,\ldots ,n_d)\in \N_0^d$, set 
\begin{equation} \label{eq:multiresstate}
\psi_n(q)=q_1^{n_1} \varphi_{n_2}(q_2) \cdots \varphi_{n_{d}}(q_{d}) 
\,\, .
\end{equation}
Then we have 
\begin{equation}\label{eq:mult-dim-eigen}
\cH\psi_{n}=K\big(-\ui\hbar(n_1+1/2),\hbar(n_{2}+1/2), \ldots  ,\hbar(n_d+1/2)\big)
\psi_n \,,
\end{equation}
and if we assume $\Im K(-\ui x_1,x_2 \ldots , x_{d}) <0$ for $x_1>0$ and  $x_2,\ldots ,x_{d}$ in a neighbourhood of $0$, we can conclude 
as before  that the resonances of $\cH$ are given by 
\begin{equation}\label{eq:resonances-ndim}
E_n =K\big(-\ui \hbar(n_1+1/2), \hbar(n_{2}+1/2), \ldots ,\hbar(n_d+1/2)\big)\,\, , \quad n\in \N_0^d\,\, .
\end{equation}

To summarize, we have shown 

\begin{thm}
Suppose $\cH=K(\hat{I},\hat{J}_2,\cdots,\hat{J}_d)$ and that $K$ satisfies the condition
\begin{equation}
\Im K(-\ui x_1,x_2,\ldots ,x_d)<0
\end{equation}
for $x_1>0$ and $x_2,\cdots x_d$ in some neighbourhood of $0$. Then the resonances in a neighbourhood of 
$0$ are given by 
\begin{equation}\label{eq:resonances-ndim-th}
E_n =K\big(-\ui \hbar(n_1+1/2), \hbar(n_{1}+1/2), \ldots ,\hbar(n_d+1/2)\big)\,\, , \quad n\in \N_0^d\,\, .
\end{equation}
and the corresponding resonance eigenstates are 
\begin{equation} \label{eq:multiresstate-th}
\psi_n(q)=q_1^{n_1} \varphi_{n_2}(q_2) \cdots \varphi_{n_{d}}(q_{d}) 
\,\, .
\end{equation}
\end{thm}

Following \eqref{eq:comp-lag1} the resonance eigenstate can be interpreted as Lagrangian states associated 
with the complex Lagrangian manifolds 
\begin{equation}\label{eq:comp-lagd}
\Lambda_{\psi_n} = \{ (q,p)\in \R^{2d}  \,:\,p_1=I_{n_1}/q_1\, ,\,\,  (p_k^2+q_k^2)=2n_k \hbar\,,\,\,k=2,\dots,d \}\,\, .
\end{equation}

\subsection{Lifetime of the activated complex}\label{sec:lifetime-act-complex}

The geometric object in classical phase space associated with the activated complex is the centre manifold, a    
$(2d-2)$-dimensional invariant submanifold. As mentioned in 
Sec.~\ref{sec:classicalrate} 
this submanifold can be considered as the phase space of a $(d-1)$ \dof invariant subsystem  
related to the supermolecule poised between reactants and products in the 
chemistry literature \cite{Pechukas76,Marcus92}.
This invariant subsystem is unstable, i.e. a trajectory with initial condition near but not in the subsystem will leave the neighbourhood of this subsystem. 

For the corresponding quantum system the 
Heisenberg uncertainty relation excludes the 
existence of a quantum analogue of the classical invariant subsystem. 
This is because in normal form coordinates the invariant manifold is defined by 
$q_1=p_1=0$ and in quantum mechanics we have the uncertainty relation
$\Delta p_1\Delta q_1\geq \hbar/2$, i.e. 
$p_1$ and $q_1$ cannot be $0$ simultaneously. 
The closest one can get to a state which initially has $q_1=p_1=0$ 
is a minimal uncertainty state which is a Gaussian of the form 
\begin{equation}
\psi_0(q_1)=\frac{1}{(\pi\hbar)^{1/4}}\ue^{-\frac{1}{\hbar}\frac{q_1^2}{2}}\,\, .
\end{equation}
In order to obtain a state which at time $t=0$ 
is localised on the centre manifold we choose
\begin{equation}\label{eq:NHIM-state}
\psi(q_1,\dots, q_d)=\frac{1}{(\pi\hbar)^{1/4}}\ue^{-\frac{1}{\hbar}\frac{q_1^2}{2}}\varphi_{n_2}(q_2)\cdots \varphi_{n_d}(q_d)
\end{equation}
for some fixed quantum numbers $n_2,\dots, n_d\in\N_0$, 
where $\varphi_{n_k}$ again denote the harmonic oscillator eigenfunctions. 

A suitable quantity for measuring the lifetime of such a state is the decay of the autocorrelation function 
\begin{equation}
\abs{\la \psi, \widehat{U}(t)\psi\ra}^2\,.
\end{equation}
\rem{
which is constant if the state $\psi$ is invariant and decays otherwise. 
}
We will compute the autocorrelation function for the case that the Hamiltonian is in quantum normal form. 
Inserting the expression \eqref{eq:NHIM-state} for $\psi$ and expanding the Gaussian into a Taylor series gives
\begin{equation}
 \begin{split}
\la \psi, \widehat{U}(t)\psi \ra 
&=\sum_{k=0}^{\infty} \frac{1}{k!}\frac{(-1)^k}{(2\hbar)^k}\frac{1}{(\pi\hbar)^{1/4}}
\la \psi_0\varphi_{n_2}\cdots \varphi_{n_d},\widehat{U}(t)q^{2k}\varphi_{n_2}\cdots \varphi_{n_d}\ra\\
&=\sum_{k=0}^{\infty} \frac{1}{k!}\frac{(-1)^k}{(2\hbar)^k}\frac{1}{(\pi\hbar)^{1/2}}
\int\ue^{-\frac{1}{\hbar}\frac{q_1^2}{2}} q_1^{2k}\,\, \ud q_1 
\ue^{-\frac{\ui}{\hbar} tH\big(-\ui\hbar(2k+1/2), \hbar(n_2+1/2),\cdots,\hbar(n_d+1/2)\big)}\,\, ,
\end{split}
\end{equation}
where we have used as well that $q^{2k}$ is a resonance state \eqref{eq:mult-dim-eigen}. 
The integral over $q_1$ gives $\int\ue^{-\frac{1}{\hbar}\frac{q_1^2}{2}} q_1^{2k}\,\, \ud q_1
=\Gamma(k+1/2)(2\hbar)^{k+1/2}$, and  we thus find 
\begin{equation}
\la \psi, \widehat{U}(t)\psi \ra 
=\bigg(\frac{2}{\pi}\bigg)^{1/2}\sum_{k=0}^{\infty} \frac{\Gamma(k+1/2)}{k!}(-1)^k
\ue^{-\frac{\ui}{\hbar} tH\big(-\ui\hbar (2k+1/2), \hbar(n_2+1/2),\cdots,\hbar(n_d+1/2)\big)}\,\, .
\end{equation}
The leading term in this sum for $t\to\infty$ is given by the smallest resonance with $k=0$. Hence, 
\begin{equation}
\abs{\la \psi, \widehat{U}(t)\psi \ra}^2 \sim 2\ue^{\frac{1}{\hbar} t2\Im H\big(-\ui\hbar/2, \hbar(n_2+1/2),\cdots,\hbar(n_d+1/2)\big)}\,,
\end{equation}
and this determines the maximal lifetime  
of a quantum state of the activated complex, i.e. a 
state initially localised on the invariant subsystem given by the centre manifold.

For small $\hbar$ the quantum normal form is dominated by its quadratic part and that gives 
\begin{equation}
\lim_{\hbar\to 0} \frac{1}{\hbar} 2\Im H\big(-\ui\hbar/2, \hbar(n_2+1/2),\cdots,\hbar(n_d+1/2)\big)=-\lambda
\end{equation}
and therefore for small $\hbar$ 
\begin{equation}
\abs{\la \psi, \widehat{U}(t)\psi \ra}^2 \sim 2\ue^{-t\lambda}\,\, .
\end{equation}
The quantum lifetime of the activated complex is  in leading order for $\hbar\to 0$ thus given by the reciprocal value of the classical Lyapunov 
exponent associated with the saddle equilibrium point.


\subsection{On the relation between the resonances of the Quantum Normal Form and the 
full system}

We have seen that the resonances of an operator in quantum normal form can be 
computed explicitly. They are obtained from the Bohr-Sommerfeld type 
quantization condition \eqref{eq:resonances-ndim}. 
In Sec.~\ref{sec:SNF} we have shown how to approximate a 
Hamilton operator near an equilibrium point of the principal symbol by an 
operator in quantum normal form.
We now want to discuss under 
which conditions this quantum normal form can be used to compute 
the resonances of the full Hamilton operator. 
This question has been studied in 
\cite{KaiKer00}
and we will mainly cite their results. 

One would expect that resonances of the full system are close to the one of the quantum normal form 
around an equilibrium point if that equilibrium point dominates the reaction, i.e., if it is the only 
equilibrium point at that energy, and all other trajectories come from infinity or can escape to infinity.  
This idea is formalized by using the trapped set of the classical Hamiltonian, whose definition we now recall.  

Let $H(q,p)$ be a Hamilton function and $\flow_H^t$ the Hamiltonian flow 
generated by it. The {\bf trapped set} at energy $E$ is defined by 
\begin{equation}
TS^E(H):=\{ (q,p)\in \R^d\times \R^d \,:\, H(q,p)=E\,,\quad |\lim_{t\to\pm \infty}\Phi_H^t(q,p)| < \infty\}\,.
\end{equation}
It consists of the trajectories which stay in some bounded region for $t\to \pm \infty$.  

\begin{thm}[\cite{KaiKer00}] \label{thm:KK}Assume $H$ satisfies the general conditions of \cite{HelSjo86}
  and 
has a equilibrium point at $z_0$ with energy $E_0$ and
$TS^{E_0}(H)=\{z_0\}$. Let $\Hqnf^{(N)}$ be the $N^{\text{th}}$ order quantum normal form of $H$ with respect to $z_0$. 
Then the resonances of $\Op[H]$ in a $\hbar^{\delta}$ neighbourhood of 
$E_0$, $1\geq \delta >0$,  are $\hbar^{\delta N}$ close to the resonances of $\Hqnf^{(N)}$.  
\end{thm}

The conditions from  \cite{HelSjo86} referred to above are conditions on $H$ which ensure 
that the resonances can be defined by a complex deformation of phase space, 
a  generalisation of the complex dilation method \cite{Simon79,Reinhardt82,Moiseyev98} which we will use in Sec.~\ref{sec:examples} 
to compute numerically exact quantum resonances. For a more recent and more 
accessible presentation see \cite{LahMar02}.

More explicitly, the main consequence of Theorem \ref{thm:KK} is that for 
every $n\in \N_0^d$, there is a resonance $E_n\in \C$ of $\Op[H]$ 
with 
\begin{equation} {\label{eq:BohrSommerfeldresonances}}
E_n =\Hqnf^{(N)}\big(-\ui \hbar(n_1+1/2), \hbar(n_{1}+1/2), \ldots ,\hbar(n_d+1/2)\big)+
O\big((\abs{n}\hbar)^{N+1}\big)\,\,   .
\end{equation}
The quantum normal form thus provides an asymptotic expansion of the resonances for small $\hbar$. 
If we want to have all resonances in a neighbourhood of $E_0$ of radius $\hbar^{\delta}$, 
then we must go in $n$ up to a size determined by $\hbar \abs{n}\sim \hbar^{\delta}$ in which case 
the error term becomes of order $\hbar^{\delta N}$. Since we are interested in the 
first few resonances only we can take $\delta=1$.

We note that the resonances \eqref{eq:BohrSommerfeldresonances} coincide with the poles of the S-matrix 
which we computed in \eqref{eq:local-S} and \eqref{eq:Smatrixfull}. 
As can be seen from \eqref{eq:local-S} the poles of the S-matrix are simply given by the poles of the gamma function 
at non-positive integers.

In cases when the trapped set is larger, e.g., when there are several equilbrium points at the same energy, 
the situation is more complicated and the structure of the set of resonance is no longer 
necessarily determined by the contributions from the individual equilibrium points. 
Instead one has to use the methods sketched in Sec. \ref{sec:global-S-matrix} to construct a global S-matrix 
which will bring the global geometry 
into play.


\subsection{Distribution of the resonance states in phase space}

We now want to study the distribution of the resonance states in phase space in terms of Husimi functions. 
Like in the case of the scattering states in Sec.~\ref{sec:husimi_scatt} 
the Husimi function of the resonance states 
\eqref{eq:multiresstate} is given by the product of the Husimi functions of  harmonic oscillator eigenfunctions $\varphi_{n_k}$ 
and the Husimi function of $\phi_{n_1}(q_1)=q_1^{n_1}$.  
We already discussed the Husimi function of the $\varphi_{n_k}$ 
in Sec.~\ref{sec:husimi_scatt}. 
The computation of the Husimi function of the $\phi_{n_1}$ is rather straightforward,
and we obtain 
\begin{equation} 
\la \psi_{p_1,q_1},\phi_{n_1}\ra =\frac{\sqrt{2\pi\hbar}}{(\pi\hbar)^{1/4}} \left(\frac{\hbar}{2}\right)^{n/2} \ui^n 
H_{n_1}\bigg(\frac{p_1-\ui q_1}{\sqrt{2\hbar}}\bigg) \ue^{\frac{\ui}{2\hbar}p_1q_1-\frac{1}{2\hbar}p_1^2}\,,
\end{equation}
where $H_{n_1}$ is the $n_1^\text{th}$ Hermite polynomial. Therefore we have 
\begin{equation} \label{eq:resonanceHussimi}
H_{\phi_{n_1}}(q_1,p_1)=\frac{1}{\sqrt{\pi\hbar}}\left(\frac{\hbar}{2}\right)^{n_1}\biggabs{H_{n_1}\bigg(\frac{p_1-\ui q_1}{\sqrt{2\hbar}}\bigg)}^2 \ue^{-p_1^2/\hbar}\,\, .
\end{equation}
Figure~\ref{fig:husreson} shows contour plots of the Husimi functions of the  first five resonance states. Due to the exponential damping in the direction of $p_1$
the Husimi functions $H_{\phi_n}$ are concentrated along $p_1=0$.
Along $p_1=0$ they increase in leading order in $q_1$ as
\begin{equation} 
H_{\phi_n}(q_1,0)\sim  \frac{1}{\sqrt{\pi\hbar}} \left(\frac{\hbar}{2}\right)^{n/2} q_1^n +
O(q_1^{n-2})\,.
\end{equation}

It follows from \eqref{eq:resonanceHussimi} that $H_{\phi_{n_1}}$  has  $n_1$ zeroes located near the origin 
on $q_1=0$.

\def\fighusreson{%
Contour plots of the Husimi functions 
$H_{\phi_{n_1}}$ in the $(q_1,p_1)$-plane for $n_1=0,\dots,5$. 
Red corresponds to low values; blue corresponds to high values. The spacing between the values of the contourlines 
is decreasing exponentially. ($\hbar=0.1\,\,.$)
}
\def\FIGhusreson{
\centerline{
\includegraphics[angle=0,width=10cm]{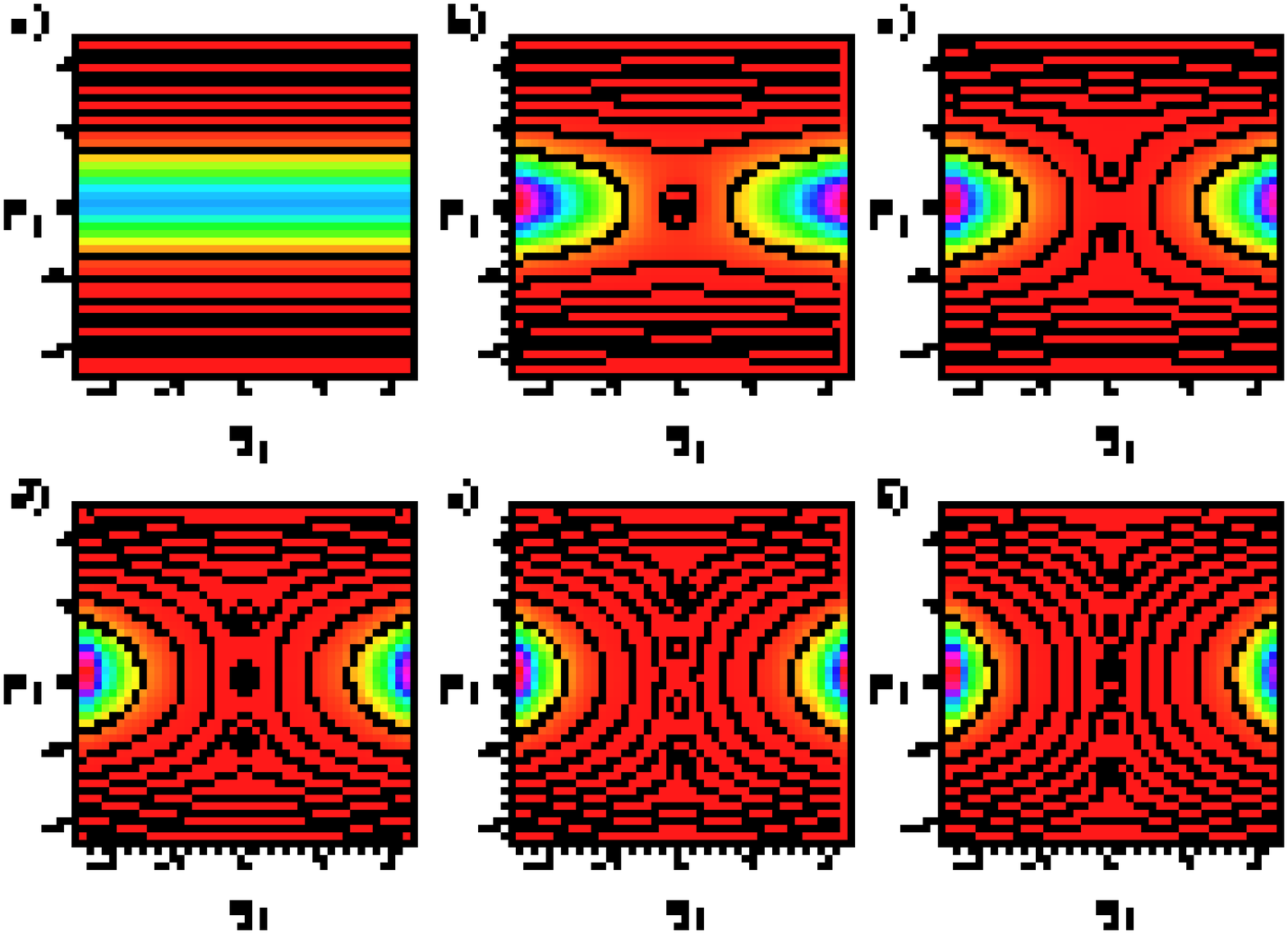}
}
}
\FIGo{fig:husreson}{\fighusreson}{\FIGhusreson}

For $n=(n_1,\dots,n_d)\in\N_0^d$ the Husimi function of a multi-dimensional scattering wavefunction $\psi_n$ defined in \eqref{eq:multiresstate} is simply given by the product of the functions defined in \eqref{eq:harmHussimi} and \eqref{eq:resonanceHussimi}, i.e.
\begin{equation}
H_{\psi_n}(q,p) = H_{\phi_{n_1}}(q_1,p_1)  H_{\varphi_{n_2}}(q_2,p_2) \cdots  H_{\varphi_{n_d}}(q_d,p_d)\,.
\end{equation}
From the distribution of the functions \eqref{eq:harmHussimi} and \eqref{eq:resonanceHussimi} it thus follows that
the resonance states $\psi_n$ are concentrated on 
the real projections of the complex Lagrangian manifolds $\Lambda_{\psi_n}$ in \eqref{eq:comp-lagd}
\begin{equation} \label{eq:lagrangereson}
\{ (q,p) \in \R^{2d}\,:\, p_1=0\,,\,\, (p_k^2+q_k^2)=2n_k \hbar\,,\,\,k=2,\dots,d \}\,.
\end{equation}

%% file: examples.tex
\section{Examples}
\label{sec:examples}

In the following we illustrate the classical and quantum reaction dynamics for 
concrete examples with one, two and three degrees of freedom. As we
will see the reaction dynamics in systems with one or two degrees of freedom
still has certain features that do not persist in the multidimensional case
(of three or more degrees of freedom). 
We will use the classical normal form to realise the phase space structures
that control classical reaction dynamics for these systems and compute the
classical flux. Likewise we will use the quantum normal form 
to compute cumulative reaction probabilities and quantum resonances.
We note that we  implemented the procedures to compute the classical and quantum normal forms
in the programming language 
{\sf C}$++$. In our object-oriented implementation the number of degrees of freedom and 
the order of the normal form can be chosen arbitrarily.


\subsection{Example with 1 \dof}
\label{sec:1Dexample}

The most frequently used systems to model one-dimensional reaction problems, 
like  the paradigm hydrogen exchange reaction
H$_2$+H $\rightarrow$ H + H$_2$, are the parabolic barrier and the Eckart potential
(see, e.g., \cite{SeidemanMiller91,SkodjeYang04}). 
The reason for choosing these model systems is that the reflection 
coefficient and the quantum 
resonances can be computed analytically for these systems.
We have already seen that the quantum normal form computation 
of the reflection coefficient and the resonances is exact for a parabolic
barrier. We therefore focus here on the Eckart barrier which provides a much
more realistic model of reactions than the parabolic barrier.

The Hamilton function for an Eckart barrier \cite{Eckart30} is given by
\begin{equation}\label{eq:ham_eckart}
H=p^2/(2m) + V_{\text{E}}(x)\,, 
\end{equation} 
where
$V_{\text{E}}$ is defined as
\begin{equation}
\label{eq:eckartpotential}
V_{\text{E}}(x) = A \frac{\exp((x+x_0)/a)}{1+\exp ((x+x_0)/a)} + B
\frac{\exp ((x+x_0)/a)}{(1+\exp ((x+x_0)/a))^2}
\end{equation}
with
\begin{equation}
\label{eq:equieckart}
x_0 = a \ln \frac{B+A}{B-A}\,.
\end{equation}
For $B>A\ge0$ the Eckart potential possesses a maximum which we shifted to
$x=0$ for convenience. The value of the potential at its maximum is
\begin{equation}
V_{\text{E}}(0) = \frac{(A+B)^2}{4B}\, .
\end{equation}
The potential monotonically decreases to 0 as $x\to -\infty$ and to  $A$
as $x\to \infty$ (see Fig.~\ref{fig:eckartpotential}a).
For $A=0$, the potential is symmetric.

The Weyl quantisation of the Hamilton function \eqref{eq:ham_eckart} gives the
Hamilton operator
\begin{equation}
\Op[H] = -\frac{\hbar^2}{2m } \frac{\partial^2}{\partial x^2} + V_{\text{E}}\,.
\end{equation}
The Hamilton function $H$ in \eqref{eq:ham_eckart} is then the principal
symbol of the Hamilton operator $\Op[H]$.

\def\figeckartpotential{%
(a) Graph of the Eckart potential $V_{\text{E}}$ defined in (\ref{eq:eckartpotential}) 
with parameters $a=1$, $B=5$ and different values of $A$. 
(b) Phase portaits for the Eckart potential with parameters $a=1$, $A=0.5$,
$B=5$ and $m=1$. The green and red lines mark the stable and unstable
manifolds of the equilibrium point $(x,p_x)=0$.
}

\def\FIGeckartpotential{
\centerline{
\raisebox{5.2cm}{a)}\includegraphics[angle=0,height=5cm]{eckart_potentials_NONL}
\raisebox{5.2cm}{b)}\includegraphics[angle=0,height=5cm]{eckart_phase_portrait}
}
}
\FIGo{fig:eckartpotential}{\figeckartpotential}{\FIGeckartpotential}


\subsubsection{Computation of the classical and quantum normal forms}

In order to compute the quantum normal form we can follow the calculation for one-dimensional potential barriers  described in 
Sec.~\ref{sec:1Dpotentialbarriers}.
Using the notation of Sec.~\ref{sec:1Dpotentialbarriers}
the coefficients of the Taylor expansion to fourth order are
\begin{equation} \label{def:lambdaeckart}
\lambda = \frac{1}{\sqrt{ 8 m a^2 B^3}} \left( B^2 - A^2 \right) \,,
\end{equation}
and
\begin{equation}
V_{30} =
-\frac{1}{16}\,A  {\frac {\sqrt {{B}^{2}-{A}^{2}}}{{B}^{7/4} }}  \left( {\frac {2  }{m a^2}} \right) ^{3/4} \,,\quad
V_{40} 
= {\frac {1}{96}}\,{\frac {2\,{B}^{2}-9\,{A}^{2}}{m\,{a}^{2}{B}^{2}}}\,.
\end{equation}
We refrain from giving the analytical expressions for the higher order terms
as the actual computation of the classical and quantum normal form 
implemented in our {\sf C}$++$ program is carried out numerically. However, we used 
the coefficients above together with Equation~\eqref{eq:opH4} to check
the numerically 
computed $4^\text{th}$ order quantum normal form.
To give the reader the opportunity to verify our results we list in
Tab.~\ref{Tab:HSCNF1D} in Sec.~\ref{sec:appendixcoefficients} of the appendix 
the coefficients of the symbol of the $10^\text{th}$ order quantum normal
form of the Eckart barrier with parameters $a=1$, $B=5$, $A=1/2$ and
$m=1$. The classical normal form can be obtained from the symbol by discarding
all terms that involve a factor  $\hbar$.

\subsubsection{Classical reaction dynamics}

Since the energy surface of a 1 \dof system is one-dimensional, the classical
reaction dynamics of 1 \dof systems is trivial. The question of whether a
trajectory is reactive or nonreactive is determined by the energy alone, i.e.,
in the case of the Eckart barrier trajectories are forward or backward
reactive if they have  energy $E>V_{\text{E}}(0)$, and they are nonreactive
localised in reactants or products if $E<V_{\text{E}}(0)$
(see Fig.~\ref{fig:eckartpotential}b). 
Fixing an energy $E>V_{\text{E}}(0)$ one can choose any point $x_{\text{ds}}\in\R$ to define
a dividing `surface' on the energy surface according to
$\{(x,p_x):x=x_{\text{ds}},\,H(x,p_x)=E\}=\{(x,p_x)=(x_{\text{ds}},\pm
\sqrt{2m(E-V_{\text{E}}(x))}\}$. This dividing `surface' consists of two
points which have $p_x>0$ and $p_x<0$ and are crossed by
all forward reactive trajectories and backward reactive trajectories,
respectively. In fact the two points can be considered to form a
zero-dimensional sphere, $S^0$ with each point forming by itself a zero-dimensional ball, $B^0$.
Note that many of the other phase space
structures that we discussed in Sec.~\ref{sec:classical} do not make sense for
the case of $d=1$ degrees of freedom. Moreover, the case of one degrees of
freedom is special because it is the only case for which 
the location of the dividing surface is not
important. 

Note that the formalism to compute the classical flux $\flux(E)$ developed in
Sec.~\ref{sec:classicalrate} does not apply either to the case $d=1$. Still it
useful to view the classical flux to be given by the step function
$\flux(E)=\Theta(E-V_{\text{E}}(0))$, i.e., 
classically, we have full transmission for $E>V_{\text{E}}(0)$ and full
reflection for $E<V_{\text{E}}(0)$.

\subsubsection{Quantum reaction dynamics}

The effect of quantum mechanical tunneling makes the
quantum reaction dynamics even of 1 \dof systems
more complicated than the corresponding classical reaction dynamics.
The quantum mechanically exact transmission coefficient $T_{\text{exact}}$ can be computed analytically  for the Eckart potential \cite{Eckart30}. One finds 
\begin{equation}
\label{eq:TEexact}
T_{\text{exact}}(E) = 1 - \frac{ \cosh [ 2\pi(\alpha-\beta) ] + \cosh [2\pi \delta] }{\cosh [ 2\pi(\alpha+\beta) ] + \cosh [2\pi \delta]}\,,
\end{equation}
where
\begin{equation}
\alpha = \frac12 \sqrt{\frac{E}{C} } \,,\quad
\beta = \frac12 \sqrt{\frac{E-A}{C} } \,,\quad
\delta = \frac12 \sqrt{\frac{B-C}{C} } \,,\quad
C= \frac{\hbar^2}{8m a^2}\,.
\end{equation}
Note that $T_{\text{exact}}(E)\rightarrow 0$ when the energy $E$ approaches the limiting  value
$A$ of the potential from above.
Figure~\ref{fig:eckart} shows the graph of $T_{\text{exact}}(E)$ versus the energy $E$. 
Following Sec.~\ref{sec:oneDSmatrix} we can compute the transmission coefficient from the
$N^{\text{th}}$ order quantum normal form $\Hqnf^{(N)}$ 
according to
\begin{equation}
T^{(N)}_{\text{QNF}}(E) = \bigg[1+\exp\bigg(-2\pi\frac{I^{(N)}(E)}{\hbar}\bigg) \bigg]^{-1}\,,
\end{equation}
where $I^{(N)}(E)$ is obtained from inverting the equation
\begin{equation}
\Hqnf^{(N)}(I^{(N)}(E))=E\,.
\end{equation}

\def\figeckart{%
Exact transmission coefficient $T_{\text{exact}}(E)$ (top panel) and resonances in the complex energy plane (bottom panel) for the Eckart potential.
The parameters $a$, $B$, $C$ and $m$ are the same as in
Fig.~\ref{fig:eckartpotential}b, and  $\hbar=0.1$.
}
\def\FIGeckart{
\centerline{
\includegraphics[angle=0,width=10cm]{eckart_as_cumul_res_NONL}
}
}
\FIGo{fig:eckart}{\figeckart}{\FIGeckart}

We illustrate the high quality of the quantum normal form computation of the
transmission coefficient in
Fig.~\ref{fig:eckarterrors}a which shows the difference between $T_{\text{exact}}$ and $T^{(N)}_{\text{QNF}}$ 
for different orders, $N$, of the quantum normal form. 
Though the quantum normal form expansion is not expected to converge, 
the difference between $T_{\text{exact}}$ and $T^{(N)}_{\text{QNF}}$  decreases as $N$
increases to the maximum value of 10 at which we stopped the quantum normal
form computation. In fact, the difference decreases 
 from the order of 1 percent for the $2^\text{nd}$ order quantum normal form to
  the order of 
10$^{-11}$ for the $10^\text{th}$ order quantum normal form.

We can also compute the quantum mechanically exact resonances
analytically. They are given by the
poles of the transmission coefficient \eqref{eq:TEexact}. We find
\begin{equation}
E_{\text{exact},n} = C \frac{\left( (\delta -i (n+\frac12))^2 + \frac{A}{4C} \right)^2}{(\delta
-i (n+\frac12))^2} \,, \qquad n=0,1,2,\dots\,.
\end{equation}
We illustrate the location of the quantum resonances in the complex energy plane
in the bottom panel of Fig.~\ref{fig:eckart}. 
Following Sec.~\ref{sec:oneDrescompâ} we can compute the resonances
from the $N^{\text{th}}$ order quantum normal form according to
\begin{equation}
E^{(N)}_{\text{QNF},n} = \Hqnf^{(N)}(-\ui\hbar (n+1/2))\,,\qquad n=0,1,2,\dots\,.
\end{equation}
For the $2^\text{nd}$ order quantum normal form this reduces to
\begin{equation}
E^{(2)}_{\text{QNF},n} = V_{\text{E}}(0) -\ui \lambda \hbar \big( n+\frac12 \big) \,, \qquad n=0,1,2,\dots\,.
\end{equation} 
As mentioned earlier the $2^\text{nd}$ order quantum normal form resonances
would be exact for a parabolic potential barrier. 
For comparison we also show the location of these resonances in the complex
energy plane in the bottom panel of Fig.~\ref{fig:eckart}. Note that the
$2^\text{nd}$ order resonances have a constant real part.
The ``bending'' of the series of exact resonances in Fig.~\ref{fig:eckart} is a 
consequence of the nonlinearity of the Eckart potential. 
The quantum normal form is able to describe this effect very accurately. The approximation of the exact
resonances by the $4^\text{th}$ order quantum normal is already so good that the error is no longer 
visible on the scale of Fig.~\ref{fig:eckart}. 
We therefore show the differences between the exact and quantum normal form
resonances for different orders of the quantum normal form
in a separate graph in Fig.~\ref{fig:eckarterrors}b.
Again, up to the maximal order shown, the accuracy of the quantum normal form increases with the order. 
As to be expected, for a fixed order of the quantum normal form $N$, 
the error of the quantum normal form increases with the quantum number $n$. 
Note that the sequence of resonances is  localised in the complex energy plane
in Fig.~\ref{fig:eckart} in such a way that the real part of the resonance closest to the real axis coincides with the position of the (smooth) step of the transmission coefficient on the (real) energy axis.

\def\figeckarterrors{%
(a) Error for the transmission coefficient of the Eckart potential computed from quantum normal forms of different orders $N$. 
(b) Errors for the the resonances of the Eckart potential computed from quantum normal forms of different orders $N$
as a function of the quantum number $n$. 
The parameters for the Eckart potential are the same as in Fig.~\ref{fig:eckart}.
}
\def\FIGeckarterrors{
\centerline{
\raisebox{5.2cm}{a)}\includegraphics[angle=0,width=7.5cm]{cummulative_eckart_as_difference_NONL}
\raisebox{5.2cm}{b)}\includegraphics[angle=0,width=7.5cm]{eckart_as_poles_difference_NONL}
}
}
\FIGo{fig:eckarterrors}{\figeckarterrors}{\FIGeckarterrors}

\subsection{Example with 2 \dof}
\label{sec:2Dexample}

We now illustrate the quantum normal form computation for a 2 \dof model system which consists of 
an Eckart barrier in the $x$-direction that is coupled to a Morse oscillator in the
$y$-direction. A Morse oscillator is a typical model for a chemical bond.
 The Hamilton function is 
\begin{equation}
\label{eq:HEckartMorse}
H = \frac{1}{2 m} \big( p_x^2+p_y^2 \big) + V_{\text{E}}(x) + V_{\text{M}}(y) + \epsilon H_c\,,
\end{equation}
where $V_{\text{E}}$ is the Eckart potential from (\ref{eq:eckartpotential}) and $V_{\text{M}}$ is the Morse potential
\begin{equation} \label{eq:Morsepot}
V_{\text{M}}(y) = D_e (\exp(-2 a_{\text{M}} y) - 2 \exp(-a_{\text{M}} y))
\end{equation}
with positive valued parameters  $D_e$ (the {\em dissociation energy}) and $a_{\text{M}}$  (see Fig.~\ref{fig:eckartasmorse}a).
For the coupling term $H_c$ we choose a so called {\em kinetic coupling} (see,
e.g., \cite{Heller95})
\begin{equation}
H_c = p_x \, p_y\, .
\end{equation}
The strength of the coupling is controlled by the parameter $\epsilon$ in \eqref{eq:HEckartMorse}.
The vector field corresponding to the Hamilton function~(\ref{eq:HEckartMorse}) has an equilibrium point at 
$(x,y,p_x,p_y)=0$.
For $|\epsilon|$ sufficiently small (for given parameters of the Eckart and Morse potentials), 
the equilibrium point is of saddle-centre stability type.
Contours of the Eckart-Morse potential $V(x,y)=V_{\text{E}}(x)+V_{\text{M}}(y)$  are shown in 
Fig.~\ref{fig:eckartasmorse}b. These indicate the bottleneck-type structure of the energy surfaces with energies slightly above the energy of the saddle-centre equilibrium point. 
Note that the relation between the saddle of
the potential $V(x,y)=V_{\text{E}}(x)+V_{\text{M}}(y)$ at $(x,y)=0$ and the
equilibrium point of Hamilton's equations at $(x,y,p_x,p_y)=0$ is complicated
by the kinetic coupling in \eqref{eq:HEckartMorse}.

\def\figeckartasmorse{%
(a) Morse potential $V_{\text{M}}(y)=D_e\,(\exp(-2\,a_{\text{M}}
\,y)-2\exp(-2a_{\text{M}} \,y))$. The potential approaches 0 for $y\rightarrow\infty$.
The parameter $D_e=V_{\text{M}}(\infty)-V_{\text{M}}(0)$ gives the 
depth of the well potential well while
$a_{\text{M}}$ determines the width of the well.
(b) Contours of the Eckart-Morse potential $V_{\text{E}}(x)+V_{\text{M}}(y)$. 
Red correspond to small values of the potential; blue corresponds to large values.
The parameters for the Eckart potential are the same as in
Fig.~\ref{fig:eckart}. The parameters for the Morse potential are $D_e=1$ and 
$a_{\text{M}}=1$.
}
\def\FIGeckartasmorse{
\centerline{
\raisebox{5cm}{a)}\includegraphics[angle=0,height=5cm]{morse}
\raisebox{5cm}{b)}\includegraphics[angle=0,height=5cm]{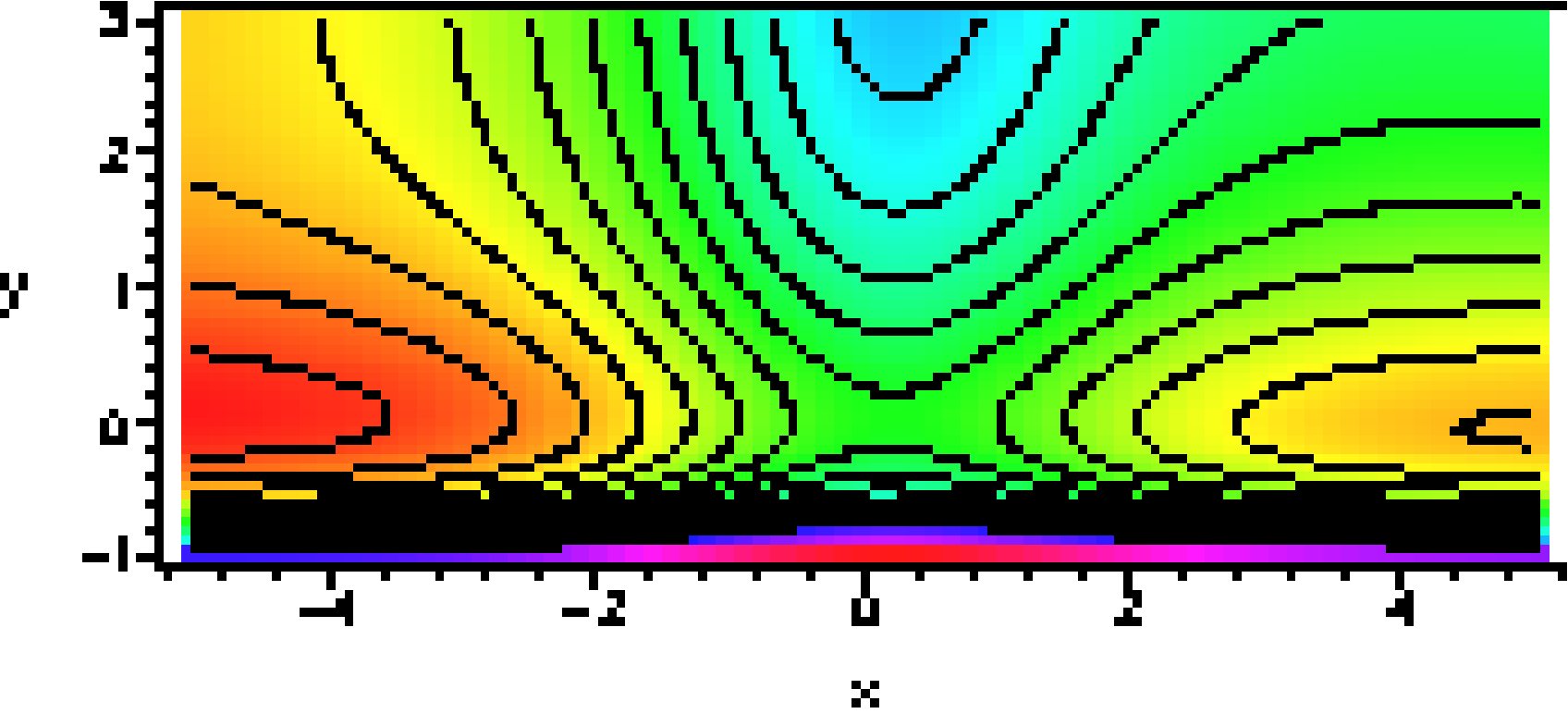}
}
}
\FIGo{fig:eckartasmorse}{\figeckartasmorse}{\FIGeckartasmorse}

The Weyl quantisation of the Hamilton function $H$ in \eqref{eq:HEckartMorse}
gives the operator
\begin{equation}
\Op[H] = -\frac{\hbar^2}{2m} \left( \frac{\partial^2}{\partial x^2} +
  \frac{\partial^2}{\partial y^2} \right) + V_{\text{E}} + V_{\text{M}}  -
  \epsilon \hbar^2  \frac{\partial^2}{\partial x \partial y}\,.
\end{equation}
The Hamilton function \eqref{eq:HEckartMorse} is the principal symbol of the
operator $\Op[H]$.


\subsubsection{Computation of the classical and quantum normal forms}

Since the equilibrium point is already at the origin of the coordinate system
we can skip the first step in the classical and quantum normal form transformation 
sequences~\eqref{eq:transformation-sequence} and \eqref{eq:transformation-sequence_quantum}, and start with the second
step which consists of simplifying the quadratic part of the Hamilton function or symbol, respectively.
To this end we follow 
Sec.~\ref{sec:examp_comp_cnf} and compute 
the matrix $J\, \Hess H$ associated with the linearisation of Hamilton's equations 
about $(x,y,p_x,p_y)=0$. This gives
\begin{equation}\label{eq:linearisationEckartMorse}
J\, \Hess H(0)
= 
\left(
\begin{array}{cccc}
0 & 0 & 1/m & \epsilon   \\
0 & 0 & \epsilon    & 1/m \\
m\lambda_{\text{E}}^2  & 0 & 0 & 0   \\
0 & -m \omega_{\text{M}}^2 & 0   & 0 \\
\end{array}
\right) \,,
\end{equation}
where $\lambda_{\text{E}}$ is defined as in \eqref{def:lambdaeckart} and 
\begin{equation}
\omega_{\text{M}} = \sqrt{ \frac{1}{m}  V''_{\text{M}}(0)} = a_{\text{M}} \sqrt{\frac{2 D_{e}}{m}}
\end{equation}
is the linear frequency of the Morse oscillator.
The matrix in \eqref{eq:linearisationEckartMorse} has eigenvalues
\begin{eqnarray}
\lambda := e_1 &=&  \frac12 \sqrt{2\lambda_{\text{E}}^2 - 2\omega_{\text{M}}^2 + 2 \sqrt{\omega_{\text{M}}^4 + 2\lambda_{\text{E}}^2 \omega_{\text{M}}^2 + \lambda_{\text{E}}^4 - 4 \epsilon m^2 \lambda_{\text{E}}^2 \omega_{\text{M}}^2 }} \,,\\
e_3 &=& -\lambda\,,\\
\ui \omega := e_2 &=& \ui \frac12 \sqrt{ 2\omega_{\text{M}}^2 - 2\lambda_{\text{E}}^2 + 2 \sqrt{\omega_{\text{M}}^4 + 2\lambda_{\text{E}}^2 \omega_{\text{M}}^2 + \lambda_{\text{E}}^4 - 4 \epsilon
m^2 \lambda_{\text{E}}^2 \omega_{\text{M}}^2 }}  \,,\\
e_4 &=&  -\ui \omega \,,
\end{eqnarray}
where as mentioned above, 
for given parameters of the Eckart and Morse potentials and $|\epsilon|$ sufficiently small, 
the eigenvalues $e_1$ and $e_3$ (and hence $\lambda$) are real, and $e_2$ and $e_4$ are purely imaginary 
(and hence $\omega$ is real). For $\epsilon \rightarrow 0$, $\lambda$ and $\omega$ converge to
$\lambda_{\text{E}}$ and $\omega_{\text{M}}$, respectively.
The corresponding eigenvectors are
\begin{equation}
\label{eq:eigenvect}
v_k = \left( e_k \big( e_k^2+\omega^2 \big), \epsilon m \lambda_{\text{E}}^2 e_k , m \lambda_{\text{E}}^2\big( e_k^2+\omega^2 \big),-\epsilon m^2\lambda_{\text{E}}^2
\omega^2 \right)^T\,, \quad k=1,2,3,4\,.
\end{equation}
Following  Sec.~\ref{sec:examp_comp_cnf} we obtain a 
real linear symplectic change of coordinates by
using the $v_k$ to define the columns of a matrix $M$ according to
\begin{equation}
M = \left( c_1 v_1, c_2 \Re v_2, c_1 v_3, c_2 \Im v_2 \right)
\end{equation}
with the coefficients $c_1$ and $c_2$ defined as
\begin{equation}
\label{eq:defd1d2}
c_1^{-2} := \la v_1, J v_3 \ra \,,\quad 
c_2^{-2} := \la \Re v_2, J \Im v_2 \ra \,.
\end{equation}
Now set
\begin{equation}
(q_1,q_2,p_1,p_2)^T = M^{-1} (x,y,p_x,p_y)^T\,.
\end{equation}
Then the Hamilton function (\ref{eq:HEckartMorse}) becomes
\begin{equation} \label{eq:HEckartMorse2}
H = V(0) + \lambda q_1 p_1 + \frac{\omega}{2} \left( q_2^2+p_2^2 \right) + \dots\,,
\end{equation}
where the neglected terms are of order greater than 2. 
The constant term is
\begin{equation}
V(0) = V_{\text{E}}(0) + V_{\text{M}}(0) = \frac{(A+B)^2}{4B} - D_e\,.
\end{equation}
The truncation of  \eqref{eq:HEckartMorse2} at order 2 is the symbol of the $2^\text{nd}$ order quantum normal form of \eqref{eq:HEckartMorse}.

The classical and quantum normal forms are then computed from the algorithm described in 
Sections~\ref{sec:examp_comp_cnf} and \ref{sec:compscnf}, respectively. 
For the parameters $a=1$, $B=5$, $A=1/2$ for the Eckart potential and $D_e=1$ and $a_{\text{M}}=1$ for the Morse potential, $\epsilon=0.3$ for the coupling strength, and $m=1$,
we list the coefficients of the symbol of the $10^\text{th}$ order quantum normal form in 
Tab.~\ref{Tab:HSCNF2D} of the Appendix.
The classical normal form can be obtained from the symbol by neglecting all terms that
involve a factor  $\hbar$.


\subsubsection{Classical reaction dynamics}

For a 2 \dof system the NHIM is a one-dimensional sphere, $S^1$,  i.e., a periodic orbit.
This is the Lyapunov periodic orbit associated with the saddle point.
As discussed in the introduction, for 2 \dof systems with time-reversal symmetry, the periodic orbit
can be used to define a dividing surface without recrossing -- the so called periodic orbit dividing surface -- from the projection of the periodic orbit to configuration space 
\cite{PechukasMcLafferty73,PechukasPollak78}. Note that, as mentioned earlier,
 such a construction in configuration space does not work for systems with 3 or more \dof \cite{WaalkensWiggins04}.

The NHIM has stable and unstable manifolds with the structure of cylinders or `tubes', $S^1\times \R$.
They inclose the forward and backward reactive trajectories as discussed in detail in, e.g., 
\cite{WaalkensBurbanksWiggins05b,WaalkensBurbanksWiggins05c}.
The flux is given by the action of the periodic orbit \cite{WaalkensWiggins04}. In the uncoupled case 
($\epsilon=0$) the periodic orbit (p.o.) is contained in the
$(y,p_y)$-plane and its action can be computed analytically. One finds
\begin{equation}
\flux(E) =  \oint_{\text{p.o.}} p_y \ud y = \frac{2\pi}{a} (\sqrt{2mD_e}-\sqrt{-2m(E-V_{\text{E}}(0))}) 
\end{equation}
for $-D_e+V_{\text{E}}(0) < E < V_{\text{E}}(0) $
and $\flux(E)=0$ (no classical transmission) for $E\le -D_e+V_{\text{E}}(0) $.


\subsubsection{Quantum reaction dynamics}

\def\figeckartmorsecoupled_cum{%
The top panel shows the
cumulative reaction probabilities $N_{\text{exact}}(E)$ (oscillatory curve) and 
$\Nweyl(E)$
for  the Eckart-Morse potential defined in the text with $\epsilon =0$. 
The bottom panel shows the (numerically) exact resonances  computed from the complex dilation method in the complex energy plane.
Circles mark resonances for the uncoupled case $\epsilon =0$ and crosses mark resonances for the strongly 
coupled case  $\epsilon = 0.3$. The parameters for the potential are the same as in 
Fig.~\ref{fig:eckartasmorse}. Again we choose $m=1$ and $\hbar=0.1$.
}
\def\FIGeckartmorsecoupled_cum{
\centerline{
\includegraphics[angle=0,width=10cm]{cum_res_eckart_as_morse_NONL}
}
}
\FIGo{fig:eckartmorsecoupled_cum}{\figeckartmorsecoupled_cum}{\FIGeckartmorsecoupled_cum}

For the uncoupled case we can  compute the cumulative transmission probability analytically. We have 
\begin{equation}
N_{\text{exact}}(E) = \sum_{n_2} T_{\text{Eckart; exact}}(E-E_{\text{Morse;}n_2})\,,
\end{equation}
where $ T_{\text{Eckart; exact}}$ denotes the transmission coefficient for the Eckart barrier given in \eqref{eq:TEexact} and
$E_{\text{Morse;}n_2}$ are the energy levels of a one-dimensional Morse oscillator
\begin{equation}
E_{\text{Morse;}n_2} = -\frac{a_{\text{M}}^2 \hbar^2}{2 m} \left( n_2+\frac12 - \frac{\sqrt{2mD_e}}{a_{\text{M}} \hbar} \right)^2\,,\qquad n_2=0,1,2,\dots\,.
\end{equation}
The graph of $N_{\text{exact}}$ in the top panel of Fig.~\ref{fig:eckartmorsecoupled_cum} shows that 
$N_{\text{exact}}$ is  ``quantised'', i.e., it increases in integer steps each time a new transition channel opens. The opening of a
Morse oscillators mode $(n_2)$ as a transition channel can be defined as the energy where 
$T_{\text{Eckart; exact}}(E-E_{\text{Morse;}n_2})=1/2$.
The quantisation of the cumulative reaction probability has been observed experimentally, e.g., in molecular isomerisation experiments 
\cite{LovejoyMoore93} and also in ballistic electron transport problems in semiconductor nanostructures 
where the analogous effect leads to a quantised conductance \cite{vanWees88,Wharam88}.
As mentioned in Sections~\ref{sec:classicalrate} and \ref{sec:smatrixmultiD}, the quantity
\begin{equation}
\Nweyl (E)=\flux(E)/(2\pi \hbar)
\end{equation}
can be interpreted as the mean number of open transmission channels at energy $E$. 
This is illustrated in the top panel of  Fig.~\ref{fig:eckartmorsecoupled_cum} which shows
$\Nweyl$ together with $N_{\text{exact}}$. 
Note the nonlinear increase of $\Nweyl(E)$ with $E$ which is an indication of the strong anharmonicity 
of the Morse oscillator.

In order to compute the cumulative reaction probability from the quantum normal form we follow the procedure described in Sec.~\ref{sec:smatrixmultiD}.
We get
\begin{equation}
N^{(N)}_{\text{QNF}}(E) = \sum_{n_2} \bigg[1+\exp\bigg(-2\pi\frac{I^{(N)}_{n_2}(E)}{\hbar}\bigg) \bigg]^{-1}\,,
\end{equation}
where $I^{(N)}_{n_2}(E)$ is obtained from inverting
\begin{equation}
\Hqnf^{(N)}(I^{(N)}_{n_2}(E),\hbar(n_2+1/2))=E\,,\quad n_2=0,1,2,\dots\,.
\end{equation}
The high quality of the quantum normal form computation of the cumulative reaction probability  is illustrated in 
Fig.~\ref{fig:eckartmorsecoupled}a which shows $|N_{\text{QNF}}(E)-N_{\text{exact}}(E)|$ versus the energy $E$ for 
quantum normal forms with $N=2$ to $N=10$.
Like in the 1 \dof example in Sec.~\ref{sec:1Dexample} we find that up to the orders shown, 
the accuracy of the quantum normal form increases with the order of the quantum normal form.
The error is of order 10$^{-10}$ for the $10^\text{th}$ order quantum normal form. 

For the coupled case $\epsilon\ne 0$ we also make a comparison of the quantum mechanically exact resonances and the resonances computed from the quantum normal form. The exact resonances cannot be computed analytically for the coupled case. To get them numerically we use the
\emph{complex dilation method} \cite{Simon79,Reinhardt82,Moiseyev98} 
whose implementation for the present system we describe in Sec.~\ref{sec:complscaling} of the Appendix.
The bottom panel in Fig.~\ref{fig:eckartmorsecoupled_cum} shows the (numerically) exact resonances for the uncoupled case 
and the strongly coupled case $\epsilon =0.3$. In both cases the resonances form a distorted lattice in the complex energy plane. 
The quantum normal form computation of the resonances is given by
\begin{equation}
E^{(N)}_{\text{QNF},(n_1,n_2)} = \Hqnf^{(N)}(-\ui\hbar (n_1+1/2),\hbar(n_2+1/2))\,,\qquad n_1,n_2=0,1,2,\dots\,.
\end{equation}
One of the benefits of the quantum normal form is that it
leads to an assignment of the resonance lattice by quantum numbers. 
The quantum number $n_1$ labels the resonances in vertical direction, and
the quantum number $n_2$ labels the resonances in horizontal direction.
Each vertical string of resonances (i.e., sequence of resonances for fixed $n_2$) 
gives rise to one quantisation step of the cumulative reaction probability.
Note that an assigment of the resonances is very difficult to obtain only from 
the exact quantum computation. 
Figure~\ref{fig:eckartmorsecoupled}b illustrates the high accuracy of the quantum normal form computation for a selection of resonances.

\def\figeckartmorsecoupled{%
(a) Errors for the cumulative reaction probability 
in the top panel of Fig.~\ref{fig:eckartmorsecoupled_cum} for different orders $N$ of the quantum normal form.
(b) Difference  $|E^{(N)}_{\text{QNF}}-E_{\text{exact}}|$ for a selection of resonances
with quantum numbers $(n_1,n_2)$ for the resonances shown in the bottom panel of Fig.~\ref{fig:eckartmorsecoupled_cum} 
for the coupled case $\epsilon = 0.3$. 
}
\def\FIGeckartmorsecoupled{
\centerline{
\raisebox{5.2cm}{a)}\includegraphics[angle=0,width=7.5cm]{cummulative_eckart_as_morse_difference_NONL}
\raisebox{5.2cm}{b)}\includegraphics[angle=0,width=7.5cm]{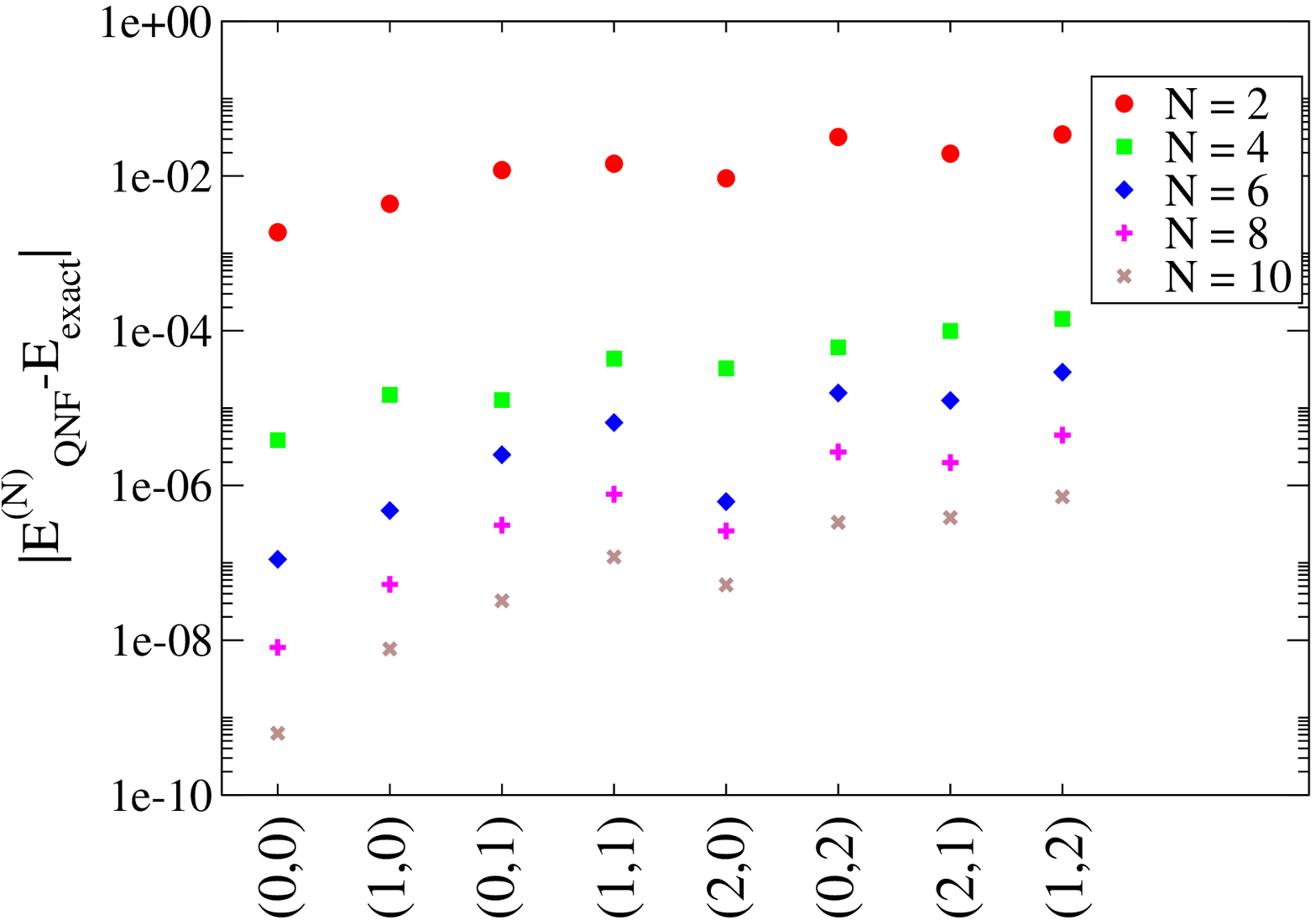}
}
}
\FIGo{fig:eckartmorsecoupled}{\figeckartmorsecoupled}{\FIGeckartmorsecoupled}

\subsection{Example with 3 \dof}
\label{sec:3Dexample}

Our final example is a 3 \dof model system consisting of an Eckart barrier in the $x$-direction that is coupled to 
Morse oscillators in the $y$-direction and in the $z$-direction. The Hamilton function is
\begin{equation} \label{eq:HEckartMorseMorse}
H = \frac{1}{2 m} \big( p_x^2+p_y^2 +p_z^2\big) + V_{\text{E}}(x) + V_{M;2}(y) + V_{M;3}(z) + \epsilon H_c\,,
\end{equation}
where $V_{\text{E}}$ is the Eckart potential from (\ref{eq:eckartpotential}) and $V_{M;k}$, $k=2,3$, are Morse potentials 
of the form \eqref{eq:Morsepot} with parameters $D_{e;k}$ and $a_{M;k}$, $k=2,3$, respectively.
For $H_c$ we choose the mutual  kinetic coupling 
\begin{equation}
H_c = p_x \, p_y\, + p_x \, p_z\, + p_y \, p_z\,  .
\end{equation}
The strength of the coupling is again controlled by the parameter $\epsilon$ in \eqref{eq:HEckartMorseMorse}.
The vector field generated by the Hamilton function  has an equilibrium point at 
$(x,y,z,p_x,p_y,p_z)=0$.
For $|\epsilon|$ sufficiently small (for given parameters of the Eckart and Morse potentials), 
the equilibrium point is of saddle-centre-centre stability type.
Figure~\ref{fig:eckartasmorsemorse} shows 
contours of the potential $V(x,y,z)= V_{\text{E}}(x) + V_{M;2}(y) + V_{M;3}(z)$ which, for energies slightly above the saddle-centre-centre equilibrium point, indicate the bottleneck-type structure of the corresponding energy surfaces. 

\def\figeckartasmorsemorse{%
Contours $V_{\text{E}}(x)+V_{M;2}(y)+V_{M;3}(z)=$const.  of the Eckart-Morse-Morse potential. 
The parameters for the Eckart potential are the same as in
Fig.~\ref{fig:eckart}. The parameters for the Morse potentials are $D_{e;2}=a_{M;2}=a_{M;3}=1$ and 
$D_{e;3}=2/3$.
}
\def\FIGeckartasmorsemorse{
\centerline{
\includegraphics[angle=0,width=8cm]{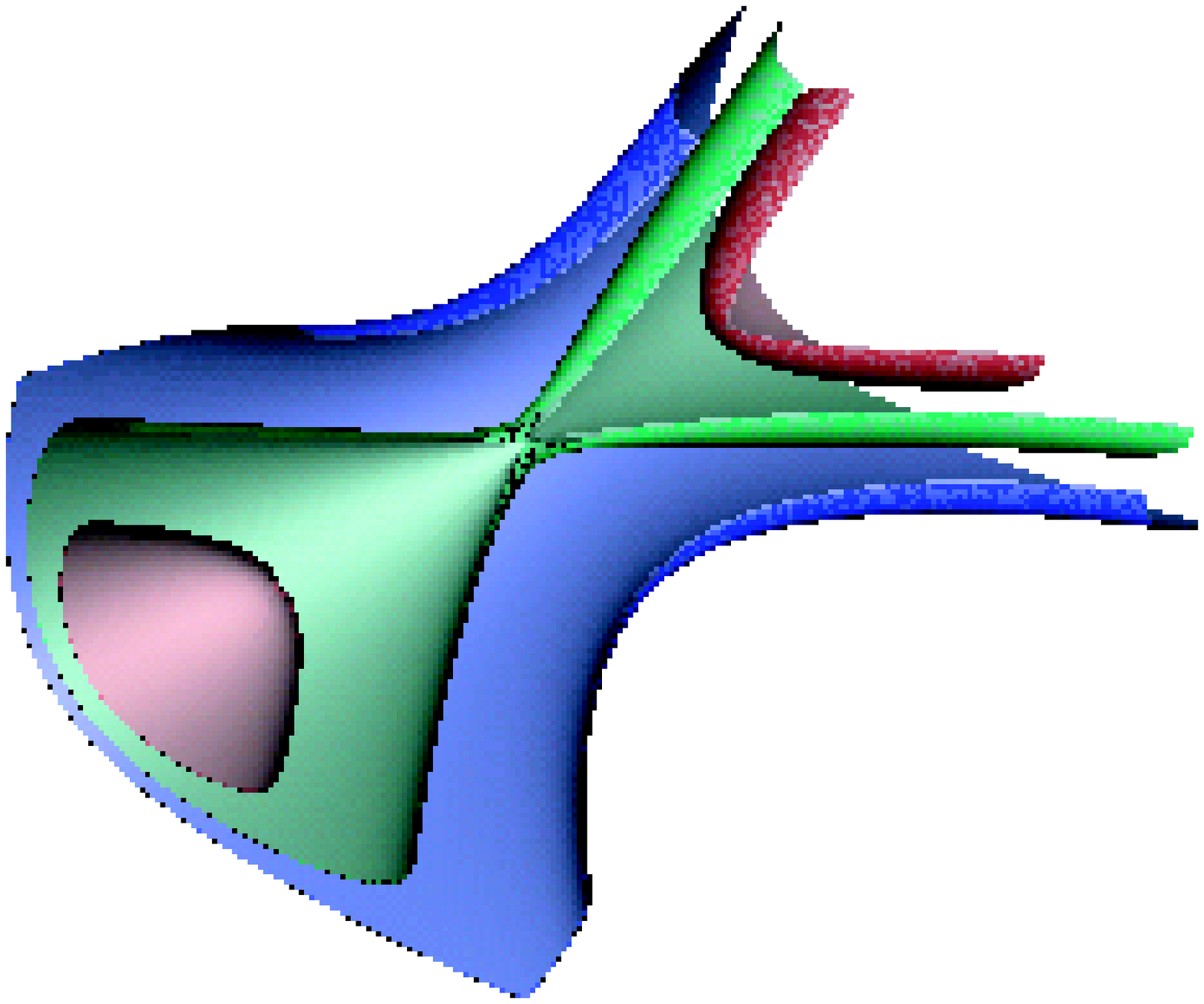}
}
}
\FIGo{fig:eckartasmorsemorse}{\figeckartasmorsemorse}{\FIGeckartasmorsemorse}

The Weyl quantisation of the Hamilton function $H$ in \eqref{eq:HEckartMorseMorse}
gives the operator
\begin{equation}
\Op[H] = -\frac{\hbar^2}{2m} \left( \frac{\partial^2}{\partial x^2} +
  \frac{\partial^2}{\partial y^2}  + \frac{\partial^2}{\partial z^2}    \right) + V_{\text{E}} + V_{\text{M;2}}  + V_{\text{M;3}}  
  - \epsilon \hbar^2  \left(  \frac{\partial^2}{\partial x \partial y}  + \frac{\partial^2}{\partial x \partial z}   + \frac{\partial^2}{\partial y \partial z}    \right) \,.
\end{equation}
The Hamilton function \eqref{eq:HEckartMorse} is the principal symbol of the
operator $\Op[H]$.

\subsubsection{Computation of the classical and quantum normal forms}

Like in Sec..~\ref{sec:2Dexample} 
the  equilibrium point is again already at the origin of the coordinate system. 
For the computation of the classical and quantum normal forms 
we therefore again start with the second step in the 
sequences~\eqref{eq:transformation-sequence} and \eqref{eq:transformation-sequence_quantum}, respectively. Following again 
Sec..~\ref{sec:examp_comp_cnf}, 
we compute the Hamiltonian matrix associated with the linearisation of Hamilton's equations 
about the equilibrium point $(x,y,z,p_x,p_y,p_z)=0$. This gives
\begin{equation}
J\,  \Hess H (0)
= 
\left(
\begin{array}{cccccc}
0 & 0 & 0 & 1/m & \epsilon & \epsilon  \\
0 & 0 & 0 & \epsilon    & 1/m & \epsilon \\
0 & 0 & 0 & \epsilon    & \epsilon & 1/m \\
m\lambda_{\text{E}}^2  & 0 & 0 & 0 & 0 & 0  \\
0 & -m \omega_{\text{M};2}^2 & 0   & 0 & 0 & 0 \\
0 & 0 & -m \omega_{\text{M};3}^2 & 0   & 0 & 0 \\
\end{array}
\right) \,,
\end{equation}
where $\lambda_{\text{E}}$ is defined in \eqref{def:lambdaeckart} and  
\begin{equation}
\omega_{\text{M};k} = \sqrt{ \frac{1}{m}  V''_{\text{M};k}(0)} =  a_{\text{M};k} \sqrt{\frac{2 D_{e;k}}{m}} \,,\quad k=2,3\,,
\end{equation}
are the linear frequencies of the Morse oscillators.
The matrix $J\,  \Hess H (0)$ has six eigenvalues, one pair of real eigenvalues of opposite signs and two pairs of imaginary eigenvalues with opposite signs.
We label them according to
\begin{equation}
e_1 = \lambda \,, \quad
e_4 = -\lambda \,, \quad
e_2 =  \ui \omega_2 \,, \quad
e_5 = -\ui \omega_2 \,, \quad
e_3 =  \ui \omega_3 \,, \quad
e_6 = -\ui \omega_3 \,,
\end{equation}
where $\lambda$, $\omega_2$ and $\omega_3$ are real positive constants that converge to
$\lambda_{\text{E}}$ and
the linear frequencies 
$\omega_{\text{M};2}$ and $\omega_{\text{M};3}$, respectively,  when
$\epsilon\rightarrow 0$\,.
We assume that the parameters $D_{e;k}$ and $a_{\text{M};k}$, $k=2,3$, are chosen such that $\omega_{2}$ and  $\omega_{3}$ are linearly independent over $\Z$.
Let us again denote the corresponding eigenvectors by $v_k$, $k=1,\dots,6$.
In order to define a real linear symplectic change of coordinates 
we use the eigenvectors $v_k$ to define the columns of a matrix $M$ according to
\begin{equation}
M = \left( c_1 v_1, c_2 \Re v_2, c_3 \Re v_3, c_1 v_4, c_2 \Im v_2,  c_3 \Im v_3 \right)
\end{equation}
with the coefficients $c_1$, $c_2$ and $c_3$ defined as
\begin{equation}
\label{eq:defd1d2d3}
c_1^{-2} := \la v_1, J v_4 \ra \,,\quad 
c_2^{-2} := \la \Re v_2, J \Im v_2 \ra \,,\quad
c_3^{-2} := \la \Re v_3, J \Im v_3 \ra\,.
\end{equation}
We choose the eigenvectors $v_1$ and $v_3$ such that  $\la v_1, J v_3\ra$ is positive 
(if $\la v_1, J v_2\ra<0$ then multiply $v_2$ by -1). As mentioned in Sec.~\ref{sec:defcompcnf} 
the coefficients $c_2^{-2}$ and $c_{3}^{-2}$ in \eqref{eq:defd1d2d3}
are automatically positive and the matrix $M$ is symplectic.
For
\begin{equation}
(q_1,q_2,q_3,p_1,p_2,p_3)^T = M^{-1} (x,y,z,p_x,p_y,p_z)^T
\end{equation}
the Hamilton function (\ref{eq:HEckartMorseMorse}) becomes
\begin{equation}\label{eq:HEckartMorseMorse2}
H = V(0) + \lambda q_1 p_1 + \frac{\omega_2}{2} \left( q_2^2+p_2^2 \right) +  
\frac{\omega_3}{2} \left( q_3^2+p_3^2 \right) + ...\,,
\end{equation}
where the neglected terms are of order greater than 2.
The constant term is
\begin{equation}
V(0) = V_{\text{E}}(0) + V_{M;2}(0)+ V_{M;3}(0) = \frac{(A+B)^2}{4B} - D_{e;2}  -  D_{e;3} \,.
\end{equation}
The truncation of  \eqref{eq:HEckartMorseMorse2} at order 2 is the symbol of the $2^\text{nd}$ order quantum normal form of 
\eqref{eq:HEckartMorseMorse}. 
The higher order classical and quantum normal forms are then computed from the algorithm described in Sections~\ref{sec:examp_comp_cnf} and \ref{sec:compscnf}. 
For the parameters $a=1$, $B=5$, $A=1/2$ for the Eckart potential and $D_{e;1}=1$, $D_{e;2}=3/2$ and 
$a_{\text{M};1}=a_{\text{M};2}=1$ for the Morse potential, $\epsilon=0.3$ for the coupling strength, 
and $m=1$,
we list the coefficients of the symbol of the $10^{\text{th}}$ order quantum normal form in Tab.~\ref{Tab:HSCNF} of the appendix.  The classical normal form is obtained from discarding terms involving a factor  $\hbar$.

\subsubsection{Classical reaction dynamics}

The NHIM is  a three-dimensional sphere, $S^3$.
In Fig.~\ref{fig:EMMnhim} we show the NHIM with the energy 0.1 above the energy of the saddle-centre-centre equilibrium point 
projected into configuration space, with the equipotential at the same energy for
reference.  Note that the projection of the NHIM to configuration space is a three-dimensional object. This can be viewed as an indication that the construction of a (in this case two-dimensional) dividing surface `in configuration space' without recrossing is not possible for a system with 3 (or more) \dof  since, as explained in detail in \cite{WaalkensWiggins04}, a dividing surface without recrossing needs to contain the NHIM (as its equator).

\def\figEMMnhim{%
The NHIM projected into configuration space. The energy
is 0.1 above the energy of the saddle-centre-centre equilibrium point.
}
\def\FIGEMMnhim{
\centerline{
\includegraphics[angle=0,width=8cm]{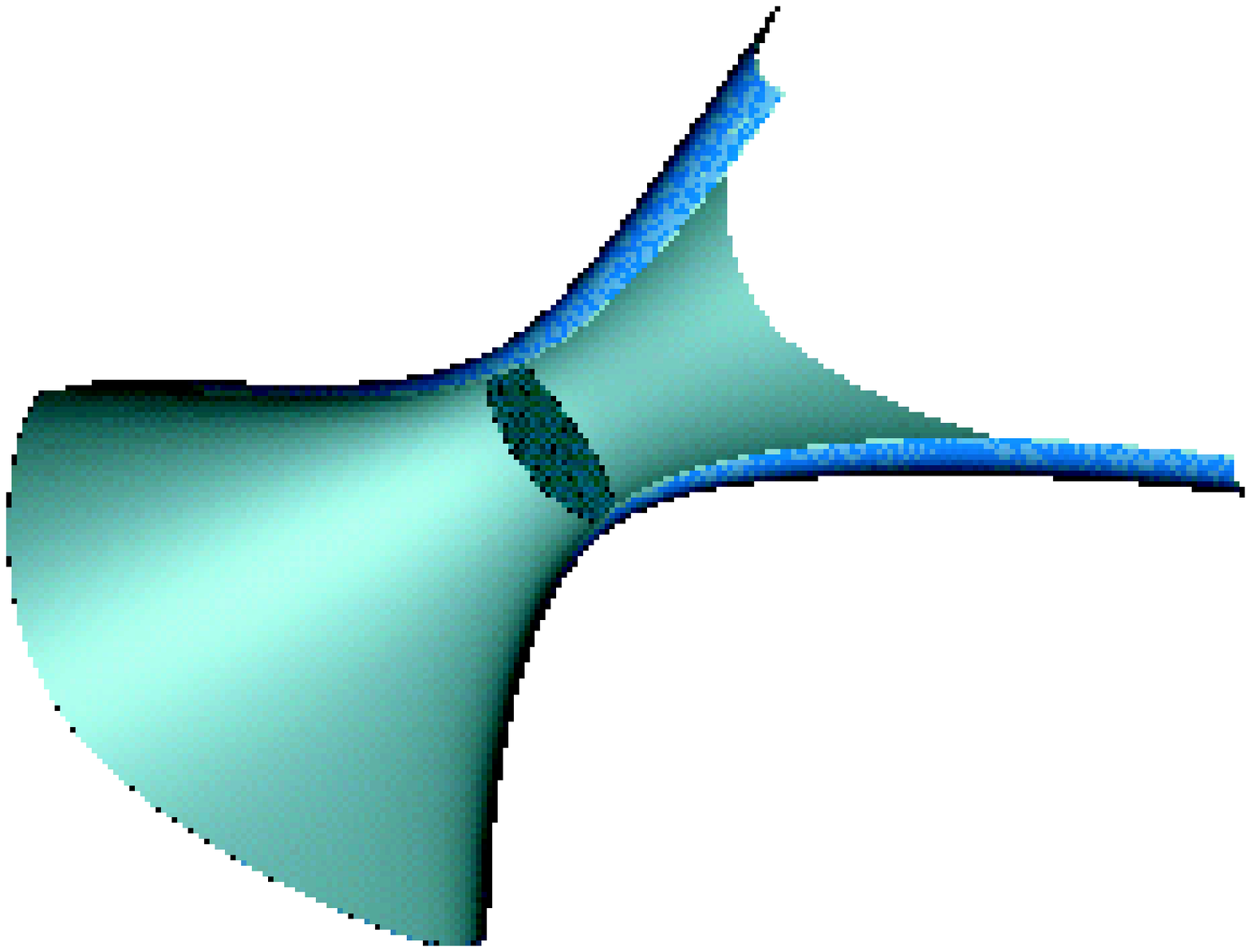}
}
}
\FIGo{fig:EMMnhim}{\figEMMnhim}{\FIGEMMnhim}

The  NHIM's  stable and unstable manifolds have the structure of spherical cylinders, $S^3\times \R$.
In Fig.~\ref{fig:EMMmanifolds} we show projections into
configuration space of local pieces of the backward branch of the
stable manifold of the NHIM, the forward branch of the stable
manifold of the NHIM, the backward branch of the unstable manifold
of the NHIM, and the forward branch of the unstable manifold of
the NHIM. Due to the time-reversal symmetry of the system
the stable and unstable manifolds
project onto each other in configuration space.
The stable and unstable manifolds  enclose the forward and backward reactive trajectories  as discussed in Sec.~\ref{sec:classical}.

\def\figEMMmanifolds{%
The stable and unstable manifolds of the NHIM projected
into configuration space. Due to the time-reversal symmetry, these manifolds project
onto each other in configuration space. The two colors represent
the forward and backward branches of the manifolds, and they are
``joined'' at the NHIM. The energy is 0.1 above the energy of the
saddle-centre-centre equilibrium point.
}
\def\FIGEMMmanifolds{
\centerline{
\includegraphics[angle=0,width=8cm]{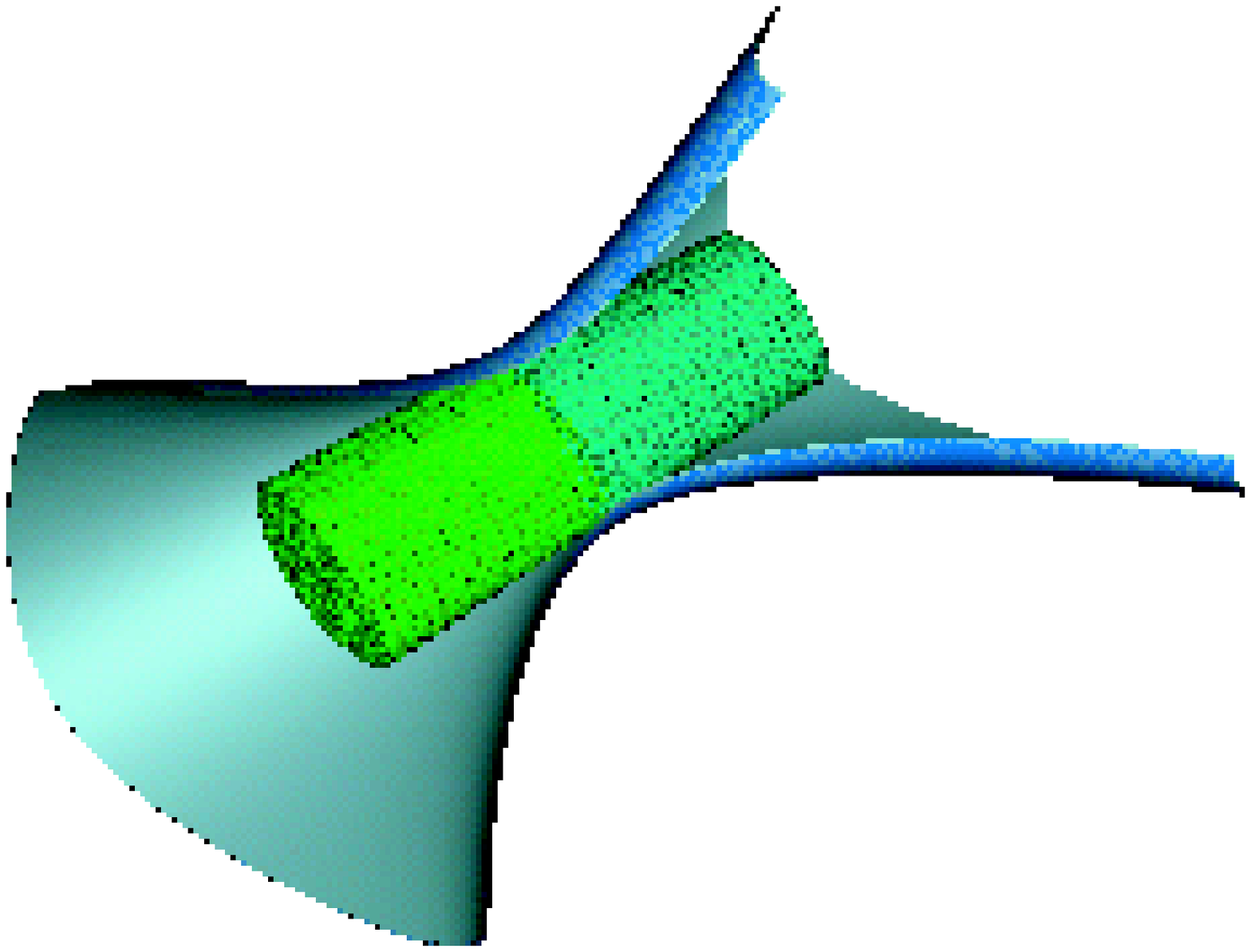}
}
}
\FIGo{fig:EMMmanifolds}{\figEMMmanifolds}{\FIGEMMmanifolds}

\def\figeckartmorsemorseesurf{%
Energy contours in the plane of the Morse oscillator actions $(J_2,J_3)$.
The morse oscillators have energies $E_y$ and $E_z$ such that $V_{\text{E}}(0)+E_y+E_z=E$ with $V_{\text{E}}(0)$ being the height of the 
one-dimensional Eckart barrier (see text).
The parameters for the potential are the same as in
Fig.~\ref{fig:eckartasmorsemorse}. The mass $m$ is 1.
}
\def\FIGeckartmorsemorseesurf{
\centerline{
  \includegraphics[angle=0,width=8cm]{morse_morse_esurf}
}
}
\FIGo{fig:eckartmorsemorseesurf}{\figeckartmorsemorseesurf}{\FIGeckartmorsemorseesurf}

The NHIM is foliated by invariant 2-tori.
According to Sec.~\ref{sec:classicalrate} the classical flux for an energy $E$ is given by 
\begin{equation}
\flux(E) = (2\pi)^2 {\cal V}(E)\,,
\end{equation}
where ${\cal V}(E)$ is the area enclosed by the energy contour in the plane of the corresponding action variables $J_2$ and $J_3$.
In the uncoupled case the 2-tori are given by the cartesian products of two circles that are contained in the 
$(y,p_y)$-plane and $(z,p_z)$-plane, 
respectively. The corresponding action variables $J_2$ and $J_3$ can be easily computed in this case. 
Let $E_y$ and $E_z$ be the energies contained in these two \dof.
Then
\begin{equation}
J_2(E_y) = \frac{1}{2\pi}\oint_{\text{p.o.}} p_y\,\ud y =  \frac{1}{a_2} (\sqrt{2mD_{e;2}}-\sqrt{-2mE_y}) \,,\qquad -D_e < E_y < 0\,,
\end{equation}
and similarly for $J_3(E_z)$.
The NHIM has energy $E=V_{\text{E}}(0)+E_y+E_z$, where $V_{\text{E}}(0)=(A+B)^2/(4B)$ is the height  of the one-dimensional Eckart barrier.
Fig.~\ref{fig:eckartmorsemorseesurf} shows some energy contours in the $(J_2,J_3)$-plane.
The fact that the energy contours are no straight lines is an indication of the strong
nonlinearity of the Morse oscillators for the energies shown.
For an energy $V_{\text{E}}(0)-D_{e;2}-D_{e;3}<E<V_{\text{E}}(0)-D_{e;3}$, the inclosed area is given by 
\begin{eqnarray}
 {\cal V}(E) &=& \int_{-D_{e;2}}^{E+D_{e;3}-V_{\text{E}}(0)} J_3(E-V_{\text{E}}(0)-E_y) \frac{\ud J_2(E_y)}{\ud E_y} \, \ud E_y   \\ \nonumber
&=& \frac{2 m \sqrt{D_{e;3}}}{a_2\,a_3} (  \sqrt{D_{e;2}} - \sqrt{V_{\text{E}}(0)-D_{e;3}-E})  \\
& & - \frac{m}{a_2\,a_3} \big(  g(E-V_{\text{E}}(0)+D_{e;3})-g(-D_{e;2}) \big)\,,
\end{eqnarray}
where 
\begin{equation}
g(E_y) := \sqrt{-E_y(E_y-E+V_{\text{E}}(0))} - \frac12 (E-V_{\text{E}}(0)) \, \arctan \bigg(\frac{E-V_{\text{E}}(0)-2E_y}{2\sqrt{-E_y(E_y-E+V_{\text{E}}(0))}}\bigg)\,.
\end{equation}
For $E\le V_{\text{E}}(0)-D_{e;2}-D_{e;3}$ the classical flux is zero.
The graph of $\Nweyl (E)= \flux(E)/(2\pi\hbar)^2$ is shown in the top panel of 
Fig.~\ref{fig:eckartmorsemorsecoupled_cum}.

\subsubsection{Quantum reaction dynamics}

In the uncoupled case the exact cumulative reaction probability $N_{\text{exact}}$ can be computed analytically. We have
\begin{equation}
N_{\text{exact}}(E) = \sum_{n_2,n_3} T_{\text{Eckart; exact}}(E-E_{\text{Morse;2},n_2}-E_{\text{Morse;3}, n_3})\,,
\end{equation}
where $ T_{\text{Eckart; exact}}$ denotes the transmission coefficient for the Eckart barrier given in \eqref{eq:TEexact} and
$E_{\text{Morse;k},n_k}$, $k=2,3$, are the energy levels of the one-dimensional Morse oscillators,
\begin{equation}
E_{\text{Morse;k},n_k} = -\frac{a_{M;k}^2 \hbar^2}{2 m} \left( n_{k}+\frac12 - 
\frac{\sqrt{2mD_{e;k}}}{a_{M;k} \hbar} \right)^2\,,\qquad n_k=0,1,2,\dots\,.
\end{equation}
The graph of $N_{\text{exact}}$ gives the oscillatory curve shown in the top panel of Fig.~\ref{fig:eckartmorsemorsecoupled_cum}.

For the quantum normal form computation of the cumulative reaction probability we get
\begin{equation}
N^{(N)}_{\text{QNF}}(E) = \sum_{n_2,n_3} \bigg[1+\exp\bigg(-2\pi\frac{I^{(N)}_{n_2,n_3}(E)}{\hbar}\bigg) \bigg]^{-1}\,,
\end{equation}
where $I^{(N)}_{(n_2,n_3)}(E)$ is obtained from inverting
\begin{equation}
\Hqnf(I^{(N)}_{(n_2,n_3)}(E),\hbar(n_2+1/2),\hbar(n_3+1/2))=E\,,\quad n_2,n_3=0,1,2,\dots\,.
\end{equation}
The high quality of the quantum normal form approximation of the cumulative reaction probability  is illustrated in 
Fig.~\ref{fig:eckartmorsemorsecoupled}a which shows $|N^{(N)}_{\text{QNF}}(E)-N_{\text{exact}}(E)|$ versus the energy $E$.

For the coupled case $\epsilon\ne 0$ we again make a comparison of the exact resonances and the resonances computed from the quantum normal form.
We again compute the (numerically) exact resonances from the
complex dilation method whose implementation is described in Sec.~\ref{sec:complscaling} of the Appendix.
The bottom panel in Fig.~\ref{fig:eckartmorsemorsecoupled_cum} shows the exact resonances for the uncoupled case 
and the strongly coupled case $\epsilon =0.3$. In both cases the resonances now form a superposition of distorted lattices. 
The quantum normal form computation of the resonances 
\begin{equation}
E^{(N)}_{\text{QNF},(n_1,n_2,n_3)} = \Hqnf^{(N)}(-\ui\hbar (n_1+1/2),\hbar(n_2+1/2),\hbar(n_3+1/2))\,,\quad n_1,n_2,n_3\in\N_0,
\end{equation}
allows one to organise the resonance by quantum numbers. 
The quantum numbers $n_1$ label the resonances in vertical direction, and
the pairs of Morse oscillator mode quantum numbers $(n_2,n_3)$ label the resonances in horizontal direction.
Here each vertical string of resonances (i.e., sequence of resonances for fixed $(n_2,n_3)$) 
gives rise to one step of the cumulative reaction probability.
In the top panel of Fig.~\ref{fig:eckartmorsemorsecoupled_cum} we mark the energies 
at which a mode $(n_2,n_3)$ opens as a transmission channel. These energies are defined in 
the same way as in Sec.~\ref{sec:2Dexample}.

Since the density of the resonances in the complex energy plane is higher for 
the 3 \dof case than it is in the 2 \dof case the
quantisation of the cumulative reaction probability is more ``washed out''.
Again note that an assigment of the resonances is very difficult to obtain only from 
the exact quantum computation. 
The resonances computed from the quantum normal form are again of a very accuracy as shown 
for a selection of resonances with quantum numbers $(n_1,n_2,n_3)$
in Fig.~\ref{fig:eckartmorsemorsecoupled}b.

\def\figeckartmorsemorsecoupled_cum{%
The top panel shows the cumulative reaction probabilities $N_{\text{exact}}(E)$ (oscillatory
curve) and $\Nweyl(E)$ (smooth curve)
for  the Eckart-Morse-Morse potential defined in the text with $\epsilon =0$.
It also shows the quantum numbers $(n_2,n_3)$ of the Morse oscillators that contribute to the quantization steps.
The bottom panel shows the resonances in the complex energy plane
marked by circles for the uncoupled case $\epsilon =0$ and by crosses for the 
strongly coupled case  $\epsilon = 0.3$.
The parameters for the Eckart potential are the same as in
Fig.~\ref{fig:eckart}. The parameters for the Morse potential are  $D_{e;2}=1$, 
$D_{e;3}=3/2$,
$a_{M;2}=1$ and $a_{M;3}=1$. Again we choose $m=1$ and $\hbar=0.1$.
}
\def\FIGeckartmorsemorsecoupled_cum{
\centerline{
  \includegraphics[angle=0,width=8cm]{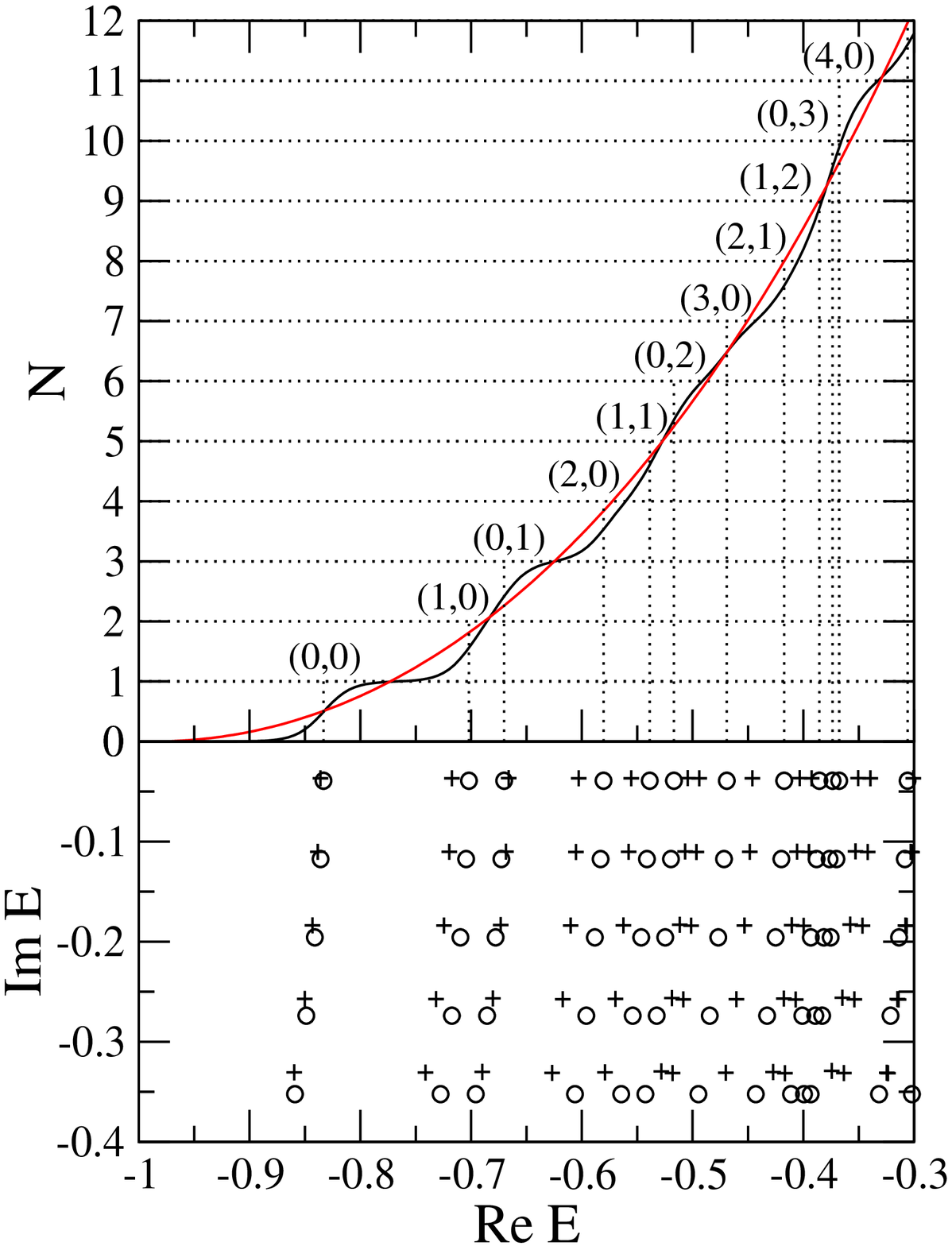}
}
}
\FIGo{fig:eckartmorsemorsecoupled_cum}{\figeckartmorsemorsecoupled_cum}{\FIGeckartmorsemorsecoupled_cum}

\def\figeckartmorsemorsecoupled{%
(a) Errors for the cumulative reaction probability in the top panel Fig.~\ref{fig:eckartmorsemorsecoupled_cum} for different orders $N$ of the quantum normal form.
(b) Errors  $|E^{(N)}_{\text{QNF}}-E_{\text{exact}}|$ for a selection of resonances with quantum numbers  $(n_1,n_2,n_3)$ 
for the coupled case $\epsilon = 0.3$ in the bottom panel of Fig.~\ref{fig:eckartmorsemorsecoupled_cum}.
}
\def\FIGeckartmorsemorsecoupled{
\centerline{
\raisebox{5.2cm}{a)}\includegraphics[angle=0,width=7.5cm]{cumulative_eckart_as_morse_morse_difference_big_NONL}
\raisebox{5.2cm}{b)}\includegraphics[angle=0,width=7.5cm]{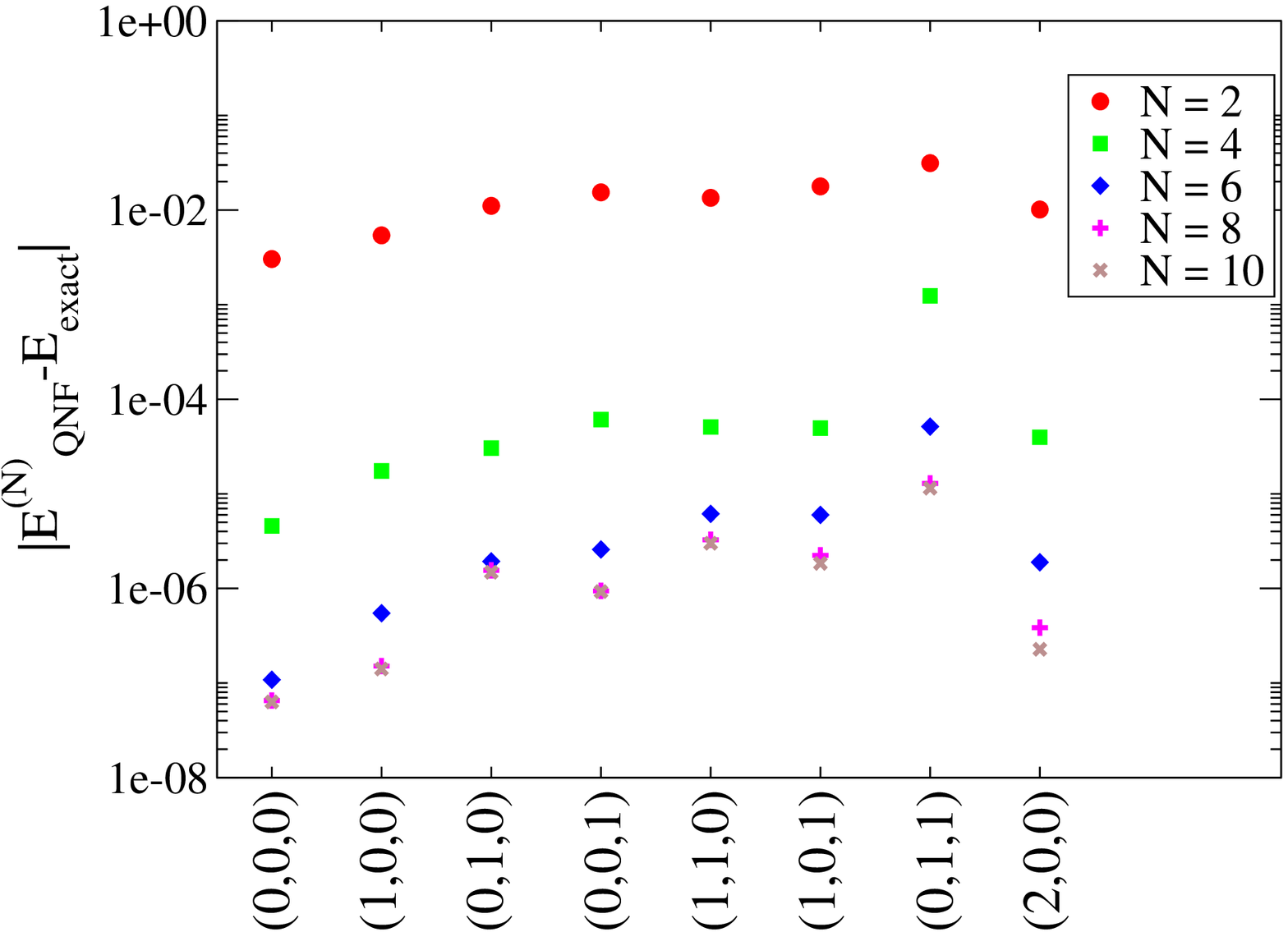}
}
}
\FIGo{fig:eckartmorsemorsecoupled}{\figeckartmorsemorsecoupled}{\FIGeckartmorsemorsecoupled}

%% file: conclusions.tex
\section{Conclusions and Outlook}
\label{sec:conclusions}


In this paper we have developed a phase space version of Wigner's
dynamical transition state theory for both classical and quantum
systems.
In the  setting of Hamiltonian classical mechanics, reaction type dynamics is induced by the presence of
a saddle-centre-$\cdots$-centre equilibrium point ('saddle' for short). 
For a fixed  energy slightly above the energy of the saddle, the energy surface has a wide-narrow-wide structure in the neighbourhood of the saddle. Trajectories must pass through this bottleneck in order to evolve from reactants to products.  
We provided a detailed study of the phase space structures which for such an energy,
exist near  the saddle and control the dynamics in the neighbourhood of the saddle. In particular we showed the existence of a dividing surface which is free of local recrossings, i.e. it  has the property that all trajectories extending from reactants to products (or vice versa) intersect this dividing surface exactly once without leaving a neighbourhood of the saddle and nonreactive trajectories 
which enter the neighbourhood from the side of reactants (resp. products) and exit the neighbourhood back to reactants (resp. products) do not intersect the dividing surface.
This dividing surface minimizes the directional flux in the sense that a (generic) deformation of the dividing surface leads to an increase of the directional flux through the dividing surface.
Such a dividing surface is a prerequisite for the computation of reaction rates from the directional flux and its construction for multi-degree-of-freedom systems was considered a major problem in transition state theory. We showed that the existence of such a dividing surface is related to the presence of a normally hyperbolic invariant manifold (NHIM) which exists near the saddle. The NHIM has the structure of a sphere of two dimensions less than the energy surface. It can be considered to form the equator of the dividing surface which itself is a sphere of one dimension less than the energy surface. This way
the NHIM divides the dividing surface into two hemispheres of which one is intersected by all trajectories evolving from reactants to products and the other is crossed by all trajectories evolving from products to reactants. 

The NHIM is the mathematical manifestation of what is referred to as activated complex in the chemistry literature.  In fact, the NHIM, which is the intersection of the centre manifold of the saddle with the energy surface of the (full) system,  can itself be viewed as the energy surface of an unstable invariant subsystem (the subsystem  given by the centre manifold). This subsystem  has one degree of freedom less than the full system and as a kind of super molecule is poised between reactants and products.  The theoretical background presented in this paper thus shows that the activated complex is not merely an heuristic concept
utilised by transition state theory, but a geometric object of a precise  significance for the dynamics.
 In particular, 
the NHIM has stable and unstable manifolds which have sufficient dimensionality to act as separatrices. They form the phase space conduits for reactions in the sense that they enclose the reactive volumes  (which consist of  trajectories evolving from reactants to products or vice versa) and separate them from the nonreactive volumes  (which consist of nonreactive trajectories). They  have  the structure of spherical cylinders (i.e., cylinders where the base is a sphere). We discussed how the centre lines of the reactive volumes enclosed by these spherical cylinders  naturally lead to the definition of a reaction path, i.e. as a kind of guiding trajectories about which other reactive trajectories rotate in phase space (observed as an oscillation when projected to configuration space) in a well defined manner.  In contrast to the usual, often heuristic definitions of a reaction path, the reaction path presented in this paper incorporates the full dynamics in a mathematically precise way.

We showed that all the phase space structures mentioned above can be realised through an efficient algorithm based on a standard Poincar{\'e}-Birkhoff normal form.  This algorithm allows one to transform the Hamilton function which describes the classical reaction dynamics to a simpler (`normal') form to any order of its Taylor expansion about the saddle point through a succession of symplectic transformations. In several examples we showed that the normal form computation truncated at a suitable order leads to a very accurate description of the dynamics near the saddle. In the generic situation where there are no resonances between the linear frequencies associated with the centre direction of the saddle the normal form is integrable and explains the regularity of the motion near the saddle which has been discovered in the chemistry literature  \cite{HindeBerry93,KomaBerry1999,Miller77}. The integrability leads to a foliation of the  neighbourhood of the saddle by invariant Lagrangian manifolds. These Lagrangian manifolds have the structure of toroidal cylinders, i.e. cylinders where the base is formed by a torus.

We showed that similar to the unfolding of the classical dynamics in the neighbourhood of a saddle point we can obtain an unfolding of the corresponding quantum dynamics.  We therefore reviewed some basic tools from the theory of micro local analysis which allow one to study properties of quantum operators in  a region of interest in the phase space of the corresponding classical system.  The main idea is to use the Weyl calculus to relate Hamilton operators to phase space functions (symbols) and vice versa. This way one can extract properties of a Hamilton operator resulting from some classical phase space region by studying 
its symbol restricted to (or `localised at')  this phase space region. In the case of reaction dynamics the region of interest is the neighbourhood of a saddle point. We showed that in the neighbourhood of a saddle  the Hamilton operator can be transformed to a simple form -- the quantum normal form -- to any order of the Taylor expansion of its symbol about the saddle by conjugating the Hamilton operator by a succession of  suitable unitary transformations. 
We showed that the quantum normal form computation can be cast into an explicit algorithm based on the Weyl calculus. This algorithm consists of two parts of which the first part takes place on the level of the symbols and is therefore very  similar in nature  to the classical normal form computation. The main difference is that the Poisson bracket involved in the symplectic transformation in the classical case is replaced by the Moyal bracket. In the second part of the quantum normal form algorithm  the symbols are quantised to obtain the corresponding quantum operators.  For this part we also developed an explicit algorithm. 

Through applications to several examples we illustrated the efficiency of the quantum normal form algorithm  for computing quantum reaction quantities like the cumulative reaction probability and quantum resonances. The cumulative reaction rate is the quantum analogue of the classical flux. Quantum resonances describe the decay of wavepackets initialised on the centre manifold. In fact, quantum mechanically, the Heisenberg uncertainty principle excludes the existence of an invariant subsystem representing the activated complex analogously to the classical case. So quantum mechanically a state initially localised on the centre manifold is unstable and will spread out. The quantum resonances  describe the lifetimes of such states.  We showed that these resonances are also related to the stepwise increase (``quantisation'') of the cumulative reaction probability as a function of energy. The dependence of the cumulative reaction probability  on the energy and also the resonances are viewed as the quantum signatures of the activated complex, and there is a huge experimental interest in these quantities \cite{SkodjeYang04}. In fact, recent advances in spectroscopic techniques allow one to study quantum scattering with unprecedented detail (see, e.g., \cite{Zare06}).  We hope that the results presented and the methods developed in this paper will contribute to the understanding and a better interpretation of such experiments.

The benefit of the quantum normal form presented in this paper is not only to give a firm theoretical framework for a quantum version of an activated complex but it moreover leads to a very efficient method for computing quantum reaction rates and the associated resonances.  
In fact, the quantum normal form computation of reaction probabilities and
resonances is highly promising since it opens the way to study
high dimensional systems for which other techniques based on the
{\em ab initio} solution of the quantum scattering problem like
the complex dilation method \cite{Simon79,Reinhardt82,Moiseyev98}
or the utilization of an absorbing potential
\cite{NeumaierMandelshtam01} do not seem feasible. We mention, that in
order to compute resonances from the complex
dilation method that are sufficiently accurate to facilitate a comparison 
with our quantum
normal form computations for the  three-degree-of-freedom example
studied in this paper
 we had to diagonalise matrices of size
$2\,500\times2\,500$, and this way we reached the limits of our numerical
computation capabilities. Furthermore, the complex dilation method
requires the ``tuning'' of the scaling angle which is not
straightforward but has to be worked out by repeating the
numerical computation for different scaling angles. In contrast to
this, the quantum normal form computation can be implemented in
a similarly transparent and efficient way as the classical normal
form.  The quantum normal form
then gives an explicit formula for the resonances from which they can
be computed directly by inserting the corresponding quantum
numbers. In particular, this leads to a direct assignment of the resonances which
one cannot obtain from the \emph{ab initio} methods mentioned
above.

We used the Weyl calculus  as a tool to systematically study several further aspects of the quantum-classical correspondence. 
One such aspect is the relation between  the quantum mechanics of reactions to the phase space structures that control classical reaction dynamics. We showed that the scattering wavefunctions are concentrated on those Lagrangian manifolds foliating the neighbourhood of a saddle 
whose toroidal base fulfill   Bohr-Sommerfeld quantisation conditions. The location of such a `quantised' Lagrangian manifold relative to the NHIM's stable and unstable manifold, i.e. the question of whether the classical trajectories on such a Lagrangian manifold are reactive or nonreactive, determines whether the scattering wavefunction corresponds to an open or a closed transmission channel. In fact, the cumulative reaction probability can be interpreted as a counting function of the number of open transmission channels (i.e., the number of quantised Lagrangian manifolds in the reactive volume of phase space) at a given energy. We showed that the Weyl approximation of this number is obtained from dividing the phase space volume of the invariant subsystem representing the activated complex enclosed by the NHIM of the given energy by elementary quantum cells, i.e. quantum cells with sidelength given by Planck's constant.    

We moreover showed that the resonance states can be viewed to be localised on Lagrangian manifolds for which in addition  to the Bohr-Sommerfeld quantisation of the toroidal base the remaining degree of freedom fulfills a complex Bohr-Sommerfeld quantisation condition. These complex Lagrangian manifolds project to the NHIM and its unstable manifolds (the direction of the decay of the resonance states) in real phase space.

Most of the theory discussed in this paper, both classically and quantum mechanically, is local in nature. 
In fact, the flux in the classical case, and the cumulative reaction probability and the associated resonances in the quantum case only require local information derived from properties of the Hamilton function or operator, respectively, in the neighbourhood of the saddle point.  This information can therefore be extracted from the classical and quantum normal form.
 Some of the classical phase space structures in the neighbourhood of the saddle where they are accurately described by the normal form are non-local in nature. This concerns the stable and unstable manifolds and the Lagrangian manifolds mentioned above. In fact they can extend to regions far away from saddle point.  This `global' information is important for the study of state specific reactivity and the control of reactions. Since these phase space structures are invariant manifolds and hence consist of trajectories they can be obtained from ``growing'' them out of the neighbourhood described by the (classical) normal form by integrating the equations of motion generated by the original Hamilton function. 
 For the classical case we used this as we mentioned in this paper  to develop an efficient procedure to determine, e.g.,  the volume of reactive initial conditions in a system.
For  the quantum case we mention the recent work by Creagh 
\cite{Creagh04,Creagh05} who developed a semiclassical theory of a reaction operator from a kind of normal form expansion about what we defined as the dynamical reaction path in this paper.
Our own future work will follow similar ideas by extending the Bohr-Sommerfeld quantised Lagrangian manifolds that carry the scattering wavefunctions to the Lagrangian structures associated with the asymptotic states of  reactants and products. The goal is to develop an efficient semiclassical procedure to compute full scattering matrices. This would not only allow one to compute state-specific reactivities but also give a clearer idea of how the quantum signatures of the activated complex are manifested in scattering experiments.

%% file: appendix.tex
\section*{Appendix}

\appendix

\section{Proof of Lemma~\ref{lem:Beals-conjug}}
\label{sec:proof-Beals-conjug}

We here provide a short sketch of the proof of Lemma~\ref{lem:Beals-conjug}.

\begin{proof}
Let  $A\in \cS_{\hbar}(\R^d\times\R^d)$
and $A'$ be the symbol of $\ue^{\frac{\ui}{\hbar}
\Op[W]}\Op[A]\ue^{-\frac{\ui}{\hbar} \Op[W]} $ with $W\in \cWl^s$.
We need to show that $A'\in
\cS_{\hbar}(\R^d\times\R^d)$. 
To this end define for $s\geq 0$,
\begin{equation}
\ccH^{s}(\R^d):=\{ \psi\in L^{2}(\R^d)\,:\,\Op[B]\psi\in
L^{2}(\R^d)\,\, \forall B\in\cW^{s'}\, \text{with}\,  0\leq s'\leq
s\}\,\, ,
\end{equation}
and let $\ccH^{-s}(\R^d)$ denote the dual of $\ccH^{s}(\R^d)$. 
Then a variant of the usual Beals characterisation of pseudo-differential
 operators 
gives that $A\in\cS_{\hbar}(\R^d\times\R^d)$ if and only if for all $s,s'\in \Z$,
\begin{equation}\label{eq:smoothing}
\Op[A]:\ccH^{s'}(\R^d)\to \ccH^{s}(\R^d)\,.
\end{equation}
This follows because \eqref{eq:smoothing} implies that for any $B_j\in \cW^{s_j}$, $j=1, \cdots ,N$  with $s_j\geq 0$, we have 
\begin{equation}
\big[\Op[B_N],\big[\Op[B_{N-1}], \cdots ,\big[\Op[B_1],\Op[A]\big]\big]\big] :L^2(\R^d)\to L^2(\R^d)\,\, ,
\end{equation} 
and this implies that $A\in\cS_{\hbar}(\R^d\times\R^d)$ (see \cite{DimSjo99}). 

For a real valued $W\in\cWl^s$, we
define $\widehat{U}(\flowparam):=\ue^{-\frac{\ui}{\hbar} \flowparam\Op[W]}$.
Then $\widehat{U}(\flowparam):L^2(\R^d)\to L^2(\R^d)$ since $\widehat{U}(\flowparam)$
is unitary. Moreover, we have that 
\begin{equation}\label{eq:sob-scale}
\widehat{U}(\flowparam):\ccH^{s}(\R^d)\to \ccH^{s}(\R^d)
\end{equation}
for all $s$. To see this, let $\psi\in \ccH^{s}(\R^d)$. Then we have
to show that $\Op[B] \widehat{U}(\flowparam)\psi\in L^2(\R^d)$ for all $B\in
\cW^{s'}$ with $s'\leq s$. But
$\Op[B]\widehat{U}(\flowparam)=
\widehat{U}(\flowparam)\widehat{U}(-\flowparam)\Op[B]\widehat{U}(\flowparam)$ and
\begin{equation}
\begin{split}
\widehat{U}(-\flowparam)\Op[B]\widehat{U}(\flowparam)-\Op[B]&=\int_0^\flowparam \frac{\ud}{\ud
\flowparam'} \bigg( \widehat{U}(-\flowparam')\Op[B]\widehat{U}(\flowparam') \bigg) \,\, \ud \flowparam' \\
&=\int_0^\flowparam
\widehat{U}(-\flowparam')\frac{\ui}{\hbar}[\Op[W],\Op[B]]\widehat{U}(\flowparam')\,\, \ud
\flowparam'\,.
\end{split}
\end{equation}
Hence 
\begin{equation}
\Op[B]\widehat{U}(\flowparam)=\widehat{U}(\flowparam)\bigg(\Op[B]+\int_0^\flowparam
\widehat{U}(-\flowparam')\frac{\ui}{\hbar}[\Op[W],\Op[B]]\widehat{U}(\flowparam')\,\, \ud
\flowparam'\bigg)\,\, .
\end{equation}
Since $W$ is localised, the commutator $\frac{\ui}{\hbar}[\Op[W],\Op[B]]$ is a bounded operator,
and therefore $\Op[B]\widehat{U}(\flowparam)\psi\in L^2(\R^d)$. 

By
\eqref{eq:sob-scale} we see then that if $\Op[A]$ satisfies
\eqref{eq:smoothing} then $\widehat{U}(-\flowparam)\Op[A]\widehat{U}(\flowparam)$
satisfies \eqref{eq:smoothing}, too, and therefore $A'\in
\cS_{\hbar}(\R^d\times\R^d)$.
\end{proof}


\section{Symbols of the quantum normal forms of the systems studied in Section~\ref{sec:examples}}
\label{sec:appendixcoefficients}

\begin{table}[H]
\caption{Nonvanishing coefficients of the symbol 
$H_{\text{QNF}}^{(10)}(\hbar,x,\xi)=\sum_{\alpha+\beta+2\gamma\le10}
h_{(\alpha,\beta,\gamma)}x^{\alpha} \xi^{\beta} \hbar^\gamma$
of the $10^{\text{th}}$ order quantum normal form
of the one \dof Eckart barrier with the potential~\eqref{eq:eckartpotential}
studied in Sec.~\ref{sec:1Dexample}. 
Recall that the nonvanishing terms in the normal form have $\alpha=\beta$.
}  \label{Tab:HSCNF1D}
\begin{center}
\begin{tabular}{|c|c|r||c|c|r|}\hline
$\alpha$    & $\gamma$ & $h_{(\alpha,\beta,\gamma)}$\hspace*{1.5cm} & 
$\alpha$    & $\gamma$ & $h_{(\alpha,\beta,\gamma)}$ \hspace*{1.5cm} \\ \hline

0 & 0&$    1.512 \,500 \,000 \,000 \,000 \,000 $ &4 & 0&$    0.000 \,625 \,000 \,000 \,000 \,000 $\\
1 & 0&$    0.782 \,663 \,720 \,891 \,674 \,056 $ &2 & 2&$    0.002 \,375 \,000 \,000 \,000 \,005 $\\
2 & 0&$    0.128 \,750 \,000 \,000 \,000 \,027 $ &0 & 4&$    0.000 \,250 \,000 \,000 \,000 \,000 $\\
0 & 2&$    0.001 \,250 \,000 \,000 \,000 \,000 $ &5 & 0&$   -0.000 \,237 \,170 \,824 \,512 \,630 $\\
3 & 0&$   -0.001 \,581 \,138 \,830 \,084 \,187 $ &3 & 2&$   -0.001 \,877 \,602 \,360 \,724 \,986 $\\
1 & 2&$   -0.012 \,155 \,004 \,756 \,272 \,212 $ &1 & 4&$   -0.001 \,098 \,767 \,960 \,437 \,415 $\\

\hline
\end{tabular}
\end{center}
\end{table}

\begin{table}[H]
\caption{Nonvanishing coefficients of the symbol 
$H_{\text{QNF}}^{(10)}=\sum_{|\alpha|+|\beta|+2\gamma\le10}
h_{(\alpha,\beta,\gamma)}x_1^{\alpha_1}x_2^{\alpha_2}\xi_1^{\beta_1}\xi_2^{\beta_2}
\hbar^\gamma $
of the $10^{\text{th}}$ order quantum normal form
of the coupled two \dof Eckart-Morse system defined in 
Equation~\eqref{eq:HEckartMorse}  in
Sec.~\ref{sec:2Dexample}. 
Recall that the nonvanishing terms in the normal form have $\alpha=\beta$.
}  \label{Tab:HSCNF2D}
\begin{center}
\begin{tabular}{|c|c|c|r||c|c|c|r|}\hline
$\alpha_1$  & $\alpha_2$   & $\gamma$ & $h_{(\alpha,\beta,\gamma)}$\hspace*{1.5cm} & 
$\alpha_1$  & $\alpha_2$   & $\gamma$ & $h_{(\alpha,\beta,\gamma)}$ \hspace*{1.5cm} \\ \hline

0 & 0 & 0 &$  0.512 \,500 \,000 \,000 \,000 \,000           $&1 & 3 & 0 &$       -\ui\,0.011 \,273 \,157 \,211 \,934 \,359   $ \\   
1 & 0 & 0 &$  0.754 \,753 \,936 \,565 \,858 \,878       $&1 & 1 & 2 &$     -\ui\,0.011 \,172 \,831 \,518 \,205 \,997  $ \\
0 & 1 & 0 &$      \ui\,1.398 \,960 \,687 \,353 \,887 \,473 $&0 & 4 & 0 &$  0.002 \,732 \,350 \,157 \,899 \,168  $ \\
2 & 0 & 0 &$  0.123 \,785 \,339 \,782 \,523 \,858              $&0 & 2 & 2 &$  0.007 \,186 \,254 \,008 \,569 \,981 $ \\  
1 & 1 & 0 &$     -\ui\,0.001 \,065 \,319 \,634 \,986 \,676  $&0 & 0 & 4 &$  0.000 \,687 \,695 \,639 \,095 \,786  $ \\
0 & 2 & 0 &$  0.502 \,213 \,521 \,058 \,802 \,562       $&5 & 0 & 0 &$ -0.000 \,214 \,239 \,042 \,469 \,975  $ \\
0 & 0 & 2 &$  0.125 \,449 \,608 \,038 \,641 \,072       $&4 & 1 & 0 &$      -\ui\,0.000 \,985 \,595 \,001 \,405 \,555  $ \\
3 & 0 & 0 &$  0.000 \,021 \,351 \,350 \,002 \,054       $&3 & 2 & 0 &$  0.003 \,423 \,967 \,215 \,023 \,733   $ \\
2 & 1 & 0 &$       \ui\,0.008 \,176 \,183 \,587 \,983 \,269 $&3 & 0 & 2 &$ -0.000 \,782 \,323 \,582 \,246 \,664   $ \\
1 & 2 & 0 &$ -0.013 \,717 \,963 \,053 \,750 \,142     $&2 & 3 & 0 &$         \ui\,0.001 \,688 \,243 \,381 \,394 \,164  $ \\
1 & 0 & 2 &$ -0.014 \,142 \,331 \,760 \,119 \,375       $&2 & 1 & 2 &$         \ui\,0.000 \,302 \,145 \,176 \,622 \,814  $ \\
0 & 3 & 0 &$       -\ui\,0.002 \,237 \,031 \,129 \,850 \,027 $&1 & 4 & 0 &$  0.003 \,334 \,954 \,065 \,262 \,960  $ \\
0 & 1 & 2 &$    -\ui\,0.002 \,154 \,732 \,890 \,857 \,193 $&1 & 2 & 2 &$  0.011 \,718 \,284 \,130 \,545 \,851       $ \\
4 & 0 & 0 &$  0.000 \,388 \,266 \,134 \,708 \,556     $&1 & 0 & 4 &$  0.000 \,106 \,782 \,020 \,240 \,749      $ \\
3 & 1 & 0 &$    \ui\,0.001 \,167 \,305 \,695 \,975 \,092 $&0 & 5 & 0 &$        \ui\,0.001 \,836 \,329 \,386 \,792 \,953  $ \\
2 & 2 & 0 &$ -0.007 \,789 \,574 \,828 \,129 \,416       $&0 & 3 & 2 &$          \ui\,0.011 \,444 \,002 \,354 \,002 \,782  $ \\
2 & 0 & 2 &$  0.000 \,318 \,492 \,327 \,523 \,421      $&0 & 1 & 4 &$          \ui\,0.004 \,314 \,246 \,915 \,341 \,055  $ \\

\hline
\end{tabular}
\end{center}
\end{table}

\begin{table}[H]
\caption{Nonvanishing coefficients of the symbol
$H_{\text{QNF}}^{(10)}=\sum_{|\alpha|+|\beta|+2\gamma\le10} h_{(\alpha,\beta,\gamma)}x_1^{\alpha_1}x_2^{\alpha_2}x_3^{\alpha_3}\xi_1^{\beta_1}\xi_2^{\beta_2}\xi_3^{\beta_3} \hbar^\gamma $
of the coupled 3 \dof Eckart-Morse-Morse system defined in Equation~\eqref{eq:HEckartMorseMorse}  in
Sec.~\ref{sec:3Dexample}. 
Recall that the nonvanishing terms in the normal form have $\alpha=\beta$.
}  \label{Tab:HSCNF}
\begin{center}
\begin{tabular}{|c|c|c|c|r||c|c|c|c|r|}\hline
$\alpha_1$  & $\alpha_2$  &$\alpha_3$ & $\gamma$ & $h_{(\alpha,\beta,\gamma)}$\hspace*{1.5cm} & 
$\alpha_1$  & $\alpha_2$  &$\alpha_3$ & $\gamma$ & $h_{(\alpha,\beta,\gamma)}$ \hspace*{1.5cm} \\ \hline

0  & 0  & 0 &  0 &$     -0.987 \, 500 \, 000 \, 000 \,000 \,000 $   &0  & 2  & 0 & 2 &$      0.152 \, 783 \, 733 \, 769 \,442 \,116 $  \\
1  & 0  & 0 &  0 &$      0.734 \, 955 \, 236 \, 108 \,148 \,115 $   &0  & 1  & 3 & 0 & $     0.310 \, 986 \, 515 \, 383 \,694 \,741  $ \\
0  & 1  & 0 &  0 &$  \ui\, 1.822 \,517 \,936 \,036 \,739 \,209    $   &0  & 1  & 1 & 2 &    $ -4.151 \, 328 \, 593 \, 608 \,719 \,646   $ \\
0  & 0  & 1 &  0 &$  \ui\, 1.267 \,290 \,444 \,967 \,990 \,459    $   &0  & 0  & 4 & 0 &   $   0.006 \, 137 \, 865 \, 049 \,515 \,079  $ \\
2  & 0  & 0 &  0 &$      0.118 \, 038 \, 678 \, 383 \,844 \,813 $   &0  & 0  & 2 & 2 & $     0.859 \, 423 \, 987 \, 882 \,411 \,768 $ \\
1  & 1  & 0 &  0 &$ -\ui\, 0.012 \,334 \,879 \,342 \,872 \,699    $   &0  & 0  & 0 & 4 &  $   -0.265 \, 855 \, 011 \, 175 \,839 \,773    $ \\
1  & 0  & 1 &  0 &$  \ui\, 0.005 \,310 \,192 \,075 \,685 \,135    $   &5  & 0  & 0 & 0 & $ -0.000 \, 210 \, 376 \, 032 \, 140 \,816    $ \\
0  & 2  & 0 &  0 &$      0.393 \, 832 \, 730 \, 618 \,103 \,493 $   &4  & 1  & 0 & 0 &$ -\ui\, 0.000 \,284 \,795 \,393 \,758 \,395 $ \\
0  & 1  & 1 &  0 &$      0.909 \, 582 \, 776 \, 314 \,433 \,320 $   &4  & 0  & 1 & 0 & $-\ui\, 0.000 \,337 \,276 \,968 \,652 \,946   $ \\
0  & 0  & 2 &  0 &$      0.173 \, 096 \, 436 \, 125 \,076 \,552 $   &3  & 2  & 0 & 0 & $    -0.000 \, 627 \, 685 \, 605 \,556 \,083 $ \\
0  & 0  & 0 &  2 &$      0.266 \, 664 \, 869 \, 446 \,484 \,871 $   &3  & 1  & 1 & 0 & $     0.003 \, 281 \, 135 \, 664 \,719 \,332  $ \\
3  & 0  & 0 &  0 &$      0.000 \, 552 \, 036 \, 804 \,498 \,563 $   &3  & 0  & 2 & 0 & $     0.000 \, 026 \, 178 \, 720 \,039 \,055  $ \\ 
2  & 1  & 0 &  0 &$  \ui\, 0.002 \,430 \,126 \,450 \,332 \,083    $   &3  & 0  & 0 & 2 &   $  -0.000 \, 809 \, 539 \, 163 \,262 \,948 $ \\
2  & 0  & 1 &  0 &$  \ui\, 0.004 \,886 \,339 \,438 \,884 \,285    $   &2  & 3  & 0 & 0 &$ -\ui\, 0.001 \,666 \,813 \,854 \,950 \,104 $ \\
1  & 2  & 0 &  0 &$     -0.000 \, 569 \, 612 \, 518 \,570 \,350 $   &2  & 2  & 1 & 0 &$ -\ui\, 0.011 \,060 \,027 \,951 \,060 \,662 $ \\
1  & 1  & 1 &  0 &$     -0.039 \, 861 \, 920 \, 250 \,395 \,527 $   &2  & 1  & 2 & 0 &$  \ui\, 0.021 \,558 \,200 \,542 \,697 \,081 $ \\
1  & 0  & 2 &  0 &$      0.005 \, 117 \, 262 \, 453 \,168 \,276 $   &2  & 1  & 0 & 2 &$ -\ui\, 0.001 \,561 \,080 \,497 \,427 \,406 $ \\
1  & 0  & 0 &  2 &$     -0.015 \, 343 \, 995 \, 286 \,930 \,709 $   &2  & 0  & 3 & 0 &$ -\ui\, 0.004 \,089 \,992 \,505 \,729 \,545 $ \\
0  & 3  & 0 &  0 &$ -\ui\, 0.063 \,077 \,949 \,720 \,773 \,535    $   &2  & 0  & 1 & 2 &$ -\ui\, 0.000 \,973 \,333 \,949 \,567 \,230 $ \\
0  & 2  & 1 &  0 &$  \ui\, 0.851 \,786 \,534 \,413 \,891 \,081    $   &1  & 4  & 0 & 0 &$      0.002 \, 350 \, 577 \, 380 \,299 \,191 $ \\   
0  & 1  & 2 &  0 &$ -\ui\, 1.430 \,298 \,863 \,449 \,648 \,912    $   &1  & 3  & 1 & 0 &$      0.165 \, 841 \, 199 \, 916 \,935 \,531 $ \\  
0  & 1  & 0 &  2 &$ -\ui\, 0.085 \,082 \,314 \,838 \,682 \,922    $   &1  & 2  & 2 & 0 &$      0.544 \, 009 \, 061 \, 075 \,099 \,235  $ \\
0  & 0  & 3 &  0 &$  \ui\, 0.243 \,714 \,959 \,199 \,355 \,560    $   &1  & 2  & 0 & 2 &$     -0.126 \, 519 \, 607 \, 942 \,211 \,641 $ \\
0  & 0  & 1 &  2 &$  \ui\, 0.066 \,628 \,105 \,873 \,760 \,135    $   &1  & 1  & 3 & 0 &$     -0.182 \, 508 \, 361 \, 502 \,375 \,351   $ \\ 
4  & 0  & 0 &  0 &$      0.000 \, 459 \, 055 \, 390 \,142 \,951 $   &1  & 1  & 1 & 2 & $     2.075 \, 443 \, 875 \, 730 \,594 \,886 $ \\ 
3  & 1  & 0 &  0 &$  \ui\, 0.002 \,154 \,242 \,685 \,324 \,458    $   &1  & 0  & 4 & 0 &  $   -0.016 \, 321 \, 552 \, 385 \,758 \,725   $ \\
3  & 0  & 1 &  0 &$  \ui\, 0.000 \,532 \,155 \,590 \,347 \,800    $   &1  & 0  & 2 & 2 &  $   -0.493 \, 860 \, 447 \, 549 \,943 \,396 $ \\
2  & 2  & 0 &  0 &$     -0.004 \, 840 \, 046 \, 450 \,845 \,588 $   &1  & 0  & 0 & 4 &$      0.100 \, 323 \, 914 \, 200 \,990 \,955$ \\ 
2  & 1  & 1 &  0 &$     -0.008 \, 696 \, 277 \, 962 \,945 \,819 $   &0  & 5  & 0 & 0 &$ -\ui\, 0.035 \,437 \,158 \,103 \,964 \,192 $ \\
2  & 0  & 2 &  0 &$     -0.001 \, 521 \, 249 \, 367 \,386 \,827 $   &0  & 4  & 1 & 0 &$ -\ui\, 1.098 \,730 \,518 \,769 \,317 \,535 $ \\
2  & 0  & 0 &  2 &$     -0.000 \, 729 \, 166 \, 304 \,792 \,555 $   &0  & 3  & 2 & 0 &$ -\ui\,16.346 \,415 \,113 \,011 \,525 \,772$ \\
1  & 3  & 0 &  0 &$ -\ui\, 0.005 \,625 \,488 \,538 \,854 \,559    $   &0  & 3  & 0 & 2 &$  \ui\, 3.112 \,191 \,195 \,399 \,140 \,660 $ \\
1  & 2  & 1 &  0 &$ -\ui\, 0.042 \,200 \,218 \,044 \,352 \,362    $   &0  & 2  & 3 & 0 & $ \ui\,25.261 \,986 \,404 \,397 \,997 \,857$ \\
1  & 1  & 2 &  0 &$  \ui\, 0.035 \,856 \,221 \,513 \,981 \,255    $   &0  & 2  & 1 & 2 & $-\ui\,79.663 \,024 \,974 \,498 \,432 \,059$ \\
1  & 1  & 0 &  2 &$ -\ui\, 0.019 \,385 \,123 \,066 \,852 \,090    $   &0  & 1  & 4 & 0 & $-\ui\, 3.564 \,332 \,805 \,428 \,819 \,594 $ \\
1  & 0  & 3 &  0 &$ -\ui\, 0.005 \,485 \,448 \,552 \,764 \,268    $   &0  & 1  & 2 & 2 & $ \ui\,91.718 \,400 \,582 \,446 \,722 \,291 $ \\
1  & 0  & 1 &  2 &$  \ui\, 0.005 \,553 \,350 \,862 \,328 \,742    $   &0  & 1  & 0 & 4 & $-\ui\, 7.653 \,275 \,405 \,441 \,236 \,619 $ \\
0  & 4  & 0 &  0 &$     -0.022 \, 779 \, 283 \, 170 \,516 \,708 $   &0  & 0  & 5 & 0 &$ -\ui\, 0.071 \,898 \,162 \,267 \,093 \,398 $ \\ 
0  & 3  & 1 &  0 &$     -0.382 \, 813 \, 075 \, 268 \,433 \,553 $   &0  & 0  & 3 & 2 &$ -\ui\, 8.612 \,377 \,204 \,404 \,908 \,782 $ \\
0  & 2  & 2 &  0 &$     -0.852 \, 347 \, 953 \, 691 \,774 \,933 $   &0  & 0  & 1 & 4 &$  \ui\, 6.544 \,597 \,476 \,333 \,204 \,031 $ \\

\hline
\end{tabular}
\end{center}
\end{table}


\section{Computation of Quantum Resonances from the Complex Dilation Method}
\label{sec:complscaling}

We here provide some details on the {\em complex dilation method} \cite{Simon79,Reinhardt82,Moiseyev98} 
that we used to numerically compute
the quantum resonances of the 2 \dof coupled Eckart-Morse system in Sec.~\ref{sec:2Dexample} 
and the 3 \dof coupled Eckart-Morse-Morse system in Sec.~\ref{sec:3Dexample}. 
We illustrate the method for the 2 \dof system. The generalisation to 3 \dof is straightforward. 

Let $\widehat{H}=\Op[H]$ be the Weyl quantisation of the Hamilton function $H$ defined in \eqref{eq:HEckartMorse}.
For an angle $\alpha\ge0$,
we define the scaled operator $\widehat{H}^\alpha$ that is obtained from the
operator $\widehat{H}$ by substituting 
for the coordinate $x$ the scaled coordinate $\exp(\ui\alpha)x$, i.e.,
\begin{equation}
\widehat{H}^\alpha = -\frac{\hbar^2}{2m} \big( \ue^{-2\ui\alpha} \frac{\partial^2}{\partial x^2} + \frac{\partial^2}{\partial y^2}\big)
+ V_{\text{E}}( \ue^{\ui\alpha}x ) + V_{\text{M}}(y) - \epsilon \hbar^2 \ue^{-\ui\alpha} \frac{\partial^2}{\partial x \partial y}\,.
\end{equation}
For $\alpha\ne 0$ this operator is no longer Hermitian. 
The effect of the complex scaling is that, for  suitable $\alpha>0$, the generalised eigenfunctions of 
$\widehat{H}$ that correspond to resonances become square-integrable after the substitution 
$x\mapsto \exp(\ui\alpha)x$, i.e., they become genuine elements of the Hilbert space $L^2(\R^2)$.
The resonances are then given by 
the eigenvalues of the operator $\widehat{H}^\alpha$ which can be computed from a standard variational principle
using a finite matrix representation 
in which $\widehat{H}^\alpha$ is expanded in terms of some truncated basis set.  

We
choose the basis set given by 
the product states
$|n_{\text{dv}},n_{\text{M}}\rangle:=|n_{\text{dv}}\rangle \otimes
|n_{\text{M}}\rangle $, where, using the Dirac notation,  the states $|n_{\text{dv}}\rangle$ and
$|n_{\text{M}}\rangle$ with quantum numbers
$n_{\text{dv}}$ and $n_{\text{M}}$ form  1D basis states in the directions of $x$ and $y$, respectively. 
For the  $y$-direction, we choose the eigenstates that correspond to the discrete part
of the spectrum of the 1D Morse
oscillator $\widehat{H}_{\text{M}} := -(\hbar^2/2m)\pa_y^2 + V_\text{M}(y)$.
The quantum number $n_{\text{M}}$ then runs from $0$ to $n_{\text{M\,max}}-1$, where
\begin{equation}
n_{\text{M\,max}} = [\frac{\sqrt{2m D_e}}{a_{\text{M}} \hbar}+\frac12]
\end{equation}
is the number of bound states of the 1D Morse oscillator.  
The matrix with elements $\langle n_{\text{M}} | \widehat{H}_{\text{M}} | n'_{\text{M}} \rangle$
    is then
diagonal with the Morse oscillator energies on the diagonal, i.e., 
\begin{equation}
\langle n_{\text{M}}| \widehat{H}_{\text{M}} |n'_{\text{M}} \rangle =
E_{\text{M}}(n_{\text{M}}) \delta_{n_{\text{M}}\, n'_{\text{M}}}\,,\quad 
E_{\text{M}}(n_{\text{M}}) = -\frac{a_{\text{M}}^2 \hbar^2}{2 m} \left( n_{\text{M}}+\frac12 - \frac{\sqrt{2mD_e}}{a_{\text{M}} \hbar} \right)^2\,.
\end{equation}
In order to compute the matrix elements of $\widehat{H}^\alpha$ with respect
to the product states we also need
the matrix elements $\langle n_{\text{M}}| \widehat{p}_y |n'_{\text{M}} \rangle$,
where $\widehat{p}_y$ is the momentum operator 
$\widehat{p}_y=-\ui \hbar \pa_y$. For the elements above the diagonal, we get (see, e.g., \cite{Spirkoetal85})
\begin{equation}
\langle n_{\text{M}}| \widehat{p}_y |n'_{\text{M}} \rangle = 
(-1)^{n_{\text{M}}-n'_{\text{M}}-1} \ui \left( \frac{b_{n_{\text{M}}} b_{n'_{\text{M}}} n_{\text{M}}! \Gamma(2\beta-n_{\text{M}})}{2\beta^2 n'_{\text{M}}! \Gamma(2\beta-n'_{\text{M}})} m D \right)^{1/2}\,, \qquad n_{\text{M}}>n'_{\text{M}}\,,
\end{equation}
where
\begin{equation}
b_{n_{\text{M}}} = 2\beta -2n_{\text{M}} -1 \,, \quad \beta = \frac{\sqrt{2mD_e}}{a_{\text{M}} \hbar}\,.
\end{equation}
The  diagonal elements vanish and the elements below the diagonal can be obtained from the elements above the diagonal,
\begin{equation}
\langle n_{\text{M}}| \widehat{p}_y |n_{\text{M}} \rangle = 0\,,\quad
\langle n'_{\text{M}}| \widehat{p}_y |n_{\text{M}} \rangle = - \langle n_{\text{M}} |\widehat{p}_y |n'_{\text{M}} \rangle\,.
\end{equation} 

In $x$-direction  we choose a so called {\em discrete value representation} \cite{LightHamiltonLill85}
which consists of a basis set $|n_{\text{dv}}\rangle$, $n_{\text{dv}}\in\Z$, for which the 
wave functions $\langle x | n_{\text{dv}} \rangle$ are localised in space on a discrete grid. 
Concretely, we choose the ``sinc'' functions
\begin{equation}
\langle x| n_{\text{dv}}  \rangle= \sqrt{\Delta x} \frac{\sin \big( \frac{\pi}{\Delta x} (x-n_{\text{dv}}  \Delta x) \big) }{\pi ( x-n_{\text{dv}}  \Delta x ) }\,,
\end{equation}
where $\Delta x$ is a positive constant (the grid spacing). The states $|n_{\text{dv}}\rangle$ are normalised and orthogonal.
The matrix elements of the kinetic energy operator $\widehat{p}_x^2/(2 m)$ are easily worked out to give
\begin{equation}
\langle n_{\text{dv}} |\frac{\widehat{p}_x^2}{2 m}| n'_{\text{dv}} \rangle = 
\left\{ 
\begin{array}{cc}
\frac16 \frac{\hbar^2\pi^2}{m\Delta x^2} & ,\,n_{\text{dv}}=n'_{\text{dv}} \\
(-1)^{n_{\text{dv}}-n'_{\text{dv}}} \frac{\hbar^2}{m \Delta x^2 (n_{\text{dv}}-n'_{\text{dv}})^2} & ,\, n_{\text{dv}} \ne n'_{\text{dv}}
\end{array}
\right.\,.
\end{equation}
Similarly, we get for the above-diagonal 
matrix elements of the momentum operator $\widehat{p}_x$ in this representation
\begin{equation}
\langle n_{\text{dv}} | \widehat{p}_x| n'_{\text{dv}} \rangle = (-1)^{n'_{\text{dv}}-n_{\text{dv}}} \ui \frac{\hbar}{(n'_{\text{dv}}-n_{\text{dv}})\Delta x}\,, \qquad n_{\text{dv}}>n'_{\text{dv}}\,.
\end{equation}
The  diagonal elements vanish and the elements below the diagonal can be obtained from the elements above the diagonal,
\begin{equation}
\langle n_{\text{dv}}| \widehat{p}_x |n_{\text{dv}} \rangle = 0\,,\quad
\langle n'_{\text{dv}}| \widehat{p}_x |n_{\text{dv}} \rangle = - 
\langle n_{\text{dv}}| \widehat{p}_x |n'_{\text{dv}} \rangle\,.
\end{equation} 
The matrix elements of the potential $V_{\text{E}}$, or more precisely the complexified potential $V^\alpha_{\text{E}}(x) = V_{\text{E}}(\exp(\ui\alpha) \,x )$, 
have to be computed from  numerical quadrature.

Using the results above, we find that the matrix elements 
$\widehat{H}^{\alpha}_{(n_{\text{dv}},n_{\text{M}};n'_{\text{dv}},n'_{\text{M}})} 
:= \langle
n_{\text{M}},n_{\text{dv}}|\widehat{H}^{\alpha}|n_{\text{dv}},n_{\text{M}}\rangle$
of the full operator $\widehat{H}^\alpha$ 
are given by
\begin{equation}
\begin{split}
\widehat{H}^{\alpha}_{(n_{\text{dv}},n_{\text{M}};n'_{\text{dv}},n'_{\text{M}})} = &  
\ue^{-2\ui\alpha} \, \langle n_{\text{dv}} | \frac{\widehat{p}_x^2}{2m} | n'_{\text{dv}} \rangle \, \delta_{n_{\text{M}}\,n'_{\text{M}}} 
+
 \langle n_{\text{dv}} | \widehat{V}_{\text{E}}^{\alpha} | n'_{\text{dv}} \rangle  \, \delta_{n_{\text{M}}\,n'_{\text{M}}} \\
&+
 E_{\text{M}}\big( n_{\text{M}} \big) \, \delta_{n_{\text{dv}}\,n'_{\text{dv}}} \, \delta_{n_{\text{M}}\,n'_{\text{M}}} 
+
 \ue^{-\ui\alpha} \, \langle n_{\text{dv}} | \widehat{p}_x | n'_{\text{dv}} \rangle  \, \langle n_{\text{M}} | \widehat{p}_y | n'_{\text{M}} \rangle \,.
\end{split}
\end{equation}
In our numerical study of the 2 \dof system we chose 
$n_{\text{M}} \in \{0,\dots,13\}$ (for our choice of parameters in Sec.~\ref{sec:2Dexample} the Morse oscillator has 14 bound states),  
$n_{\text{dv}}\in \{-50,\dots,50\}$, and $\Delta_x=0.1$. This led to a matrix of size $1414\times1414$.  
In our numerical study of the 3 \dof system we chose 
$n_{\text{dv}}\in \{-25,\dots,25\}$, $\Delta_x=0.16$,
$n_{\text{M};2} \in \{0,\dots,6\}$ and $n_{\text{M};3} \in \{0,\dots,6\}$
(for our choice of parameters in Sec.~\ref{sec:3Dexample} the 1D Morse oscillators have 14 and 17 bound states, respecively).
This led to a matrix of size $2499\times2499$. 
We computed the eigenvalues of these matrices using the function {\sf eigs} in {\sf Matlab}.
For both systems we chose the scaling angle to be $\alpha=1.2$.